\documentclass[12pt]{report}
\usepackage[margin=1in,footskip=0.5in]{geometry}
\usepackage{titlesec}
\usepackage{titletoc}
\usepackage{appendix}

\makeatletter

\makeatletter

\titlecontents{chapter}%
 [0em]%
 {\addvspace{2em}}%
 {\bfseries\@chapapp\ \thecontentslabel\quad}%
 {\hspace{-0em}}%
 {\hfill\contentspage}%
 [\addvspace{0pt}]%

\g@addto@macro\appendices{%
  \addtocontents{toc}{\protect\renewcommand{\protect\@chapapp}{\appendixname}}%
}

\makeatother

\usepackage{lipsum}

\usepackage{comment}
\usepackage{graphicx}
\usepackage{amsmath,amsfonts,amssymb,amsthm,amstext, amscd,extarrows,mathtools,graphicx,subfigure,setspace}
\usepackage{cite}
\usepackage{verse}
\usepackage{sectsty}
\usepackage{lipsum}
\usepackage{blindtext}
\usepackage{epsfig}
\usepackage[section]{placeins}
\usepackage{slashed}
\usepackage{setspace}
\usepackage{color}
\usepackage{caption}
\usepackage{amsmath}
\usepackage[mathscr]{eucal}
\numberwithin{equation}{section}
\usepackage{hyperref}
\usepackage[font=small,labelfont=bf]{caption}
\chapterfont{\centering}
\makeatother

\newcommand{\be}{\begin{equation}}
\newcommand{\bea}{\begin{eqnarray}}
\newcommand{\eea}{\end{eqnarray}}
\newcommand{\ba}{\begin{array}}
\newcommand{\ea}{\end{array}}
\newcommand{\ee}{\end{equation}}
\newcommand{\bes}{\begin{equation*}}
\newcommand{\beas}{\begin{eqnarray*}}
\newcommand{\eeas}{\end{eqnarray*}}
\newcommand{\bas}{\begin{array*}}
\newcommand{\eas}{\end{array*}}
\newcommand{\ees}{\end{equation*}}

\numberwithin{equation}{section}

\newcommand{\mrd}{\mathrm{d}}

\begin{document}

\onehalfspacing
\vfill

\begin{titlepage}
\vspace{10mm}

\begin{center}

\vspace*{25mm}
\vspace*{1mm}

{\setstretch{1.0}

{\LARGE  \textbf{Beyond AdS Space-times,\\}}
{\LARGE   \textbf{ New Holographic Correspondences and \\ }}
\vspace*{0.25cm}
{\LARGE   \textbf{ Applications }}
 \vspace*{0.5cm} }
 
 by

  \vspace*{0.5cm}
{ Mahdis Ghodrati\\}

 \vspace*{2cm}

{\setstretch{1.0}
{A dissertation submitted in partial fulfillment\\}
{of the requirements for the degree of\\}
{Doctor of Philosophy\\}
{(Physics)\\}
{ in the University of Michigan\\}
{2016}}
\end{center}

 \vspace*{3cm} 
 
\begin{verse}
Doctoral Committee:\\  
{\setstretch{1.0}
 \vspace*{0.4cm}
 \ \ \ \ \ \ \ \ Professor Leopoldo A. Pando Zayas, Chair\\
 \ \ \ \ \ \ \ \ Professor Finn Larsen\\
 \ \ \ \ \ \ \ \ Professor James T. Liu\\
 \ \ \ \ \ \ \ \ Professor Jon M. Miller\\
\ \ \ \ \ \ \ \ Professor Gregory Tarle \\}
\end{verse}
\end{titlepage}

\pagenumbering{roman}

\vspace*{10mm}
{ \textbf{\LARGE \centerline {Abstract}} }
\addcontentsline{toc}{chapter}{Abstract}
\setcounter{page}{2}
 \vspace*{0.5cm}

The AdS/CFT correspondence conjectures a mathematical equivalence between string theories and gauge theories. In a particular limit it allows a description of strongly coupled conformal field theory via weakly coupled gravity. This feature has been used to gain insight into many condensed matter (CM) systems. such as high temperature superconductors, superfluids or quark-gluon plasma. However, to apply the duality in more physical scenarios, one needs to go beyond the usual AdS/CFT framework and extend the duality to non-AdS situations which is the aim of this dissertation.

To describe Lifshitz and hyperscaling violating (HSV) phenomena in CM one uses gauge fields on the gravity side which naturally realize the breaking of Lorentz invariance. These gravity constructions often contain naked singularities. In this thesis, we construct a resolution of the infra-red (IR) singularity  of the HSV background. The idea is to add squared curvature terms to the Einstein-Maxwell dilaton action to build a flow from $\text{AdS}_4$ in the ultra violate (UV)  to an intermediating HSV region and then to an $\text{AdS}_2 \times {\text{R}}^2$ region in the IR. This general solution is free from the naked singularities and would be more appropriate for applications of HSV in physical systems.

We also study the Schwinger effect (the creation of particles and anti-particles in the presence of strong electric or magnetic fields) by using the AdS/CFT duality. We present the phase diagrams of the Schwinger effect and also the ``butterfly shaped-phase diagrams" of the entanglement entropy for four different confining supergravity backgrounds. Comparing different features of all of these diagrams could point out to a potential relation between the Schwinger effect and the entanglement entropy which could lead to a method of measuring entanglement entropy in the laboratory.

Finally, we study the ``new massive gravity" theory and the different black hole solutions it admits. We first present three different methods of calculating the conserved charges. Then, by calculating the on-shell Gibbs free energy we construct the Hawking-Page phase diagrams for different solutions in two thermodynamical ensembles.  As the massive gravity models are dual to dissipating systems, studying the Hawking-Page diagrams could point out to interesting results for the confinement-deconfinement phase transitions of the dual boundary theories. 

So this dissertation discusses various generalizations of the AdS/CFT correspondence of relevance for cases which violate Lorentz symmetry.

 \pagebreak

 \vspace*{10mm}
{ \textbf{\LARGE \centerline {Acknowledgements}} }
\addcontentsline{toc}{chapter}{Acknowledgements}
  \vspace*{0.5cm}
 
Firstly, I would like to thank my advisor Professor Leopoldo Pando Zayas for the many informative discussions we had, for providing me the opportunity to attend many conferences and workshops, for the great courses he offered and also for letting me to work independently and with my own pace and methods. 

 I am very thankful to Professor Finn Larsen who trusted me from the beginning of my graduate studies and was always supportive during these years. He let me to change my field of study after two years. With his flexibility I could also travel a lot between Iran and US to have progress in my work which I am deeply thankful for.
 
 I would like to thank Professor James Liu and Professor Ratindranath Akhoury who first showed me the beauty of doing independent research in theoretical physics and made me determined to follow this path.
 
I would like to thank Dr. Mohsen Alishahiha as the head of school of Particles and Accelerators at IPM who during these years was a main source of encouragement and support. He is leading a great group of physicists and provides wonderful opportunities for the young researchers whenever he is able to.

I am very thankful to Dr. Shahin Sheikh Jabbari for the great leadership of school of Physics of IPM and specially the string theory group. Specially I would like to thank him for the great courses he offered, his informative lectures and for the wonderful weekly meetings we had at IPM.

I am deeply grateful to Dr. Mohammadi Mozaffar, my first collaborator, who patiently guided me through all the details and disciplines of working in theoretical physics.

Thanks to all my friends working in physics specially Kamal Hajian, Zahra Rezaei, Hajar Ebrahim, Saeedeh Sadeghian, Ali Naseh, Ali Seraj, Arash Arabi, Arya Farahi, Pedro Lisbao, Alejandro Lopez, Joshua Gevirtz, Tim Olson, John Kearney and Uri Kol who I had very enjoyable physics dissuasions with.

I would like to thank my father Zia, my sister Laya and my grandmother Pari for staying with me through any difficulty and for their warm supports during all these years of my life.

Finally and specially I would like to thank my mother Monireh, my greatest friend and the biggest source of love, encouragements and support in my life. Without her help and support I could not carry on in anything in my life. I would like to dedicate this work to her.

\pagebreak

\tableofcontents

\pagebreak 

\cleardoublepage
\addcontentsline{toc}{chapter}{\listfigurename}
\listoffigures

\pagebreak


 \chapter{ Introduction} 
\pagenumbering{arabic}

One of the goals of human beings from the early stages of civilization has been to find connections between different phenomena seen in the universe, and then using these connections explaining the world around us. Using these connections, human got the power to actually predict many events in nature.  The laws of physics developed as the result of finding connections between different phenomena. Unifying different theories, made the physical models simpler, more beautiful and more powerful. For example, one of the most successful example in this regard is the theory of electricity and magnetism discovered by James Clerk Maxwell in the nineteen century. This theory unifies two different forces of nature: electricity and magnetism.
This unification lead to important technical improvements and the possibility to predict new phenomena such as the propagations of electromagnetic waves at the speed of light.

This mindset of unifying different models and theories to build a simpler, more general, master theory has paved the way for many other brilliant discoveries in physics. As the results of such attempts, three forces of nature, the electromagnetic, the weak and the strong interaction were merged into a Grand Unified Theory (GUT). To reconcile general relativity theory of graviton with the quantum filed theory,  sting theory has been developed.

The main idea of string theory, which first was constructed to explain aspects of the strong interaction, is to replace point-like particles with one-dimensional strings. The different vibrational states of these strings correspond to different particles. As one of these particles is the graviton, the mediator of gravitational force, the string theory is actually a quantum theory of gravity, capable of unifying general relativity with quantum field theory. 

String theory turned out to be a very rich theoretical framework which generates many interesting discoveries even in pure mathematics such as progresses in non-commutative geometry, K-theory, homology, cohomology, homotopy and so on. However, still many theoretical and physical aspects of string theory remained unknown. 

This theory possesses many exotic features, such as extra dimensions: 26 dimensions for bosonic strings and 10 dimensions for superstring theory are required. For reducing the number of dimensions to $4d$, one needs to compactify these extra dimensions. As there are many different ways of compactifying, string theory has a huge landscape of vacuua with different physical constants. If one wants to obtain our own universe with our specific physical constants, the $6d$ geometry that is being compactified should be a Calabi-Yau manifold.

Within String theory there are also other objects such as D-branes which have specific masses and charges. The end points of open strings as they propagate in the world-volume lie on D-branes with the Dirichlet boundary condition. One can also study the dynamics of strings and D-branes, considering different mathematical limits, such as near horizon limit or assume large number of D-branes and then derive interesting results in string theory. The duality between string theory on the background of asymptotically Anti-deSitter space-times and the conformal invariant field theories, (the AdS/CFT correspondence), which were first derived by Maldacena in 1997 is an example of having such a mindset in studying string theory which led to exciting progresses in many areas of high energy physics.

Anti-de Sitter (AdS) space generally is a mathematical model of space-time which is closely related to hyperbolic space where in the two dimensional case can be viewed as a disk where its boundary is infinitely far from any point in the interior region. Then if we stack these disks like a cylinder, the whole $3d$ AdS space-time can be constructed where time runs on its vertical direction. Similar to this view one can imagine the AdS space in higher dimensions.

The prototypical example of the AdS/CFT correspondence is a duality between string theory on the background of $\text{AdS}_5 \times S^5$ and the conformal field theory on the boundary of the Anti-deSitter space-time. In the original paper, Maldacena considered a stack of D-branes in string theory and then considered the low energy limit and showed that the field theory on the D-branes decouples from the bulk. He demonstrated that in the near horizon regions of these D-branes, there is an extra  supersymmetry leading to a super conformal group. Specifically he found that the large $N$ limit of $4d$ $\mathcal{N}=4$ super-Yang-Mills at the conformal point has a specific sector in its Hilbert space which is isomorphic to type IIB strings on the background of AdS. More generally he showed that string theory on various AdS space times is dual to various conformal field theories.

Some typical examples of AdS/CFT are the correspondence between type IIB string theory on $\text{AdS}_5 \times S^5$ and $4d$, $\mathcal{N}=4$ supersymmetric Yang-Mills theory,  M-theory on $\text{AdS}_7 \times S^4$, (2,0)-theory in six dimensions, and M-theory on $AdS_4 \times S^7$ which is equivalent to the ABJM superconformal field theory in three dimensions.

This duality is the most successful realization of the holographic principle, a property of string theory and quantum gravity which states that all the information in a volume of a space-time can be encoded in its boundary which has one dimension lower. The calculability power of AdS/CFT comes from the fact that it connects a strongly coupled gauge theory on one side to a weakly coupled gravity theory on the other side where perturbative techniques can be implemented rather easily.

Until now the AdS/CFT duality was mostly implemented for deriving information about strongly coupled gauge theories. Different problems such as in nuclear and condensed matter physics were attacked by modeling them with black hole solutions in the weakly coupled gravitational backgrounds. This thesis is based on this approach to AdS/CFT as well. In the future, however, it is anticipated that the other direction of the duality will be used to gather more information about the quantum gravity.

Attempts to apply string theory and AdS/CFT to condensed matter physics have led to a program called AdS/CMT which mostly deals with exotic states of matter such as superconductors and superfluids.
One of the main successes in this regard was describing the transition of superfluids to insulators by using higher dimensional black holes in the bulk. The other success was to find a lower bound in the ratio of shear viscosity $\eta$ to the volume density of entropy $s$ as $\frac{\eta}{s} \approx \frac{\hbar}{4\pi k}$ which later has been tested at the Relativistic Heavy Ion Collider (RHIC). In this thesis we also try to build some models of bulk geometry which can be used to study some exotic phases of matter and specifically strange metals which are a form of non-Fermi liquid systems. 

One should note that condensed matter problems are mostly non-relativistic, demanding a non-relativistic, Lorentz violating geometry as their bulk dual to describe them. Therefore, in the bottom-up gauge/gravity duality, the Lifshitz and Hyperscaling violating (HSV) geometries have been introduced to model specific phases of matters such as high temperature superconductors and strange metals. In AdS/CFT the symmetries in the boundary CFT and of the bulk gravity should be the same. Therefore, since in condensed matter physics the systems are usually non-relativistic, the Lifshitz or more generally the HSV geometries could be used as the bulk gravity dual where Lorentz invariance is broken. Therefore, different properties of these geometries could point out to specific properties of different exotic phases of matter. For example the creation of black hole in these backgrounds is dual to confinement/de-confinement phase transitions in the dual boundary theory.

In applying this duality, however, some subtleties might arise which have been investigated in different works. For instance, in using Lifshitz and HSV, one main problem, as noted in the literature, was that the these geometries have a naked singularity in the IR limit \cite{Copsey:2012gw} \cite{Copsey:2010ya}. This is an unfavorable property for a metric which is built to model condensed matter systems, as in the dual CFT part, there would not be such a singularity or any corresponding physical effects. More generally, the presence of naked singularities signals a break down in the applicability of general relativity. Therefore, one ought to look for a way to resolve these singularities of Lifshitz and HSV in order to be able to use them for modeling different problems in condensed matter physics.

 In chapter \ref{chap2} of this thesis which is based on the work in \cite{Ghodrati:2014spa}, we construct a solution to resolve this issue.  First, we add higher derivative gravity correction terms to the Einstein-Maxwell dilaton action of the following form
 
 \begin{gather} \label{eq:action1}
S=\int{d^{d+2}x\sqrt{-g}\bigg(R+V(\phi)-\frac{1}{2}(\partial \phi)^2-f(\phi)F_{\mu \nu}F^{\mu \nu}+g(\phi)(\alpha_1 R_{\mu \nu \rho \sigma}R^{\mu \nu \rho \sigma}+\alpha_2 R_{\mu \nu} R^{\mu \nu}+\alpha_3 R^2)\bigg)},
\end{gather}

where it has the hyperscaling violating metric solution in the form below
\begin{eqnarray}\label{eq:Ansatz}
ds^2_{d+2}=r^{-2\frac{\theta}{d}}\left(-r^{2z}dt^2+\frac{dr^2}{r^{2}}+\Sigma \ r^2 d\vec{x_i}^2\right), \ \ \ i=1,2..,d.
\end{eqnarray}

 Then for the four dimensional case, by changing the weight of higher derivative correction terms, $g(\phi)$, we could find a flow which interpolates between $\text{AdS}_4$ in the UV of the theory, to an intermediate HSV region and then to an $\text{AdS}_2 \times S^2$ in the IR. Finding this flow therefore, could resolve the IR singularity of the HSV.  Also by using the null energy and causality conditions we put constraints on the parameters of the model and narrow it down to physical regions. Specifically we find the allowed regions for the strange metal system. 

\begin{figure}[h!]
\centering
\begin{minipage}{.55\textwidth}
  \centering
  \includegraphics[width=.9\linewidth]{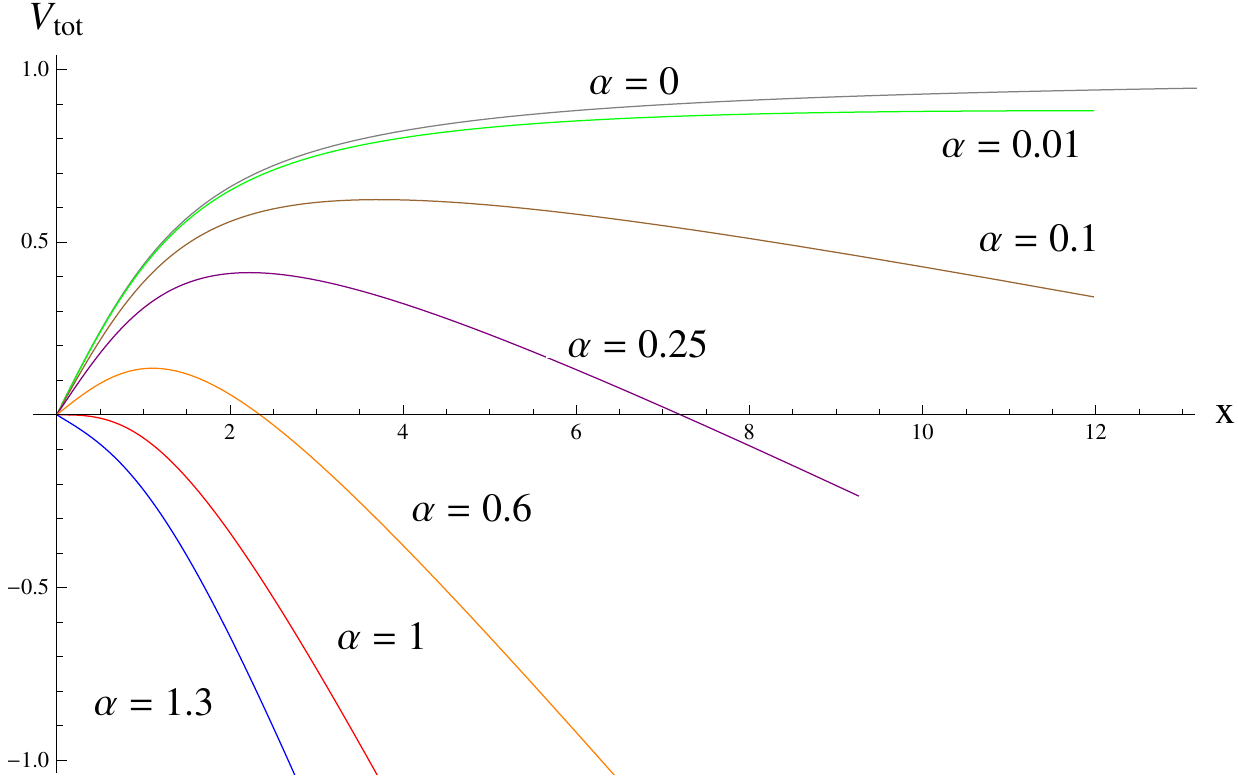}
  \captionof{figure}{ Schwinger phase diagram for\\ a conformal geometry. }
  \label{fig:AdS1}
\end{minipage}%
\begin{minipage}{.5\textwidth}
  \centering
  \includegraphics[width=.9\linewidth]{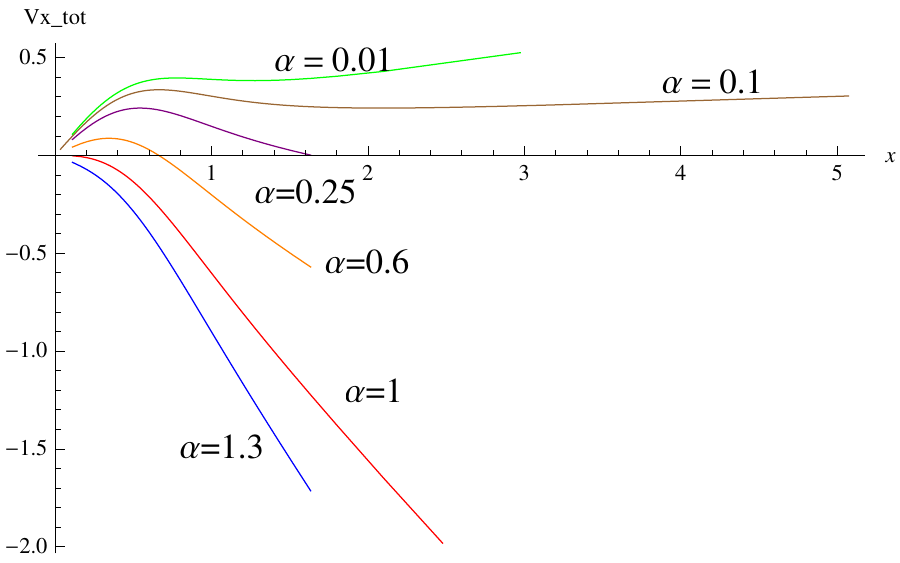}
  \captionof{figure}{Schwinger phase diagram for \\ a confining geometry. }
  \label{fig:WQCD1}
\end{minipage}
\end{figure}

\begin{figure}[h!]
\centering
\begin{minipage}{.55\textwidth}
  \centering
  \includegraphics[width=.9\linewidth]{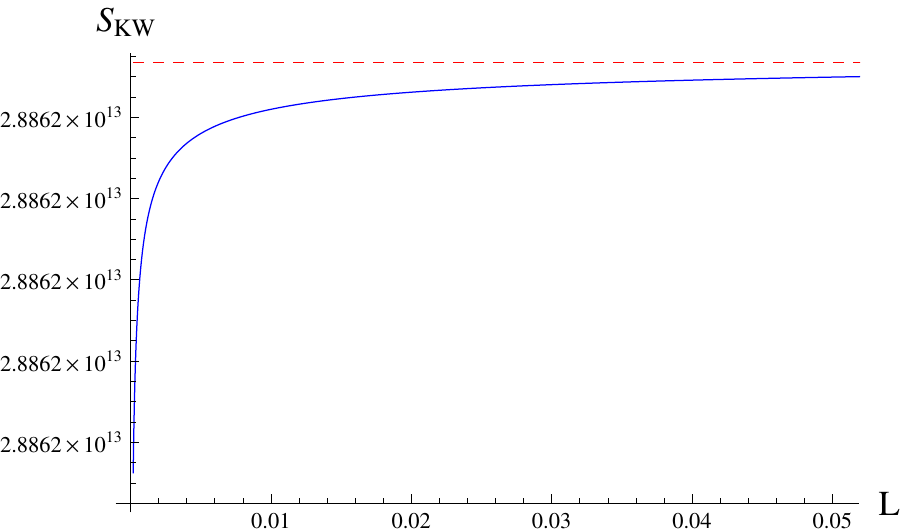}
  \captionof{figure}{Entangement entropy phase diagram \\ for a conformal geometry. }
  \label{fig:KW-EE1}
\end{minipage}%
\begin{minipage}{.55\textwidth}
  \centering
  \includegraphics[width=.9\linewidth]{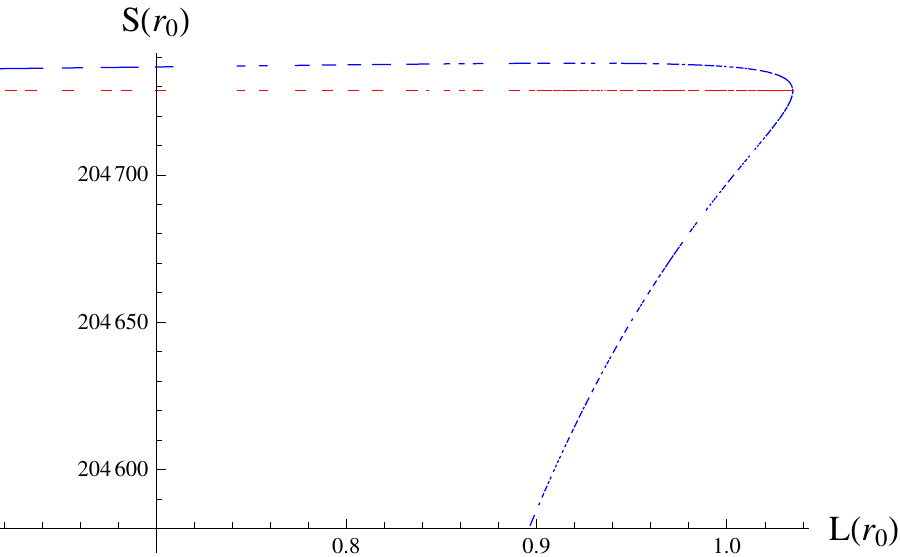}
  \captionof{figure}{Entangement entropy phase diagram \\ for a confining geometry. }
  \label{fig:KTSS32}
\end{minipage}
\end{figure}

In chapter \ref{chap3} of this thesis, which is based on the work in \cite{Ghodrati:2015rta}, we use the gauge/gravity duality to search for a relation between the Schwinger effect (or particle pair creation rate in the presence of an electric field) and the entanglement entropy. Schwinger pair creation effect is a phenomenon occurring when in a strong electric or magnetic field the imaginary virtual particles become real and an electric current is being created. The entanglement entropy (EE) is a measure of how the quantum information is stored in a quantum state and in the holographic dual systems it is encoded in geometric features of the bulk geometry. One would think that in a system with higher entanglement entropy the virtual particles would be more entangled and could find each other easier and as a consequence the rate of creation of particle pairs from the strong electric or magnetic field would be higher. Therefore, a relationship between these two effects is plausible. One way to try to find such a relation is to look at phase diagrams and check how one phase turns to another phase for each effect and each background. If a relationship is present, it should show itself in the phase transitions as well. The phase diagrams for each effect for a confining and a conformal geometry are shown in figures (\ref{fig:AdS1}), (\ref{fig:WQCD1}), (\ref{fig:KW-EE1}) and (\ref{fig:KTSS32}).

 We find the Schwinger effect phase diagrams and the entanglement entropy phase diagrams for several confining models: Witten-QCD, Maldacena-Nunez, Klebanov-Strassler and Klebanov-Tseytlin and also the conformal models of Klebanov-Witten and AdS space. We then compare the rate of phase transitions for both effects for all these models and also compare them with each other. We show that the phase transitions have a higher rate in WQCD and Klebanov-Tseytlin relative to Maldacena-N\'{u}\~{n}ez and Klebanov-Strassler backgrounds.  
 
 This could point out to a hidden relation between these two effects which needs to be clarified and quantified better in the future works. Quantifying this relation could have many applications, such as in cosmology or for measuring the entanglement entropy of condensed matter systems in the lab. As entanglement entropy is a non local quantity which depends on the correlation functions between different regions of space, measuring it in the laboratory would be a very difficult task and no precise method of measuring it currently exists. Thus, quantifying the relation between Schwinger pair creation rate and entanglement entropy could be a breakthrough in measuring entanglement entropy.

Next, as another direction of research, one could note that there are many attempts to generalize AdS/CFT to other cases. Examples are lower dimensional dualities which do not need M-theory or String theory such as $3d$ gravity and Liouville field theory. Also for generalizing and using the duality in cosmology, Andrew Strominger first introduced the dS/CFT correspondence \cite{Strominger:2001pn}. The Kerr/CFT correspondence was first introduced in \cite{Guica:2008mu} to study real astrophysical black holes using the correspondence. There is also another duality closely related to AdS/CFT which first conjectured in \cite{Klebanov:2002ja} that connects higher spin gauge theories with $O(N)$ vector models. 

As an example of generalizing AdS/CFT to other geometries, in chapter \ref{chap4} of this thesis we specifically study the warped $\text{AdS}_3$ solution of a massive gravity theory. In the context of gauge gravity duality, a massive gravity theory could be used to study a system with momentum dissipation in condensed matter physics. Thus, studying the phase diagrams of different solutions in this massive theory of gravity can relate to some dissipating phases of matter. 

In chapter \ref{chap4}, which is based on \cite{Ghodrati:2016ggy}, we study the Hawking-Page phase transitions between different solutions of a chiral massive gravity, named the Bergoshoeff- Hohm-Townsend (BHT) theory which is of the following form

\begin{gather}\label{eq:action}
S=\frac{1}{16 \pi G_N} \int d^3x \sqrt{-g} \Big[ R-2\Lambda+\frac{1}{m^2} \Big( R^{\mu\nu} R_{\mu\nu}-\frac{3}{8} R^2 \Big)  \Big].
\end{gather}

We first review how one derives the conserved charges for this specific theory which are: mass, angular momentum and entropy. We review the Abbott, Deser and Tekin (ADT) formalism in section \ref{sec:ADTcharge}, the $SL(2,R)$ reduction method in \ref{sec:chapSL} and the recently proposed method of calculating charges, the solution phase space method, for  the higher curvature theories in Appendix \ref{sec-Higher d} \cite{Hajian:2015xlp, Ghodrati:2016vvf}.

\begin{figure}[h!] 
  \centering
    \includegraphics[width=0.3\textwidth]{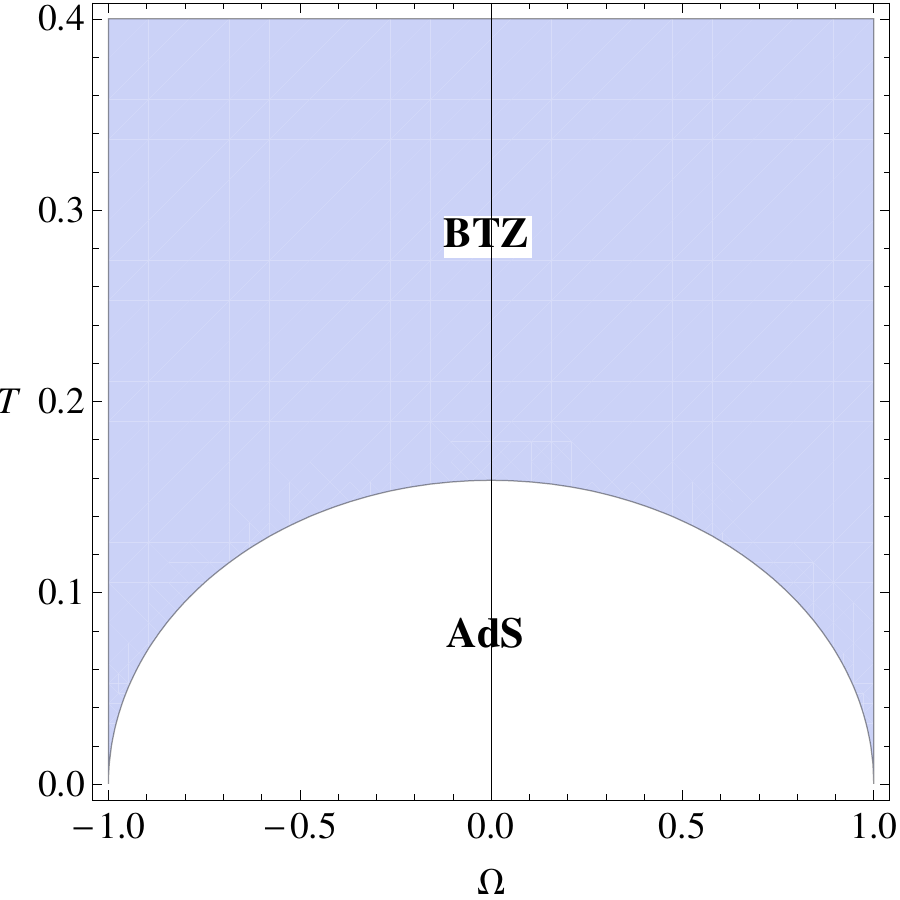}
       \caption{The phase diagram of BTZ black hoel in the grand-canonical ensemble. }
       \label{fig:phasegrand}
\end{figure}

Then using the conserved charges we derive the Gibbs free energy for each solution and then by calculating the Hessian of the free energy we derive the stability conditions. Next, we derive the phase diagrams for each solution of this theory and for two different thermodynamical ensembles we determine the regions of the parameters where a black hole is being formed. In the grand canonical ensemble which is the physical ensemble for our solution, the phase diagram is symmetric as one expects for a non-chiral theory and it is in the form shown in Figure (\ref{fig:phasegrand}).

In the non-local/quadratic ensemble, as we will present, the phase diagram is not symmetric and therefore, one expects it would not be a physical ensemble to describe the black hole phase transitions. 

We also study the modular invariance properties, the inner horizon thermodynamics of the black hole solutions and then the entanglement entropies of the warped AdS solution in the BHT massive gravity.

In the last chapter, which is based on un-published work, we use the gauge/gravity duality in the top-down approach. It is worth mentioning that in the top-down approach one chooses a background that is already a solution of string or M-theory and then studies the dual CFT. However, in the bottom-up approach one considers the symmetries or interactions of the theory, and tries to build a bulk gravity background for a specific field theory. 

More precisely, in chapter \ref{chap5}, we consider the $3+1$-charge sector of the $4d$ $U(1)^4$ gauged supergravity model which is 
\begin{equation} \label{eq:lagmain}
\mathcal{L}_{(3+1)}=R-\frac{1}{2}(\partial \vec{\phi})^2+24g^2 \text{cosh} \frac{\phi}{\sqrt{3}}-\frac{3}{4} e^{\frac{\phi}{\sqrt{3}}} F^2-\frac{1}{4} e^{-\sqrt{3} \phi}f^2.
\end{equation}

We study the $3+1$-charged black brane solution of the above Lagrangian in the following form: 

\begin{equation} \label{metric}
ds^2=-H^{-\frac{1}{2}} f(r) dt^2+H^\frac{1}{2}(\frac{dr^2}{f(r)}+r^2d\Omega_{2,k}^2),
\end{equation}
where
\begin{eqnarray}
H=H_1H_2 H_3 H_4,\ \ \ \ \ \
H_I=1+\frac{\mu\ \text{sinh}^2\beta_I}{k\ r}=1+\frac{q_I}{r}, \ \ \ \ \ \
f(r)=k-\frac{\mu}{r}+ \frac{r^2}{L^2} H,
\end{eqnarray}

 Using this solution we search for the Fermi surfaces and the superconducting phases in the dual theory from a top-down approach. 
In this chapter, we study how the dilaton field and chemical potentials behave under different limits of turning off the charges and then we show that there exist a gap in the near horizon region of this solution which corresponds to a gap in the dual field theory or possibly a superconducting phase. Then, we solve the Dirac equation for this solution. Finally, we uplift the theory to $5d$ and then to $11d$ and show that a warped $\text{AdS}_3$ solution exists in the near horizon limit of the black brane solutions.

All in all, we examine different setups in generalizing the AdS/CFT correspondence, suitable for various applications in condensed matter physics, and during these examinations, we discuss a variety of problems, relations and diagrams which we discuss in further details.

\chapter{Hyperscaling violating solution in coupled dilaton-squared curvature gravity}\label{chap2}

In this chapter we introduce Lifshitz and Hyperscaling violating geometries which have been studied extensively  (for example see: \cite{Maldacena:1997re, Gath:2012pg, Kachru:2008yh, Lu:2012xu, Taylor:2008tg, Taylor:2015glc, Dong:2012se, Dey:2013vja, Fonda:2014ula, Cremonini:2012ir, Edalati:2013tma, Balasubramanian:2008dm, Son:2008ye, Herzog:2008wg, Guica:2010sw, Narayan:2011az, Donos:2010tu }) as in the context of gauge/gravity duality {\cite{Maldacena:1997re}, they can be used to study phases of matter in condensed matter physics such as non-Fermi liquids.

The Lifshitz metric can be written as
\begin{eqnarray}
ds^2_{d+2}=-r^{2z}dt^2+\frac{dr^2}{r^{2}}+\sum\limits_{i=1}^d  r^2 d\vec{x_i}^2 ,
\end{eqnarray}
and has the following scaling symmetry
\begin{eqnarray}
r \to  \lambda^z r, \ \ \ \ \ \  x_i \to \lambda x_i .
\end{eqnarray}
These types of metrics can be derived from gravity coupled to massive gauge fields \cite{Kachru:2008yh}, and they do not respect the relativistic scaling symmetry.  The Lifshitz metric is a generalization of the AdS metric where for the special case of $z=1$ is ${\mathrm{AdS}_{d+2}}$ and when $z \to \infty$ becomes an ${\mathrm{Ads_2 \times \mathbb{R}^{d}}}$ geometry.

One can generalizes the Lifshitz metric to a hyperscaling violating background by adding a non-zero exponent $\theta$ in the following form
\begin{eqnarray}
ds^2_{d+2}=r^{-\frac{2\theta}{d}}(-r^{2z}dt^2+\frac{dr^2}{r^{2}}+\sum\limits_{i=1}^d d\vec{x_i}^2).
\end{eqnarray}
Now the scaling symmetries become
\begin{eqnarray}
t \to \lambda^z t, \ \ \ \  r \to \lambda^{-1} r,\ \ \ \  x_i \to \lambda x_i,\ \ \ \  ds_{d+2} \to \lambda^{\frac{\theta}{d}} ds_{d+2}.
\end{eqnarray}
This metric is also not Lorentz invariant and can be derived by adding a dilaton field to the Einstein-Maxwell action with a particular dilatonic potential.

The Lifshitz geometry is a possible candidate for describing the behaviors of strange metals, holographically \cite{Hartnoll:2009ns}. Also in order to investigate a system with Fermi surface, we can consider hyperscaling violating geometries in a specific range of parameters for its gravitational dual \cite{Dong:2012se}.

Although these two space-times have constant and, therefore, finite scalar curvature invariants, they both have null singularity in the ${\mathrm{IR}}$ which makes the infrared region incomplete \cite{Knodel:2013fua, Horowitz:2011gh, Lei:2013apa, Bhattacharya:2012zu, Copsey:2012gw, Harrison:2012vy, O'Keeffe:2013nha, Copsey:2010ya}. On the field theory side this might suggest that the solution cannot be trusted in the IR unless these singularities are resolved.

One resolution of the singularity was argued in \cite{Knodel:2013fua}. As they have suggested, for the 4-dimensional Lifshitz metric in Einstein-Weyl gravity, one can construct numerically a flow from $\mathrm{AdS_4}$ in the UV to an intermediate Lifshitz region and then to $\mathrm{AdS_2 \times \mathbb{R}^2}$ in the IR. As $\mathrm{AdS}$ space-times are free from singularity, constructing this flow can resolve the IR singularities.

In this chapter we generalize the solutions in \cite{Knodel:2013fua} to non-zero $\theta$ and study the effects of squared curvature terms on the solutions of hyperscaling violating backgrounds. 
In section \ref{sec:solution} , we derive the analytical solution by coupling the higher derivative terms to the dilaton field by a $\mathrm{g(\phi)}$ function. Letting $\mathrm{\theta}=0$ in our results, the solution of  \cite{Knodel:2013fua} can be re-derived. We study how our solution is being renormalized in $z$ by the effects of higher derivative gravity.
In section \ref{sec:cases}, we study the regions of $d$ and $\theta$ which can lead to physical solutions by satisfying the imposed constraints, most importantly, null energy condition and stability of the solution. We study several different special cases in the parameter region and investigate constraints in various physical limits such as for the strange metals.

In section \ref{sec:resolve}, we consider a four dimensional metric Ansatz and then study the perturbations around $\mathrm{AdS_2 \times \mathbb{R}^2}$ in the IR and $\mathrm{AdS_4}$ in the UV which can support a hyperscaling violating solution in the intermediate region. We investigate the allowed parameter space region for constructing numerical flow and then for some specific initial values, analytically estimate the cross over parameters from each region of the parameter space to the next one for the complete and free from singularity solution. 

\section{General Solution}\label{sec:solution}

The action which gives a hyperscaling violating solution corrected by squared curvature terms is 
\begin{flalign} \label{eq:action1}
S=\int{d^{d+2}x\sqrt{-g}\bigg(R+V(\phi)-\frac{1}{2}(\partial \phi)^2-f(\phi)F_{\mu \nu}F^{\mu \nu}+g(\phi)(\alpha_1 R_{\mu \nu \rho \sigma}R^{\mu \nu \rho \sigma}+\alpha_2 R_{\mu \nu} R^{\mu \nu}+\alpha_3 R^2)\bigg)} .
\end{flalign}
As discussed in\cite{Shaghoulian:2013qia}, one way to derive a hyperscaling violating solution and fix the exponents is that the higher derivative terms should be coupled to the scalar field $\phi$, by multiplying these terms to a $g(\phi)$ function.

For deriving hyperscaling violating solution, the Ansatz metric is
\begin{eqnarray}\label{eq:Ansatz}
ds^2_{d+2}=r^{2\alpha}\left(-r^{2z}dt^2+\frac{dr^2}{r^{2}}+\Sigma \ r^2 d\vec{x_i}^2\right), \ \ \ i=1,2..,d.
\end{eqnarray}
We have set the AdS radius to one; $L=1$.
Here $\alpha=-\frac{\theta}{d}$, where $\theta$ and z are hyperscaling violation exponent and dynamical exponent respectively.

Taking the variation of the action and neglecting the surface terms, the Einstein's equations can be derived,

\begin{eqnarray}\label{Eineq}
E_{\mu \nu}=T_{\mu \nu} ,
\end{eqnarray}

where\footnote{We use $R_{[\mu |\beta | \nu] \alpha}=\frac{1}{2}(R_{\mu \beta  \nu \alpha}-R_{\nu \beta  \mu \alpha})$.}
\begin{eqnarray}\label{Emunu}
E_{\mu \nu}&=&R_{\mu \nu}-\frac{1}{2}g_{\mu \nu}R+g(\phi)\Big(-\frac{1}{2}g_{\mu \nu}\big(\alpha_1 R_{\alpha \beta \rho \lambda}R^{\alpha \beta \rho \lambda}+\alpha_2 R_{\alpha \beta}R^{\alpha \beta}+\alpha_3 R^2\big)+2\alpha_1 R_{\mu \alpha \beta \rho}R_\nu^{\alpha \beta \rho}+\nonumber\\
&&2\alpha_2 R_{\mu \alpha}R_\nu^\alpha+ 2\alpha_3 R R_{\mu \nu}\Big)+g_{\mu \nu}\big(\alpha_2 \nabla_\alpha \nabla_\beta g(\phi)R^{\alpha \beta}+2\alpha_3 \Box g(\phi)R\big)+4\alpha_1 \nabla^\alpha \nabla^\beta g(\phi)R_{[\mu |\beta | \nu] \alpha}\nonumber\\
&&+\alpha_2 \Big(-2\nabla_\alpha \nabla_\nu g(\phi)R_\mu^\alpha+\Box g(\phi)R_{\mu \nu}\Big)-2\alpha_3 \nabla_\mu \nabla_\nu g(\phi)R ,
\end{eqnarray}

and
\begin{eqnarray}\label{Tmunu}
T_{\mu \nu}&=&\frac{1}{2}\partial_\mu \phi \partial_\nu \phi +2 f(\phi)\left(F_{ \mu} ^\rho F_{ \nu \rho}-\frac{1}{4}g_{\mu \nu} F_{ \rho \sigma}F^{ \rho \sigma}\right)+\frac{1}{2}g_{\mu \nu}\left(V(\phi)-\frac{1}{2}\partial^\rho \phi \partial_\rho \phi \right).
\end{eqnarray}

These equations need to be supplemented with the Maxwell and scalar equations of motion,

\begin{flalign}\label{eq:max}
\nabla_\mu \left(f(\phi)F^{\mu \nu}\right)=0,
\end{flalign}
\begin{flalign}
\Box \phi-f'(\phi)F_{\mu \nu}F^{\mu \nu}+g'(\phi)\left(\alpha_1 R_{\mu \nu \rho \sigma}R^{\mu \nu \rho \sigma}+\alpha_2 R_{\mu \nu} R^{\mu \nu}+\alpha_3 R^2\right)=&-\frac{dV(\phi)}{d\phi},
\end{flalign}
where $V(\phi)=V_0 e^{\gamma \phi}, f(\phi)=e^{\lambda \phi}, g(\phi)=e^{\eta \phi}$ and prime denotes the derivative with respects to the argument.

Using the metric Ansatz, the Maxwell equation \ref{eq:max} leads to
\begin{eqnarray}\label{maxsol}
F_{ rt}=\rho e^{-\lambda \phi}r^{(2-d)\alpha+z-d-1}.
\end{eqnarray}
For solving the Einstein's equations more easily, we combine the various components of the energy-momentum tensor in the following way

\begin{eqnarray}\label{Tcomp}
T_t^t-T_r^r&=&-\frac{1}{2} r^{2-2 \alpha } (\partial_r \phi)^2, \nonumber\\
T_x^x-T_t^t&=&2 \rho^2 r^{-2d (\alpha+1) } e^{-\lambda \phi} , \nonumber\\
T_r^r+T_x^x&=&V_0 e^{\gamma \phi}.
\end{eqnarray}
Then, considering a logarithmic function for the dilaton as $\phi(r)=\phi_0+\beta \log r$ implies that $\eta=\frac{2\alpha}{\beta}$ and then one finds

\begin{eqnarray*}
\frac{r^{2 \alpha }(E_t^t-E_r^r)}{d(1+\alpha )}&=&1-z-\alpha+2e^{\eta  \phi _0}\Big[2\alpha _1\big(z(1-z)(1+d+\alpha (d-2))+z^3+\alpha+\nonumber\\&& \alpha ^2(1-z-\alpha )-1\big)+(z+\alpha -1)\big(\alpha _2\left(d+z^2+(d+z)\alpha \right)+\nonumber\\&&\alpha _3\left(\left(1+\alpha ^2\right)d(d+1)+2(d+z)(z+\alpha (d+1))\right)\big)\Big],
\end{eqnarray*}
\begin{gather}
\frac{r^{2\alpha }(E_x^x-E_t^t)}{(z-1)(z+d(1+\alpha ))}=1-2e^{\eta  \phi _0}\Big(2\alpha _1(1+z^2-\alpha ^2-z(d+(d-1)\alpha ))- \nonumber \\ \alpha _2(d+z^2+d \alpha +z \alpha )
-\alpha _3(2z^2+d(d+1)(1+\alpha )^2+2z(d+(d+1)\alpha ))\Big),
\end{gather}

\begin{gather}
r^{2\alpha}(E_r^r+E_x^x)=(d+z+d \alpha -1) (d+z+d \alpha )+e^{\eta  \phi _0} (1+\alpha )\times \nonumber \\
\Big(2\big(d^2 (1+\alpha )^2 (3 \alpha -1)+d ((\alpha-3)(\alpha^2-1)+2z(\alpha(4\alpha+3)-(2+(z-2)z)))\nonumber\\ +2 z (1-\alpha +z (z+2 \alpha -1))\big) \alpha _1-\big(d z (2 z^2-2 (1+\alpha )^2-z (3+\alpha ))\nonumber \\  -d^3 (\alpha -1) (1+\alpha )^2-2 z^2 (z+\alpha )+d^2 (1+\alpha ) (z (2+z)-2 z \alpha -\alpha ^2-3)\big) \alpha _2\nonumber\\-(d^2 (1+\alpha )^2+2 z (z+\alpha )+d (1+\alpha ) (1+2 z+\alpha )) (d (d-3+2 z+(d-1) \alpha )-2 z) \alpha _3\Big).
\end{gather}

Now changing the basis of $\alpha$ corrections, we will write the solution in $\alpha_{GB}$, $\alpha_R$ and $\alpha_W$ basis which is related to the previous one by
\begin{eqnarray}
\alpha_1&=&\alpha_{GB}+\alpha_W, \nonumber\\
\alpha_2&=&-4\alpha_{GB}-\frac{4}{d}\alpha_W, \nonumber\\
\alpha_3&=&\alpha_{GB}+\frac{2}{d(d+1)}\alpha_W+\alpha_R.
\end{eqnarray}
\\
In this new basis the general Lagrangian of the theory is \cite{Knodel:2013fua},
\begin{eqnarray}
\mathcal{L}=\alpha_{W} C_{\mu\nu\rho\sigma}C^{\mu\nu\rho\sigma}+\alpha_{GB} G+\alpha_{R}R^2 ,
\end{eqnarray}
where $C$ is the Weyl tensor and $G$ is the Gauss-Bonnet combination with the following definitions
\begin{eqnarray}
C_{\mu\nu\rho\sigma}=R_{\mu\nu\rho\sigma}-\frac{1}{d-1}(g_{\mu[\rho} R_{\sigma_]\nu}-g_{\nu[\rho}R_{\sigma]\mu})+\frac{1}{d(d-1)}g_{\mu[\rho}g_{\sigma]\nu}R,
\end{eqnarray}
and 
\begin{eqnarray}
G=R_{\mu\nu\rho\sigma}R^{\mu\nu\rho\sigma}-4R_{\mu\nu}R^{\mu\nu}+R^2.
\end{eqnarray}
By combining the above equations, considering $\alpha=-\theta/d$ and also using equation \ref{maxsol} and after some algebra, one finds the solution in the new basis
\begin{eqnarray*}
ds^2&=&r^{-2\frac{\theta}{d}}\left(-r^{2z}dt^2+\frac{dr^2}{r^{2}}+r^2 d\vec{x}^2\right), \\
F_{rt}&=& \rho \hspace*{2mm}e^{\frac{2}{\beta}(d-\theta+\theta/d)\phi_0}r^{d+z-\theta-1} , \\
e^{\phi}&=&e^{\phi_0}r^\beta, 
\end{eqnarray*}

which the parameters of the solution, $\beta$, charge and potential are
\begin{eqnarray} \label{eq:beta}
\beta^2 &=&2(d-\theta)(z-1-\frac{\theta}{d})+\frac{4(d-\theta)}{d^3}e^{\eta \phi _0}\Big(\frac{2 z d^3(d-1) (z-1) (d-z-\theta +2)}{1+d}\alpha_W+\nonumber\\ &&d (d (1-z)-\theta ) \left(2z(d(d+z-\theta)-\theta)+(d+1)(\theta-d)^2\right)\alpha_R+\nonumber\\ &&(1-d) (d-\theta ) \left(d^2(d-2)(z-1)-d ((4+d) z-4) \theta +(2+d) \theta ^2\right)\alpha_{GB}\Big), \nonumber\\
\end{eqnarray}

\begin{gather}\label{eq:rho}
\rho^2e^{-\lambda \phi_0}=\frac{(z-1)(d+z-\theta)}{2}\Big(1+\frac{e^{\eta \phi _0}}{d^2 (d+1)}\big(2d(d+1)( 2z(\theta -d(d+z-\theta ))-\nonumber\\ (d+1)(\theta -d)^2)\alpha_R
+4 d z\text{  }(1-d)(\theta +d (z+\theta -d-2))\alpha_W+\nonumber\\ 2 (1-d^2) (d^2(d-2-2\theta)+(d+2) \theta ^2)\alpha_{GB}\big)\Big), \nonumber\\
\end{gather}

\begin{gather} \label{eq:V}
V_0e^{\gamma\phi _0}=(d+z-\theta )(d+z-\theta -1)+\frac{(d-\theta)e^{\eta \phi_0}}{d^3}\Big(\frac{4z d^2(1-d)(z-1) (d (z-2)-z+\theta )}{d+1}\alpha_W\nonumber\\
+d (d (3-d)+(1-d)(2 z-\theta )) (((\theta -d)^2-2z \theta )(d+1)+2d z(d+ z))\alpha_R\nonumber\\+(1-d) \big( d^2(d-2) (d^2+2 (z-1) z+d (4z-3))-d \big(d(d-2)(1+3 d)+2 (4 d^2+d-2) z\nonumber\\
+2 (2+d) z^2\big) \theta +d (5 d-6+3 d^2+4 (d+3) z\big)\theta^2-(2+d (d+5)) \theta ^3)\alpha_{GB}\Big),\nonumber\\
\end{gather}

\begin{eqnarray}
\gamma=-\eta=\frac{2\alpha}{\beta},\ \ \ \  \  \ \  \alpha=-\frac{\theta}{d}, \hspace*{1cm}\lambda=\frac{2}{\beta}(\theta-d-\frac{\theta}{d}).\ \ \ \ \ 
\end{eqnarray}

One can check that the scalar equation is satisfied accordingly and does not imply any further relation.
Therefore, the action \ref{eq:action1} with the Ansatz \ref{eq:Ansatz} admits hyperscaling violating solution with an electric gauge potential and a logarithmic scalar field in the form $\phi=\phi_0+\beta \  \mathrm{log} \ r$.

For a consistency check, we may notice that if we let $\theta=0$ and set the dimension $d^\prime=d-1$, these solution will exactly match the Lifshitz solution found in  \cite{Knodel:2013fua} and for their case of $d=1$ and 2, the factor of Gauss-Bonnet combination does not contribute to the equation of motion as we expect from the theory of Gauss-Bonnet gravity. However, for our general case that we coupled the dilaton with higher derivative terms, the theory is no longer the simple Gauss-Bonnet gravity and it would be no longer necessary that the factor of $\alpha_{GB}$ be zero in these specific dimensions and in our solution, in the general case of $\theta \ne 0$, this factor is not zero for $d=1,2$.

As it is obvious from the above solution, the scalar field $\phi$ runs logarithmically and causes the coupling functions  $f(\phi)=e^{\lambda\phi_0} r^{\lambda\beta}$, (which couples the dilaton to the gauge field, and also $g(\phi)$ which couples the dilaton to the higher derivative terms), run from weak coupling in the IR ($r \to 0$) to the strong coupling in the UV when $(r\to \infty)$.

The function $g(\phi)$ coupled to the squared curvature terms, changes the usual hyperscaling violating solution in the Einstein gravity in a non trivial way. This term induces corrections of order $z^3$ and $z^4$ in the solution as one can check that $\beta^2$ which is zero at Lifshitz solution, now is corrected by order of $z^3$, the electric charge $\rho^2$ is renormalized by order of $z^4$ and the potential $V_0$ by order of $z^3$. Also the maximum order of $z$ inducing by Gauss-Bonnet factor is 2 in all the quantities and $\alpha_R$ and $\alpha_W$ can also induce corrections of order $z^3$ and $z^4$. 

One should notice, there would not be a physical solution for any arbitrary $d$, $z$ and $\theta$. In \cite{O'Keeffe:2013nha} three constraints of null energy condition, causality $(z > 1)$, and $0<d-\theta<d$ have been assumed to derive the range of physical regions for the parameters of the theory $\alpha_W$, $\alpha_{GB}$, $\alpha_R$. In the next section, we consider the specific cases of the solution by assuming different combinations of $\alpha_{GB}, \alpha_R$, $\alpha_W$  terms and plot the ranges of $z$ and $\theta$ for each case coming from the conditions of $\beta^2\ge 0$, i.e. scalar field solution should not be oscillatory, and also $\rho^2 \ge 0$ (NEC) coming from only studying the gravity side. Also as has been pointed out in \cite{Hoyos:2010at} for $\theta=0$, based on null energy condition, one should remove the part of $z<1$ which has beed done with a red line in our figures.

Other constraints that can be imposed on the region of parameters, is the stability of the thermodynamics. For the general hyperscaling violating metric Ansatz, there should be the following relation $\frac{d-\theta}{z}>0$, to satisfy the positivity of specific heat \cite{Dong:2012se}. We don't consider this constraint here for plotting the physical region. Considering this constraint also plus the other two, for the full corrected gravity theory when all $\alpha$ corrections are present gives a smaller region of $\theta<d$ and $0<z<1$ which is both stable and physical. 

\section{Specific Cases of the Solution}\label{sec:cases}

In the following sections, by letting the different combinations of higher derivative terms to vanish, we investigate several special cases of the solution obtained before. Assuming the remaining factors of each $\alpha_R$, $\alpha_W$ or $\alpha_{GB}$ get positive values, to compare the qualitative behaviors in different limits, we plot the allowed regions for the parameters $d$, $z$ and $\theta$. 

\subsection{Hyperscaling Violating Solution in Einstein-Weyl gravity }

For the case of Einstein-Weyl gravity we have $\alpha_R=\alpha_{GB}=0$, and then we can read the hyperscaling violating solution in this setup. 
So, Weyl solution is as follows
\begin{flalign}\label{nec}
\beta^2&= 2 (d-\theta ) \left(z-1-\frac{\theta }{d}\right)+\text{$\alpha_W$}\frac{8 e^{\eta  \text{$\phi $0}}z\text{  }(z-1)(d-1)(d-\theta ) (d-z-\theta +2)}{d+1}. 
\end{flalign}

 Notice, in order to have a solution, the right hand side of Eq. (\ref{nec})  must be positive and this is leading to a constraint similar to the NEC. If we set $\alpha_W=0$, the case with no higher derivative gravity, then $(\theta-d)(\theta-d z +d)\ge 0$, which is the NEC in pure hyperscaling violating solution.
 Also, the electric charge and the constant of scalar potential are 
\begin{flalign}
\rho^2 e^{-\lambda \phi_0}  =\frac{1}{2} (z-1) (d+z-\theta ) \left(1+\text{$\alpha_W$} \frac{4 e^{\eta  \text{$\phi $0}} z(\theta +d (z+\theta -d-2))(1-d)}{d (d+1)}\right),
\end{flalign}

\begin{flalign}
V_0 e^{\gamma \phi_0} =& (z+d-\theta -1) (z+d-\theta )+\text{$\alpha_W$}\frac{4e^{\eta  \text{$\phi $0}}z (z-1)(z(d-1)-2d+\theta )(d-\theta )(1-d)  }{d (d+1)}.
\end{flalign}

Again, letting $\theta=0$ and $d^\prime=d -1$ in here, these equations will match the results of  \cite{Knodel:2013fua}. So this solution is the generalization of Lifshitz to HSV for the Einstein-Weyl gravity. 

Now we would like to study the allowed regions of parameters.  To have a meaningful physical solution, generally the three constraints mentioned above should be satisfied. Also, one needs to make sure that the choices for $\alpha_W$ does not lead to a blow up in the potential. Here we just choose $\alpha_W=1$, so for now, we don't need to worry about this issue. However, in the next section we should consider this condition as well.
 
The allowed region from the constraint of having physical solution is plotted in Figure (\ref{fig:weyl}). The initial value for the dilaton is assumed to be zero, also $\alpha_W=1$ and $d=4$.  One can notice specifically that the range of $0<z<1$ with general $\theta$ is not included in the physical space. 
\begin{figure}[]
\centering
\parbox{15cm}{
\centering
\includegraphics[scale=0.55]{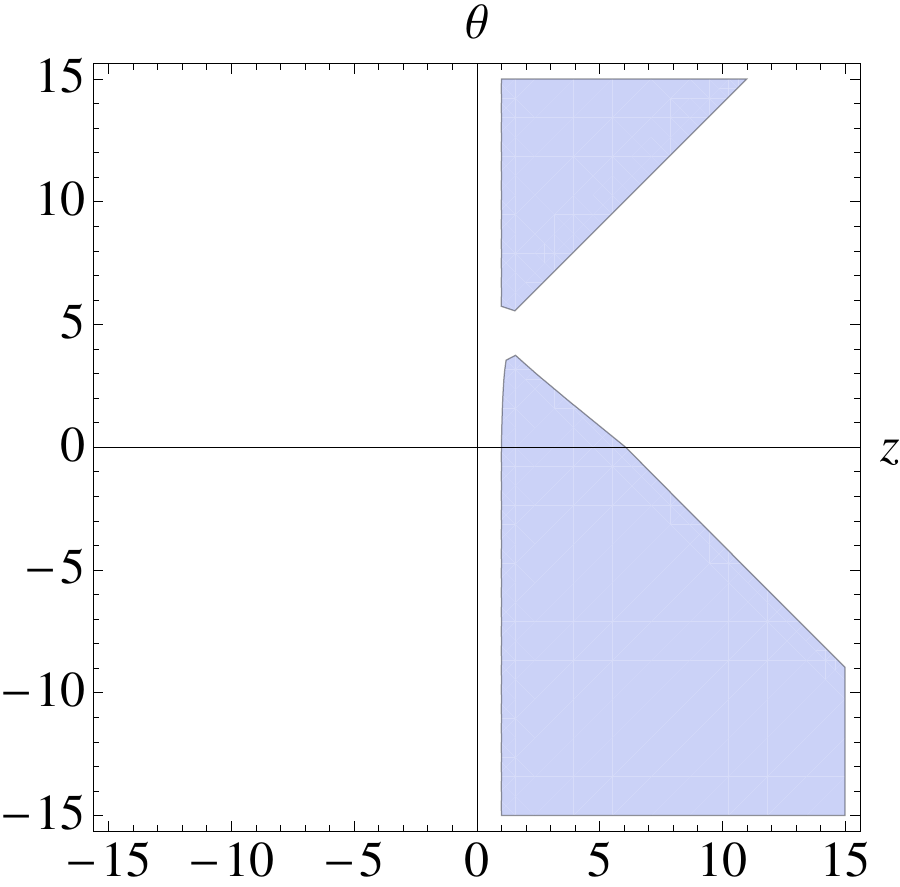}
\caption{Allowed region for  $z$, $\theta$ from both constraints of $\beta^2 \ge 0$ and $\rho^2 \ge 0$ for the Einstein-Weyl gravity, assuming $(\alpha_W=1,\ \phi_0=0,  d=4).$ \label{fig:weyl} }
\label{fig:2figsA4}}
\qquad
~ ~ ~ ~ ~ ~ ~ ~ ~ ~

\end{figure}

Also if we let $d=1$, only the first term in $\beta^2$, $\rho^2$ and $V$ will remain; i.e., the factor of $\alpha_W$ is zero in all of them. So in this case we will reach to the pure hyperscaling violating solutions of previous works \cite{Alishahiha:2012qu}. This means that the higher derivative corrections cannot induce any correction to hyperscaling violating and also to Lifshitz space-times in three dimension, i.e.,  $d+2=3$.

\subsection {Pure gravitational field with higher derivative terms $(\rho=0)$}

We now study the case where $\rho=0$  and matter field is decoupled, so a pure gravitational field is being recovered. There are two ways that this can happen. One is $z=1$ which relates to pure $\mathrm{AdS_{d+2}}$, as a pure $\mathrm{AdS}$ background describes bulk without matter field.
The second possibility is when:

\begin{flalign} \label {eq:alpha}
\alpha_W=& -\frac{e^{-\eta  \text{$\phi $0}}d (1+d) }{4 z(d-1)\left(d^2+2d-d z-\theta -d \theta \right)}.
\end{flalign}

If we let $\theta=0$ we will get the Lifshitz result\cite{Knodel:2013fua},
\begin{flalign} 
\alpha_W=-\frac{d+1}{4 z (d-1) (2+d-z)}.
\end{flalign}

In this case $\rho^2=0$ and

\begin{flalign}
\beta^2=&-\frac{2 \theta (d+1) (d-\theta )^2  }{d \left(d^2-\theta -d (-2+z+\theta )\right)},
\end{flalign}

\begin{flalign}
V_{0} e^{\gamma \phi_0}= &(d+z-\theta -1) (d+z-\theta )-\frac{(z-1) (d-\theta ) (d (z-2)-z+\theta )}{\theta +d (z+\theta -d-2)} \ .
\end{flalign}

If we consider the case where $\theta=0$ corresponding to a Lifshitz metric, and also $d=d^\prime+1$ instead of $d=d^\prime+2$, we will get $\beta^2=0$. Therefore, the dilaton is a constant value. and equation (2.24) of \cite{Knodel:2013fua} corresponds to a pure gravitational Lifshitz solution in this limit. However, for the hyperscaling violating space-times with $\theta \ne 0$, for the limit of $\rho \to 0$, $\beta^2$ is no longer zero and therefore the dilaton would not be necessarily a constant and the purely gravitational solution cannot be recovered unless $\theta=d$. The reason behind this situation is that even without the charge, the potential of the dilaton can source the running of the dilaton and breaks the scale invariance of the theory. This is one difference between Lifshitz and hyperscaling violating solution in this limit.  

For conformal gravity where $\alpha_W \rightarrow \infty$, there are two possible values for the dynamical exponent $z$ which are the roots of the denominator of $\alpha_W$ in equation (\ref{eq:alpha}). One of them is $z=0$, where in this case $\alpha_W \to \infty$, $\rho^2=0$ and 
\begin{flalign}
\beta^2=&-\frac{2 \theta }{d}\frac{(d+1) (d-\theta )^2 }{ d (d+2)- \theta (d+1)},
\end{flalign}
\begin{flalign}
V_{0} e^{\gamma \phi_0}= &\frac{(d+1) (d-\theta )^3}{d (d+2)-\theta (d+1) }.
\end{flalign}

The other case is $z=d+2-\theta -\frac{\theta }{d}$. At this $z$, generally both $V$ and $\beta^2$ will blow up. One case which can make both of them well behaved is when $d=\theta$, then $z=1$ and we will again end up with an AdS space-time and pure gravity theory.  The other case is when $\theta=0$ corresponding to a Lifshitz background. Then from the condition on $\beta^2$, $z=d+2$ and from the condition on $V^2$, $d=2$. Thus, for the case of $z=4,  d=2, \theta=0$ conformal gravity has a solution, consistent also with  \cite{Knodel:2013fua}. Therefore, at this limit, hyperscaling violating background has not any further well behaved solution.    

Again, we would like to see which ranges give us a physical solution in this case where $\rho^2 \to 0$. We should satisfy the constraint of $\beta^2 \ge 0$ and $d-\theta >0$ which corresponds to stability and also we need to make sure that the parameters in these regions do not lead to blowing up of the potential $V(\phi)$. Figure (\ref{fig:rho}) shows the allowed region for this limit.

\begin{figure}[!ht] 
\centering
\parbox{17cm}{
\centering
\includegraphics[scale=0.55]{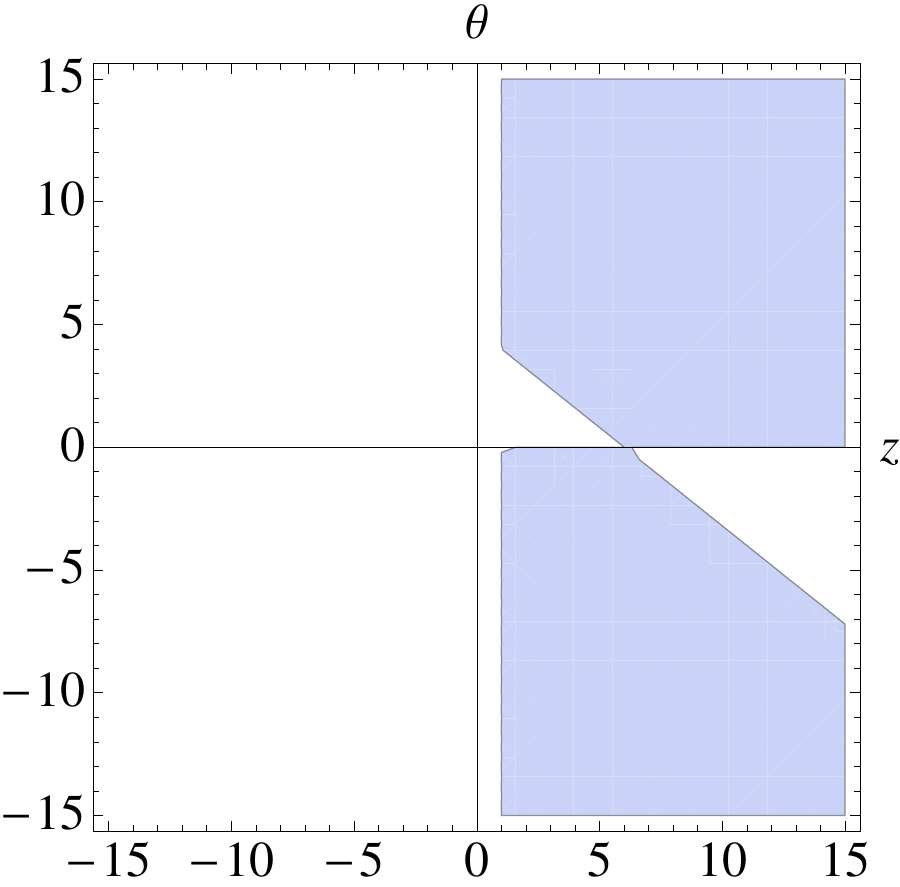}
\caption{Allowed region for  $z$, $\theta$ from the constraint of $\beta^2 \ge 0$, for the case of $\rho^2=0, d=4.$ \label{fig:rho}  }
\label{fig:2figsA5}}
\qquad

~ ~ ~ ~ ~ ~ ~ ~ ~ ~ 
\end{figure}

\subsection{Hyperscaling Violating Solution in Gauss-Bonnet gravity }

If we assume $\alpha_W=0$ and $\alpha_R=0$, the solution in the Gauss-Bonnet gravity is
\begin{gather}
\beta^2= 2 (d-\theta ) \left(z-\frac{\theta }{d}-1\right)+ \nonumber\\ \text{$\alpha_{GB}$}\frac{4 e^{\eta  \text{$\phi $0}}  (1-d)(d-\theta )^2 \left((d-2) d^2 (z-1)-d ((4+d) z-4) \theta +(d+2) \theta ^2\right)}{d^3},
\end{gather}

\begin{flalign}
\rho^2 e^{-\lambda \phi_0}=\frac{1}{2} (z-1) (d+z-\theta ) \left(1+ \text{$\alpha_{GB}$}\frac{e^{\eta  \text{$\phi $0}}2 \left(1-d^2\right)\left(d^2 (d-2 \theta -2)+(d+2) \theta ^2\right)}{d^2 (d+1)}\right),
\end{flalign}

\begin{gather}
V_0 e^{\gamma \phi_0}=(d+z-\theta -1) (d+z-\theta )+\nonumber \\  \frac{e^{\eta  \text{$\phi $0}} \text{$\alpha_{GB}$}}{d^3}(1-d)(d-\theta ) ((d-2) d^2 (d^2+2 (z-1) z+d (4 z-3))  \nonumber \\  -d ((d-2) d (1+3 d)+2 (4 d^2+d-2) z+2 (d+2) z^2) \theta \\ \nonumber +d (3 d^2+5d-6+4 (d+3) z) \theta ^2-(2+d (d+5)) \theta ^3).
\end{gather}
\begin{figure}[!ht]
\centering
\parbox{15cm}{
\centering
\includegraphics[scale=0.55]{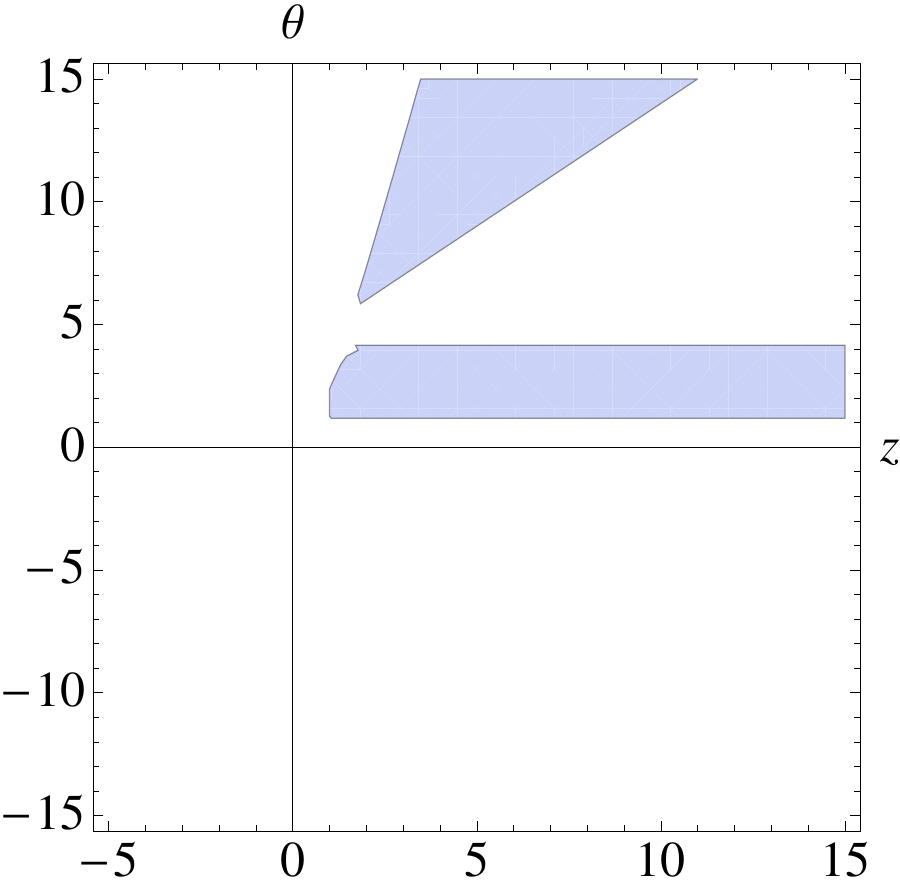}
\caption{Allowed region for  $z$, $\theta$ from the constraint of $\beta^2 \ge 0$, $d=4$ for Gauss-Bonnet case where $\alpha_W=\alpha_R=0$ and $\alpha_{GB}=1.$ \label{fig:gauss} }
\label{fig:2figsA2}}
\end{figure}

As can be seen from Figure (\ref{fig:gauss}), ($d=4$), two distinct regions are allowable. Notably, the region of $z>1$ with $ 1<\theta < d=4$ is present in the allowed region. Particular changes happen at $z=1$, $\theta=1$ and $\theta=d=4$. For Lifshitz solution where $\theta=0$, the region of $-d=-4<z<1$, is physical. Also for AdS case where $z=0$, $-d=-4<\theta<1$ is within the allowable ranges. It is worth mentioning that changing the dimension, does not change these qualitative behaviors much.

\subsection{Hyperscaling Violating Solution in  $\mathrm{R^2}$  gravity }

We then study the solution for the case where $\alpha_W=\alpha_{GB}=0$. In this case the solution is 

\begin{gather} \nonumber \\
\beta^2=2 (d-\theta ) \left(z-\frac{\theta }{d}-1\right)+ \nonumber \\ \text{$\alpha_R$}\frac{4}{d^2} e^{\eta  \text{$\phi $0}}\text{  }(d-\theta ) (d (1-z)-\theta ) \left(2 z (d (d+z-\theta )-\theta )+(1+d) (\theta -d)^2\right),
\end{gather}

\begin{flalign}
\rho^2 e^{-\lambda \phi_0}=(z-1) (d+z-\theta ) \left(\frac{1}{2}+ \text{$\alpha_R$}\frac{e^{\eta  \text{$\phi $0}}}{d}\text{  }\left(-(1+d) (\theta -d)^2+2 z (-d (d+z-\theta )+\theta )\right)\right),
\end{flalign}

\begin{flalign}
V_0 e^{\gamma \phi_0}&=(d+z-\theta -1) (d+z-\theta )+\nonumber \\ & \text{$\alpha_{R}$} \frac{e^{\eta  \text{$\phi $0}}}{d^2}((3-d) d+(1-d) (2 z-\theta )) (d-\theta ) \left(2 d z (d+z)+(1+d) \left(-2 z \theta +(\theta -d)^2\right)\right).
\end{flalign}

\begin{figure}[]
\centering
\parbox{15cm}{
\centering
\includegraphics[scale=0.55]{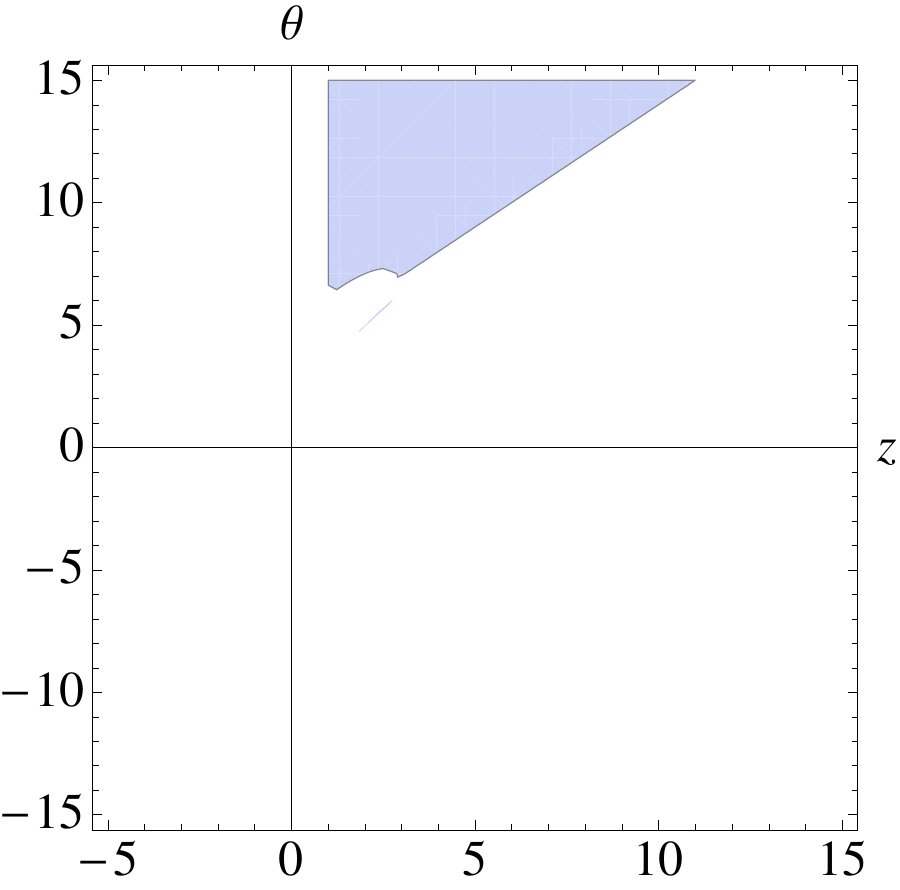}
\caption{Allowable region for  $z$, $\theta$ from both constraint of $\beta^2 \ge 0$ and $\rho^2 \ge 0$ for the $R^2$ gravity, assuming $(\alpha_R=1,\ \phi_0=0,  d=4).$  }
\label{fig:2figsA3}}
\qquad
\end{figure}
As can be noticed, in $R^2$ theory, the region of $z>0$, $1< \theta <4$ is not present anymore.


\subsection{Hyperscaling Violating Solution in Gauss-Bonnet and $\mathrm{R^2}$  gravity }

In this section and the following sections \ref{sec:sub3} and \ref{sec:sub4}, we will turn off only one alpha correction and keep the other two to study the different combinations of $R^2$ corrections.

For the case where $\alpha_W=0$, the solution is

\begin{gather}
\beta^2=2 (d-\theta ) (z-\frac{\theta }{d}-1) \nonumber \\+ \frac{4}{d^3} e^{\eta  \text{$\phi $0}} (d-\theta ) (\text{$\alpha _{GB}$}(1-d)(d-\theta )((d-2) d^2 (z-1)-d (-4+(d+4) z) \theta +(d+2) \theta ^2)+\nonumber \\ \frac{\text{$\alpha_ {R}$}}{d^2} (d (1-z)-\theta ) (2 z (d (d+z-\theta )-\theta )+(d+1) (\theta -d)^2)),
\end{gather}

\begin{flalign}
\rho^2 e^{-\lambda \phi_0}=&\frac{1}{2} (z-1) (d+z-\theta ) \bigg(1+\frac{e^{\eta  \text{$\phi $0}}}{d^2 (1+d)} (2 \text{$\alpha_{GB}$} (1-d^2)(d^2 (d-2 \theta -2)+(d+2) \theta ^2)+\nonumber \\& 2 \text{$\alpha_{R}$}\text{  }d (d+1)(-(1+d) (\theta -d)^2+2 z (-d (d+z-\theta )+\theta )))\bigg),
\end{flalign}

\begin{flalign}
V_0 e^{\gamma \phi_0}&=(d+z-\theta -1) (d+z-\theta )+\nonumber \\&\frac{1}{d^3}e^{\eta  \text{$\phi $0}} (d-\theta ) (\text{$\alpha_{GB}$}(1-d)((d-2) d^2 (d^2+2 (z-1) z+d (4 z-3))\nonumber \\& -d ((d-2) d (1+3 d)+2 (4 d^2+d-2) z+2 (d+2) z^2) \theta \nonumber \\&+ d (3 d^2+4 (3+d) z+5d-6) \theta ^2-(2+d (5+d)) \theta ^3) \nonumber \\&+ \text{$\alpha_R$} d(2 z-\theta +d (3-d-2 z+\theta ))(2 d z (d+z)+(1+d) ((d-\theta )^2-2 z \theta ))).
\end{flalign}
\begin{figure}[]
\centering
\parbox{15cm}{
\centering
\includegraphics[scale=0.55]{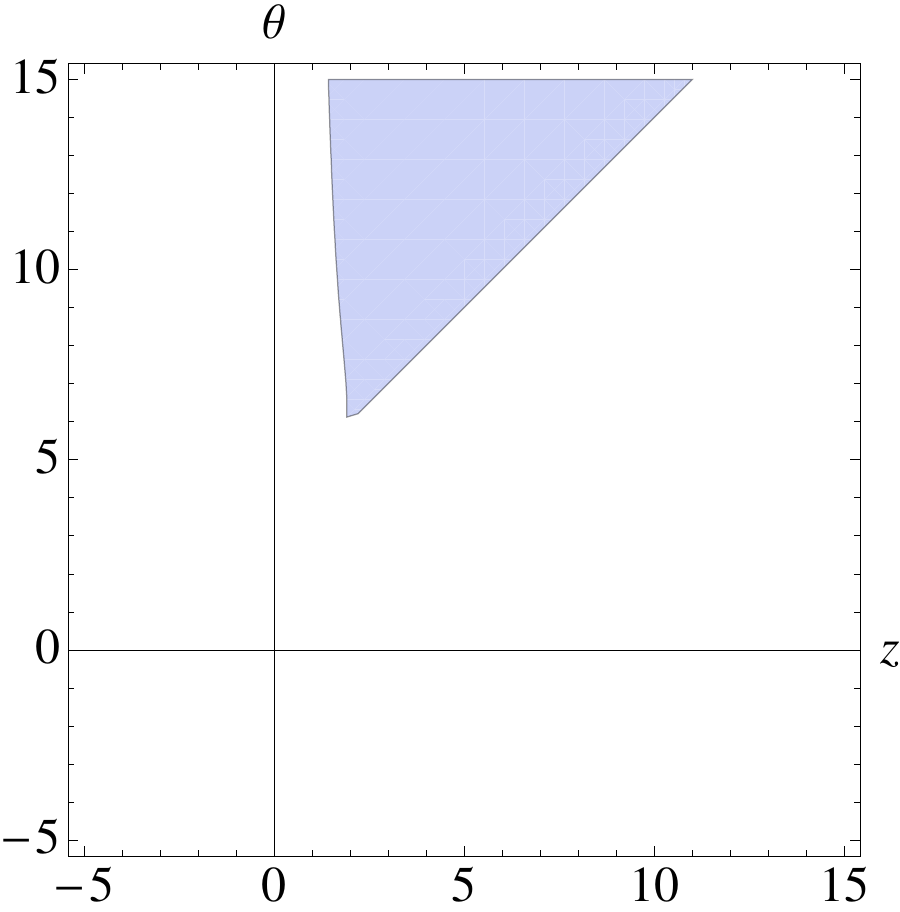}
\caption{Allowable region for  $z$, $\theta$ from both constraints of $\beta^2 \ge 0$ and $\rho^2 \ge 0$ for the $R^2$ and Gauss-Bonnet gravity, assuming $(\alpha_R=1,\alpha_{GB}=1,\alpha_W=0\ \phi_0=0,  d=4).$  }
\label{fig:2figsA6}}
\qquad
~ ~ ~ ~ ~ ~ ~ ~ ~ ~ 

\end{figure}

\subsection{Hyperscaling Violating Solution in Gauss-Bonnet and Weyl gravity }\label{sec:sub3}
We then study the solution for the case where $\alpha_R=0$. In this case the solution will be 

\begin{flalign}
\beta^2=&2 (d-\theta ) (z-\frac{\theta }{d}-1)+4 e^{\eta  \text{$\phi $0}} \frac{(d-\theta )}{d^3} (\text{$\alpha _W$}\frac{2 (d-1) d^3 (z-1) z\text{  }(d-z-\theta +2)}{1+d}+\nonumber \\&\text{$\alpha_{GB}$}(1-d)(d-\theta ) ((d-2) d^2 (z-1)-d (-4+(d+4) z) \theta +(d+2) \theta ^2)),
\end{flalign}

\begin{flalign}
\rho^2 e^{-\lambda \phi_0}=&(z-1) (d+z-\theta ) (\frac{1}{2}+e^{\eta  \text{$\phi $0}} (1-d)(\text{$\alpha_{GB}$} \frac{1 }{d^2 } (d^2 (d-2 \theta -2)+(d+2) \theta ^2)+\nonumber \\& \text{$\alpha_{W}$}\frac{2\text{  }}{d (1+d)}z (\theta +d (z+\theta -d-2)))),
\end{flalign}

\begin{flalign}
V_0 e^{\gamma \phi_0}&=(d+z-\theta -1) (d+z-\theta )+\frac{1}{d^3}e^{\eta  \text{$\phi $0}} (d-\theta ) (\text{$\alpha_{W}$}\frac{4 (1-d) d^2 (z-1) z(d (z-2)-z+\theta )}{1+d}\nonumber \\&+ \text{$\alpha_{GB}$}(1-d) ((d-2) d^2 (d^2+2 (z-1) z+d (4 z-3)) \nonumber \\& -d ((d-2) d (3d+1)+2 (4d^2+d-2) z+2 (d+2) z^2) \theta \nonumber \\&+ d (3d^2+5d+4 (d+3) z-6) \theta ^2-(2+d (d+5)) \theta ^3)).
\end{flalign}
\begin{figure}[]
\centering
\parbox{15cm}{
\centering
\includegraphics[scale=0.55]{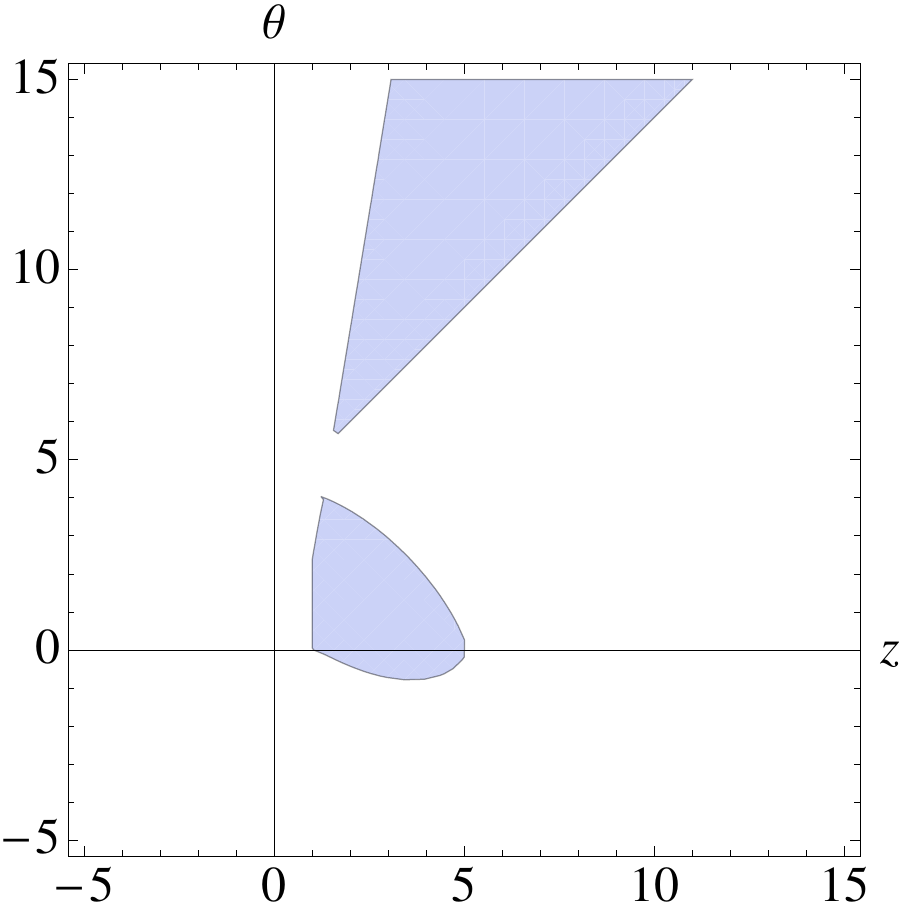}
\caption{Allowable region for  $z$, $\theta$ from both constraints of $\beta^2 \ge 0$ and $\rho^2 \ge 0$ for the  Gauss-Bonnet and Weyl gravity, assuming $(\alpha_W=\alpha_{GB}=1,\alpha_R=0\ \phi_0=0,  d=4).$  }
\label{fig:2figsA7}}
\qquad
~ ~ ~ ~ ~ ~ ~ ~ ~ ~ 

\end{figure}

One may notice that in all solutions of hyperscaling violating a term of $d-\theta$ exist which can indicate again that the degrees of freedom of the theory effectively are in $d_{\mathrm{eff}}= d-\theta$, which can also be seen from the scaling relation of the entropy \cite{Dong:2012se} as it scales with the exponent containing $d-\theta$ rather that just $\theta$.   
 
\subsection{Hyperscaling Violating Solution in $R^2$ and Weyl gravity }\label{sec:sub4}
For the case where $\alpha_{GB}=0$, the solution will be 

\begin{flalign}
\beta^2=&2 (d-\theta ) (z-\frac{\theta }{d}-1)+4 e^{\eta  \text{$\phi $0}} (d-\theta ) ( \text{$\alpha_{W}$}\frac{2 (d-1) (z-1) z (2+d-z-\theta )}{1+d}\nonumber \\& +\frac{1 }{d^2}\text{$\alpha_R$} (d (1-z)-\theta ) (2 z (d (d+z-\theta )-\theta )+(d+1) (\theta -d)^2)),
\end{flalign}

\begin{flalign}
\rho^2 e^{-\lambda \phi_0}=&\frac{1}{2} (z-1) (d+z-\theta ) (1+e^{\eta  \text{$\phi $0}} (4 \text{$\alpha_W$} \frac{(1-d) }{d (1+d)} z (\theta +d (z+\theta -d-2))\nonumber \\&+2 \frac{\text{$\alpha_R$}}{d }(-(d+1) (\theta -d)^2+2 z (-d (d+z-\theta )+\theta )))),
\end{flalign}

\begin{flalign}
V_0 e^{\gamma \phi_0}&=(d+z-\theta -1) (d+z-\theta )+e^{\eta  \text{$\phi $0}} (d-\theta ) (\text{$\alpha $w} \frac{4 (1-d) (z-1) z (d (z-2)-z+\theta )}{(1+d)d}\nonumber \\&+\frac{\text{$\alpha_R$}}{d^2}\text{  }((3-d) d+(1-d) (2 z-\theta )) (2 d z (d+z)+(1+d) (-2 z \theta +(\theta -d)^2))).
\end{flalign}
\begin{figure}[ht!]
\centering
\parbox{15cm}{
\centering
\includegraphics[scale=0.55]{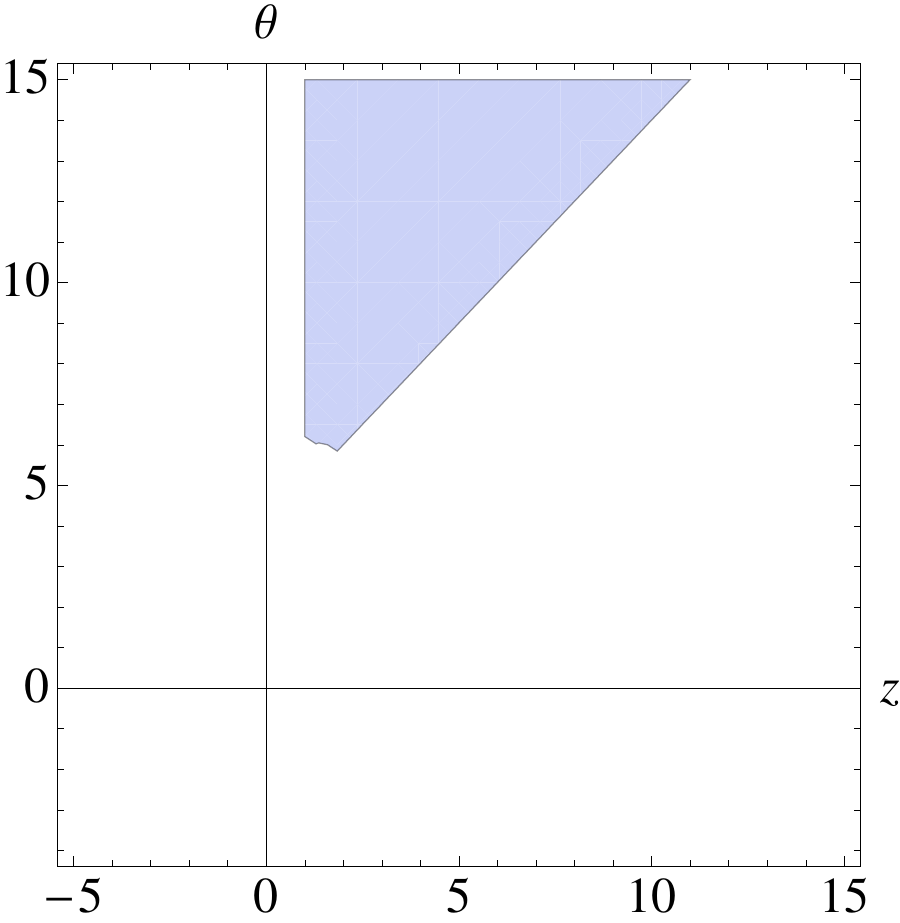}
\caption{Allowable region for  $z$, $\theta$ from both constraints of $\beta^2 \ge 0$ and $\rho^2 \ge 0$ for the  $R^2$ and Weyl gravity, assuming $(\alpha_W=\alpha_{R}=1,\alpha_{GB}=0\ \phi_0=0,  d=4). $  }
\label{fig:2figsA8}}
\qquad
~ ~ ~ ~ ~ ~ ~ ~ ~ ~

\end{figure}

\begin{figure}[]
\centering
\parbox{15cm}{
\centering
\includegraphics[scale=0.55]{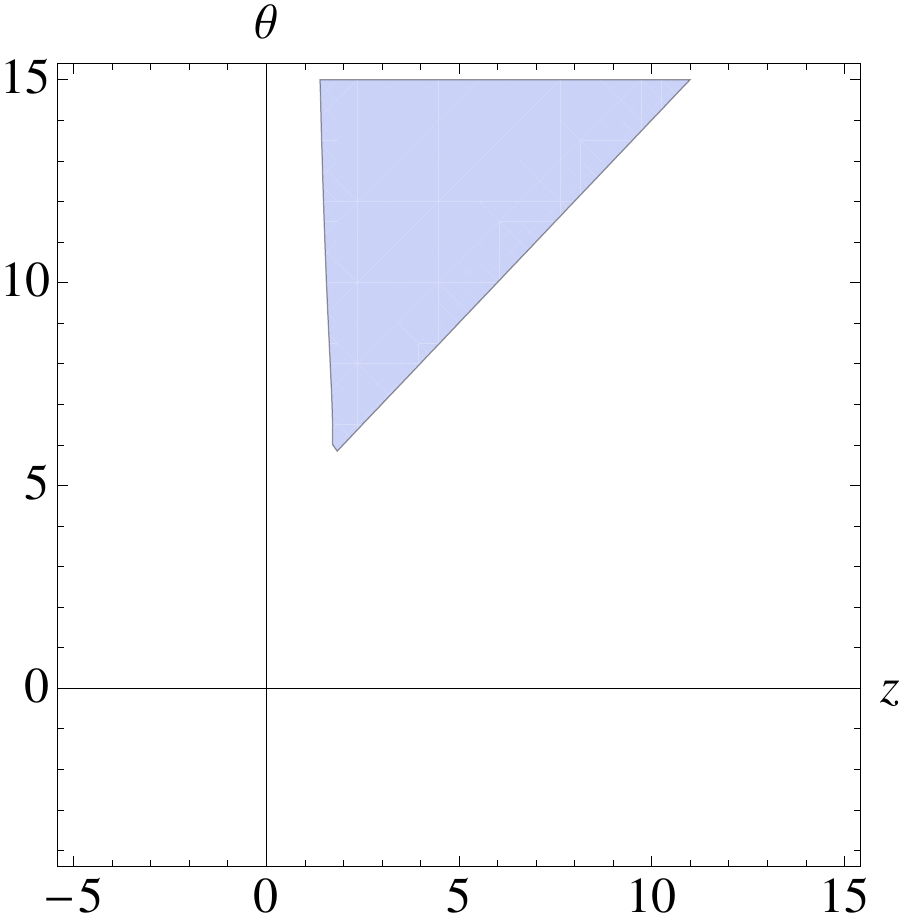}
\caption{Allowable region for  $z$, $\theta$ from both constraints of $\beta^2 \ge 0$ and $\rho^2 \ge 0$ for the most general higher derivative gravity theory, assuming $(\alpha_W=\alpha_{GB}=\alpha_{R}=1, \phi_0=0,  d=4).$  }
\label{fig:2figsA9}}
\qquad
~ ~ ~ ~ ~ ~ ~ ~ ~ ~

\end{figure}

We can compare the physical region of $\theta$ and $z$ for each combination of $\alpha$ corrections. For instance, one can notice that Einstein-Weyl gravity is the least strict theory while on the physical region, $R^2$ is the most strict one. The Gauss-Bonnet gravity has specifically an approximate region of $z>1$ and $1< \theta <d$ in its physical solution that is not present in the $R^2$ gravity, and Einstein-Weyl theory can contain two more regions which placed in the second and forth quarter of space coordinate which are not present in the other two theories. 

As discussed in \cite{Dong:2012se}, the point of $z=1$ and $\theta=-\frac{1}{3}$, is also special as it corresponds to $N$ D2-branes, in type $IIA$ supergravity. It can be seen that this point is not within the physical region for $\alpha_{GB}$ gravity corrected term, Fig~(\ref{fig:gauss})  and $\alpha_{GB}+\alpha_W$ terms, Fig~(\ref{fig:2figsA9}) .

If we also consider the constraint of $0<d-\theta <d$, corresponding to stability of the solution, for $d=4$, all the regions of $\theta>d$ would be eliminated from the plots. However, here we kept the thermodynamically unstable regions within the physical regions for the higher derivative corrected solution. 

Another special limit is when $\theta=d$, where in the squared curvature corrected gravity, there is no region that the right hand side of the equation for $\beta^2$ is positive and so there is no region that dilaton is non-oscillatory in this limit.

Also one can notice that changing the dimension of the theory, $d$, does not change the general qualitative behavior much, and in each $d$ the allowed region in each theory approximately behaves the same.  
One may also fix $\theta$, $z$ and $d$ and plot the regions for $\alpha_W$, $\alpha_{GB}$ and $\alpha_R$ as has been done in \cite{O'Keeffe:2013nha}.

\subsection{The allowable regions for Strange Metals}

One special case where the degrees of freedom effectively live in one dimension, where $d-\theta=1$, corresponds to strange metals, a from of non-Fermi liquid or non-zero temperature phase of correlated electrons in solids, whose resistivities increase linearly with $T$ rather than with $T^2$ \cite{Musso:2014efa}. High temperature superconductors is one example of such materials. These metals recently have been studied using hyperscaling violating backgrounds in the context of holography and AdS/CMT as the Bardeen-Cooper-Schrieffer (BCS) theory failed in studying them and also they can be put among the strongly correlated fermion systems. \cite{Dey:2013vja}. 

Without higher derivative corrections, from NEC,  the relation of $z \ge 2-\frac{1}{\theta+1}$ gives the allowed region for the strange metals \cite{Dong:2012se}. When all the $\alpha$ corrections are turned on, considering higher gravity corrections, gives more complicated relations for our solution.  The allowed region of parameters is shown in Figure~(\ref{fig:metal} ). One can see that the range of $-2< \theta <-1$ and $z  \ge 3$ is within the allowable region in the full squared curvature corrected theory, in agreement with the results without any higher derivative gravity term. 
\begin{figure}[!ht]
\centering
\parbox{14cm}{
\centering
\includegraphics[scale=0.55]{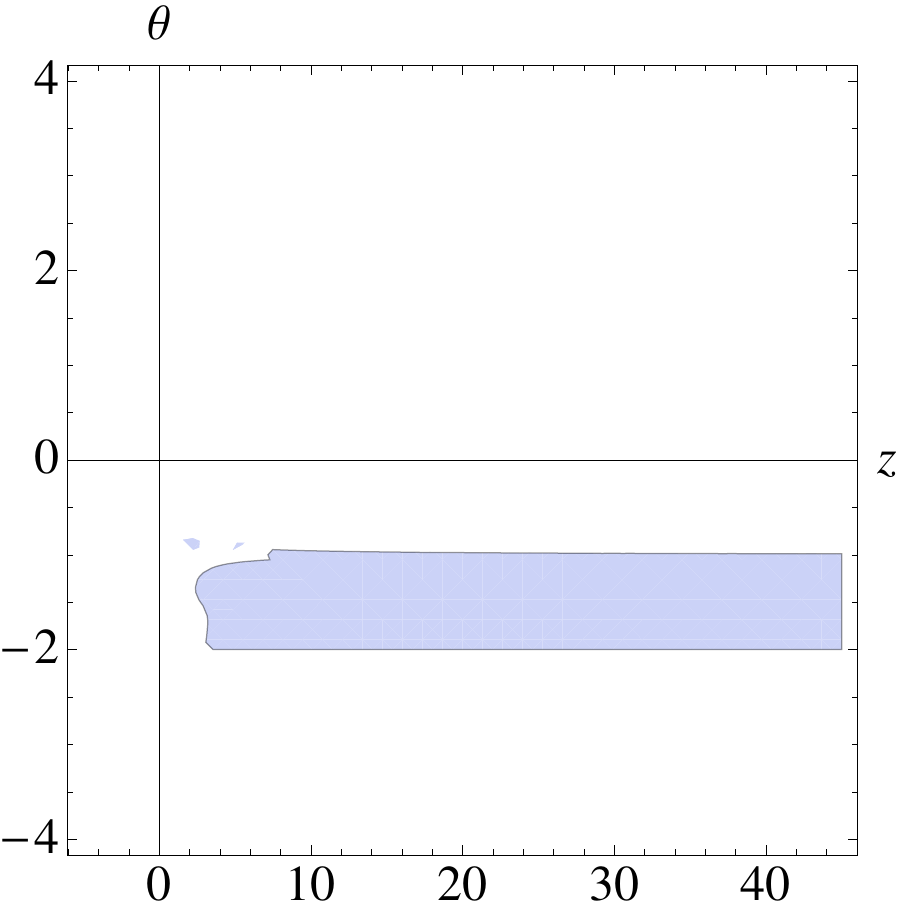}
\caption{Allowable region for strange metals where $d=\theta+1$, all $\alpha$ corrections are turned on and $\alpha_R=\alpha_{GB}=\alpha_{W}=1$. \label{fig:metal} }
\label{fig:2figsA2}}
\end{figure}

However, when only $\alpha_{GB}$ is on, all the region of $z>1$ with any $\theta$ is acceptable and so it is the least strict theory with biggest allowable region. On the other hand, $\alpha_R$ is the most strict theory as we also saw in the previous section, and only approximately a small region of $-2< \theta <4$, $-1<z <1$ is allowable.

\section{Resolving the Singularity}\label{sec:resolve}
In this section we consider adding only a Weyl correction term to the action, coupled to the dilaton, and investigate the singularity resolution. By choosing an appropriate $g(\phi)$ function, this term can lead to corrections to effective potential in the deep IR which stabilizes the dilaton at a finite value $\phi_I$ and the geometry would be free from singularity in this regime. In the UV limit of the theory as Weyl tensor vanishes, this correction would have no effect. 

The action that we consider is
\begin{eqnarray}
S=\int{d^{4}x\sqrt{-g}(R+V(\phi)-\frac{1}{2}(\partial \phi)^2-f(\phi)F_{\mu \nu}F^{\mu \nu}+g(\phi)C_{\mu \nu \rho \sigma}C^{\mu \nu \rho \sigma})},
\end{eqnarray}

where $g(\phi)=\frac{3}{4}\left(c_0e^{\eta \phi}+c_1\right)$ and $\alpha_W=\frac{3}{4}c_0$. 
The metric Ansatz for constructing the flow is
\begin{eqnarray}\label{allsol}
ds^2&=&a^2(r)\left(-dt^2+dr^2+b^2(r)\left(dx^2+dy^2\right)\right).
\end{eqnarray}

We would call this new form of Ansatz, the (a-b) gauge. The field parametrization in this new gauge is
\begin{eqnarray}
F_{rt}&=&\frac{Q}{f(\phi)b^2},
\end{eqnarray}

and the Einstein equations are

\begin{eqnarray}\label{e1}
E_{rr}&=&\frac{3 a'^2}{a^2}+\frac{b'^2}{b^2}+\frac{4 a' b'}{a b}+\frac{4g}{3a^2}\left(\frac{ b'^4}{2b^4}+\frac{ b''^2}{2b^2}-\frac{b'b^{(3)}}{ b^2}+\frac{ b'g' \phi '}{ g b^2}\left(\frac{ b'^2}{ b}-b''\right)\right),\nonumber\\
E_{tt}&=&-2\left(\frac{a''}{a}+\frac{b''}{b}\right)+\frac{a'^2}{a^2}-\frac{b'^2}{b^2}-\frac{4a'b'}{ab}-\frac{4g}{3a^2}\times\nonumber\\
&&\left(\frac{(b'b^{(3)})'}{b^2}-\frac{2b'^2b''}{b^3}-\frac{b''^2}{2b^2}+\frac{b'^4}{2b^4}+\frac{g'\phi''+\phi '^2 g''}{gb}\left(b''-\frac{b'^2}{b}\right)+\frac{g' \phi '}{gb}\left(2b^{(3)}-\frac{b'^3}{b^2}-\frac{b' b''}{b}\right)\right),\nonumber\\
\frac{E_{ii}}{b^2}&=&\frac{2 a''}{a}+\frac{b''}{b}-\frac{a'^2}{a^2}+\frac{2 a' b'}{a b}-\frac{4g}{3a^2}\times\nonumber\\
&&\left(\frac{ b^{(4)}}{2b}+\frac{ b'^4}{2b^4}-\frac{ b'^2 b''}{ b^3}+\frac{ g' \phi '}{g b}\left( b^{(3)}-\frac{ b'b''}{ b}\right)+\frac{1}{2g b}\left(b''- \frac{b'^2 }{b}\right)\left(\phi '^2g''+ g'\phi ''\right)\right),
\end{eqnarray}

with the right hand side
\begin{eqnarray}\label{e2}
T_{rr}&=&-\frac{Q^2}{b^4 a^2f }+\frac{1}{4} \phi '^2+\frac{a^2}{2} V, \nonumber\\
T_{tt}&=&\frac{Q^2}{b^4 a^2f }+\frac{1}{4} \phi '^2-\frac{a^2}{2} V, \nonumber\\
\frac{T_{ii}}{b^2}&=&\frac{Q^2 }{b^4 a^2f } -\frac{1}{4}\text{  }\phi '^2+\frac{a^2}{2} V, 
\end{eqnarray}

and the scalar equation is
\begin{eqnarray*}
\phi ''+2\left(\frac{a'}{a}+\frac{b'}{b}\right)\phi '+\frac{2 Q^2 f'}{b^4 a^2 f^2}+\frac{4g'}{3b^2a^2}\left(\frac{b'^4}{b^2}-\frac{2 b'^2 b''}{ b}+ b''^2\right)+a^2V'=0.
\end{eqnarray*}

As we can see from this equation, the derivative of the function $g(\phi)$ coupled to the metric function $b(r)$, affects the effective potential of the dilaton.

Next, we consider an initial solution of $\mathrm{AdS}_2\times \mathbb{R}^2$ in the $\mathrm{IR}$
\begin{eqnarray*}
a(r)=\frac{1}{r},\hspace*{1cm}b(r)=b_Ir,\hspace*{1cm}\phi(r)=\phi_I.
\end{eqnarray*}

Then from \eqref{e1} and \eqref{e2} we can derive
\begin{eqnarray*}
V(\phi_I)=1,\hspace*{1cm}\frac{Q^2}{b_I^4}=\frac{f(\phi_I)}{2}\left(1-\frac{4}{3}g(\phi_I)\right),
\end{eqnarray*}

and the scalar equation becomes
\begin{eqnarray*}
\frac{f'(\phi_I)}{f(\phi_I)}\left(1-\frac{4}{3}g(\phi_I)\right)+\frac{4}{3}g'(\phi_I)+V'(\phi_I)=0.
\end{eqnarray*}

Finally, by considering $f(\phi)=e^{\lambda \phi}, V(\phi)=V_0e^{\gamma \phi}$, we have the following solution in the IR
\begin{eqnarray*}
V_0 e^{\gamma_1 \Phi_I}&=&1,\nonumber\\
\frac{Q^2}{b_I^4}&=&\frac{e^{\lambda \Phi_I}}{2}\frac{(c_1-1)\eta-\gamma}{\lambda-\eta}, \nonumber\\
\Phi_I&=&\frac{1}{\eta}\log \left(\frac{\lambda(1-c_1)+\gamma}{c_0(\lambda-\eta)}\right).
\end{eqnarray*}

\subsection{IR Perturbations}
We now consider perturbations around the above IR, $\mathrm{AdS}_2 \times \mathbb{R}^2$ solution,
\begin{eqnarray*}
a(r)=\frac{1}{r}+\delta a(r),\hspace*{1cm}b(r)=b_Ir+\delta b(r),\hspace*{1cm}\phi(r)=\phi_I+\delta \phi(r).
\end{eqnarray*}

For simplicity we can take $b_I=1$. Then from the Einstein and dilaton equations, there are four coupled perturbative equations with maximum order of 4. 
\begin{flalign}
rr: r^2 \delta b^{(3)}+\frac{3 }{2g_I}\left(r^2 \delta a\right)'-2 \delta b'\frac{f_I}{f_I'g_I}\left(g_I'+\frac{3}{4}V_I'\right)-\frac{g_I'}{g_I}(r\delta \phi)'&= 0,
\end{flalign}
\begin{flalign}
tt:\left(r \delta  b^{(3)}\right)'+\frac{3}{2 g_I}\frac{\left( r^3 \delta  a'\right)'}{r^2}-\frac{g_I'}{g_I}\frac{(r \delta  \phi ')'}{r}+\frac{ g_I'}{g_I}\frac{\delta  \phi }{r^2}-2\frac{f_I}{f_I'g_I}\left(g_I'+\frac{3}{4}V_I'\right)\left(\frac{\delta  b'}{r}\right)'&=0, 
\end{flalign}
\begin{flalign}
ii: r^2 g_I \delta  b^{(4)}-3\left(r^2 \delta  a'\right)'-\frac{6 \delta  b}{r^2}-r g_I' \delta  \phi ''+2\frac{\delta  \phi }{r}\left(g_I'+\frac{3 V_I'}{4}\right)-2r^2\left(\frac{3}{4}+g_I\right)\left(\frac{\delta  b'}{r^2}\right)'&=0, 
\end{flalign}
and the scalar equation is
\begin{flalign}
  \delta  \phi ''-\frac{8}{3}r g_I'\left(\frac{ \delta  b'}{r^2}\right)'+\frac{4}{r}\left(\delta  a+\frac{\delta  b}{r^2} \right)V_I'+\frac{\delta  \phi }{r^2}\left(\frac{2f_I'}{ f_I}\left(\frac{4}{3}g_I'+V_I'\right)+\frac{ f_I''}{ f_I}-\frac{4 g_I f_I''}{3f_I}+\frac{4g_I''}{3 }+ V_I''\right)&=0.
\end{flalign}
One can read $\delta b^ {(4)} $ from ($ii$) and replace it in $(tt)$ and therefore end up with 3 equations with maximum order of 3.

Now considering the IR perturbations
\begin{eqnarray*}
\delta a(r)\sim r^{\nu-1},\hspace*{1cm}\delta b(r) \sim r^{\nu+1},\hspace*{1cm}\delta \phi(r)\sim r^{\nu},
\end{eqnarray*}

we can derive the following solutions for $\nu$
\begin{gather}
\nu_{1,2}=-1,\hspace*{2cm} \nu_3=2,\nonumber\\
\nu_{4,..,7}=\frac{1}{2}\pm\frac{1}{2}[1-\frac{2}{ 3( \gamma - c_1 \eta + \lambda _1)}[u_0\pm(24(\gamma +\eta -c_1 \eta ) (\gamma +\lambda -c_1 \eta ) (c_1 \eta  \lambda - \nonumber\\(\lambda +\gamma ) (\eta +\gamma ))+u_0^2)^{\frac{1}{2}}]]^{\frac{1}{2}}, \nonumber\\
u_0=(\gamma +(1-c_1) \eta ) (-2+3 (\gamma +\lambda )^2)+3c_1 \eta ^2(\gamma +(1-c_1) \lambda ).
\end{gather}

\subsection{ Hyperscaling Violating Solution in the (a-b) gauge} \label{ab}
Now we need to write the hyperscaling violating solution that we have found in section \ref{sec:solution} in a new gauge as \eqref{allsol} in order to match all three regions in a similar form of a metric.
Changing variable in \eqref{allsol}, in the way that $z\tilde{r}=r^{-z}, \tilde{x}=z^{1-\frac{1}{z}}x$, leads to
\begin{eqnarray}
ds^2=\tilde{L}^2\tilde{r}^{\frac{\theta}{z}-2}\left(-dt^2+d\tilde{r}^2+\tilde{r}^{2-\frac{2}{z}}d\tilde{x}^2\right),
\end{eqnarray}
where $\tilde{L}^2=L^2z^{\frac{\theta}{z}-2}$.\\

The hyperscaling violating solution in this gauge is
\begin{eqnarray}\label{eq:newsolg}
ds^2&=&\tilde{L}^2 r^{(1-\tilde{z})\theta-2}\left(-dt^2+dr^2+r^{2\tilde{z}}dx^2\right), \nonumber\\
\phi(r)&=&\phi_0+ (\tilde{z}-1)\beta \log r ,   \nonumber\\
F_{rt}&=& \rho e^{-\lambda \phi_0}r^{2(\tilde{z}-2)-(\tilde{z}-1)\theta} , \nonumber\\
\beta^2&=&\frac{(\theta-2)(2\tilde{z}+(\tilde{z}-1)\theta)}{\tilde{z}-1}+\frac{2c_0e^{\eta \phi_0}}{\tilde{L}^2}\frac{\tilde{z}(\theta-2)(3-4\tilde{z}+(\tilde{z}-1)\theta)}{\tilde{z}-1} , \nonumber\\
\rho^2 e^{-\lambda \phi_0}&=&\frac{\tilde{L}^2}{2}\tilde{z}(3-2\tilde{z}+(\tilde{z}-1)\theta)+\frac{c_0e^{\eta \phi_0}}{4}\tilde{z}(3-2\tilde{z}+(\tilde{z}-1)\theta)(6-8\tilde{z}+3(\tilde{z}-1)\theta) , \nonumber\\
V_0 e^{\gamma \phi_0}&=&\frac{(3-2\tilde{z}+(\tilde{z}-1)\theta)(2-\tilde{z}+(\tilde{z}-1)\theta)}{\tilde{L}^2}+\frac{c_0e^{\eta \phi_0}}{2\tilde{L}^4}\tilde{z}(\tilde{z}-1)(\theta-2)(3-4\tilde{z}+(\tilde{z}-1)\theta),\nonumber\\
\end{eqnarray}
where $\eta=-\gamma=\frac{-\theta}{\beta}, \lambda \beta=\theta-4$ and $z=\frac{1}{1-\tilde{z}}$.

\section{Allowed Regions for the Numerical Solution}
To search for the numerical solution, we should consider some constraints on our extensive parameters.
In order to have an acceptable solution in the IR, from the terms for $Q^2$ and $\phi_I$, we get two constraints between $\lambda$, $\eta$, $\gamma$, $c_0$ and $c_1$ which are:
$\mathrm{\frac{ \left(\gamma +\eta -\eta  c_1\right) }{(\eta -\lambda )}}>0$ and $\mathrm{\frac{\lambda(1-c_1) +\gamma }{c_0(\lambda-\eta)} }>0$. If we consider two cases of $\mathrm{c_1<1+\frac{\gamma}{\lambda}  }$ or $\mathrm{c_1>1+\frac{\gamma}{\lambda}  }$, for both cases one can demonstrate that $\mathrm{c_0<0 }$. Thus, we consider a negative value for $c_0$.

Then we would like to find all the acceptable regions for $\lambda$, $\eta$ and $c_1$. The conditions that we will impose, similar to  \cite{Knodel:2013fua} are:
 \\ \\ 
 1) $\mathrm{\lambda, \eta, \gamma >0}$. \\ \\
 2) There should be a region where $V_{\mathrm{eff}}(\phi_0)=0$. So the argument of the logarithm in the equation for $\phi_0$ should be positive.  \\ \\
3) $\mathrm{g(\phi_0) >0}$ which means $\mathrm{\frac{3}{4} (c_0 e^{\eta\phi_0}+c_1)>0}.$\\ \\
 4) $\mathrm{Q^2 >0}$, therefore, $\mathrm{\frac{ \left(\gamma +\eta -\eta  c_1\right) }{(\eta -\lambda )}}>0.$\\ \\
 5) The perturbations should not be oscillatory, which means all the $\nu$'s that we have found should be real parameters, which put constraints on the terms which are under the radical.\\ \\ 
6) At least one of the $\nu$ should be negative and therefore one of the dilaton perturbation should be irrelevant.\\ 

Using these conditions we can specify different regions of the parameter space in the following figures. Similar to  \cite{Knodel:2013fua} which separated the parameters regions for the Lifshitz metric, we will do the same for the hyperscaling violating metric. So the green region is the allowed region, red is when $g(\phi)<0$, yellow is when $g(\phi)>0$, but $\nu$ is imaginary or all $\nu$'s are irrelevant, and grey is when any of the conditions 1, 2 or 3 is violated.

We found that for  $c_1< 1 $ and $\gamma >1$, there is no green region. But for $c_1> 1 $ for both cases, $\gamma >1$ and $\gamma <1$, green regions do exist. From figures (\ref{fig:13}) -~(\ref{fig:16}), one can notice that specifically the region of  $0< \tilde{z} <1$ is green, which indicates that $z>1$, consistent with causality condition of hyperscaling violating solution. So the green region is the most restricted region for the parameter space coming from the condition where the background is non-singular.

\begin{figure}[ht!] \label{Fig:figs}
\centering
\parbox{5.5cm}{
\includegraphics[scale=0.6]{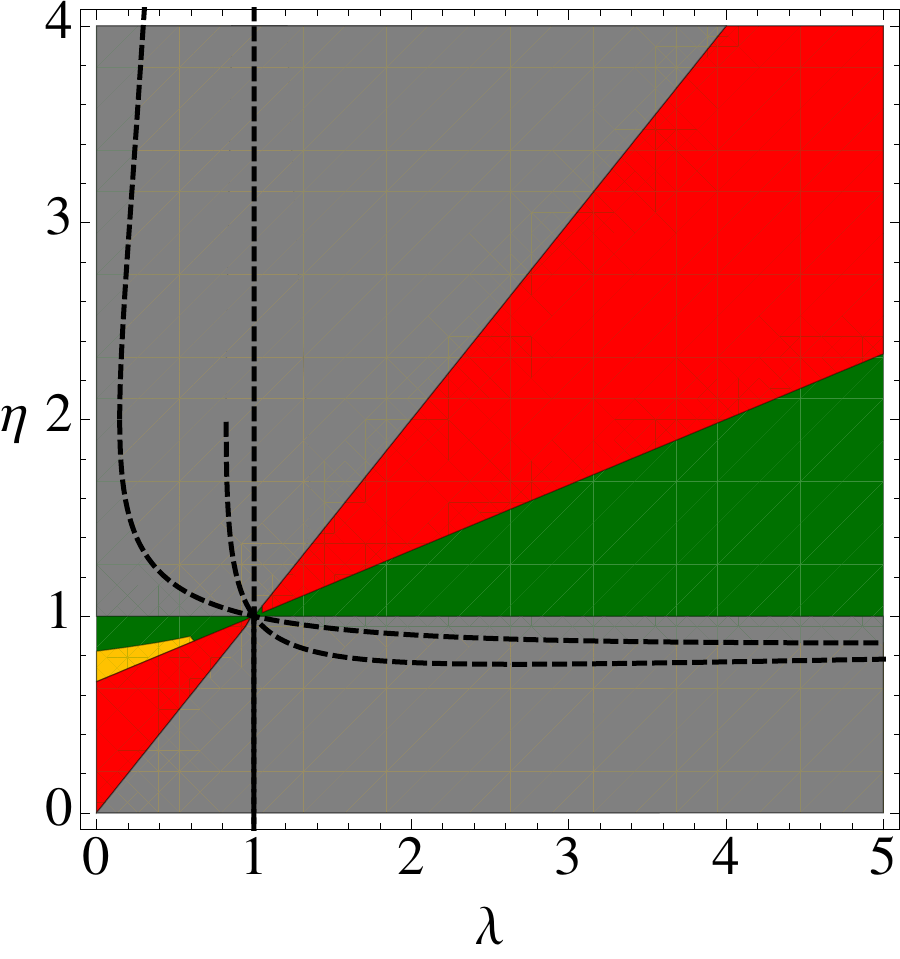}
\caption{Plot of $\eta$ v.s $\lambda$ for $\gamma=2$ and $c_1=3$. \label{fig:9} }
\label{fig:2figsA9}}
\qquad
~ ~ ~ ~ ~ ~ ~ ~ ~ ~ 
\begin{minipage}{6cm}
\includegraphics[scale=0.6]{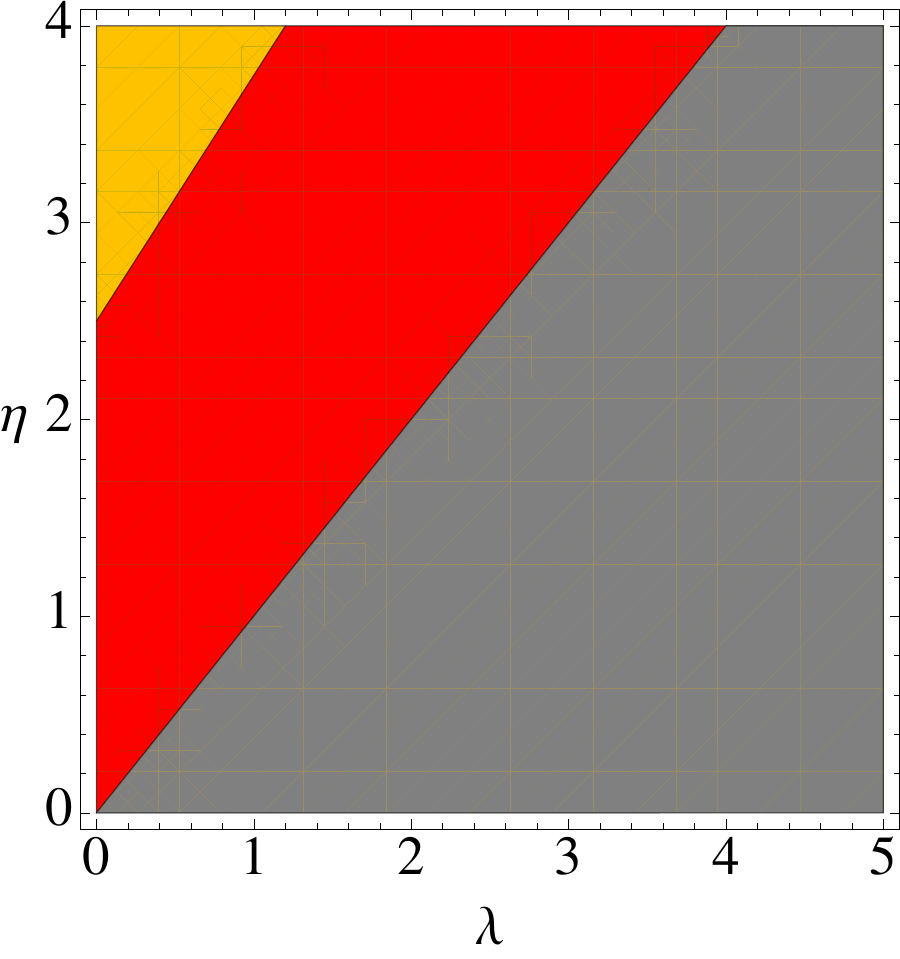}
\caption {Plot of $\eta$ v.s $\lambda$ for $\gamma=2$ and $c_1=0.8$.  \label{fig:10}   }
\label{fig:2figsB9}
\end{minipage}

\centering
\parbox{6cm}{
\includegraphics[scale=0.6]{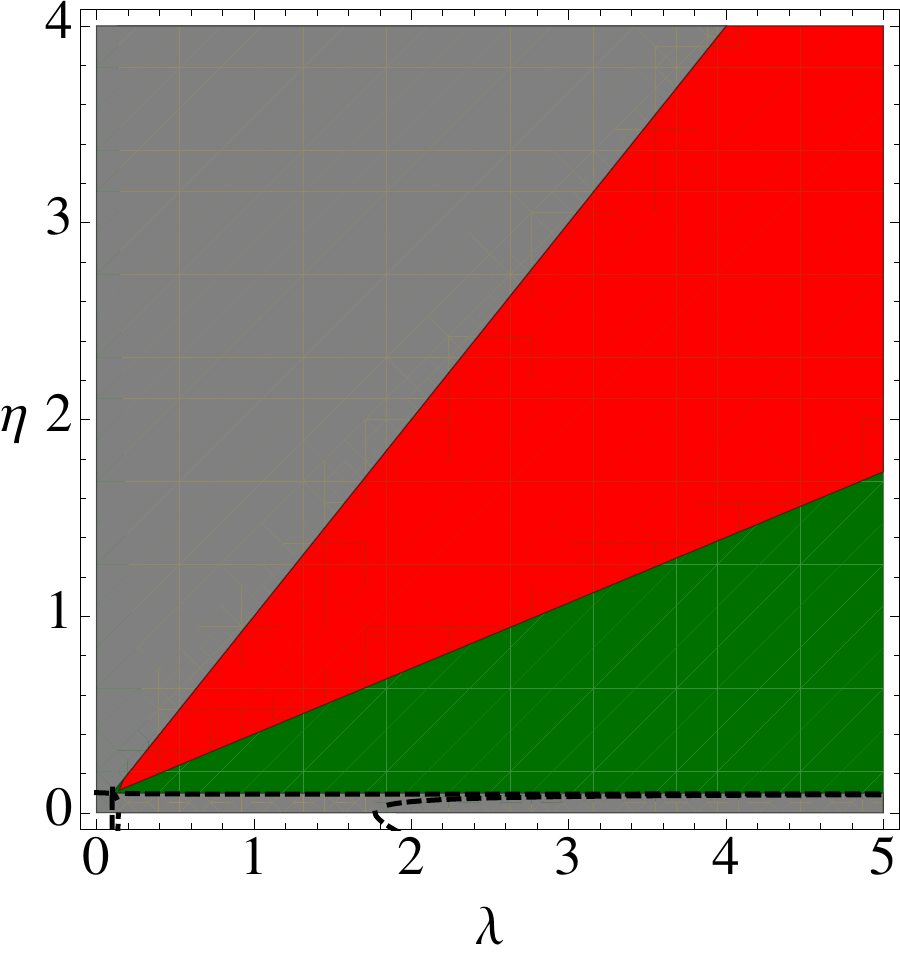}
\caption{ Plot of $\eta$ v.s $\lambda$ for $\gamma=0.2$ and $c_1=3$. \label{fig:11}  }
\label{fig:2figsA9}}
\qquad
~ ~ ~ ~ ~ ~ ~ ~ ~ ~ 
\begin{minipage}{6cm}
\includegraphics[scale=0.6]{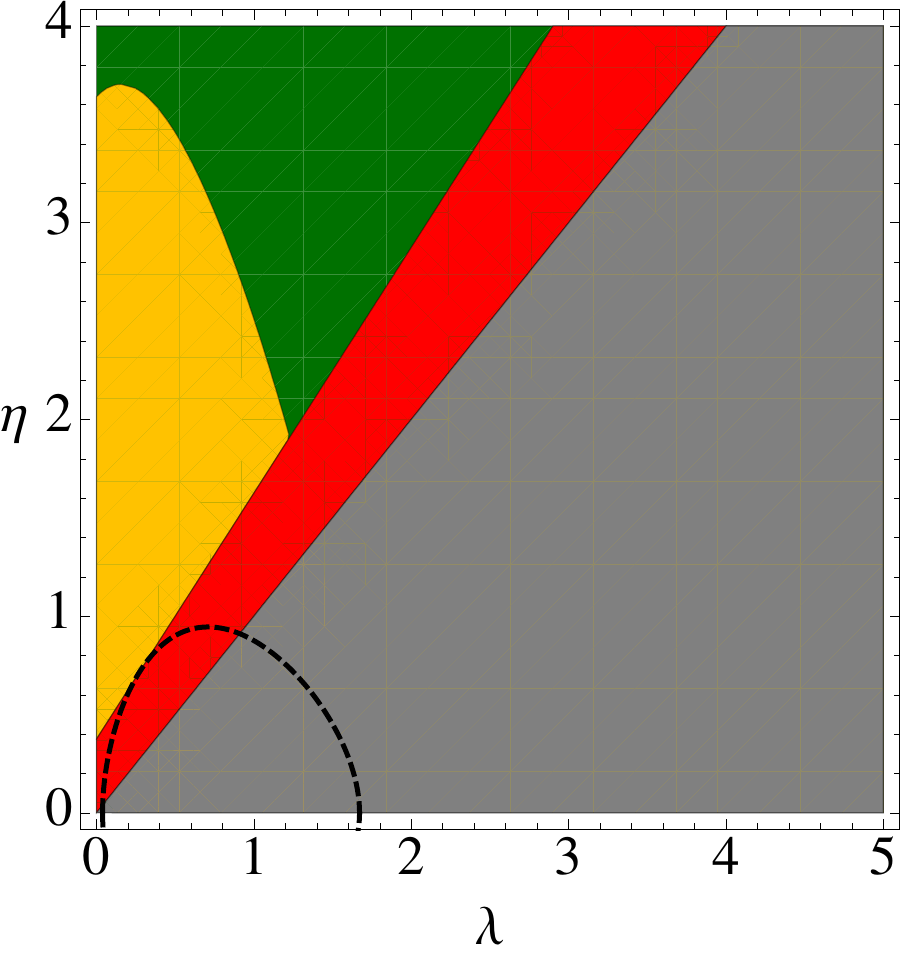}
\caption{Plot of $\eta$ v.s $\lambda$ for $\gamma=0.2$ and $c_1=0.8$.  \label{fig:12} }
\label{fig:2figsB9}
\end{minipage}

\end{figure}

\begin{figure}[]
\centering
\parbox{5.5 cm}{
\includegraphics[scale=0.6]{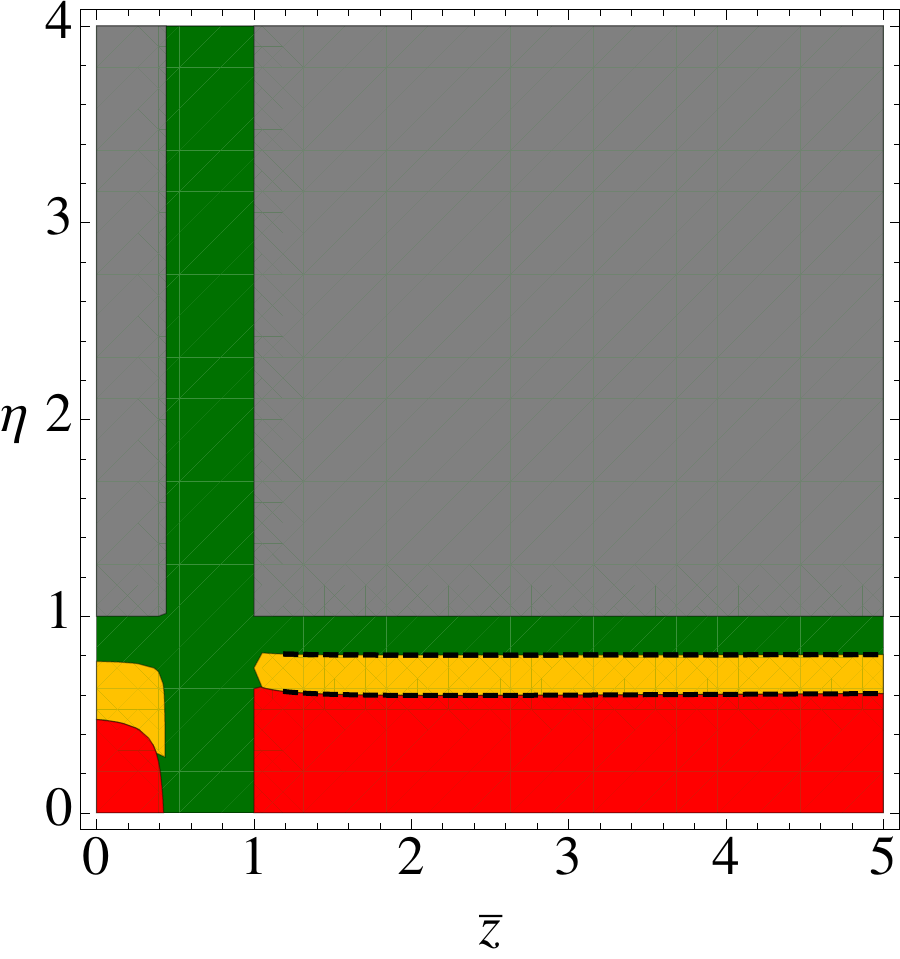}
\caption{Plot of $\eta$ v.s $\tilde{z}$ for $\gamma=2$ and $c_1=3$. \label{fig:13} }
\label{fig:2figsA9}}
\qquad
~ ~ ~ ~ ~ ~ ~ ~ ~ ~ 
\begin{minipage}{5.5cm}
\includegraphics[scale=0.6]{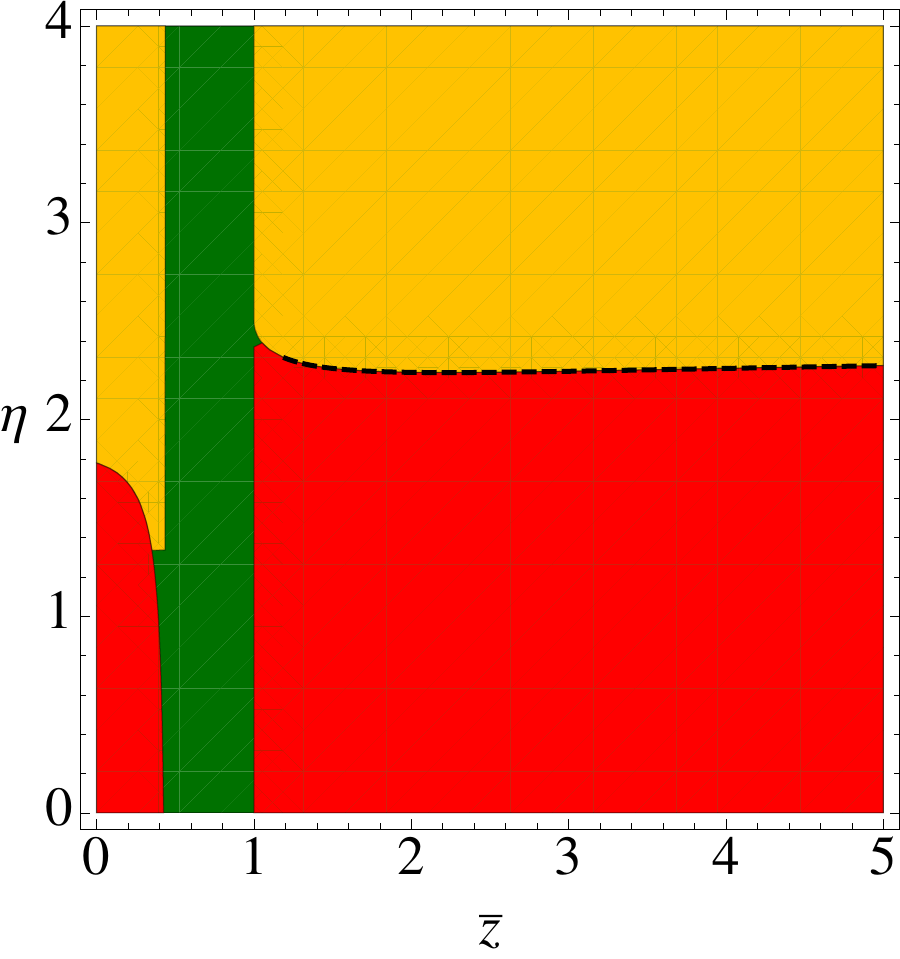}
\caption{Plot of $\eta$ v.s $\tilde{z}$ for $\gamma=2$ and $c_1=0.8$. \label{fig:14}    }
\label{fig:2figsB9}
\end{minipage}

\end{figure}

\begin{figure}[ht!]
\centering
\parbox{5.5cm}{
\includegraphics[scale=0.6]{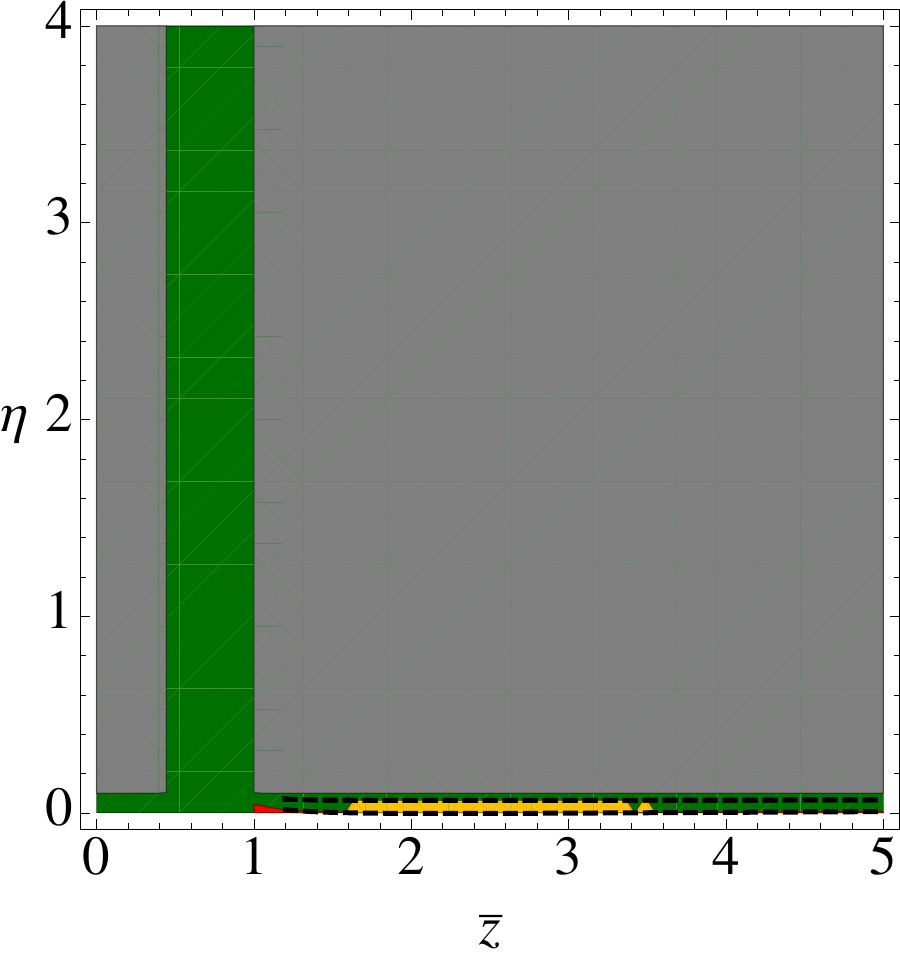}
\caption{Plot of $\eta$ v.s $\tilde{z}$ for $\gamma=0.2$ and $c_1=3$.  \label{fig:15} }
\label{fig:2figsA9}}
\qquad
~ ~ ~ ~ ~ ~ ~ ~ ~ ~ 
\begin{minipage}{5.5cm}
\includegraphics[scale=0.6]{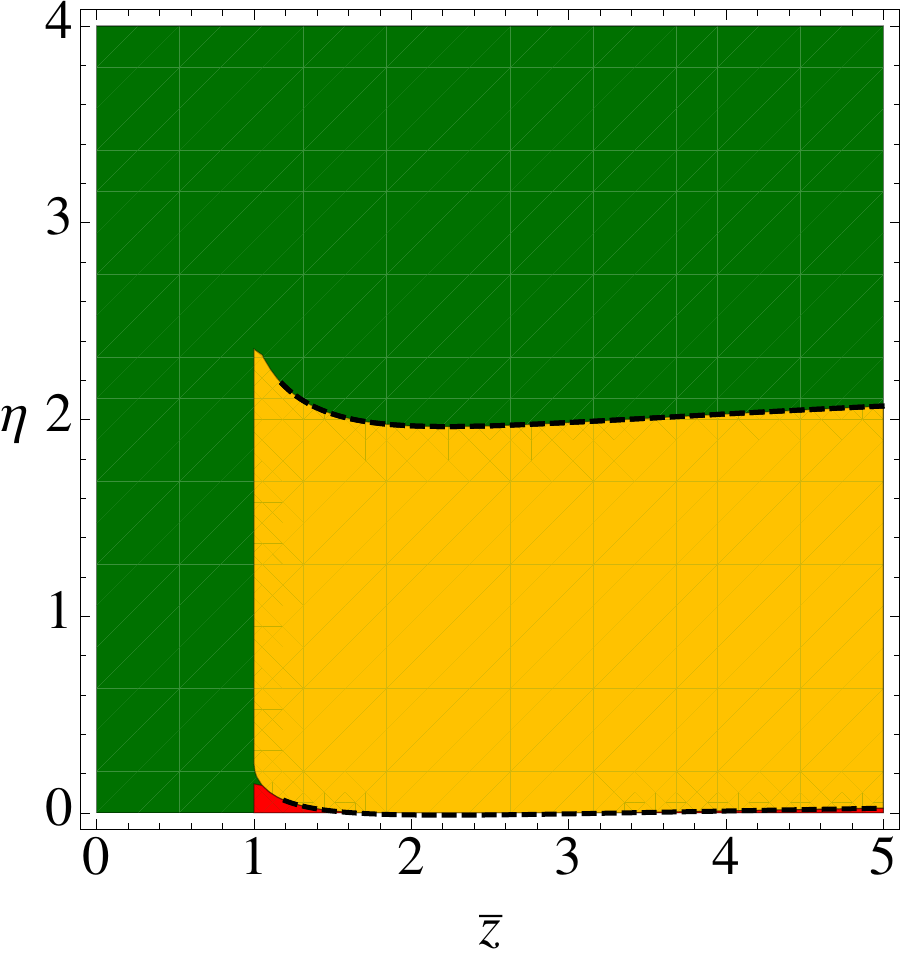}
\caption{Plot of $\eta$ v.s $\tilde{z}$ for $\gamma=0.2$ and $c_1=0.8$. \label{fig:16}   }
\label{fig:2figsB9}
\end{minipage}
\end{figure}
 
For plotting $\eta$ versus $z$, we have used the following equation that we have derived in section (\ref{ab}),

\begin{equation}
\lambda =\frac{\theta -4}{\left(\frac{\text{   }L^2 (\theta -2)(-\theta +z (2+\theta ))+2\text{  }c_0z (\theta -2)(3+z (-4+\theta )-\theta )}{L^2 (z-1)}\right){}^{\frac{1}{2}}}.
\end{equation}

For plotting figures~(\ref{fig:9}) -~(\ref{fig:16}), we made the following assumptions
\begin{gather*}
\phi_0=0, \ \ \ L=1, \ \ \ c_0=-2, \ \ \  \theta=3.
\end{gather*}

\subsection{ Crossover Estimations}
We can analyze the physics of the flow by doing several estimations. The flow is from $\text{AdS}_4$ in the $\text{UV}$ to $\text{AdS}_2 \times \mathbb{R}^2$ in the $\text{IR}$ where each of these two regions has a constant dilaton. There is an intermediate hyperscaling violating region with solution \eqref{eq:newsolg} where the dilaton flows logarithmically based on the relation $\phi(r)=\phi_0+ (\tilde{z}-1)\beta \log r $. One can approximately say that the exponential potential $V(\phi)$ is responsible for the HSV region, $f(\phi)\ F^2$ is responsible for the $\text{AdS}_4$, and $g(\phi)$ times the higher derivative terms is responsible for the emergence of $\text{AdS}_2 \times \mathbb{R}^2$ in the  $\text{IR}$. Using this we can estimate the $r$ and $\phi$ for each cross over. The cross over from $\text{AdS}_4$ in the UV to HSV happens at $\phi_U$ and $r_U$, and when $f(\phi_U)\ g^{rr} g^{tt} (F_{rt})^2 \sim V(\phi_U)$. Using this,
\begin{gather}   
r_U=\Big(\frac{-1}{\tilde{L}^4} \ \frac{V_0 e^{\gamma \phi_0}}{\rho^2 e^{-\lambda \phi_0}}\Big)^{\frac{-1}{4(\theta(\tilde{z}-1)+2) }}, \nonumber\\
\phi_U=\phi_0-\frac{\beta (\tilde{z}-1)} {4 \theta(\tilde{z}-1)+8} \ \text{log}\Big(\frac{-V_0 e^{\gamma \phi_0}}{\tilde{L}^4  \ \rho^2 e^{-\lambda \phi_0}}\Big).
\end{gather}

For the first estimation, we take $\alpha_R=\alpha_W=\alpha_{GB}=L=1 \phi_0=0$, $z=3, \ \theta=4$, leading to $\tilde{z}=\frac{2}{3}$ and $\tilde{L}=0.693$. Also $c_0=-2$ and $c_1=3$, as in figures (\ref{fig:9}) , (\ref{fig:11}). This gives
\begin{gather}
r_U \simeq 0.32, \ \ \ \ \ \ \ 
\phi_U \simeq -6.45.
\end{gather}

After choosing the specific parameters, then $r_U$ and $\phi_U$ are fixed for the UV region. The crossover from HSV region to $\text{AdS}_2 \times \mathbb{R}^2$ in the $\text{IR}$ occurs at $\phi_I$ and $r_I$, when the higher derivative correction terms become comparable to the exponential potential. So $g(\phi_I)h \sim V(\phi_I)$, where 
\begin{gather}
h= \alpha_W C_{\mu\nu\rho\sigma} C^{\mu\nu\rho\sigma}+\alpha_{GB} G+\alpha_R R^2,
\nonumber
\end{gather}

and
\begin{gather}
g(\phi_I)= \frac{3}{4}(c_0 e^{\eta \phi_I}+c_1).
\end{gather}

For the assumed values we can calculate $h$, which gives, $h(r_I)=88 \ r_I^8$. This leads to the equation
\begin{gather}
r_I^8 (-2r_I^{\frac{4}{3}}+3)=-0.0292 r_I^{\frac{-4}{3}},
\end{gather}

which gives
\begin{gather}
r_I \simeq 1.356,  \ \ \ \
\phi_I \simeq 1.758.
\end{gather}

As $r_I$ is bigger than $r_U$, the cross over from $\text{AdS}_4$ to HSV happens after the cross over from HSV to $\text{AdS}_2 \times \mathbb{R}^2$ and so the RG flow can exist. Choosing a bigger $c_1$, which makes the effect of higher derivative terms more important and with all other parameters constant, one can easily make $r_I$ and $\phi_I$ bigger. As for example for $c_1=300$, we will get $r_I=43$. So one arbitrarily can increase the intermediate HSV region. The dilaton in $\text{AdS}_4$ and $\text{AdS}_2 \times \mathbb{R}^2$ is also constant with a bigger value in the UV. 

One can choose other parameter values and check whether for those cases, a flow could exist. There are some values that $r_U$ is bigger than $r_I$, or some other singularities can happen as indicated in the conditions for the existence of numerical flow in the previous section. For resolving those singularities, other methods should be implied. 

It would be much better to actually construct the numerical flow and see explicitly the interpolations between  $\text{AdS}_4$, intermediate hyperscaling violating region and $\text{AdS}_2 \times \mathbb{R}^2$. However, due to the extensive parameters and the high sensitivity of the numerical solution to the initial values and parameters, a numerical flow could not be built explicitly here and will be done in future works. Shooting method similar to \cite{Knodel:2013fua} can be used to build the numerical flow, however for the hyperscaling violating case, it would be more difficult.

\chapter{Schwinger effect and entanglement entropy in confining geometries}\label{chap3}

The Schwinger effect in quantum field theory \cite{Schwinger:1951nm} is the creation of pairs of particles in the presence of a strong electric or magnetic field. In the description of the confinement of quarks via a gluonic flux tube, it is intuitively clear that when the strength of electric field reaches the value of the string tension between the quark and antiquark, it can break the string that is attaching them and so the virtual pairs can become on shell and a current can be created. 

As holography is a powerful tool in studying strongly correlated systems, one would like to study quark-gluon plasmas using holography. Semenoff and Zarembo \cite{Semenoff:2011ng} first studied the Schwinger effect using holography by considering the fact that the dual picture of two moving quarks is a string attaching them. Also the authors in \cite{Hubeny:2014kma} and \cite{Hubeny:2014zna}, using the Nambu-Goto action on a probe brane and by calculating the free energy and using the first law of thermodynamics, calculated the entanglement entropy of a quark and an antiquark which are accelerating in an electric field in the $\text{AdS}_5$ background. A similar calculation in a different setup was also done in \cite{Lewkowycz:2013laa}. Giving that QCD, the theory of strong interaction, is actually in the strongly coupled regime, it makes sense to try Semenoff and Hubeny's calculations in the confining backgrounds too. 

On the other hand, entanglement is a measure of quantum correlation between two or more parts of a system. There are some ideas on how to measure entanglement entropy in the lab by using the fluctuations of a current which is flowing through a quantum point contact as the probe, \cite{Klich:2008un} \cite{Klich:2011eg}, which still are not quite successful experimentally.  A plausible conjecture is that there might be a relation between the Schwinger effect and the entanglement entropy. Finding such a relation could be a breakthrough since the Schwinger pair creation rate could also act as an entanglement meter in condensed matter systems. One should notice that the AdS/CFT has successfully constructed supergravity backgrounds that have been claimed to be dual to field theories such as QCD models. These backgrounds display confinement (characteristics such as Wilson loop area law, or chiral symmetry breaking).  
So to explore such a potential relation between Schwinger effect and entanglement entropy, in the first step, we study the phase diagrams of Schwinger effect and of entanglement entropy for four confining geometries which were the Witten-QCD (WQCD), the Maldacena-Nunez (MN), the Klebanov-Strassler (KS) and the Klebanov-Tseytlin (KT) backgrounds which are dual to $\mathcal{N}=1$ field theories. 

For calculating the entanglement entropy of accelerating particles one can use the method in\cite{Hubeny:2014zna}. For doing so, first one needs to find the string world-sheet profile in these confining backgrounds. For the case of $\text{AdS}$ geometry, due to the large amount of symmetries, the corresponding PDE equation of motion is simple and has been explicitly solved in \cite{Xiao:2008nr}. Mikhailov also found a simple linear relation \cite{Mikhailov:2003er} to find the string world-sheet profile based on the position of the quark and antiquark on the boundary that can only be used for the $\text{AdS}_5$ background. For other geometries such as the confining backgrounds there is no Mikhailov-like equation and for finding the string profile one needs to solve several more difficult PDE equations analytically which for our supergravity background geometries we present them in sec (\ref{sec:one}). We just do the similar calculation for the Minkowski background and we find the free energy of the quarks and antiquarks in a flat background in sec (\ref{sec:two}).

In order to look for a relationship between the phase transitions of Schwinger effect and of the entanglement entropy, in section (\ref{sec:three}) we study the phase transitions in an electric potential similar to the procedures in \cite{Sato:2013dwa, Sato:2013hyw, Sato:2013iua, Sato:2013pxa, Kawai:2015mha, Kawai:2013xya}. By using Dirac-Born-Infeld (DBI) action, we calculate the critical electric fields in these geometries where the potential becomes catastrophically unstable and we then find the phase diagrams numerically. Interestingly the phase diagrams of all of these geometries are very similar where three different phases can be detected and this has been predicted in \cite{Sato:2013hyw} as a universal feature of all confining geometries. We also compare these diagrams with the conformal case of Klebanov-Witten in which only two phases can be detected.

In section (\ref{sec:four}), we look at the phase diagrams of the entanglement entropy of a strip, similar to the calculation in \cite{Klebanov:2007ws, Kol:2014nqa, Faraggi:2007fu}. Klebanov, Kutasov and Murugan found a generalization of Ryu-Takayanagi relation for the nonconformal geometries \cite{Klebanov:2007ws}. Then in \cite{Kol:2014nqa}, the authors presented the plots of the phase diagram for geometries constructed by Dp brane compactified on a circle which are the generalization of the Witten-QCD model. They have found a butterfly shape and a double valuedness in the phase diagram of these confining geometries. In this chapter, we additionally present the phase diagrams of the Maldacena-Nunez, Klebanov-Strassler and Klebanov-Tseytlin plus the Klebanov-Witten and Witten-QCD models by similarly calculating the length of the connected regions and the entanglement entropy of the connected and disconnect solutions. We also observe the similar butterfly shape in the phase diagrams. In addition, we find that if in a specific geometry the phase transition of Schwinger effect is fast and dramatic, this would also be the situation for the entanglement entropy phase transition. This is actually the case for WQCD and KT.  On the other hand, if the phase transition for the Schwinger effect is mild, the phase transition of entanglement entropy would also be mild and this is the case for MN and KS. One can also compare these features with the conformal case of Klebanov-Witten and AdS, as a limit of a mild transition. 

In sec (\ref{sec:five}), we study the Schwinger effect in the presence of a magnetic field in addition to the electric field and study its effects on the pair creation rate. By adding a probe D8-brane, in \cite{Hashimoto:2013mua} and \cite{Hashimoto:2014yya} the authors studied the imaginary part of the Euler-Heisenberg effective Lagrangian, the rate of pair creation, the critical electric field and the effect of the parallel and perpendicular components of the magnetic field on the rate of pair creation in the background of Sakai-Sugimoto and the deformed Sakai-Sugimoto models. Similarly we calculate the DBI action and the imaginary part of the Euler-Heisenberg effective Lagrangian in our geometries. We find that in all of our supergravity confining geometries the parallel magnetic field would increase the rate of pair creation while the perpendicular magnetic field would decrease it which should be a general feature of all confining backgrounds. We expect that the universality, within the AdS/CFT correspondence of our findings might have relevance for practical systems.

\section{The string profile in confining geometries}\label{sec:oneprofile}
In this section we present the PDE equations of the string profiles which can be used in finding the entanglement entropy of accelerating quarks moving on specific trajectories.

As in \cite{Xiao:2008nr}, by starting from the Nambu-Goto action for the $\text{AdS}_5$ geometry,
\begin{gather}
ds^2=R^2 [\frac{du^2}{u^2}-u^2 dt^2+u^2(dx^2+dy^2+dz^2)],
\end{gather}
and then by assuming the static gauge of $(\tau, \sigma)=(t,u)$ and the embedding coordinate of $X^\mu=(t,u,x(t,u),0,0)$, one can find the determinant of the induced metric as
\begin{gather}
\sqrt{-g} =R^2 \sqrt{1-\dot{x}^2+u^4 x'^2},
\end{gather}
and the equation of motion as
\begin{gather}
\frac{\partial}{\partial u} \left(\frac{u^4 x'}{\sqrt{-g}}\right)-\frac{\partial}{\partial t}\left(\frac{\dot{x}}{\sqrt{-g}}\right)=0.
\end{gather}
Solving these PDE equations, in general is a difficult task, but for the case of $\text{AdS}$ which enjoys $SO(4,2)$ group isometries, it has a simpler form. The author in \cite{Xiao:2008nr} could find the solution and therefore the string profile as a function of $t$ and $u$ as
\begin{gather}
x=\pm \sqrt{t^2+b^2-\frac{1}{u^2}}.
\end{gather}
If due to the potential of an electric field, a heavy quark and an antiquark accelerate on a specific trajectory, the classical solution from the Nambu-Goto action would be a world-sheet that is a part of $\text{AdS}_2$ and the locus of
\begin{gather}
u^2+(x^1)^2-(x^0)^2=\frac{M^2}{E^2},
\end{gather}
where $M$ is the mass of the quark and $E$ is the electric field. From this relation, the world-sheet event horizon can be read as $u_E=\frac{M}{E}$ (here, the world-sheet event horizon is defined as an event horizon on the induced metric). As in \cite{Hubeny:2014zna}, for a specific trajectory such as a hyperbola, one can read the metric near the quark trajectory.

Alternatively, for finding the induced metric on the world-sheet, similar to \cite{Hubeny:2014kma}, one can use the Mikhailov relation between the embedding coordinate $X^M(\tau,u)$ and the boundary quark position $x^\mu(\tau)$, \cite{Mikhailov:2003er}, as
\begin{gather}
X^\mu(\tau,u)=u { \dot{x}}^\mu (\tau)+{x^\mu} (\tau), \ \ \ \ \ \ X^M(\tau,u)=(X^\mu(\tau,u),u).
\end{gather}

Then, as in \cite{Hubeny:2014zna}, one can find the proper area between the probe brane and the event horizon which is proportional to the free energy. By knowing the Unruh temperature that the quarks would feel in the accelerated reference frame in $\text{AdS}$ as $T_U=\frac{E}{2\pi M}$, and by using the first law of thermodynamics one can read the entropy. If one can assume the semiclassical and heavy quark limit in the problem, all of those entanglements are due to the ``entanglement entropy" and so  for the AdS the EE where found to be $s=\sqrt{\lambda}$ \cite{Hubeny:2014zna}, where $\lambda$ is the 't Hooft coupling constant.  

Now we look at the equations for the $\mathcal{N}=1$ supergravity solutions that are dual to the confining geometries. One should notice that for this calculation, one first needs the string world-sheet profile, the induced metric and the Unruh temperature of the accelerating particles in each background which we do not present here.

\subsection{Witten-QCD}
Among all the three confining geometries mentioned, the Witten QCD model is the most similar background to the AdS metric. In the string frame, its metric and dilaton field are \cite{Witten:1997ep},
\begin{gather}
ds^2=(\frac{u}{R})^{3/2}\left(\eta_{\mu\nu} dx^\mu dx^\nu+\frac{4R^3}{9u_0} f(u) d\theta^2\right)+\left(\frac{R}{u}\right)^{3/2} \frac{du^2}{f(u)}+R^{3/2} u^{1/2} d\Omega_4^2,\nonumber\\
f(u)=1-\frac{u_0^3}{u^3}, \ \ \ \ \ \ \ \ R=(\pi Ng_s)^{\frac{1}{3}} {\alpha^\prime}^{\frac{1}{2}},\ \ \ \ \ \
e^\Phi=g_s \frac{u^{3/4}}{R^{3/4}}.
\end{gather}
If we consider the static gauge, and the embedding coordinate as: \\
$X^\mu=(t,u,x(t,u),0,0,\theta(t,u),0,0,0,0)$, then the equations would be complicated. So we assume $\theta$ is a constant, and therefore the determinant of the induced metric simplifies to
\begin{gather}
\sqrt{-g}=\sqrt{u^3 \left(\frac{ 1-\dot{x}^2}{u^3-{u_0}^3}+\frac{{x'}^2}{R^3 }\right)},
\end{gather}
which leads to the following equation of motion,
\begin{gather}
\left(\frac{u^3-u_0^3}{R^3 u^3} \right)\frac{\partial}{\partial u}\left(\frac{u^3 x'}{\sqrt{-g}}\right)-\frac{\partial }{\partial t}\left(\frac{\dot{x}}{ \sqrt{-g}}\right)=0,
\end{gather}
which is quite similar to the AdS case. If one finds the analytical solution of this PDE equation, similar to the AdS case, one can follow the procedures of \cite{Hubeny:2014zna} and find the entanglement entropy of heavy accelerating quarks in this model. 

\subsection{Maldacena-Nunez}
The MN metric is obtained by a large number of D5-branes wrapping on $S^2$ \cite{Maldacena:2000yy}. In the string frame the metric and the fields are \cite{Basu:2012ae}
\begin{gather}
ds^2_{10}=e^\phi [ -dt^2+{dx_1}^2+{dx_2}^2+{dx_3}^2+e^{2h(r)}({d\theta_1}^2+\sin^2 \theta_1 {d\phi_1}^2)+dr^2+\frac{1}{4}(\omega^i-A^i)^2 ],
\end{gather}
where
\begin{gather}
A^1=-a(r)d\theta_1, \ \ \ \ \ \ \ \ A^2=a(r)\sin \theta_1 d\phi_1, \ \ \ \ \ \ \ \ 
A^3= -\cos \theta_1d\phi_1,
\end{gather}
and the $\omega^i$ s parametrize the compactification 3-sphere which are
\begin{gather}
\omega^1 =\cos \psi d\theta_2+\sin \psi \sin\theta_2 d \phi_2,\ \ \ \ \ \ \ \ \ 
\omega^2 =-\sin \psi d \theta_2+\cos \psi \sin \theta_2 d\phi_2,\nonumber\\
\omega^3=d\psi+\cos \theta_2 d\phi_2,
\end{gather}
and also the other parameters of the metric are
\begin{gather}
a(r)=\frac{2r}{\sinh 2r}, \ \ \ \ 
e^{2h}= r\coth {2r}-\frac{r^2}{{\sinh 2r}^2}-\frac{1}{4},\ \ \ \ 
e^{-2\phi} =e^{-2\phi_0} \frac{2 e^h}{\sinh 2r}.
\end{gather}

For the case of the Maldacena-Nunez model, we assume the embedding coordinate as $(t,r, x(t,r))$ and all other coordinates will set to be zero. Then one would get
\begin{gather}
\sqrt{-g}=e^\phi \sqrt{(1-\dot{x}^2 )(1+x'^2)}.
\end{gather}
So the equation of motion of the string profile in this background is
\begin{gather}
\frac{\partial}{\partial r} \left( \frac{e^{2\phi} x'  (1-\dot{x}^2 )}{\sqrt{-g } }\right)-\frac{\partial}{\partial t} \left( \frac{ e^{2\phi} \dot{x} (1+x'^2) }{\sqrt{ (x'^2+1)(\dot{x}^2-1) } }  \right)=0.
\end{gather}

Again by solving this equation analytically one can find the string profile and then the entanglement entropy of heavy accelerating quarks.

\subsection{Klebanov-Tseytlin}
The Klebanov-Tseytin metric is a singular solution which is dual to the chirally symmetric phase of the Klebanov-Strassler model which has D3-brane charges that dissolve in the flux \cite{Bena:2012ek}.
 Although this metric is singular, but still we can extract the information we are looking for from analyzing it.
 
The metric is
\begin{gather}
ds_{10}^2=h(r)^{-1/2} \left[ -dt^2+d \vec{x}^2 \right]+h(r)^{1/2} \left[ dr^2+r^2 ds_{T^{1,1} }^2 \right].
\end{gather}
Here $ds_{T^{1,1} }^2$ is a base of a cone with the definition of
\begin{gather}
ds_{T^{1,1} }^2=\frac{1}{9} (g^5)^2+\frac{1}{6} \sum_{i=1}^4 (g^i)^2.
\end{gather}
It is the metric on the coset space $T^{1,1}=(SU(2) \times SU(2))/U(1)$.
Also $g^i$s are some functions of the angles $\theta_1, \theta_2, \phi_1, \phi_2, \psi$ as
\begin{gather}
g^1=(-\sin \theta_1 d\phi_1-\cos \psi \sin \theta_2 d\phi_2 +\sin \psi d\theta_2)/ \sqrt{2}, \nonumber\\ 
g^2=(d\theta_1-\sin \psi \sin \theta_2 d\phi_2-\cos \psi d\theta_2) / \sqrt{2},\nonumber\\
g^3=(-\sin \theta_1 d\phi_1+\cos \psi \sin \theta_2 d \phi_2-\sin\psi d \theta_2) \sqrt{2}, \nonumber\\ 
g^4=(d\theta_1+ \sin \psi \sin \theta_2 d\phi_2+\cos \psi d\theta_2) /\sqrt{2},\nonumber\\
g^5=d\psi+\cos \theta_1 d\phi_1+\cos \theta_2 d\phi_2,
\end{gather}
and also
\begin{gather}
h(r)=\frac{L^4}{r^4} \ln{\frac{r}{r_s}}, \ \ \ \ \ \ \ \ \ \  L^4=\frac{81}{2} g_s M^2 \epsilon^4.
\end{gather}
In this frame, the asymptotic flat region has been eliminated. Also, $r=r_s$ is where the naked singularity is located. We can hope to extract sensible information from this metric.

For the case of KT the embedding is $(t, \tau, x(t,\tau))$ with all other coordinates zero. Then, 
\begin{gather}
\sqrt{-g}=\sqrt{(1-\dot{x}^2)(\frac{{x'}^2}{H}+\frac {\epsilon^2}{9} e^ {\frac{2\tau}{3}}) },
\end{gather} 
and the equation of motion of the string profile is
\begin{gather}
\frac{\partial}{\partial \tau} \left(\frac{x' (1-\dot{x}^2 )}{H\sqrt{-g}} \right) -\frac{\partial}{\partial t} \left(\frac{\dot{x}  (\frac{x'^2}{H}+\frac{\epsilon^2}{9} e^{\frac{2\tau}{3} } ) }{\sqrt{-g}}  \right)=0.
\end{gather}

\subsection{Klebanov-Strassler}
The Klebanov-Strassler (KS) metric which is known also as warped deformed conifold is obtained by a collection of $N$ regular and $M$ fractional D3-branes  \cite{Klebanov:2000hb}. 

The metric is
\begin{gather}
ds_{10}^2=h^{-\frac{1}{2}}(\tau) dx_\mu dx^\mu+h^{\frac{1}{2}}(\tau) {ds_6}^2,
\end{gather}
and $ds_6^2$ is the metric of the deformed conifold which is
\begin{gather}
{ds_6}^2=\frac{1}{2} \epsilon^{\frac{4}{3}} K(\tau) \Big[ \frac{1}{3 K^3(\tau)} (d\tau^2+ (g^5)^2 )+\cosh^2 (\frac{\tau}{2}) [(g^3)^2+(g^4)^2]+\sinh^2 (\frac{\tau}{2}) [(g^1)^2+(g^2)^2] \Big].
\end{gather}
The parameters of the metric are
\begin{gather}
K(\tau)=\frac{(\sinh(2\tau) -2\tau)^{\frac{1}{3}}} {2^{\frac{1}{3}} \sinh \tau },\ \ \ \ \ \ \ \ \ h(\tau)=(g_s M \alpha')^2 2^{2/3} \epsilon^{-8/3} I(\tau),\nonumber\\  I(\tau)=\int_\tau^\infty dx \frac{x \coth x-1}{\sinh^2 x} (\sinh(2x)-2x)^{\frac{1}{3}}, 
\end{gather}
and
\begin{gather}
g^1=\frac{1}{\sqrt{2}} [ -\sin \theta_1 d\phi_1-\cos \psi \sin\theta_2 d\phi_2+\sin \psi d\theta_2],\nonumber\\
g^2=\frac{1}{\sqrt{2}}[d\theta_1 -\sin \psi \sin\theta_2 d\phi_2-\cos\psi d\theta_2],\nonumber\\
g^3=\frac{1}{\sqrt{2}}[-\sin\theta_1 d\phi_1+\cos \psi \sin \theta_2 d\phi_2-\sin\psi d\theta_2],\nonumber\\
g^4=\frac{1}{\sqrt{2}}[d\theta_1 +\sin \psi\sin\theta_2 d\phi_2+\cos \psi d\theta_2],\nonumber\\
g^5=d\psi+\cos \theta_1 d\phi_1+\cos\theta_2 d\phi_2.
\end{gather}

For the case of Klebanov-Strassler similar to the KT, the embedding is $(t, \tau, x(t,\tau))$ with all other coordinates zero. Then 
\begin{gather}
\sqrt{-g}=\sqrt{(1-\dot{x}^2) \left(\frac{x'^2}{h} +\frac{\epsilon^{\frac{4}{3}} }{6 K^2(\tau)} \right)},
\end{gather}
and the equation of motion of the string profile is
\begin{gather}
\frac{\partial}{\partial \tau} \left(\frac{x' (1-\dot{x}^2) }{h\sqrt{-g}} \right) -\frac{\partial}{\partial t} \left(\frac{\dot{x}}{\sqrt{-g}}  (\frac{x'^2}{h} +\frac{\epsilon^{\frac{4}{3}} }{6 K^2(\tau)} ) \right)=0.
\end{gather} 

\subsection{Klebanov-Witten}
The Klebanov-Witten solution is similar to the KT throat solution but with no logarithmic warping \cite{Kaviani:2015rxa}. Unlike the other four mentioned metrics, it is a conformal nonconfining geometry. We study this background to compare our results with the conformal case.  

The metric is
\begin{gather}
ds^2=h^{-\frac{1}{2}} g_{\mu\nu} dx^\mu dx^\nu+h^{\frac{1}{2}} (dr^2+r^2 ds_{T^{1,1}} ^2),
\end{gather}
where
\begin{gather}
h=\frac{L^4}{r^4}, \ \ \ \ \ \  \text{and} \ \ \ \ L^4=\frac{27\pi}{4} g_s N (\alpha')^2.
\end{gather}
The embedding is $(t, r, x(t,r))$ with all other coordinates zero. Then 
\begin{gather}
\sqrt{-g}=\sqrt{(1-\dot{x}^2)(1+\frac{x'^2}{h} )},
\end{gather} 
and the equation of motion of the string profile is
\begin{gather}
\frac{\partial}{\partial r} \left(\frac{x' (1-\dot{x}^2 )}{h\sqrt{-g}} \right) -\frac{\partial}{\partial t} \left(\frac{\dot{x} (1+\frac{x'^2}{h} ) }{\sqrt{-g}}   \right)=0.
\end{gather} 

Therefore, again solving this equation would give the string profile and the induced metric near the accelerating particles' trajectory in the background of the conformal KW model.

\section{The free energy of accelerating $q \bar{q} $ in the Minkowski background} \label{sec:two}

In \cite{Hubeny:2014zna} by minimizing the Nambu-Goto action and by using the solution of the PDE for accelerating particles in $\text{AdS}_5$, the authors found the metric near the quark and antiquark trajectory as
\begin{gather}
ds_g^2=\sqrt{\lambda} \alpha' \left[ -\left(\frac{1}{u^2}-\frac{E^2}{M^2} \right) d\tau^2-\frac{2}{u^2} d\tau du\right],
\end{gather}
then calculating the proper area between the probe brane and the event horizon yields
\begin{gather}
S_N=-2\left[M-\epsilon M -\frac{M}{2} \right] \tau_P.
\end{gather}
Knowing the Unruh temperature of AdS, $T_U=\frac{E}{2\pi M}$ and the free energy, $\frac{1}{2} \epsilon =M-\frac{M}{2} -\sqrt{\lambda} T_U$, the entanglement entropy of the accelerating quark and antiquark have been found to be $s=\sqrt{\lambda}$.

Now the flat geometry can be the UV limit of the Hard-Wall and Witten-QCD model. Therefore, for the flat metric
\begin{gather}
ds^2=\sqrt{\lambda} \alpha' (-dt^2+dx_1^2+dx_2^2+dx_3^2+dx_4^2+du^2),
\end{gather}
one can repeat the calculation of Semenoff and Hubeny \cite{Hubeny:2014zna}. So one would have
\begin{gather}
\gamma_{\tau \tau}=\dot{x}^2-1, \ \ \ \gamma_{uu}=1+x'^2, \ \ \ \ \gamma_{\tau u}=x' \dot{x},
\end{gather}
and then the equation of motion is
\begin{gather}
\partial_u\left(\frac{x'}{\sqrt{ 1-\dot{x}^2+x'^2} }\right)-\partial_t \left(\frac{\dot{x} }{\sqrt{1-\dot{x}^2+x'^2} } \right)=0.
\end{gather}

The solution of this PDE can be found as $x(t,u)=\sqrt{t^2+b^2-u^2}$, where as in the previous case, for the accelerating quark and antiquark, the constant is $b=\frac{M}{E}$. From this solution one can see that the world-sheet event horizon is at $u_E=\frac{M}{E}$.
Now by using this solution, the components of the induced metric can be found as
\begin{gather}
\gamma_{\tau \tau}=\frac{u^2-b^2}{t^2+b^2-u^2},\ \ \ 
\gamma_{uu}=\frac{t^2+b^2}{t^2+b^2-u^2},\ \ \ 
\gamma_{\tau u}=\frac{t^2+b^2}{t^2+b^2-u^2}.
\end{gather}

Similar to the conditions that have been applied to derive the induced metric of the work of Semenoff, one can similarly reach to the following induced metric
\begin{gather}
ds^2=\sqrt{\lambda} \alpha' \left[\left(1+\frac{u^2 E^2}{M^2} \right)du^2-\frac{2E^2}{M^2}u^2 d\tau du\right].
\end{gather}

The D3-brane is located at $u_M=\frac{\sqrt{\lambda} }{2\pi M}$.
So the proper area between the probe brane and the event horizon is 
\begin{gather}
S_N=2\big[ -\frac{\sqrt{\lambda}}{2\pi} \int_{-\frac{\tau_p}{2}}^{\frac{\tau_p}{2}} d\tau \int_{u_M}^{u_E} \frac{E^2 u^2}{M^2} du+\frac{M}{2}\big] \nonumber\\
=2\tau_p\big [-\frac{\sqrt{\lambda}}{6\pi}\left(\frac{M}{E}-\frac{E^2}{M^2} (\frac{\sqrt{\lambda}}{2\pi M})^3 \right)+\frac{M}{2} \big].
\end{gather}
Assuming that the Unruh temperature of the accelerated frame is $T_U=\frac{E}{2\pi M}$, then the free energy is,
\begin{gather}
\frac{\epsilon}{2}=\left(-\frac{\sqrt{\lambda}}{12 \pi^2 T}+\frac{T^2 \lambda^2}{12 \pi^2 M^3}+\frac{M}{2}\right).
\end{gather}

So knowing the solution of any of the above PDE equations in any confining geometries can similarly lead to the proper area between the probe brane and the event horizon that yields the free energy of the accelerating $q \tilde{q}$ in those geometries. In a similar way the holographic Schwinger effects were also studied in other geometries such as in de Sitter space \cite{Fischler:2014ama}.

\section{Potential analysis of the confining geometries}\label{sec:three}

One would think that for searching for any possible relationship between the entanglement entropy and the Schwinger pair creation rate, it is interesting to first find the phase diagrams and the phase transitions for both quantities in a few different confining backgrounds.  

In \cite{Sato:2013hyw} the authors demonstrated some general features in the phase diagram of Schwinger effect in all confining geometries and then in \cite{Sato:2013dwa}, they have studied the ``AdS soliton geometry" as a special case. In these papers, the authors studied the potential of a confining background and then they plotted the total potential, $V_{\text{tot}} (x)$ versus the distance between the quark and antiquark $x$. Their setup and the world-sheet configuration for the quark and antiquark potential is shown in Figs. (\ref{fig:test13a}) and (\ref{test14b}). {The plot of WQCD is shown in Fig. (\ref{WQCD}).

\begin{figure}[h!]
\centering
\begin{minipage}{.52\textwidth}
  \centering
  \includegraphics[width=.9\linewidth]{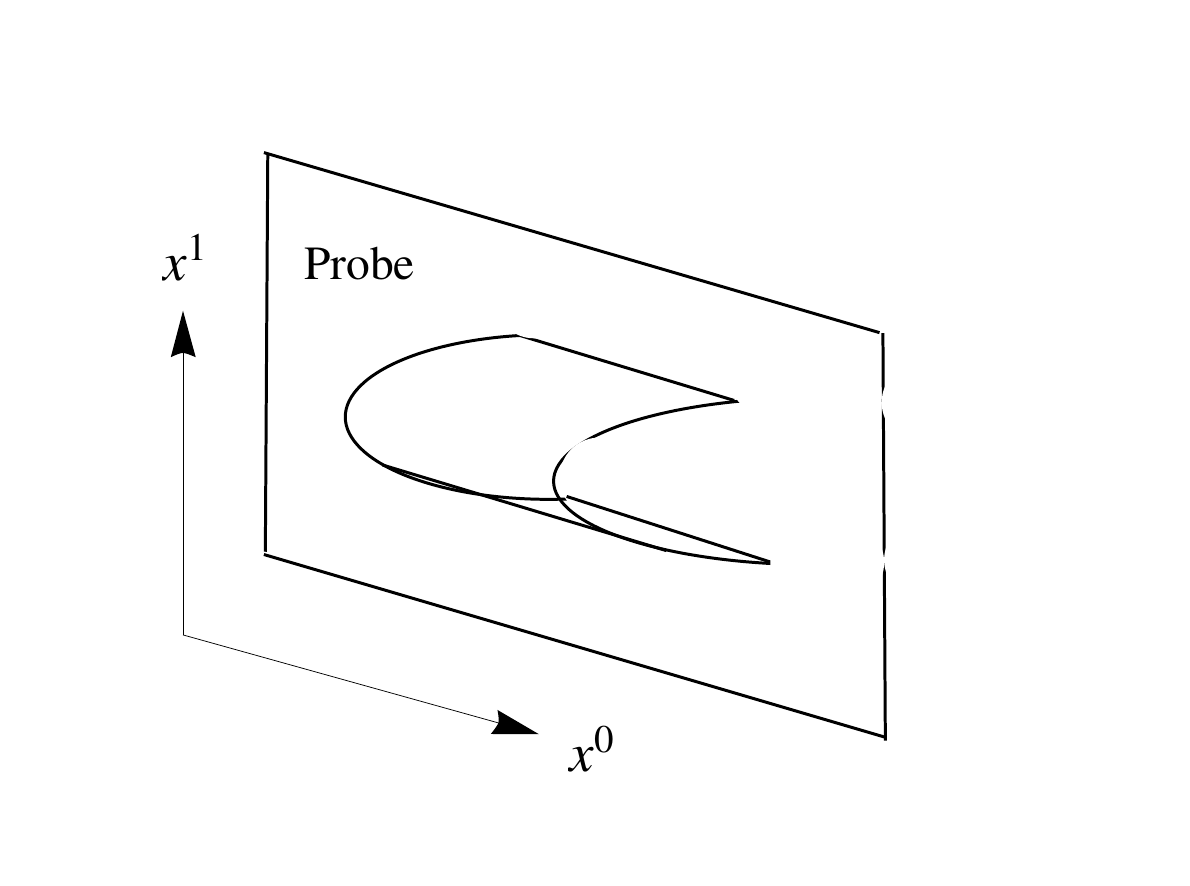}
  \captionof{figure}{World-sheet configuration in 3D.}
  \label{fig:test13a}
\end{minipage}%
\begin{minipage}{.51\textwidth}
  \centering
  \includegraphics[width=.9\linewidth]{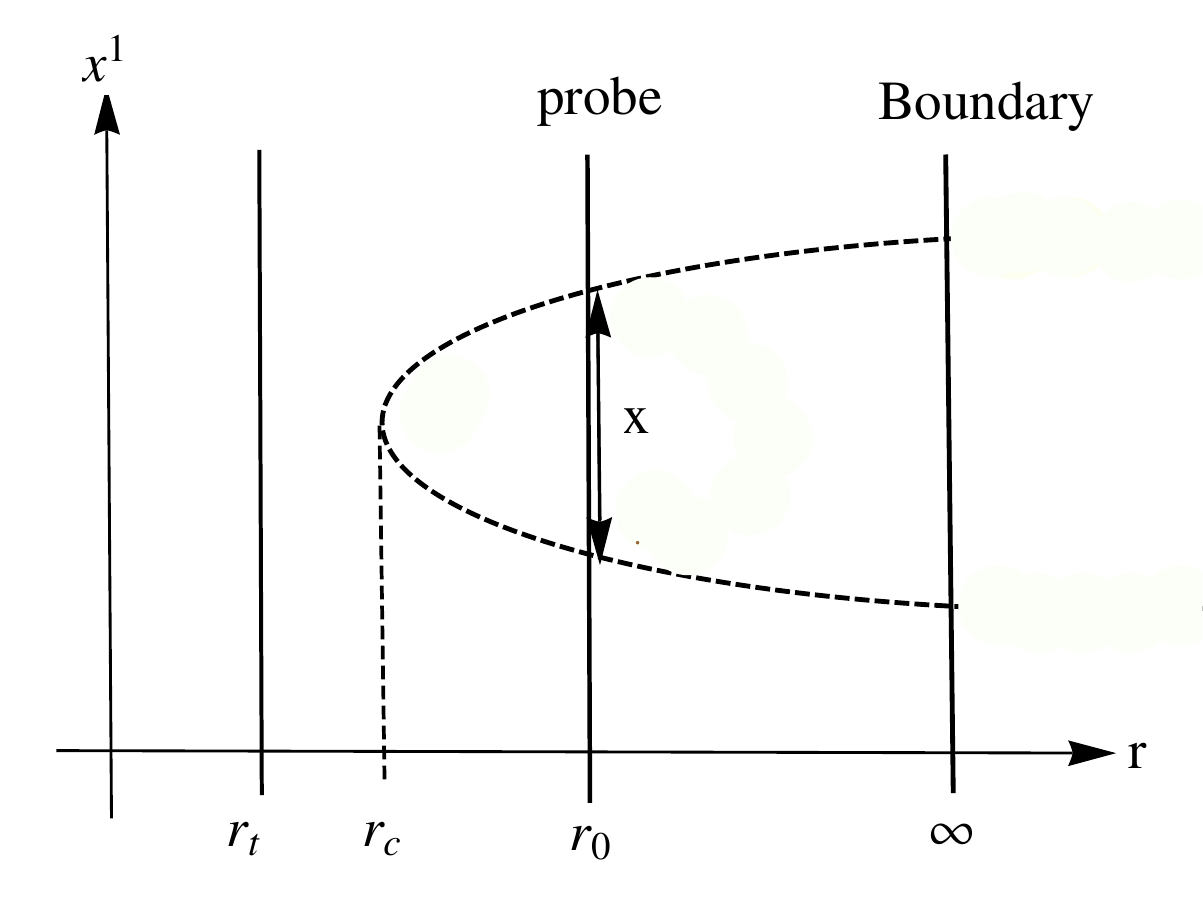}
  \captionof{figure}{World-sheet configuration in 2D.}
  \label{test14b}
\end{minipage}
\end{figure}
To compare with our diagrams, their plot of the phase diagram for the AdS soliton geometry \cite{Sato:2013hyw} is reproduced in Fig. (\ref{fig:soliton1}).

Now in the following sections, we do a similar calculation for our class of supergravity confining geometries and then numerically we find the phase diagrams.

\subsection{Witten-QCD}

First for the Witten-QCD geometry which is similar to Sakai-Sugimoto model, if we assume a probe D3-brane (which gives the practical spectrum for us) is located at $u=u_0$ and then we assume the following Ansatz for the Wilson loop,

\begin{gather}
x^0=\tau, \ \ \ \ x^1=\sigma, \ \ \ u=u(\sigma), \ \ \ \theta=\theta(\sigma),
\end{gather}
the NG action is
\begin{gather}
\mathcal{L}=\sqrt{-g}= \sqrt{ \left(\frac{u}{R}\right)^3 \left(1+\frac{4 R^3}{9 u_t}  {\theta'} ^2 \left(1-\frac{{u_t}^3}{u^3}\right)+\frac{{u'}^2}{1-\frac{u_t^3}{u^3}}\left (\frac{R}{u}\right)^3 \right)   }.
\end{gather}
As the Lagrangian does not depend on $\sigma$, the following Hamiltonians are conserved,
\begin{gather}
H_u=\frac{\partial \mathcal{L}}{\partial(\partial_\sigma u)}\partial_\sigma u -\mathcal{L}, \ \ \ \ H_\theta=\frac{\partial \mathcal{L}}{\partial(\partial_\sigma \theta)}\partial_\sigma \theta -\mathcal{L}.
\end{gather}
Then there should exist two different minimal surfaces that satisfy the following relations,
\begin{gather}
\frac{du}{d\sigma}=0, \ \ \text{at} \ \ u=u_c \ \ (u_t<u_c<u_0 ),  \ \ \ \ \ \ \frac{d\theta}{d\sigma}=0, \ \ \text{at} \ \ \theta=\theta_c \ \ (\theta_t<\theta_c<\theta_0 ).
\end{gather}
However, by assuming $\theta'=0$, we consider a rectangular Wilson loop which makes the calculation simpler. So
\begin{gather}
\mathcal{L}=\sqrt{\frac{1}{1-\frac{u_t^3}{u^3} } \left(\frac{du}{d\sigma}\right)^2+\frac{u^3}{R^3} },
\end{gather} 
and then from the conservation of $H_u$ we would get
\begin{gather}
\frac{du}{d\sigma}=\frac{1}{R^{\frac{3}{2}}} \sqrt{(u^3-{u_t}^3) \left( \frac{u^3}{u_c^3}-1  \right) }.
\end{gather}
By integrating the above equation one derives the length of the string between the quark and antiquark in WQCD as
\begin{gather}
x=2  R^{\frac{3}{2}} \int_{u_c}^{u_0} \frac{ du }{ \sqrt{ (u^3-u_t^3)\left( (\frac{u}{u_c})^3-1  \right) }}.
\end{gather}
Now, by defining the following dimensionless quantities,
\begin{gather}
y=\frac{u}{u_c}, \ \ \ \ a=\frac{u_c}{u_0}, \ \ \ \ \ b=\frac{u_t}{u_0},
\end{gather}
one can simplify $x$ as
\begin{gather}
x= \frac{ 2 R^{\frac{3}{2}}}{(u_0 a)^{\frac{1}{2}} } \int_{1}^{\frac{1}{a}} \frac{ dy }{ \sqrt{ (y^3-1)\Big(y^3-(\frac{b}{a})^3 \Big) }}.
\end{gather}
The sum of the potential and static energy is
\begin{gather}
V_{PE+SE}=2T_F\int_0^{\frac{x}{2}} d\sigma \mathcal{L}=2T_f u_0 a \int_1^{\frac{1}{a}} dy \frac{y^3}{\sqrt{(y^3-1)\Big (y^3-\frac{b^3}{a^3} \Big )}}.
\end{gather}
For the large $x$ limit, ($a\to b$), the sum of the potential and static energy is
\begin{gather}
V_{PE+SE}=T_F\frac{({u_0 b}) ^{\frac{3}{2} }}{R^{\frac{3}{2}}}x+2 T_F u_0 b  \left( \frac{1}{b}-1 \right).
\end{gather}
So from the first term which is the quark and antiquark potential, we can read the confining string tension as
\begin{gather}
\sigma_{\text{st}}= T_F {\left(\frac{u_t}{R} \right)^{ \frac{3}{2} }}.
\end{gather}
This matches with the result coming from the relation $\sigma_{st}=\frac{g{(u_t)}}{2\pi \alpha'}$.
Also the second term gives the static mass of the quark and antiquark,
\begin{gather}
2T_F (u_0-u_t)=2m_W.
\end{gather}

Then from the DBI action, one can read the critical electric field for this geometry as
\begin{gather}
E_c= T_F  \Big( \frac{u_0}{R} \Big)^{\frac{3}{2}}.
\end{gather}
Now, we can define the dimensionless parameter $\alpha=\frac{E}{E_c}$. So the total potential energy is
\begin{gather}
V_{\text{tot}}=V_{\text{PE}+\text{SE}}-Ex=\nonumber\\
2T_Fu_0 a \int_1^{\frac{1}{a}} dy \frac{y^3}{ \sqrt{(y^3-1) (y^3-\frac{b^3}{a^3} )}}-\frac{2T_F u_0 \alpha}{a^{\frac{1}{2}}} \int_1^{\frac{1}{a}} \frac{dy}{ \sqrt{(y^3-1) (y^3-\frac{b^3}{a^3} )}}.
\end{gather}

\begin{figure}[h!]
  \centering
    \includegraphics[width=0.6\textwidth]{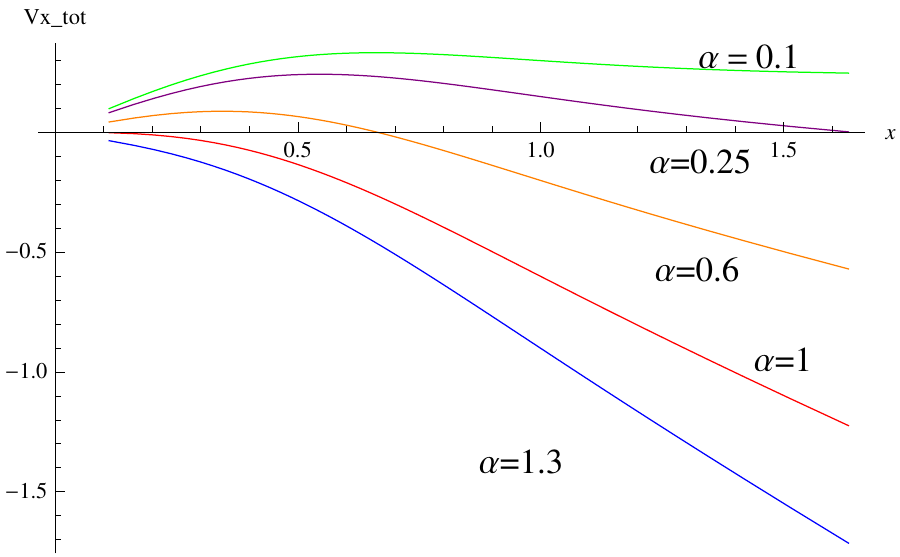}
       \caption{The plot of total potential versus $x$ for the Witten QCD model when $b=0.5$ and $2T_F=u_0=1$. }
   \label{WQCD}
\end{figure}

The plot of WQCD is shown again in Fig. (\ref{wQCD}) . By comparing the diagrams below, one can see that the form of plots is similar with slight differences in small $x$ limit.  One can see that in both of them, there exist three phases, one with stable potentials  and no pair creation, one with exponentially suppressed potentials with tunneling pair creation and one with catastrophically unstable potentials with exponential pair creation. 

As it has been demonstrated analytically for a general background in \cite{Sato:2013hyw} this behavior is universal in all confining geometries. However, there are still some minor differences between the phase diagrams of different confining backgrounds which here we aim to detect and then compare with the phase diagrams of the entanglement entropy of these backgrounds. 

\begin{figure}[h!]
\centering
\begin{minipage}{.5\textwidth}
  \centering
  \includegraphics[width=.9\linewidth]{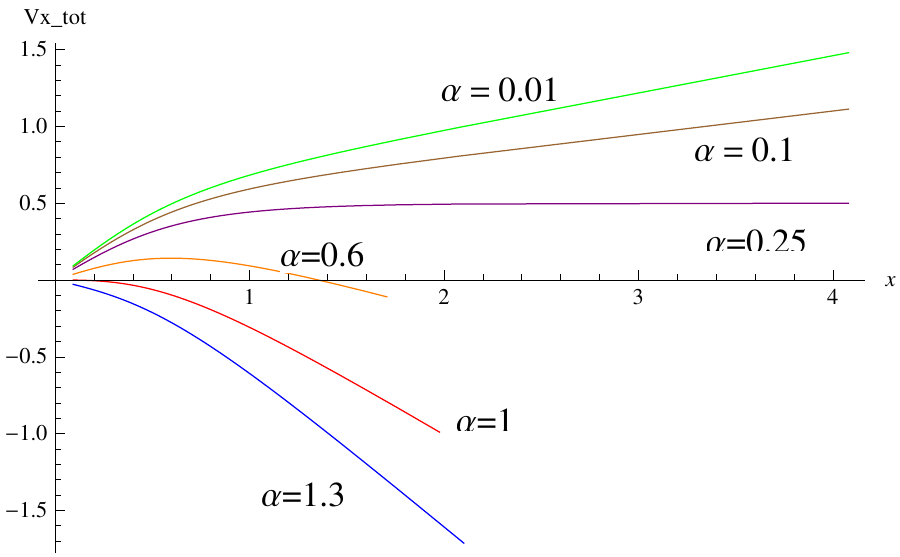}
  \captionof{figure}{AdS soliton.} \label{fig:soliton1}
  \label{fig:test15}
\end{minipage}%
\begin{minipage}{.5\textwidth}
  \centering
  \includegraphics[width=.9\linewidth]{wQCD}
  \captionof{figure}{Witten QCD.}
  \label{wQCD}
\end{minipage}
\end{figure}

From these plots one can see that in the Witten QCD diagram, for $\alpha=0.01$ there is no zero other than the origin and no Schwinger effect can occur. For $\alpha=0.1$ the potential becomes flat.  For $\alpha=0.25$ or 0.6, there is a barrier in the potential which, as can be seen from the diagrams, is different from the AdS soliton case as it has a bigger curvature in smaller $x$. In this phase the Schwinger pair creation can only occur by tunneling through this barrier and the rate of pair creation is strongly suppressed. For larger $E$, where for instance $\alpha=1.3$, the potential becomes catastrophically unstable and the Schwinger effect occurs and therefore a current can be created. In this phase the probability of pair creation would not be any more exponentially suppressed. 
The curvatures are bigger for the diagram of the Witten QCD model and a small bump can be seen around $x=0.5$ which is not present for the AdS soliton case. Also by comparing all the diagrams, one can see that in Witten-QCD and KT models, the phase transition happens faster and in a more dramatic way relative to the other backgrounds.

\subsection{Maldacena-Nunez}

Now we repeat this calculation for the Maldacena-Nunez background which potentially can show the instantons effects.

We assume $x^0=\tau, \ x^1=\sigma, \ r=r(\sigma)$, so the determinant of the induced metric and therefore the Lagrangian is $ \mathcal{L}=e^\phi \sqrt{r'^2+1}$.

Since
\begin{gather}
\frac{\partial \mathcal{L}}{\partial(\partial_\sigma r)} \partial_\sigma r-\mathcal{L}
\end{gather}
is a constant, we would get
\begin{gather}
r'=\sqrt{\frac{e^{2\phi}}{e^{2\phi({r_c})}}-1}.
\end{gather}
So, $\mathcal{L}=\frac{e^{2\phi}}{e^{\phi_c}}$. Also from the DBI action one can find the critical electric field as
\begin{gather}
E_c=T_F \frac{e^{2\phi({r_0})} }{e^{\phi({r_c})}}.
\end{gather}
We assume that the probe brane is located at $r=r_c=1$, and for the sake of similarity to the previous calculations we define $a=\frac{r_c}{r_0}$, therefore $r_0=\frac{1}{a}$.
One should also notice that the MN geometry ends when $e^\phi$ becomes undefined. This happens at the roots of $e^h=r \coth(2r) -\frac{r^2}{\sinh[2r^2]}-\frac{1}{4}$ where its plot is shown in Fig.(\ref{fig:roots1}).

\begin{figure}[h!] 
  \centering
    \includegraphics[width=0.3\textwidth]{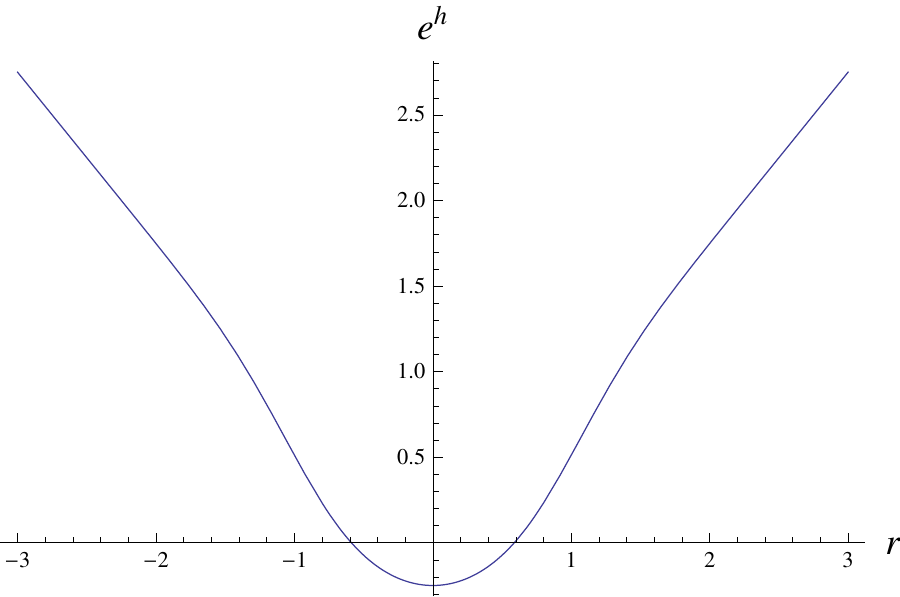}
       \caption{The Plot of $e^h$ versus $r$ . }
       \label{fig:roots1}
\end{figure}
One can see that the positive root is approximately located at $r_t=0.6$ and it is where the geometry ends.

Now, by defining 
\begin{gather}
x=2\int_{r_c}^ {r_0} \frac{dr}{ \sqrt{\frac{e^{2\phi}}{ e^{2\phi_c}} -1 }},
\end{gather}
and the total potential as
\begin{gather}
V_{\text{tot}}=2T_F \int_{r_c}^{r_0} \frac{e^{2\phi} }{e^{2\phi_c}} \frac{ dr}{\sqrt{\frac{e^{2\phi} }{e^{2\phi_c}  }-1 } } -2E_c \alpha  \int_{r_c}^{r_0} \frac{dr}{\sqrt{\frac{e^{2\phi} }{e^{2 \phi_c} }-1 }},
\end{gather}
 one can find the phase diagram in Fig. (\ref{fig:MNa}). 
 
\begin{figure}[h!] 
  \centering
    \includegraphics[width=0.6\textwidth]{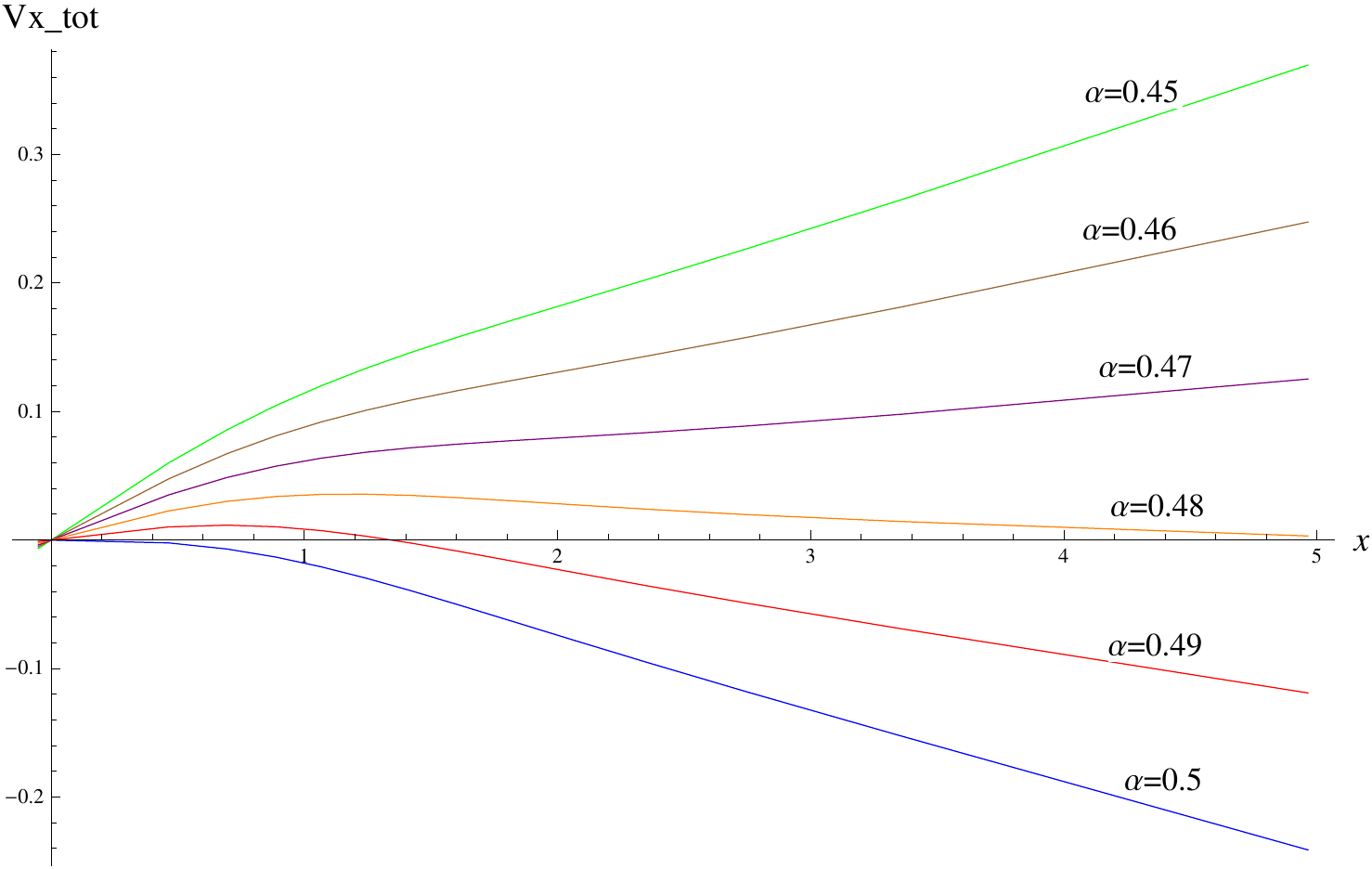}
       \caption{The plot of total potential versus $x$ for the Maldacena-Nunez model. Here $\phi_0=0$, $r_0=T_F=1$. }
  \label{fig:MNa}
\end{figure}
By expanding the potential $V_{\text{PE}+\text{SE}}$ for large $x$, i.e, $r \to r_c$, one can find the string tension and quark mass as
\begin{gather}
\sigma_{st}=e^{\phi_c} T_F, \ \ \ \ \ \  2m_W=2T_F e^{\phi_c} (r_0-r_t).
\end{gather}
This result for the string tension matches with the other result coming from relation $\frac{g(r_t)}{2\pi \alpha'}$.
One can see that still there exist three different phases. For $\alpha$ until around 0.47 no Schwinger effect would occur. For $\alpha$ between 0.47 and 0.48, Schwinger effect would occur as a tunneling process, and for $\alpha$ larger than 0.48, the potential becomes unstable. 

\subsection{Klebanov-Strassler}

Now for the KS background we assume $x^0=t, \ x^1=\sigma,\  \tau=\tau(\sigma)$.
So the components of the induced metric are
\begin{gather}
\gamma_{tt}=-h^{-\frac{1}{2}}(\tau), \ \ \ \ \ \gamma_{\sigma \sigma}=h^{-\frac{1}{2}}(\tau)+\frac{\tau'^2 h^{\frac{1}{2}}(\tau) \epsilon^{\frac{4}{3}}}{6 K^2(\tau) }.
\end{gather}
Therefore,
\begin{gather}
\mathcal{L}= \sqrt{h^{-1}(\tau)+\frac{\tau'^2 \epsilon^{\frac{4}{3}}}{6K^2(\tau) } }.
\end{gather}
Again, knowing that $\frac{\partial \mathcal{L}}{\partial(\tau')}-\mathcal{L}$ is constant, one can derive $\tau'$ as
\begin{gather}
\tau'=\frac{\sqrt{6}} {\epsilon^{\frac{2}{3}}} \frac{K(\tau)}{h(\tau)} \sqrt{h(\tau_c)-h(\tau) }.
\end{gather}
Also, the critical electric field is 
\begin{gather}
E_x= T_F h^{-\frac{1}{2}} (\tau_0).
\end{gather}
At $\tau=\tau_c$ one would have $\tau'_c=0$. We assume $\tau_0=1$, so one can find
\begin{gather}
x=2 \int_1^{\frac{1}{a}} \frac{\epsilon^{\frac{2}{3}} h(ya) }{\sqrt{6} K(ya) } \frac{1}{ \sqrt{h(a) -h(ya)} },
\end{gather}
and the total potential as
\begin{gather}
V_{tot}=\frac{2 \epsilon T_F}{\sqrt{6}} \int_1^{\frac{1}{a}} \frac{ \sqrt{h(a)} }{K(ya)} \frac{1}{\sqrt{h(a)-h(ya) } }\nonumber\\
-\frac{2\epsilon^{\frac{2}{3}} T_F \alpha}{\sqrt{6} } \int_1^{ \frac{1}{a}} \frac{h(ya) }{K(ya) } \frac{h^{-\frac{1}{2}} (1)}{\sqrt{h(a)-h(ya) }}.
\end{gather}
The plot of KS phases for different $\alpha$ is shown in Fig. (\ref{fig:KSa}).

\begin{figure}[h!]
  \centering
    \includegraphics[width=0.6\textwidth]{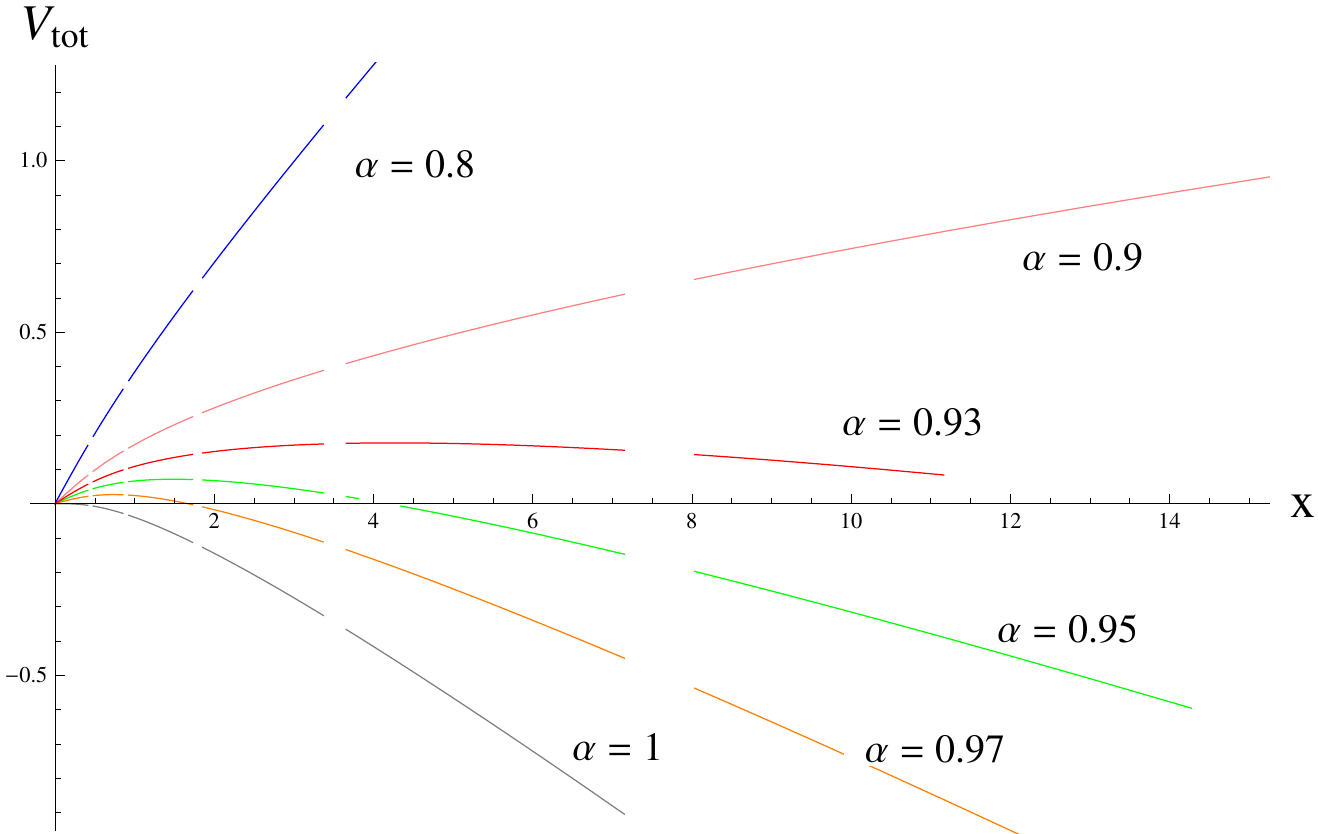}
       \caption{The plot of total potential versus $x$ for the Klebanov-Strassler model, assuming $\tau_0=\beta=T_F=\epsilon=1$. }
  \label{fig:KSa}
\end{figure}
The numerical calculation of this case would take more time to be performed but the final phase diagram in general, is similar to the three other confining geometries. The parts that are missing in the plot are due to some singularities in the numerical calculations of the integrals. Again, three phases for different $\alpha$ can be detected and the phase transitions are mild.

\subsection{Klebanov-Tseytlin}
The calculation can be repeated for the KT background.
Taking  $x^0=t, \ x^1=\sigma,\  r=r(\sigma)$, the components of the induced metric are
$\gamma_{tt}=h^{-\frac{1}{2}}, \ \gamma_{\sigma \sigma}=-(h^{-\frac{1}{2}}+r'^2 h^{\frac{1}{2}} )$ and 
 so $\mathcal{L}= \sqrt{h^{-1} +r'^2}$.
 
Again using the conservation of the Hamiltonian and knowing  that at $r=r_c$, $r'=0$, then we find
\begin{gather}
r'=\frac{\sqrt{h(r_c)-h(r)} }{h(r)}.
\end{gather}

After taking the integral of this relation and by defining $y=\frac{r}{r_c} $ , $a=\frac{r_c}{r_0}$ and $b=\frac{r_s}{r_0}$, one can find

\begin{gather}
x=9 \sqrt{2} M \epsilon ^2 \sqrt{g_s } \int_{1} ^{\frac{1}{a}}\frac{dy}{y^4 \sqrt{\ln \left(\frac{a}{b}\right) -\frac{1}{y^4} \ln \left(\frac{ya}{b} \right) } } ,\nonumber\\
V_{PE+SE}=2T_F\int_{r_0}^{r_s} \mathcal{L} d\sigma=2T_F a r_0 \sqrt{\ln \frac{a}{b}} \int_1^{\frac{1}{a}} \frac{dy}{\sqrt{\ln \frac{a}{b}-\frac{1}{y^4}\ln \frac{ay}{b}  } }.
\end{gather}
From the DBI action the critical electric field is
\begin{gather}
E_{c}=T_F h_0^{-\frac{1}{2}}= T_F \frac{r_0^2}{L^2} \left(\ln \frac{r_0}{r_s} \right)^{-\frac{1}{2}}.
\end{gather}
By defining $\alpha=\frac{E}{E_{c}}$, the total potential is, 
\begin{gather}
V_{\text{tot}}=2T_F a r_0 \sqrt{\ln \frac{a}{b}} \int_1^{\frac{1}{a}} \frac{dy}{\sqrt{\ln \frac{a}{b}-\frac{1}{y^4}\ln \frac{ay}{b}  } }- \frac{2\alpha T_F r_0^2}{\sqrt{ \ln\frac{1}{b}} }  \int_{1} ^{\frac{1}{a}}\frac{dy}{y^4 \sqrt{\ln \frac{a}{b}-\frac{1}{y^4} \ln \frac{ya}{b}  } }.
\end{gather}

Now by simplifying this relation by assuming $b=0.4$ and  $r_0=M=T_F=1$, one can find the plot of $V_{\text{tot}}$ versus $x$ which is shown in Fig. (\ref{fig:KTa}).

\begin{figure}[h!]
  \centering
    \includegraphics[width=0.6\textwidth]{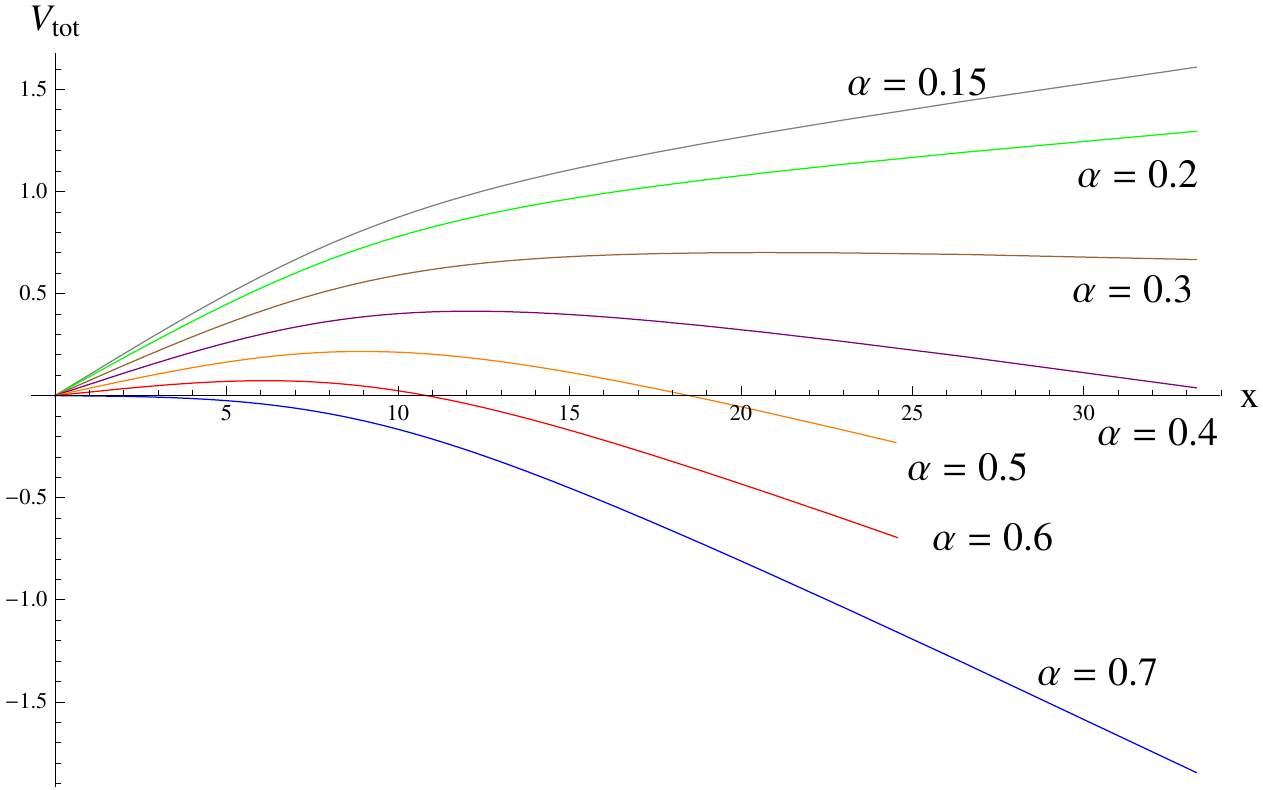}
       \caption{The plot of total potential versus $x$ for the Klebanov-Tseytlin model for $b=0.5$ and $r_0=M=T_F=1$ . }
        \label{fig:KTa}
\end{figure}

The Klebanov-Tseytlin model is a limit of Klebanov-Strassler and there is no wonder that the phase diagrams are very similar. Again three phases can be seen for different $\alpha$s. However, in these two cases the exact numerical values of $\alpha$ do not correspond to each other,  since we did not match the numerical constants.

Also, note that for this particular question that we were interested in, we can neglect the naked singularity of KT while we gain some technical advantages.

\subsection{Klebanov-Witten}

Now to compare our results of the phase diagrams of confining geometries with the conformal backgrounds, we also study the Klebanov-Witten geometry as an example of conformal background.  The phase diagram is identical to the phase diagram of the AdS case which has been studied in \cite{Sato:2013iua}.
Assuming $ x^0=t, \ x^1=\sigma, \ r=r(\sigma)$,  where $h=\frac{L^4}{r^4}$, then the Lagrangian is $\mathcal{L}= \sqrt{h^{-1} +r'^2}$, and as the Hamiltonian is conserved, one would get
\begin{gather}
r'=\frac{r^2}{L^2}\sqrt{ \frac{r^4}{r_c^4} -1},
\end{gather}
where $\frac{dr}{d\sigma}=0$ at $r=r_c$. So $\mathcal{L}=\frac{r^4}{L^2 r_c^2}$ and
from the DBI action the critical electric field is  $E_x=\frac{T_F r_0^2} {L^2}$. By integrating $r'$ and by defining $y=\frac{r}{r_c}$, $a=\frac{r_c}{r_0}$ and $b=\frac{r_t}{r_0}$, (notice that $r_t=0$ here) the distance between the quark and antiquark and the potential are
\begin{gather}
x=\frac{2L^2}{r_0 a} \int_1^{\frac{1}{a}} \frac{dy}{y^2 \sqrt{y^4-1} }, \\
V_{PE+SE}= 2T_F r_0 a  \int_1 ^{\frac{1}{a}} \frac{y^2 dy}{ \sqrt{y^4-1} }.
\end{gather}
For large $x$, as $a\to b=0$, the potential vanishes which is a feature of non-confining (conformal) backgrounds. Also, $2m_W=2T_F r_0 (1-a)$ and for $a=b=0$, it gives $2m_W=2T_F r_0$. 
Then we get,
\begin{gather}
V_{tot}=2T_F r_0 a \int_1 ^{\frac{1}{a}} \frac{y^2 dy}{\sqrt{y^4-1} }-\frac{2T_F r_0 \alpha }{a}  \int_1^{\frac{1}{a}} \frac{dy}{y^2 \sqrt{y^4-1} }.
\end{gather}
The phase diagram of KW and AdS are similar and is shown in Fig. (\ref{fig:11a}).

\begin{figure}[h!]
  \centering
    \includegraphics[width=0.6\textwidth]{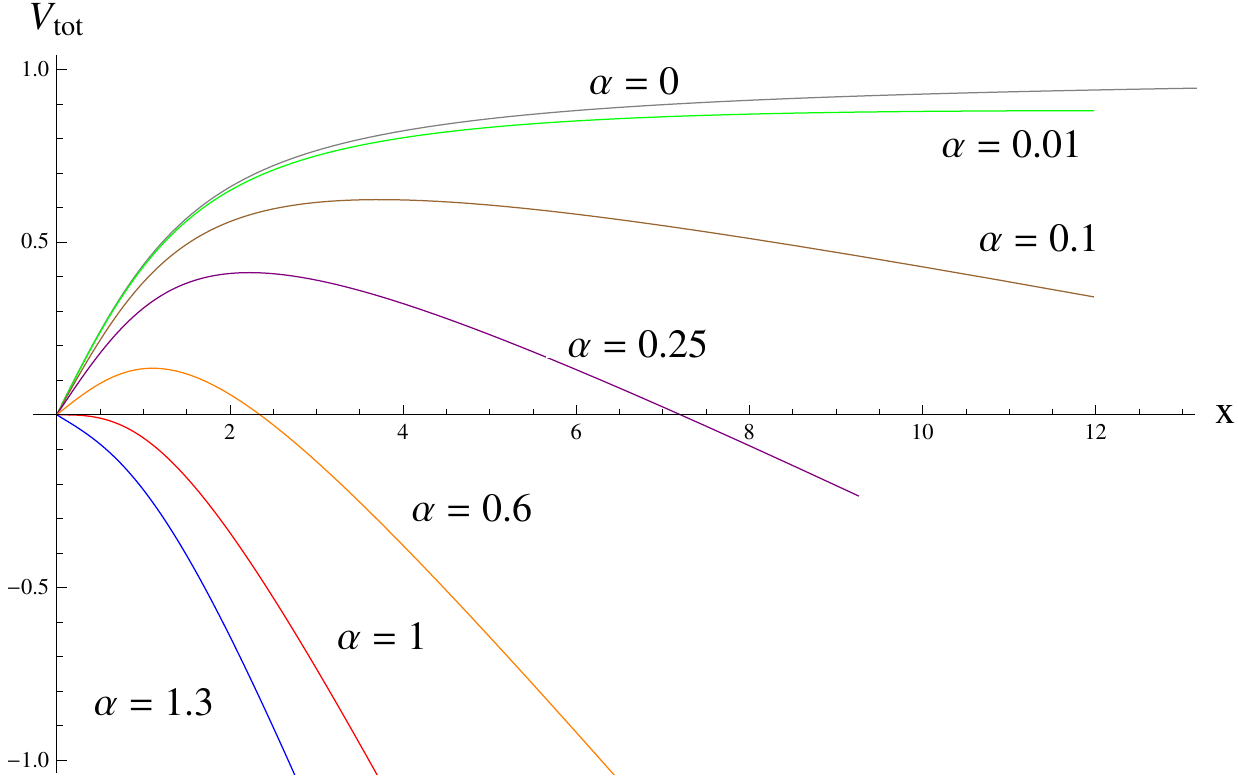}
       \caption{The Schwinger phases of $\text{AdS}_5$ and Klebanov-Witten. }
        \label{fig:11a}
\end{figure}

Unlike the other four confining geometries, as the KW and AdS are conformal geometries, there are only two phases present. Even for a very small $\alpha$ (or electric field), there is always a zero at larger $x$ and so the pair creation would happen by a tunneling process there. The other phase is the unstable one for a larger $\alpha$.

\section{Entanglement entropy of a strip in confining geometries}\label{sec:four}
Now we would like to study the phase diagrams of entanglement entropy in these confining geometries  and then compare the phase transition of EE with the phase transition of Schwinger effect and compare the diagrams of different geometries. One way is to use the method in \cite{Hubeny:2014zna} for calculating the entanglement entropy of accelerating quark and antiquark. But as we have mentioned in Sec. \ref{sec:oneprofile} , one first needs to solve several partial differential equations analytically. Other methods could be using the ideas in \cite{Caputa:2014vaa} or \cite{Wong:2013gua},  to study the entanglement entropy of local operators or localizes excited states in each background. But here, to find the entanglement entropy of a strip in each confining geometry, we follow the calculations of \cite{Kol:2014nqa} and \cite{Klebanov:2007ws}.  Also, the entanglement entropy of multiple strips can be calculated similar to calculations in \cite{Ben-Ami:2014gsa}.

So as in \cite{Klebanov:2007ws}, based on Klebanov-Kutasov-Murugan (KKM) suggestion, the generalization of Ryu-Takayanagi conjecture for the non-conformal theories is
\begin{gather}
S=\frac{1}{G_N^{(10)}} \int_\gamma d^8 \sigma e^{-2\phi} \sqrt{G_{{ind}}^{(8)}} .
\end{gather}
The authors showed that the entanglement entropy can be found by minimizing this action over all surfaces that ends on the boundary of the entangling surface. There are actually two solutions which satisfy these conditions. One of them corresponds to a disconnected region and the other one is a connected surface. In \cite{Wong:2013gua}, the authors showed that always only one of the two possible configurations would dominate and would be the physical solution. 

So if one writes the gravitational background in the general following form 
\begin{gather}
ds^2=\alpha(\rho) [ \beta(\rho) d\rho^2+dx^\mu dx_\mu]+g_{ij} d\theta^i d\theta^j, \ \ \ \  (\mu=0,1,...d),\ \ (i=d=2,...,9),
\end{gather}
then the volume of the internal manifold is $V_{\text{int}}=\int d\vec{\theta} \sqrt{\text{set} [g_{ij}]}$.

 One can also define another useful function $H(\rho)$ as
\begin{gather}
H(\rho)=e^{-4\phi} V_{\text{int}}^2 \alpha^d.
\end{gather}
For confining geometries, this function is a monotonically increasing one while $\beta(\rho)$ is a monotonically decreasing function.
Using the KKM equation \cite{Klebanov:2007ws}, \cite{Kol:2014nqa}, the EE for the connected solution is
\begin{gather}
S_C(\rho_0)=\frac{V_{d-1}}{2 G_N^{(10)}} \int_{\rho_0} ^\infty d\rho \sqrt { \frac{\beta(\rho) H(\rho)} {1-\frac{H(\rho_0)}{H(\rho)} } },
\end{gather}
the EE of the disconnected solution is given by
\begin{gather}
S_D(\rho_0)= \frac{V_{d-1}}{2G_N^{(10)}} \int_{\rho_\Lambda}^\infty d\rho \sqrt{ \beta(\rho) H(\rho) }.
\end{gather}
and the length of the line segment of the connected solution is
\begin{gather}
L(\rho_0)=2\int_{\rho_0}^\infty d\rho \sqrt{ \frac{\beta(\rho)}{\frac{H(\rho)}{H(\rho_0)}-1 } }.
\end{gather}
The difference of the connected and disconnected solution is finite, so $S$ is defined as
\begin{gather}
S(\rho_0)=\frac{2G_N^{(10)}}{V_{d-1}} (S_C-S_D)=\int_{\rho_0} ^\infty d\rho \sqrt { \frac{\beta(\rho) H(\rho)} {1-\frac{H(\rho_0)}{H(\rho)} } }-\int_{\rho_\Lambda}^\infty d\rho \sqrt{ \beta(\rho) H(\rho) }.
\end{gather}

Now we study $L$ and $S$ for our specific geometries.

\subsection{Witten-QCD background}

For the case of Witten QCD background similar to \cite{Kol:2014nqa}, one can define the functions $\alpha(u), \beta(u)$ and $H(u)$ as
\begin{gather}
\alpha(u)=\left(\frac{u}{R}\right)^{\frac{3}{2}}, \ \ \ \ \ \beta(\rho)=\left(\frac{R}{u}\right)^3 \frac{1}{f(u)},\ \ \ \ \ \ H(u)=\left( \frac{8\pi^2}{3}\right)^2 \frac{4R^6 u^5 f(u)}{9 u_0 g_s^4} ,
\end{gather}
where $f(u)=1-\frac{u_t^3}{u^3}$. From the functions of $\beta$ and $H$, it can be seen that as it has been suspected, $\beta(u)$ is monotonically decreasing, $H$ is monotonically increasing, $H$ shrinks to zero at $u=0$ and $\beta(u)$ diverges at $u=u_t=1$.

The EE for the connected and disconnected solutions are respectively 
\begin{gather}
S_{C}(u_0)=\frac{V_2}{ {G_N}^{(10)}} \frac{8\pi^2 R^{\frac{9}{2}}}{9g_s^2 \sqrt{u_t} }  \int _{u_0}^{\infty} du \sqrt{ \frac{u^2}{1-\frac{u_0^5}{u^5} \frac{f(u_0)}{f(u)}  } },\nonumber\\
S_D(u_0)=\frac{V_2}{G_N^{10}} \frac{8\pi^2 R^{\frac{9}{2}} }{9 g_s^2 \sqrt{u_t}} \int_{u_\Lambda}^\infty u du.
 \end{gather}
 Also, the length of the line segment of the connected solution is
\begin{gather}
 L(u_0)=2\int_{u_0}^\infty du \sqrt { \frac{(\frac{R}{u})^3 \frac{1}{f(u)} }{\frac{f(u) u^5}{f(u_0) u_0^5} -1} }.
 \end{gather}
 Their plots are shown in Figs. (\ref{fig:WQCDL}) and (\ref{fig:WQCDS}).

\begin{figure}[h!]
\centering
\begin{minipage}{.5\textwidth}
  \centering
  \includegraphics[width=.9\linewidth]{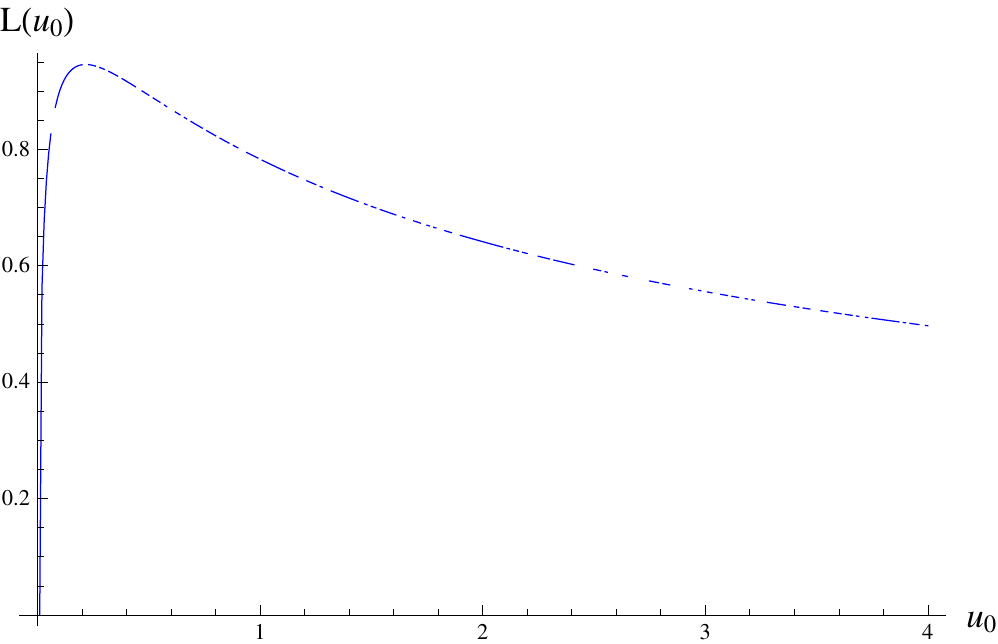}
  \captionof{figure}{Plot of $L(u_0)$ vs. $u_0$ for the \\ WQCD model. }
  \label{fig:WQCDL}
\end{minipage}%
\begin{minipage}{.5\textwidth}
  \centering
  \includegraphics[width=.9\linewidth]{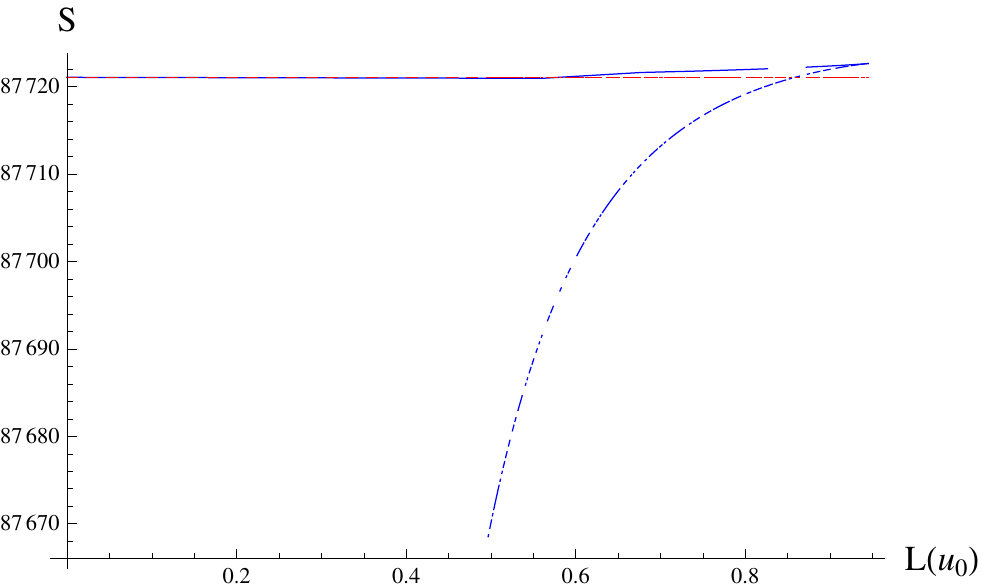}
  \captionof{figure}{Plot of $S$ vs. $u_0$ for the WQCD\\ model. The blue line is the connected solution \\ and the dashed red line is the disconnected solution. }
  \label{fig:WQCDS}
\end{minipage}
\end{figure}
Note that here, we take $u_t=u_{\Lambda}=1$.
 
Due to the peak in the plot of $L$ and also the butterfly shape and the double valuedness in the plot of $S$, one can deduce that a phase transition and therefore a confinement phase exists. This behavior of phase transition and the shape of the peak are similar to the instability of $V$ and the phase transition in the previous section. One can see that for this case, the phase transition is more dramatic relative to the other geometries. This dramatic phase transition in smaller $x$ was also seen in the Schwinger phase transition in WQCD relative to the other geometries. So there might be a deeper relationship between these two quantities.

\subsection{Klebanov-Tseytlin}
For the KT case, the volume of the internal part, $ds_{T^{1,1}}$, is $\frac{16 \pi^3}{27}$. So for the KT background the functions are
\begin{gather}
\alpha(r)=h(r)^{-\frac{1}{2}},  \ \ \ \ \  \beta(r)=h(r), \ \ \ \ \  V_{\text{int}}=\frac{16 \pi^3}{27} r^5 h(r)^{\frac{5}{4}}, \ \ \ \ \ 
H(r) = \left( \frac{16}{27} \right)^2 \pi^6 r^{10} h(r).
\end{gather}

Again $\beta$ and $H$ show the expected monotonically decreasing and increasing behaviors respectively. Then the length of the connected region and the EEs are

\begin{gather}
L(r_0)= 9\sqrt{2}   M \sqrt{g_s} \epsilon^2  \int_{r_0}^\infty dr \frac{ \sqrt{\frac{r}{r_s}} }{r^2 \sqrt{\frac{r^6}{{r_0}^6} \frac{\ln {\frac{r}{r_s} } }{\ln \frac{r_0}{r_s} } -1} }  ,\nonumber\\
S_C(r_0)=\frac{12 V_2 \pi^3 M^2 g_s  \epsilon^4}{G_N^{(10)}} \int_{r_0}^\infty dr \frac{r \ln \frac{r}{r_s} }{\sqrt{1-\frac{r_0^6}{r^6}\frac{\ln \frac{r_0}{r_s} }{\ln \frac{r}{r_s}}  } },\nonumber\\
S_D(r_0)=\frac{12V_2 \pi^3  M^2  g_s \epsilon^4 }{G_N ^{ (10)}} \int_{r_\Lambda} ^\infty dr \ r \ln \frac{r}{r_s}.
\end{gather}
The plots for the $M=\epsilon=g_s=1$ and $r_s=0.5$ is shown below in Figs. (\ref{fig:KTL2}) and (\ref{fig:KTSS2}).

\begin{figure}[h!]
\centering
\begin{minipage}{.5\textwidth}
  \centering
  \includegraphics[width=.9\linewidth]{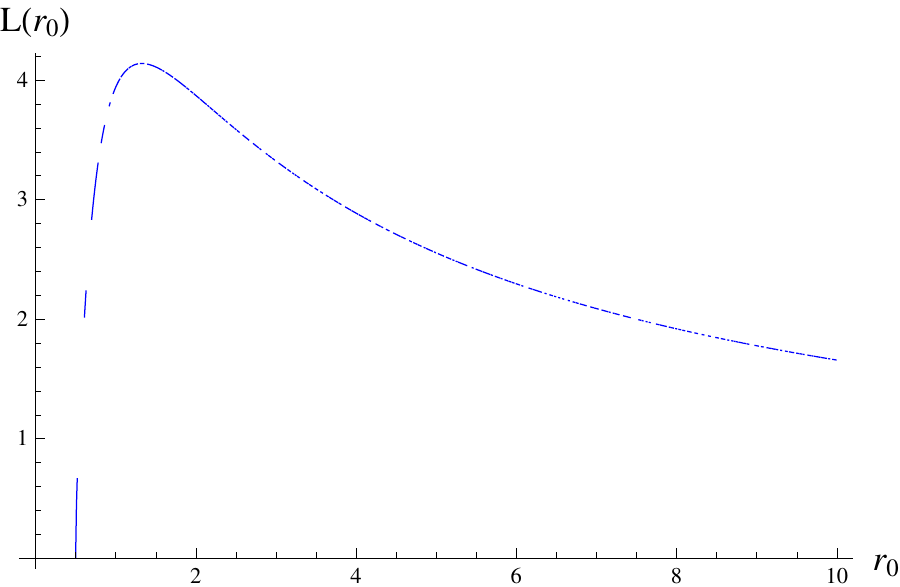}
  \captionof{figure}{Plot of $L(r_0)$ vs. $r_0$ for KT,\\ $r_s=0.5$. }
  \label{fig:KTL2}
\end{minipage}%
\begin{minipage}{.5\textwidth}
  \centering
  \includegraphics[width=.9\linewidth]{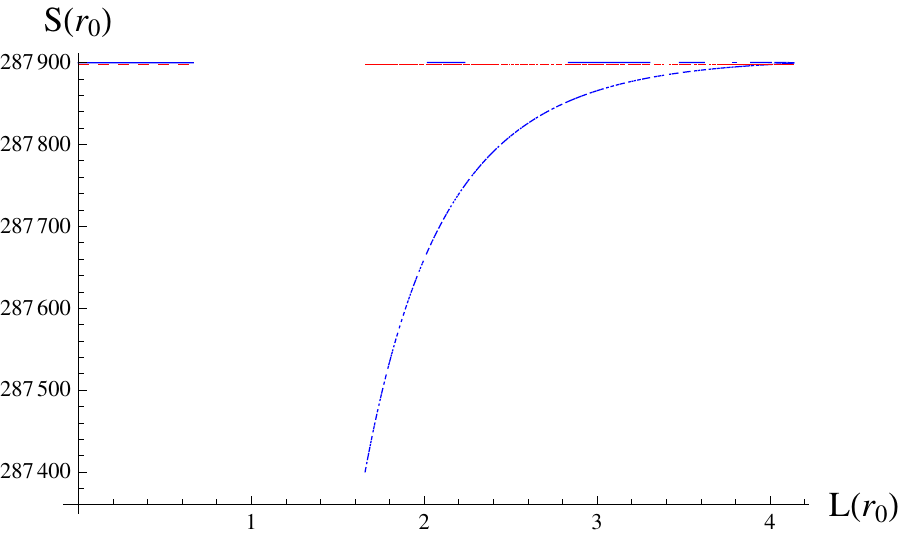}
  \captionof{figure}{Plot of $S(r_0)$ vs $L$ for KT, $r_s=0.5$. }
  \label{fig:KTSS2}
\end{minipage}
\end{figure}
As one can see, the form of the plots is very similar to the KS geometry if the naked singularity is placed at small $r$, as here it is actually set to $r_s=0.5$. However, increasing $r_s$ causes the red dashed line to go down and the butterfly shape of the diagram gets a flatter curvature as is shown in Figs (\ref{fig:KTL2}), (\ref{fig:KTSS2}), (\ref{fig:KTL3}) and (\ref{fig:KTSS3}).

\begin{figure}[h!]
\centering
\begin{minipage}{.5\textwidth}
  \centering
  \includegraphics[width=.9\linewidth]{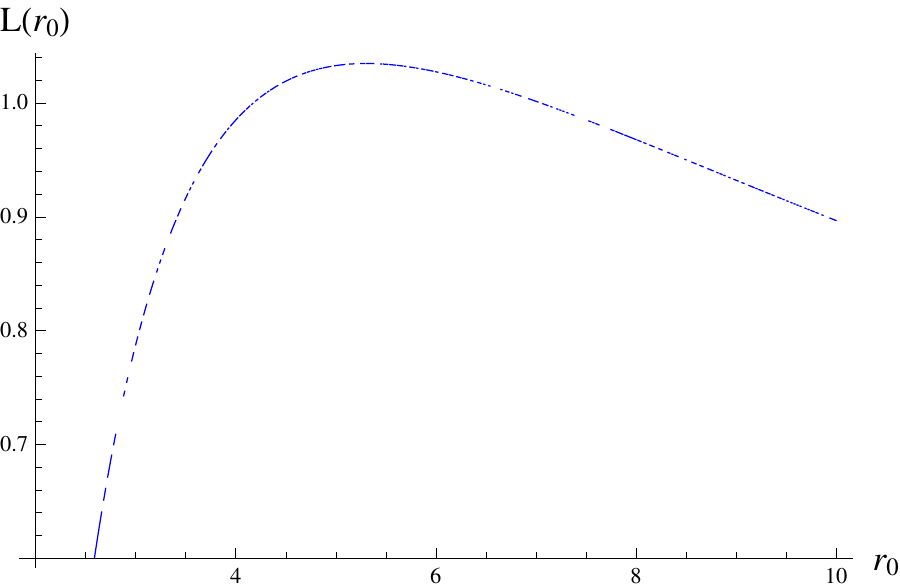}
  \captionof{figure}{Plot of $L(r_0)$ vs. $r_0$ for KT,\\ $r_s=2$. }
  \label{fig:KTL3}
\end{minipage}%
\begin{minipage}{.5\textwidth}
  \centering
  \includegraphics[width=.9\linewidth]{KTSS3}
  \captionof{figure}{Plot of $S(r_0)$ vs. $L(r_0)$ for KT, $r_s=2$. }
  \label{fig:KTSS3}
\end{minipage}
\end{figure}

\subsection{Klebanov-Strassler}

For the KS background, the functions are
\begin{gather}
\alpha=h^{-\frac{1}{2}}(\tau), \ \  \ \ \ \ \ \ \ \ \ \beta(\tau)=\frac{h(\tau) \epsilon^{4/3}}{6 K^2(\tau)},\nonumber\\
H(\tau)=\frac{8\pi^6}{3} \epsilon^{20/6} h(\tau) K^2(\tau) \sinh^4(\tau).
\end{gather}
Again, $\beta$ is a monotonically decreasing function and $H$ is a monotonically increasing function. 
Now, using $\beta$ and $H$, one can study the entanglement entropy of a strip in this geometry.
So
\begin{gather}
L(\tau_0)=\frac{2^{\frac{5}{6}} \epsilon^{\frac{2}{3}} }{\sqrt{3}} \int_{\tau_0}^\infty d\tau \frac{\sinh(\tau)}{(\sinh(2\tau) -2\tau)^{\frac{1}{3}}}\sqrt{ \frac{h(\tau)} { \frac{\sinh(\tau)^2 }{\sinh(\tau_0)^2} \left( \frac{\sinh(2\tau) -2\tau}{\sinh(2\tau_0) -2\tau_0}  \right)^{\frac{2}{3} } -1 } },\nonumber\\
S_C(\tau_0)=\frac{V_2 \pi^3 \epsilon^4}{3G_N^{(10)}} \int_{\tau_0}^\infty d\tau \frac{h(\tau) \sinh^2(\tau)}{ \sqrt{1- \left( \frac{\sinh(\tau_0) }{\sinh(\tau) }  \right)^2 \left( \frac{\sinh(2\tau_0) -2\tau_0 }{\sinh(2\tau)-2\tau }  \right)^{\frac{2}{3}}  } },\nonumber\\
S_D(\tau_0)=\frac{V_2 \pi^3 \epsilon^4}{3 G_N^{(10)}} \int_{\tau_\Lambda}^\infty d\tau h(\tau) \sinh^4 \tau.
\end{gather}

\begin{figure}[h!]
\centering
\begin{minipage}{.5\textwidth}
  \centering
  \includegraphics[width=.9\linewidth]{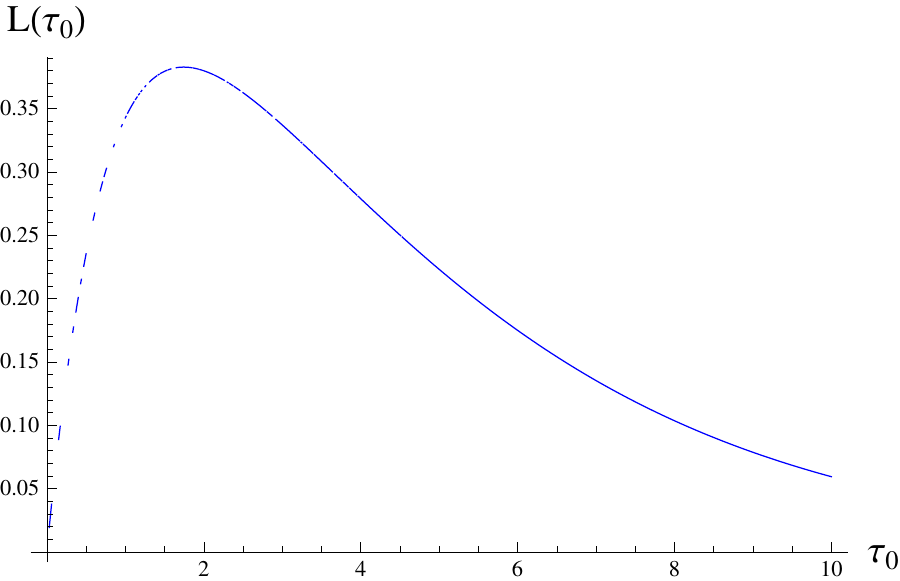}
  \captionof{figure}{Plot of $L(u_0)$ vs. $u$ for KS. }
  \label{fig:KSL2}
\end{minipage}%
\begin{minipage}{.5\textwidth}
  \centering
  \includegraphics[width=.9\linewidth]{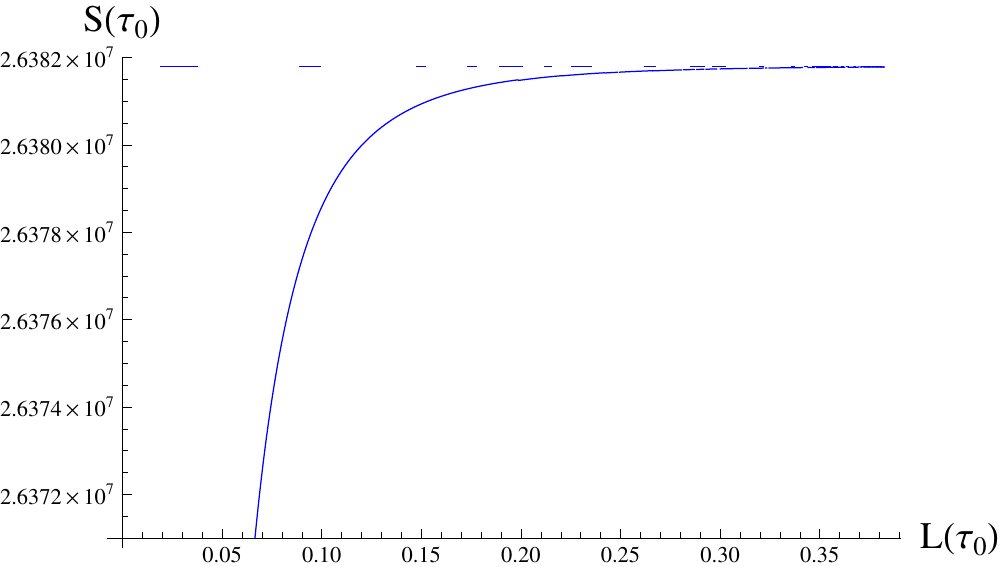}
  \captionof{figure}{Plot of $S(u_0)$ vs. $u_0$ for KS. }
  \label{fig:KSE2}
\end{minipage}
\end{figure}

The plot of the length of the connected solution $L$, and the entanglement entropy, $S$ is shown in Figs. (\ref{fig:KSL2}) and (\ref{fig:KSE2}). The entanglement entropy of the KS case with dynamical flavors were also studied in \cite{Georgiou:2015pia}. From the behavior of $L$ and the butterfly shape of $S$ one can detect the confining phase. One can see however that the phase transition is milder relative to the Witten QCD and KT cases. This milder phase transition was also seen for the Schwinger effect phase transition.

\subsection{Maldacena-Nunez background}
For the Maldacena-Nunez background, the functions are
\begin{gather}
\alpha(r)=e^{\phi}, \ \ \ \ \ \ \beta(r)=1, \ \ \ \ \ \ \ V_{\text{int}}=8\pi^3 e^{2h} e^{\frac{5}{2} \phi},\ \ \ \ \         H(r)=4\pi^6 e^{8\phi_0} (\sinh 2r)^4.
\end{gather}

Unlike the other backgrounds that we study here, for the Maldacena-Nunez metric, the function $\beta$ is a constant and is not monotonically decreasing. Also $H(r)$ is not monotonically increasing.
The $L(r_0)$ and EE functionals are

\begin{gather}
L(r_0)=\int_{r_0}^\infty dr \frac{2}{\sqrt{\frac{(\sinh 2r)^4}{(\sinh 2r_0)^4}-1 } },\nonumber\\
S_C(r_0)=\frac{V_2 \pi^3 e^{4\phi_0} }{G_N^{(10)}} \int_{r_0}^\infty dr \frac{(\sinh 2r)^2}{\sqrt{1-\frac{(\sinh 2r_0)^4}{(\sinh 2r)^4}} },\nonumber\\
S_D(r_0)=\frac{V_2 \pi^3 e^{4 \phi_0} }{G_N^{(10)}} \int_{r_0}^\infty dr (\sinh 2r)^2. 
\end{gather}
Their plots are shown in Figs. (\ref{fig:LMN}) and (\ref{fig:SMN}).

\begin{figure}[h!]
\centering
\begin{minipage}{.55\textwidth}
  \centering
  \includegraphics[width=.9\linewidth]{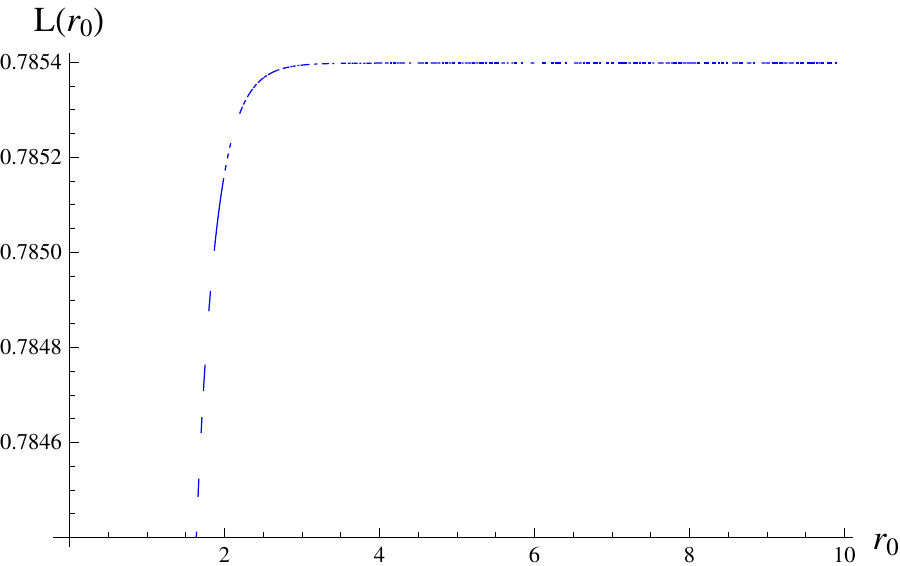}
  \captionof{figure}{Plot of $L(r_0)$ vs. $r_0$\\ for the MN model. }
  \label{fig:LMN}
\end{minipage}%
\begin{minipage}{.55\textwidth}
  \centering
  \includegraphics[width=.9\linewidth]{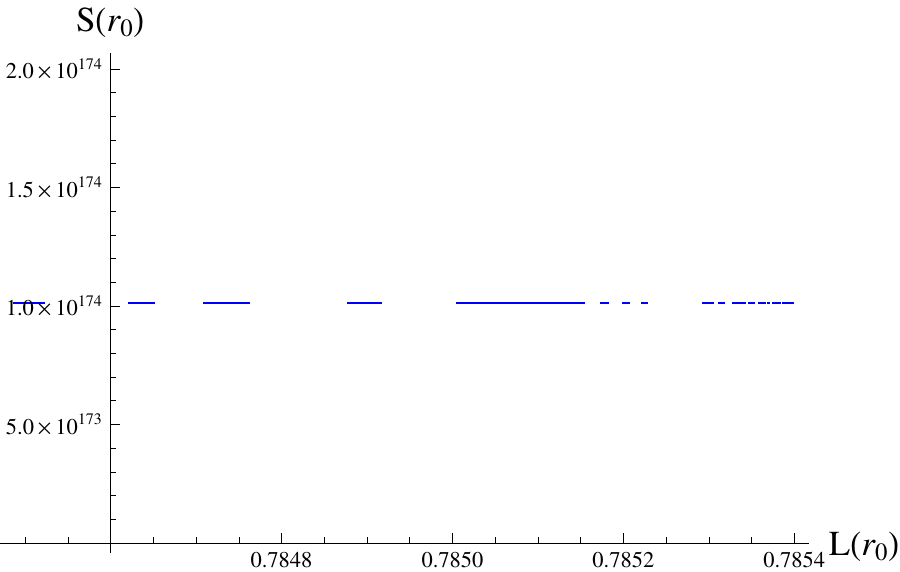}
  \captionof{figure}{Plot of $S$ vs. $L(r_0)$ \\ for the MN model. }
  \label{fig:SMN}
\end{minipage}
\end{figure}

\subsection{Klebanov-Witten}
As the Klebanov-Witten geometry is not confining, it would be interesting to compare the behavior of the functions $\beta$ and $H$ and also $L$ and $S$ with the confining geometries studied above. 

As one can see in Figs. (\ref{fig:LKW}) and (\ref{fig:KWEE}), there is no phase transition in the plot of $S$ and the true solution is the connected one. Also, there is no peak in the plot of $L$ which again specifies that KW is indeed a conformal geometry. This can be compared with the diagram of Fig. (\ref{fig:11a}).

\begin{figure}[h!]
\centering
\begin{minipage}{.5\textwidth}
  \centering
  \includegraphics[width=.9\linewidth]{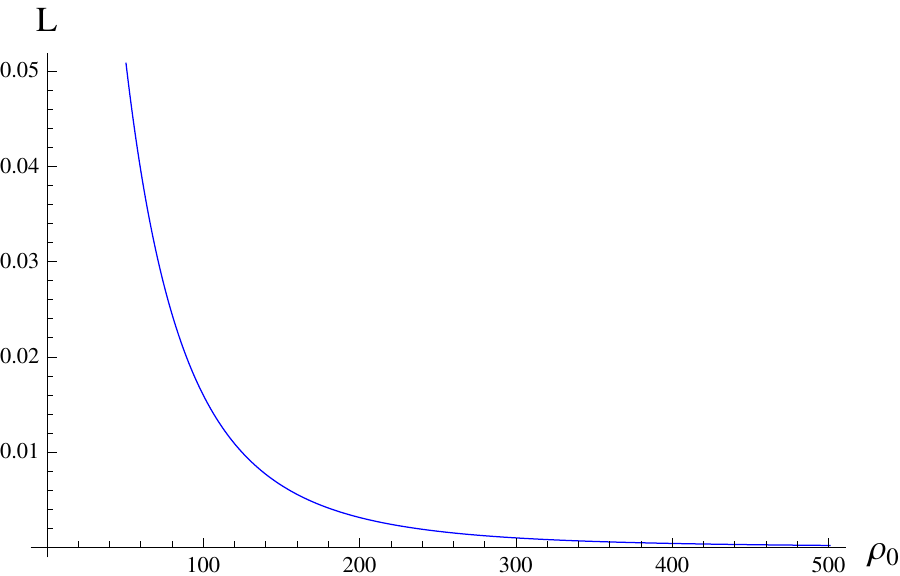}
  \captionof{figure}{Plot of $L$ vs. $\rho$ for KW. }
  \label{fig:LKW}
\end{minipage}%
\begin{minipage}{.5\textwidth}
  \centering
  \includegraphics[width=.9\linewidth]{KW-EE}
  \captionof{figure}{Plot of $S$ vs. $L$ for KW. }
  \label{fig:KWEE}
\end{minipage}
\end{figure}

\section{The critical electric field in the presence of magnetic field}\label{sec:five}
Now in this section, we assume that in addition to the electric field, a parallel and a perpendicular magnetic field components are also present. By using the Euler-Heisenberg Lagrangian, we then study the critical electric field which would lead to the Schwinger pair creation in our four confining geometries. We see that similar to the Sakai-Sugimoto and deformed Sakai-Sugimoto models, the parallel component would increase the pair creation rate and the perpendicular component would decrease it.

 \subsection{Maldacena-Nunez}
For the MN metric we have
\begin{gather}
\sigma _{\text{string}-\text{MN}}=\frac{e^{\phi _0}}{2 \pi  \alpha ' \sqrt{2}}\frac{\sqrt{ \text{sinh}\left(2r_*\right)}}{\left(r^* \text{coth}\left(2r_*\right)-\frac{r_*{}^2}{\sinh^2 \left(2r_*\right)}-\frac{1}{4}\right){}^{\frac{1}{4}}}.
\end{gather}
Based on the assumption that a geometry has an IR wall and the probe D-branes would hit the wall, by calculating the imaginary part of the DBI action with a constant field strength, the authors in \cite{Hashimoto:2014yya} showed that the critical electric field, where the Euler-Heisenberg Lagrangian becomes imaginary and therefore the Schwinger effect start to take place, is at $E_c$ which is,
\begin{gather} \label{critical E}
E_{c}=\sigma_{\text{string}} \sqrt {\frac{{\sigma^2}_{\text{string}}+|\vec{B}|^2 }   {\sigma_{\text{string} }^2+|\vec{B}_{||}|^2 }  }.
\end{gather}
Here $\sigma_{\text{string}}=\frac {g (r_{*} ) } {2 \pi \alpha}$ and $r_{*}$ is where the DBI action vanishes.
By calculating Eq. (\ref{critical E}),  one can get
\begin{gather}
E_{c-MN}=\frac{e^{\phi _0} \sqrt{\sinh  \left(2 r_*\right)}}{2 \pi  \alpha ' \left(4 \coth \left(2 r_*\right)r_* -1-4 \text{csch}^2\left(2 r_*\right) r_*^2\right){}^{\frac{1}{4}} } \times \nonumber\\  \left(\frac{e^{2 \phi _0} \sinh \left(2 r_*\right)+4\pi ^2 \alpha '^2B^2\text{  } \sqrt{4 \coth \left(2 r_*\right) r_*-1-4 \text{csch}^2\left(2 r_*\right) r_*^2} }{e^{2 \phi _0} \sinh \left(2 r_*\right)+4\pi ^2 \alpha '^2 {B_{||}}^2 \sqrt{4 \coth \left(2 r_*\right) r_*-1-4 \text{csch}^2\left(2 r_*\right) r_*^2} }\right)^{\frac{1}{2}}.
\end{gather}
For $\vec{B_{||}}=\vec{B}=0$ the critical $E$ is
\begin{gather}
E_{cr-MN}=\frac{e^{\phi _0} \sqrt{\sinh  \left(2 r_*\right)}}{2 \pi  \alpha ' \left(4 \coth \left(2 r_*\right)r_* -1-4 \text{csch}^2\left(2 r_*\right) r_*^2\right){}^{\frac{1}{4}} }.
\end{gather}

We can rederive this result by finding the imaginary part of the Euler-Heisenberg action and thus, we can find an indicator for the universality of the equation (\ref{critical E}) for the confining geometries.

We then calculate the DBI action by finding the induced metric of MN on D3-brane or D7-brane profile. The D3-brane profile is
\begin{gather}
ds^2=(1+\frac{R^4}{r^4})^{-\frac{1}{2}}(-dt^2+d {\vec {x}}^2)+(1+\frac{R^4}{r^4})^{\frac{1}{2}} (dr^2+r^2 d{\Omega_5}^2), 
\end{gather}
and the D7-brane profile is
\begin{gather}
ds_{10}^2=\frac{r^2}{R^2} dx_\mu dx^\mu +\frac{R^2}{r^2} ds_{(6)}^2,\nonumber\\
ds_{(6)}^2=dr^2+\frac{r^2}{3} \big(\frac{1}{4} ({w_1}^2 +{w_2}^2 ) +\frac{1}{3}{w_3}^2+ (d\theta-\frac{1}{2} f_2)^2 +(\sin \theta d\phi -\frac{1}{2} f_1)^2 \big),
\end{gather}
where $R^4=\frac{27}{4} \pi g_s N_c {\ell_s}^4$ and $R$ is the $\text{AdS}_5$ radius.

Now, we calculate the DBI action by finding the induced metric on the D7-brane. So,
\begin{gather}
\mathcal{L}_{\text{MN}-\text{D7}}=-T_7\int  d^2x\text{  }\text{d$\Omega $}_5 E^{-\phi _0}\int _{r_{\text{kk}}}^{\infty }dr \frac{(r \coth (2r)-\frac{r^2}{\sinh (2r)}-\frac{1}{4})^{\frac{1}{4}}}{\sqrt{\frac{1}{2} \sinh (2r)}}\times \nonumber\\ 
\Bigg(\frac{2 e^{4 \phi _0} \sinh ^4(2 r)}{1-8 r^2-\cosh (4 r)+4 r \sinh (4 r)}- \nonumber\\ \frac{(2 \pi  \alpha )^2 e^{2 \phi _0} (F_{01}{}^2-F_{12}{}^2-F_{13}{}^2-F_{23}{}^2) \sinh (2 r)}{\sqrt{4 r \coth (2 r)-4 r^2 \text{csch}^2(2 r)-1}}-  (2 \pi  \alpha )^4F_{01}{}^2 F_{23}{}^2 \Bigg)^{\frac{1}{2}}.
\end{gather}

For $F_{01}=E_1, F_{23}=B_1, F_{13}=B_2, F_{12}=B_3$ we derive the Lagrangian as
\begin{gather}
\mathcal{L}_{\text{MN}-\text{D7}}=-T_7\int  d^2x\text{  }\text{d$\Omega $}_5 E^{-\phi _0}\int _{r_{\text{kk}}}^{\infty }dr \frac{\left(r \coth (2r)-\frac{r^2}{\sinh (2r)}-\frac{1}{4}\right)^{\frac{1}{4}}}{\sqrt{\frac{1}{2} \sinh (2r)}}\times  \nonumber\\ \sqrt{\frac{2 e^{4 \text{$\phi $0}} \sinh ^4(2 r)}{1-8 r^2-\cosh (4 r)+4 r \sinh (4 r)}+\frac{(2 \pi  \alpha )^2e^{2 \text{$\phi $0}} \left({\vec{B}}^2-E_1{}^2\right)\text{  }\sinh (2 r)}{\sqrt{4 r \coth (2 r)-4 r^2 \text{csch}^2(2 r)-1}}-(2 \pi  \alpha )^4 B_1{}^2 E_1{}^2}.
\end{gather}
By solving the $E$ which makes this Lagrangian imaginary, we find the same $E$ as Eq. (\ref{critical E}). This way the universality of this relation can be checked again.

\begin{figure}[h!]
  \centering
    \includegraphics[width=0.5\textwidth]{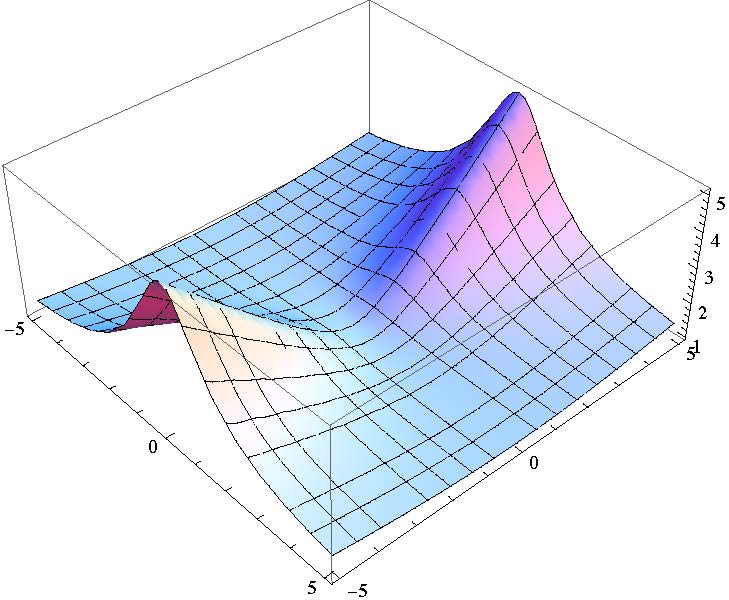}
       \caption{The $E_{cr}$ versus parallel and perpendicular magnetic field components for $\alpha=1$, $\phi_0=0$ and $r=2$.}
       \label{fig:critical}
\end{figure}

From Fig. \ref{fig:critical} one can see that the behavior of the critical $E$ versus the magnetic fields is very similar to the general figure as shown in \cite{Hashimoto:2014yya}. 

Additionally, the imaginary part of the Lagrangian versus the perpendicular and parallel magnetic field is shown in Fig. 24, the imaginary part of the Lagrangian versus the parallel electric field and perpendicular magnetic field is shown in Fig. 25 and the imaginary part of the Lagrangian versus parallel electric field and parallel magnetic field is shown in Fig. 26.

\begin{figure}[h!]
  \centering
    \includegraphics[width=0.5\textwidth]{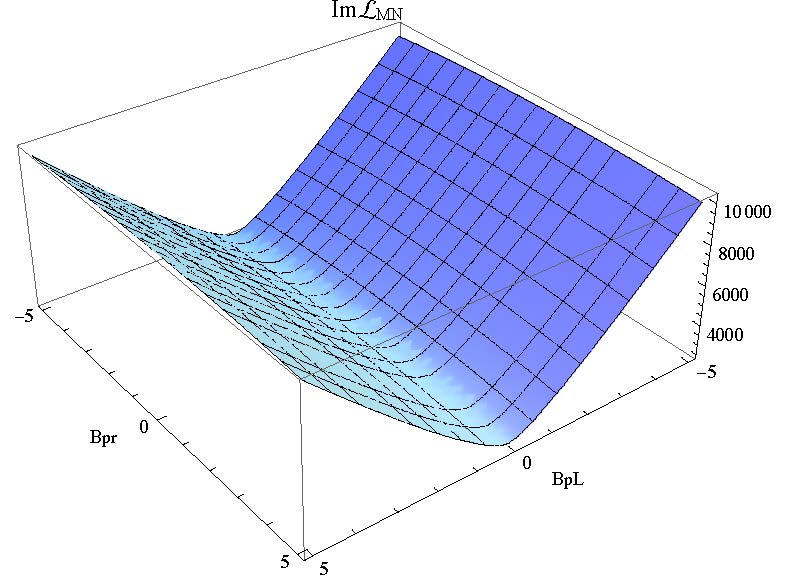}
       \caption{The Im$\mathcal{L}$ for the MN background, vs. parallel and perpendicular magnetic fields for $\alpha=1$, $\phi_0=0$ and $E=10$. (We have normalized the fields.) }
\end{figure}

One can specifically check that for this background, by increasing the parallel magnetic field the imaginary part of the EH Lagrangian would increase which leads to an increase in the pair creation rate, but increasing the perpendicular magnetic field decreases the rate. 

Now we look at the electric field dependence. For the case of $B_{||}=0$, we have
\begin{gather}
\text{Im} \mathcal{L} _{perp.B} = \int_{0.05}^5  dr \left(r \coth (2 r)-r^2 \text{csch}(2 r)-\frac{1}{4}\right)^{\frac{1}{4}} \times \nonumber\\  \sqrt{\frac{4 \sinh ^3(2 r)}{8 r^2+\cosh (4 r)-4 r \sinh (4 r)-1}-\frac{2 (2\pi  \alpha )^2 \left(B_{\text{pr}}{}^2-E_{\|}{}^2\right)\text{  }}{\sqrt{4 r \coth (2 r)-4 r^2 \text{csch}^2(2 r)-1}}},
\end{gather}
\begin{figure}[h!]
  \centering
    \includegraphics[width=0.5\textwidth]{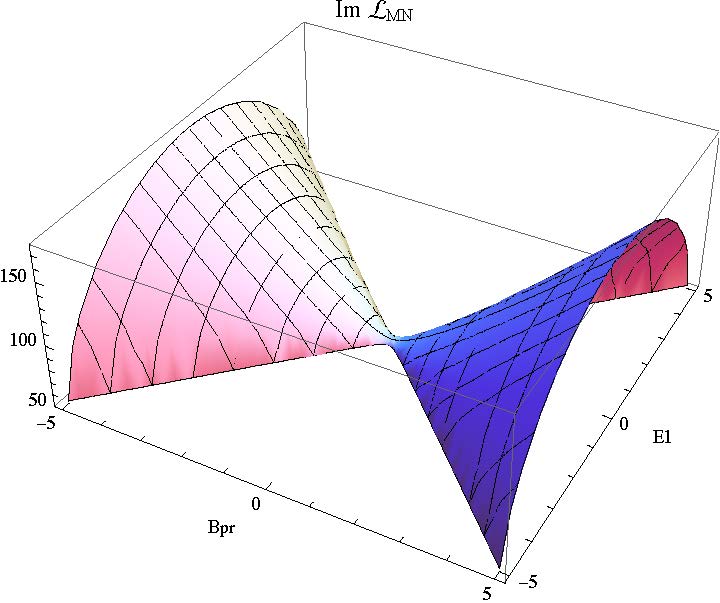}
       \caption{The Im$\mathcal{L}_{MN}$ versus electric field and perpendicular magnetic fields for  $B_{||}=0$, $\alpha=1$, $\phi_0=0$ and $r=2$  .}
\end{figure}
and for the case of $B_{\perp}=0$, we have
\begin{gather}
\text{Im}\mathcal{L} _{para.B} = \int_{0.05}^5  dr \left(r \coth (2 r)-r^2 \text{csch}(2 r)-\frac{1}{4}\right)^{\frac{1}{4}}\times \nonumber\\  \sqrt{\frac{2(2\pi \alpha )^4 B_{\|}{}^2 E_{\|}{}^2}{\sinh (2 r)}+\frac{4 e^{4 \phi _0} \sinh ^3(2 r)}{8 r^2+\cosh (4 r)-4 r \sinh (4 r)-1}-\frac{2(2\pi \alpha)^2 e^{2 \phi _0} \left(B_{\|}{}^2-E_{\|}{}^2\right)}{\sqrt{4 r \coth (2 r)-4 r^2 \text{csch}^2(2 r)-1}}}.
\end{gather}

\begin{figure}[h!]
  \centering
    \includegraphics[width=0.5\textwidth]{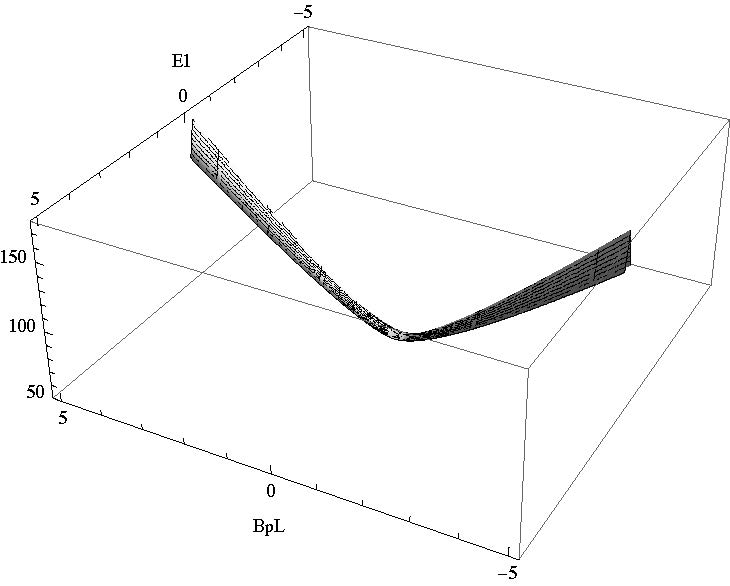}
       \caption{The Im$\mathcal{L}_{MN}$ versus electric field and parallel magnetic fields for $B_{\perp}=0$,  $\alpha=1$, $\phi_0=0$ and $r=2$.}
\end{figure}

\subsection{D3 probe brane in MN background}
The pullback of the Maldacena Nunez metric (which consists of N D5-brane) on D3-brane world volume is $e^\phi \eta_{\mu\nu}$. So the Lagrangian is
\begin{gather}
\mathcal{L}_{\text{MN}-\text{D3}}=-T_3\int  d^4x\text{  }\surd (e^{-4 \phi _0} (-(2\pi  \alpha )^4 e^{4 \phi _0} F_{01}{}^2 F_{23}{}^2+(-1+4 r \coth (2 r)) \text{csch}^2(2 r)-\nonumber \\4 r^2 \text{csch}^4(2 r) -  (2\pi  \alpha )^2 e^{2 \text{$\phi $0}} (F_{01}{}^2-F_{12}{}^2-F_{13}{}^2-F_{23}{}^2) \text{csch}(2 r) \times \nonumber \\\sqrt{-1+4 r \coth (2 r)-4 r^2 \text{csch}^2(2 r)})).
\end{gather}

There is no integral of ``r" for the D3-brane case, so by inserting D3-brane in the geometry as the probe, we can study the Schwinger effect in different $r$. First, we study the potential in MN geometry when all the electric and magnetic fields are off which is shown in Fig. (\ref{fig:curve}).

\begin{figure}[h!]
  \centering
    \includegraphics[width=0.5\textwidth]{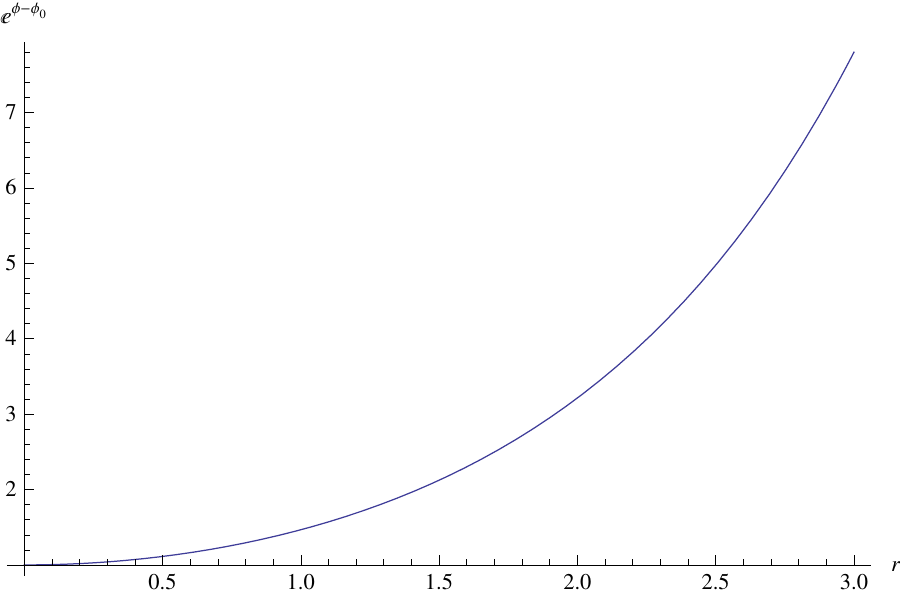}
       \caption{The potential on the D3-brane when all the fields are off.}
       \label{fig:curve}
\end{figure}
As one can see, due to the effects of $F_5$ and gravitational force, all the D3-branes are being pulled to the tip of $r = 0$ when they are inserted in MN and KS geometries. 

Now we investigate the potential when the fields are on.
If we take $F_{01}=E_1, F_{12}=B_3, F_{23}=B_1 , F_{13}=B_2$ then the potential is
\begin{gather}
V= \Big (e^{-4 \phi _0} (-(2\pi  \alpha )^4 B_1{}^2 e^{4 \phi _0} E_1{}^2+\text{csch}(2 r) ((2\pi  \alpha )^2 e^{2 \phi _0} (B_1{}^2+B_2{}^2+B_3{}^2-E_1{}^2) \times \nonumber\\ \sqrt{-1+4 r (\coth (2 r)-r \text{csch}^2(2 r))}+\text{csch}(2 r) (-1+4 r (\coth (2 r)-r \text{csch}^2(2 r))))) \Big ){}^{\frac{1}{2}}.
\end{gather}

\begin{figure}[h!] 
  \centering
    \includegraphics[width=0.5\textwidth]{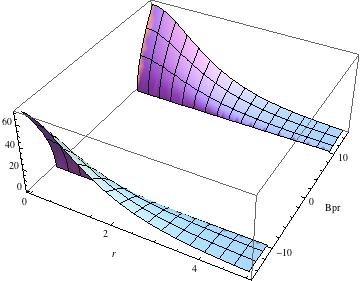}
       \caption{The potential on the D3 brane for $E_{||}=10, B_{||}=0, \alpha=1, \phi_0=0$. }
       \label{fig:pot}
\end{figure}
This is shown in Fig. (\ref{fig:pot}). One can see that the potential is looking like two walls which have higher slopes near the IR region ($ r\to 0$). In the UV ($r\to \infty$) the potential is zero and has zero slope. Increasing $E_{||}$ would increase the slope of the potential and the pair creation rate.

\begin{figure}[h!]
  \centering
    \includegraphics[width=0.5\textwidth]{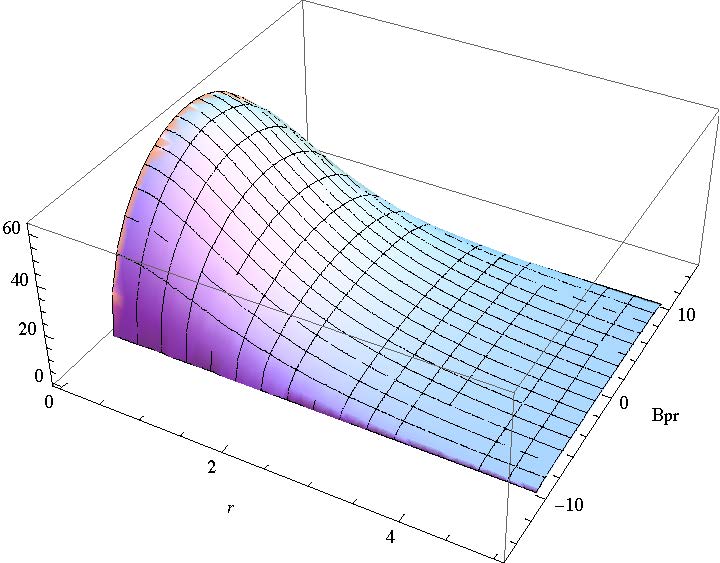}
       \caption{The imaginary part of the Lagrangian related to the pair creation for $E_{||}=10, B_{||}=0,  \alpha=1, \phi_0=0$. }
        \label{fig:Impot}
\end{figure}

Again, from Fig. (\ref{fig:Impot}) one can notice that the pair creation happens with higher rate near the origin, at $r \to 0$, and at bigger $r$ it would decrease. Increasing the perpendicular magnetic field would decrease the pair creation rate, until at any $r$ make it zero at a specific value of the perpendicular magnetic filed.

Now we turn off the perpendicular magnetic field and study the effect of the parallel component of the magnetic field.  As it can be seen from Figs. (\ref{pairz2}), (\ref{interB2}), the parallel magnetic field generally increases the whole pair creation rate, however at $B_{||}=0$ the pair creation is mainly happening at small $r$. Increasing $B_{||}$ increases the pair creation in the UV and decreases the area where the pair creation is happening in the IR until making it zero there, but in total, increasing $B_{||}$ would increase the imaginary part of the Lagrangian and therefore the rate of the Schwinger effect.

\begin{figure}[h!]
  \centering
    \includegraphics[width=0.5\textwidth]{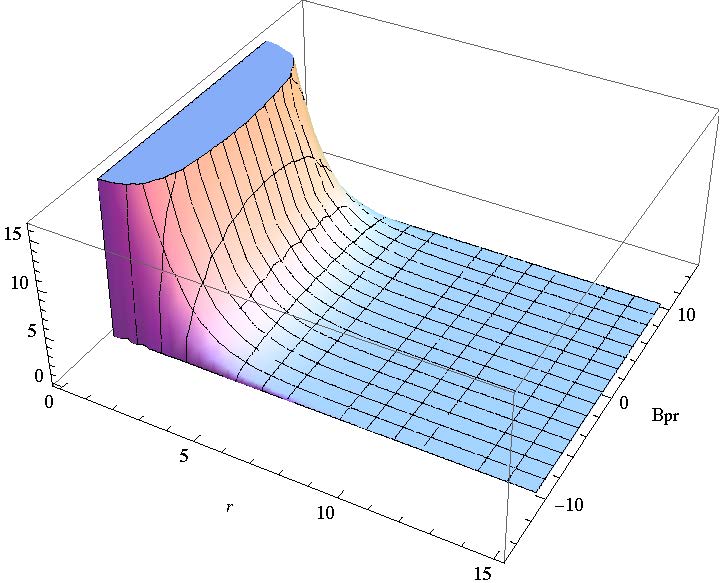}
       \caption{The imaginary part of the Lagrangian for $E_{||}=10, B_{||}=0, \alpha=1, \phi_0=0$. }
       \label{pairz2}
       \end{figure}
       \begin{figure}[h!]
  \centering
    \includegraphics[width=0.5\textwidth]{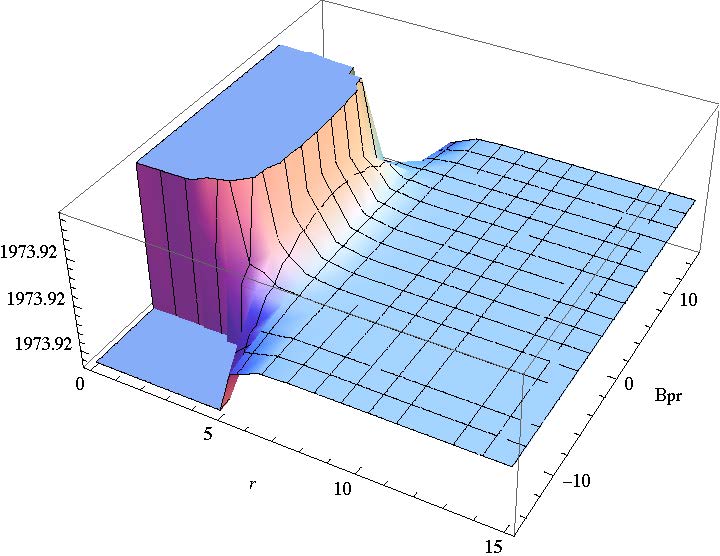}
       \caption{The imaginary part of the Lagrangian for $E_{||}=10, B_{||}=3, \alpha=1, \phi_0=0$. }
       \label{interB2}
\end{figure}

It would be interesting to notice that when $E_{||}=0$, even with a strong $B_{||}$, no pair creation happens. Only at small $r$, in the IR region, increasing the perpendicular magnetic field would increase the potential. 

One should notice that heavier states with higher charges lie in the IR as the hadrons' wave functions fall as $r^{-\Delta}$ and $\Delta \sim J$. This can be one reason that we see such a behavior in the IR. One should note that Wilson loops also find it more favorable to lie at the end of the space in the infrared region. 

The plot of $- \text{Det}(F_{MN} + g_{MN})$ versus $r$ and $B_{\perp}$, for a constant $E = 10$, and a zero parallel magnetic field, Fig. (\ref{fig:Potz}), shows that there exists a critical $r$ ( for $B_{\perp} =0$ is around $r= 1.2$) where a hole is forming.  For confining theories there is a critical $r_0$ where $\frac{\partial g_{tt} }{ \partial r} \big |_{r=r_0}=0$ which for our specific geometry gives $r=1.118$ which is close to what we have seen from the figures. Increasing $E$ would increase the radius of this hole.

     \begin{figure}[h!]
  \centering
    \includegraphics[width=0.5\textwidth]{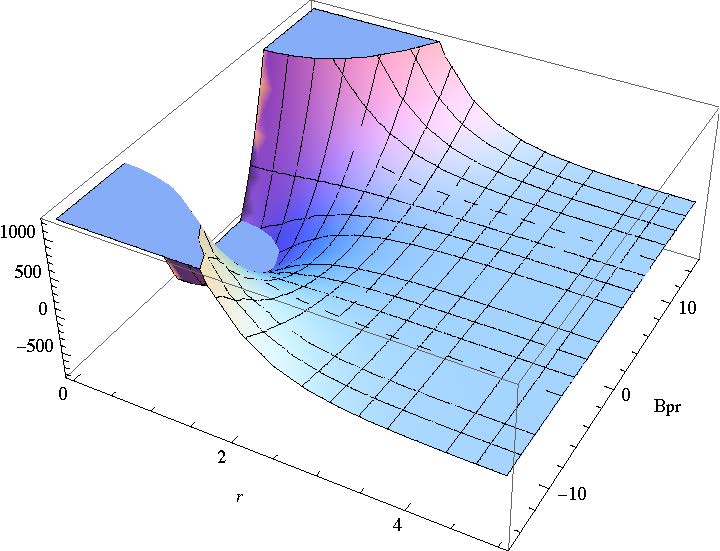}
       \caption{The potential for $E_{||}=10$ and $B_{||}=0$ showing the hole in the IR region.}
       \label{fig:Potz}
\end{figure}

\subsection{Klebanov-Strassler}
The Klebanov-Strassler (KS) metric is 
\begin{gather}
ds_{10}^2=h^{-\frac{1}{2}}(\tau) dx_\mu dx^\mu+h^{\frac{1}{2}}(\tau) {ds_6}^2.
\end{gather}
The form of this metric also looks like (\ref{eq:form}), so the D-branes hit the IR wall as it is a confining geometry. 

The Klebanov-Strassler string tension is
\begin{gather}
\sigma _{\text{string}-\text{KS}}=\frac{h^{\frac{-1}{2}}\left(\tau _*\right)}{2 \pi  \alpha '}.
\end{gather}
Using this and the Eq. (\ref{critical E}), we can find the critical $E$ as 
\begin{gather}
E_{cr}=\frac{1}{2 \pi  \alpha '}\sqrt{\frac{\frac{1}{h(\tau )}+\left( 2\pi  \alpha ' B\right)^2 }{1+\left(2\pi  \alpha 'B_{\|}\right){}^2 h(\tau )}}.
\end{gather}
For $\vec{B}=\vec{B_{||} }=0$, as it was obvious from $E_{cr}=\frac{g(r_*)}{2 \pi \alpha^\prime} $, we would have  $E_{cr}=\frac{h^{-\frac{1}{2} }(\tau)}{2 \pi \alpha^\prime}$.

The behavior of the function $h(\tau)$ is shown in Fig. (\ref{fig:hh}). It shows that after $\tau >10$ it is practically zero. So for showing the numerical plot we do not actually need to take the integral of the Lagrangian to $\tau=\infty$, but rather we take the integral from zero to $\tau=10$.
\begin{figure}[h!] \label{fig:h}
  \centering
    \includegraphics[width=0.5\textwidth]{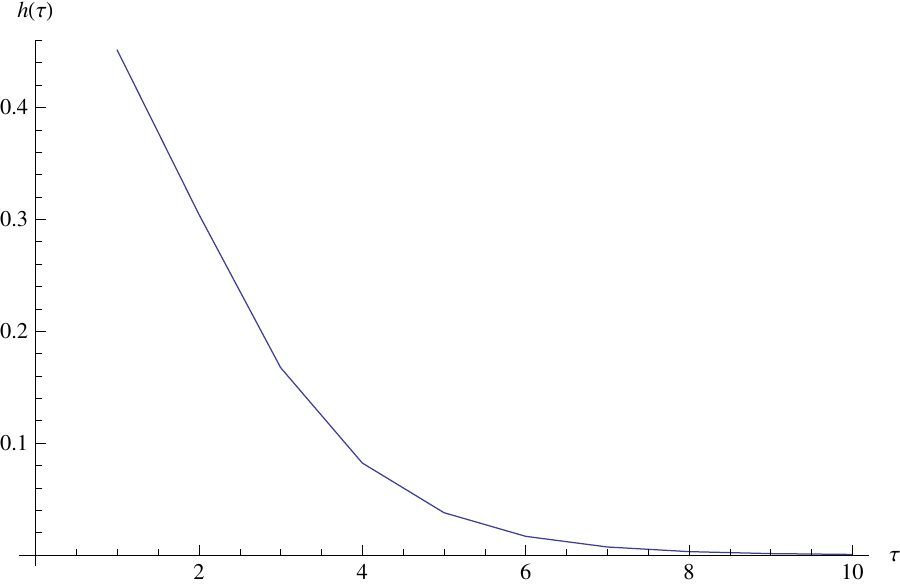}
       \caption{The behavior of the function $h(\tau)$ vs $\tau$ for the Klebanov-Stressler metric. }
       \label{fig:hh}
\end{figure}
So the Lagrangian for the constant field strength is 
\begin{gather}
\mathcal{L}_{\text{KS}}=-T_7\int  d^4x\text{  }\text{d$\Omega $}_3 e^{-\phi _0} \times \nonumber\\  \int _{\tau _{\text{kk}}}^{\infty }d\tau \sqrt{\frac{1+(2\pi \alpha)^2 \left(B_{||}{}^2+B_{\perp}{}^2-E_{\|}{}^2\right)  h(\tau )-(2\pi \alpha)^4 B_{\text{pL}}{}^2 {E_{||}}^2  h(\tau )^2}{h(\tau )^2}}.
\end{gather}

\begin{figure}[h!] \label{fig:h}
  \centering
    \includegraphics[width=0.5\textwidth]{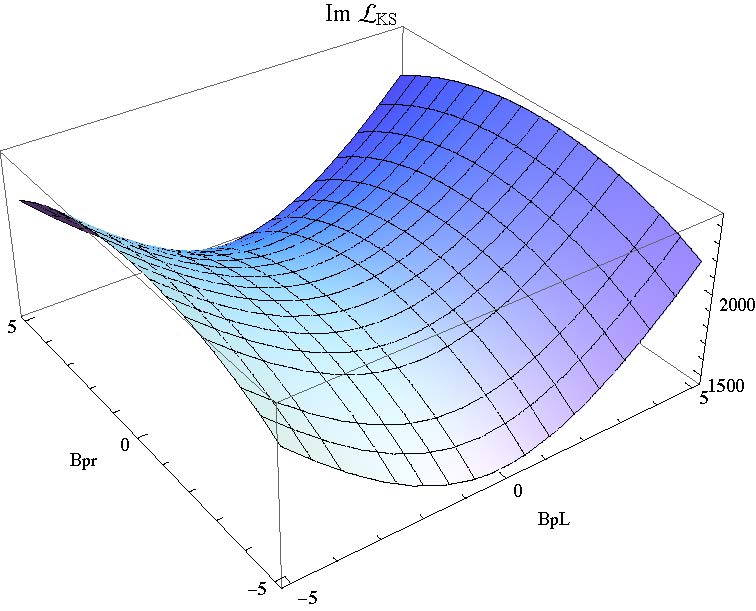}
       \caption{ The Im $\mathcal{L}_{KS}$ vs $B_{\perp}$ and $B_{||}$ of the KS background for $\tau=9.5$, $h(\tau)= 0.000798174$, and $E_{||}=10$, $\alpha=1$. }
       \label{fig:IMKS}
\end{figure}

Again as in Fig. (\ref{fig:IMKS}) one can check that for the KS background, by increasing the parallel magnetic field, the imaginary part of the Lagrangian would increase leading to an increase in pair creation and vice versa for the perpendicular magnetic field, so by increasing the perpendicular magnetic field, the imaginary part of the Lagrangian would decrease leading to a more stable phase with lower rate of pair creation. The plot of imaginary part of the Lagrangian versus the parallel electric field and parallel magnetic field is shown in Fig. (\ref{IMKS2}).

\begin{figure}[h!] \label{fig:h}
  \centering
    \includegraphics[width=0.5\textwidth]{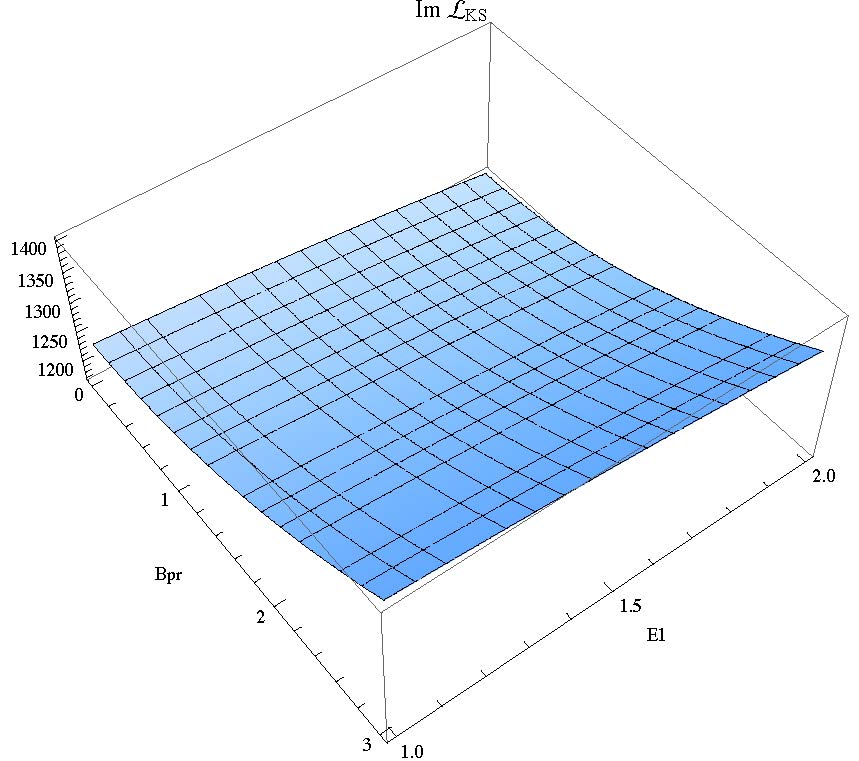}
       \caption{The Im $\mathcal{L}_{KS}$ vs. $B_{\perp}$ and $E$ of the KS background for $\alpha=1$, $\tau=9.5$, $h(\tau)= 0.000798174$. }
       \label{IMKS2}
\end{figure}

\subsection{Witten-QCD}

The Witten-QCD metric is
\begin{gather}
ds^2=\left(\frac{u}{R}\right)^{\frac{3}{2}}(\eta_{\mu\nu} dx^\mu dx^\nu)+\left(\frac{R}{u}\right)^{\frac{3}{2}}\frac{du^2}{f(u)}+\left(\frac{u}{R}\right)^{\frac{3}{2}}\frac{4R^3}{9u_0} f(u) d\theta^2+R^{\frac{3}{2}} u^{\frac{1}{2}} d\Omega_4^4.
\end{gather}
Since the generic form of the confining geometries is 
\begin{gather} \label{eq:form}
ds^2=g(r) \eta_{\mu\nu} dx^\mu dx^\nu+f(r) dr^2 +h(r) [\text{internal space}],
\end{gather}
in the Witten QCD model, the internal geometry mixes with the radial coordinate $u$, but still it is confining and the probe D-branes hit the IR wall. 

For WQCD, $\sigma_{st}=\frac{1}{2\pi \alpha'} (\frac{u_*}{R})^{\frac{3}{2}}$ which is consistent with our calculation of the potential in the previous sections.

Assuming $T_F=\frac{1}{2\pi \alpha'}$, from the above equation one would get
\begin{gather}
E_c=T_F\left (\frac{u_{*}}{R}\right)^{\frac{3}{2}} \sqrt{\frac{T_F^2 (\frac{u_*}{R} )^3+\vec{B}^2  }{T_F^2 (\frac{u_*}{R} )^3+\vec{B_{||}}^2} },
\end{gather} 
and when the magnetic field is off, this leads to the familiar result $E_c=\sigma_{st}$. Now the DBI action in the D5-brane background including a constant electromagnetic field is
\begin{gather}
S_{D5}^{\text{DBI}} =-T_5 \int d^4 x du e^{-\phi} \sqrt{-\text{det}( P[g]_{ab}+2\pi \alpha' F_{ab})}=\nonumber\\
-\frac{2}{3} \frac{T_5 g_s}{R^{\frac{3}{4}}} \int_{u_{KK}} ^\infty du \sqrt{ \frac{u^6+4(B_\perp^2+B_\parallel^2-E^2)\pi^2 R^3 u^3 \alpha^2-16 B_\parallel^2 E^2 \pi^4 R^6 \alpha^4 }{f R^3 u_0} }.
\end{gather}

Again ,one can see that in this model too, by increasing the parallel magnetic field the imaginary part of the Lagrangian and therefore the rate of pair creation would increase, while increasing the perpendicular magnetic field would decrease it.

\subsection{Klebanov-Tseytlin}
The critical electric field in the presence of a magnetic field in the KT background is
\begin{gather}
E_c=T_F \frac{L^2}{r_*^2} \sqrt{\ln \frac{r_*}{r_s} } \sqrt{ \frac{T_F^2 \frac{L^4}{r_*^4 } \ln \frac{r_*}{r_s} + |\vec{B}|^2  }{T_F^2 \frac{L^4}{r_*^4} \ln \frac{r_*}{r_s} +| \vec{B}|^2}  },
\end{gather}
and the DBI action in the D5-brane background is 
\begin{gather}
S_{D5}^{DBI}= -\frac{T_5}{L^2} \int_{r_{KK}}^\infty dr  \sqrt{ \frac{r^8 +4(B_\perp^2+B_\parallel^2-E^2) L^4 \pi^2 r^4 \alpha^2 \ln \frac{r}{r_s} -16 B_\parallel^2 E^2 L^8 \pi^4 \alpha^4 (\ln(\frac{r}{r_s})  )^2 }{r^2  \ln \frac{r}{r_s} } },
\end{gather}
where again as our general result, one can check that increasing the parallel magnetic field would increase the rate of pair creation and increasing the perpendicular magnetic field decreases the rate.

\chapter{Phase transitions in BHT Massive Gravity} \label{chap4}

Understanding quantum field theories with momentum dissipation in the context of the AdS/CFT correspondence is a crucial step toward applications to realistic condensed matter systems. So, in this regard, holographic massive gravity theories (HMGs) have been exploited. Using these models one can study different field theory features such as DC resistivity, relaxation rate or the effect of dissipation or disorder on the generic phase transitions in strongly correlated systems \cite{Adams:2014vza}.

There are various massive gravity models with multiple geometrical solutions and their corresponding field theory duals. One of these theories is ``Topological Massive Gravity'' (TMG), which is the Einstein action plus a parity breaking Chern-Simons term. Recently in \cite{Detournay:2015ysa}, the Hawking-Page phase transitions between the $\text{AdS}_3$ and BTZ solutions, and warped $\text{AdS}_3$ and warped BTZ black hole solution of TMG were investigated and the Gibbs free energies, local and global stability regions and the phase diagrams were presented.

Another rich theory is the parity preserving Bergshoeff-Hohm-Townsend (BHT) or the ``New Massive Gravity'' (NMG), which in addition to the thermal warped $\text{AdS}_3$ and warped BTZ black hole, has many different solutions as well. The aim of this chapter is, similar to \cite{Detournay:2015ysa}, we study the Hawking-Page phase transitions between different solutions of NMG and therefore learn more about the properties of the dual CFTs. Particularly, we study the phase transitions between the thermal AdS and BTZ black holes, the warped AdS and warped BTZ black holes in two different ensembles, the Lifshitz black hole and the new hairy black hole and their corresponding vacua.

Another motivation is to extend the AdS/CFT duality to more general geometries. One would think that for doing so, the most direct way is to perturbatively deform the $\text{AdS}_3$ manifold to a  warped $\text{AdS}_3$ geometry \cite{Rooman:1998xf, Israel:2003cx, Detournay:2010rh}, and then study the dual field theory. The initial works on this extension were done in \cite{Israel:2004vv}, where the authors studied the magnetic deformation of $S^3$ and the electric/magnetic deformations of $\text{AdS}_3$ which still could remain a solution of string theory vacua. Then in \cite{Anninos:2008fx, Azeyanagi:2012zd, Song:2011sr}, the dual field theories were studied. In \cite{Song:2011sr}, the dual of warped $\text{AdS}_3$ was suggested to be the IR limit of the nonlocally deformed 2D D-brane gauge theory or the dipole CFT. Constructing this duality could lead to more information about the properties of these new field theories and also some properties of the dual bulk geometries, for instance the nature of the closed time-like curves (CTCs). 

The Bergshoeff-Hohm-Townsend (BHT) gravity has both warped AdS and warped BTZ black hole solutions. The deformed $\text{AdS}_3$ preserves the $SL(2,R) \times U(1)$ subgroup of $SL(2,R) \times SL(2,R)$ isometries. The obtained space-times called null, time-like or space-like warped $\text{AdS}_{3}$ (WAdS$_{3}$) corresponding to the norm of $U(1)$ killing vectors, where the time-like WAdS$_{3}$ is just the G\"{o}del spacetime \cite{Rooman:1998xf, Reboucas:1982hn}. 

There have been several proposals generalizing the AdS/CFT correspondence. Those include: AdS/CMT (Condensed Matter Theory), AdS/QCD, dS/CFT, flat space holography, Kerr/CFT, etc \cite{Ghodrati:2015rta, Ghodrati:2014spa}. However, the dual CFT of these theories are not completely known. The advantages of WCFTs are that they posses many properties of CFTs and they can be derived from string theory and low-dimensional gravity theories and hence for studying them known CFT techniques could be deployed.

The specific properties of this new class of WCFTs were studied in \cite{Detournay:2012pc} and their entanglement entropies were first studied in \cite{Anninos:2013nja} holographically and in a more recent work, in \cite{Castro:2015csg}, by using the Rindler method of WCFT. To further study this WAdS/WCFT duality, one could study other properties such as the instabilities of the solutions and the Hawking-Page phase transitions \cite{Hawking:1982dh}. As the phase transitions from the thermal AdS or WAdS, to BTZ or warped BTZ black hole is dual to confining/deconfining phase transitions in the dual field theory, these models could be used in theories similar to QCD or condensed matter systems with dissipations.

The plan of this chapter is as follows. First, in section (\ref{sec:ADTcharge}), we review two methods of finding the conserved charges for any solution of NMG which are the ADT formalism and the $SL(2,R)$ reduction method. Mainly we use the general formulas from $SL(2,R)$ reduction method to calculate the conserved charges for any solution of NMG in different ensembles. Then in section (\ref{sec:AdSPhase}), by finding the free energies, we discuss the phase transitions between the vacuum $\text{AdS}_3$ and BTZ black hole solutions. We discuss the thermodynamics and local and global stability regions. In section (\ref{sec:sec2}), we calculate the free energies of warped $\text{AdS}_3$ vacuum and warped BTZ black hole solutions in quadratic/non-local ensemble and in section (\ref{sec:grand}), we discuss the Hawking-Page phase diagrams in grand canonical ensemble. We calculate the free energy of the $\text{WAdS}_3$ by three different methods and by doing so we could find a factor in the modular parameter which extends the result of \cite{Kraus:2005vz} for calculating the free energy of $\text{WAdS}_3$ solutions in NMG. Then we present the phase diagrams of these solutions. In section (\ref{sec:hairy}), we discuss the free energy and phase transitions of the Lifshitz and the new hairy black hole solutions. We also discuss the inner horizon thermodynamics in section (\ref{sec:inner}) and in section (\ref{entanglement}), we discuss the entanglement entropy of the vacuum solutions corresponding to the $\text{WCFT}_2$ dual of $\text{WAdS}_3$ in NMG.

\section{The Bergshoeff-Hohm-Townsend Theory }\label{sec:intro}

The Bergshoeff-Hohm-Townsend (BHT) or the new massive gravity (NMG) is a higher-curvature extension of the Einstein-Hilbert action in three dimensions which is diffeomorphism and parity invariant. In the linearized level, it is equivalent to the unitary Pauli-Fierz action for a massive spin-2 field \cite{Bergshoeff:2009hq}.

The action of NMG is
\begin{gather}\label{eq:action}
S=\frac{1}{16 \pi G_N} \int d^3x \sqrt{-g} \Big[ R-2\Lambda+\frac{1}{m^2} \Big( R^{\mu\nu} R_{\mu\nu}-\frac{3}{8} R^2 \Big)  \Big],
\end{gather}

where $m$ is the mass parameter, $\Lambda$ is a cosmological parameter and $G_N$ is the three-dimensional Newton constant. In the case of $m \to \infty$, the theory reduces to the Einstein gravity and in the limit of $m \to 0$, it is just a pure fourth-order gravity.  

The equation of motion following from the action is
\begin{gather}
R_{\mu \nu}-\frac{1}{2} R g_{\mu\nu} +\Lambda g_{\mu\nu}+\frac{1}{m^2}K_{\mu\nu}=0,
\end{gather}
with the explicit form of the tensor $K_{\mu\nu}$ as in \cite{Bergshoeff:2009hq},

\begin{gather}
K_{\mu\nu} = \nabla^{2}R_{\mu\nu}-\frac{1}{4}\left(
\nabla_{\mu}\nabla_{\nu}R+g_{\mu\nu}\nabla^{2}R\right)-4R^{\sigma}_{\mu}R_{\sigma\nu}
+\frac{9}{4}RR_{\mu\nu}+\frac{1}{2}g_{\mu\nu}\left
(3R^{\rho\sigma}R_{\rho\sigma}-\frac{13}{8}R^{2}\right).\nonumber\\
\end{gather}

The boundary terms of NMG which make the variational principle well-defined are \cite{Hohm:2010jc}\begin{gather}
S_{\text{Boundary}}=\frac{1}{16 \pi G} \int_\sigma d^3x \sqrt{-g} \left( f^{\mu \nu} (R_{\mu \nu}-\frac{1}{2} R g_{\mu \nu})-\frac{1}{4} m^2 (f_{\mu \nu} f^{\mu \nu}-f^2) \right),
\end{gather}
where $f_{\mu \nu}$ is a rank two symmetric tensor in the following form
\begin{gather}
f_{\mu \nu}=\frac{2}{m^2} (R_{\mu \nu}-\frac{1}{4} R g_{\mu \nu}).
\end{gather}

This theory admits a plethora of solutions, such as the vacuum $\text{AdS}_3$, warped $\text{AdS}_3$, BTZ black hole, asymptotic warped AdS black hole, Lifshitz, Schr$\ddot{\text{o}}$dinger and so on \cite{Bergshoeff:2009hq, AyonBeato:2009nh}. We construct the phase diagrams between several of these solutions by comparing the \textit{on-shell} free energies. 

By constructing the \textit{off-shell} free energies, one could even find all the states connecting any two solutions and therefore, create a picture of continuous evolutions of the phase transitions; similar to the work in \cite{Zhang:2015wna}, who, in the new massive gravity theory, studied the continuous phase transitions between the BTZ black hole with $M \ge 0$ and the thermal AdS soliton with $M=-1$. 

In the next section, we review how one can calculate the conserved charges and we employ several general formulas for the solutions of NMG which could be used to find the on-shell Gibbs free energies. In section (\ref{sec:AdSPhase}), we study the vacuum $\text{AdS}_3$ and BTZ solutions of NMG, the free energies and the phase diagrams. Then in section (\ref{sec:sec2}) and (\ref{sec:grand}) we discuss the warped solutions and in section (\ref{sec:hairy}) we study the new hairy black hole solution of this theory.

 \section{Review of calculating conserved charges in BHT}\label{sec:ADTcharge}
\label{sec:charges}
In three dimensions, the conserved charges associated to a Killing vector $\xi$ are
\begin{gather}
\delta Q_\xi [\delta g, g]=\frac{1}{16 \pi G} \int_0^{2\pi} \sqrt{-g} \epsilon_{\mu \nu \varphi} k_\xi^{\mu \nu} [\delta g,g] d\varphi.
\end{gather}
 
As calculated in \cite{Nam:2010ub} for BHT, the Abbott-Deser-Tekin (ADT) formalism would result in 
\begin{gather}
k_\xi ^{\mu \nu} =Q_R^{\mu \nu} +\frac{1}{2m^2} Q_K^{\mu \nu},
\end{gather}
where
\begin{gather}
Q_K^{\mu \nu} =Q_{R_2} ^{\mu \nu}-\frac{3}{8} Q_{R^2}^{\mu \nu},
\end{gather}
and the term for each charge is 
\begin{equation} \label{eq:x}
Q_R^{\mu \nu} \equiv \xi _\alpha \nabla^ {[ \mu} h^{ \nu ] \alpha}-\xi^{[\mu }\nabla_\alpha h^{\nu ] \alpha}-h^{\alpha [ \mu } \nabla _\alpha \xi^{\nu ]}+\xi ^{[\mu} \nabla^{\nu ]} h+\frac{1}{2} h \nabla^{[\mu} \xi ^{\nu ]},
\end{equation}

\begin{equation}
Q_{R^2} ^{\mu \nu}=2 R Q_R^{\mu \nu}+4\xi ^{[\mu} \nabla^{\nu ]} \delta R+2 \delta R \nabla ^{[\mu} \xi^{\nu]}-2\xi^{[ \mu} h^{\nu ] \alpha} \nabla_\alpha R,
\end{equation}
where
\begin{gather}
\delta R \equiv -R^{\alpha \beta} h_{\alpha \beta}+ \nabla^\alpha \nabla^\beta h_{\alpha \beta} -\nabla^2 h,
\end{gather}
and 
\begin{align}
Q_{R_2}^{\mu \nu}  & = \nabla^2 Q_R^{\mu \nu}+\frac{1}{2} Q_{R^2}^{\mu \nu} -2 Q_R^{\alpha [\mu} R_\alpha^{\nu ]}-2 \nabla^\alpha \xi^\beta \nabla_\alpha \nabla ^{[\mu} h_\beta ^{\nu ]}-4 \xi ^\alpha R_{\alpha \beta} \nabla^{[ \mu} h^{\nu ] \beta}-R h_\alpha^{[ \mu} \nabla^{\nu ]} \xi ^\alpha \nonumber\\ &
+2 \xi^{[ \mu} R_\alpha^{\nu ]} \nabla_\beta h^{\alpha \beta}+2 \xi_\alpha R^{\alpha [ \mu} \nabla_\beta h^{\nu ] \beta}+2 \xi^\alpha h^{\beta [ \mu} \nabla_\beta R_\alpha^{\nu ]}+2 h^{\alpha \beta} \xi^{[ \mu} \nabla_\alpha R_\beta^{\nu ]}  \nonumber\\ &
-(\delta R+2 R^{\alpha \beta} h_{ \alpha \beta} ) \nabla^{[\mu} \xi^{\nu ]}-3 \xi^\alpha R_\alpha^{[ \mu} \nabla^{\nu ]}h-\xi^{[ \mu} R^{\nu ] \alpha} \nabla_\alpha h. &
\end{align}

  For the three dimensional case with coordinates $(t, r, \phi)$, the mass and angular momentum in three dimensions are \cite{Nam:2010ub}
 
 \begin{gather}
 M=\frac{1}{4G} \sqrt{- \text{det} \ g} Q^{r t} (\xi_T) \Big |_{r\to \infty},  \ \ \ \ \ \ \ J=\frac{1}{4G} \sqrt{- \text{det} \ g} Q^{r t} (\xi_R) \Big |_{r\to \infty},
 \end{gather} 
where
 \begin{gather}
 \xi_T=\frac{1}{L} \frac{\partial}{ \partial t}, \ \ \ \ \ \ \   \ \ \ \  \xi_R= \frac{\partial}{ \partial \phi}.
 \end{gather}
 
 \subsection{The $SL(2,R)$ reduction method}\label{sec:chapSL}
 
One can also derive the charges by the $SL(2,R)$ reduction method which changes the metric to a $SO(1,2)$ form. 

For doing so one should write the metric in the form of \cite{Nam:2010ub}
 \begin{gather}
 ds^2=\lambda_{ab} (\rho) dx^a dx^b+\frac{d\rho^2}{\zeta^2 U^2(\rho) }, \ \ \ \ \ x^a=(t,\phi).
 \end{gather}
 Since there is a reparametrization invariance with respect to the radial coordinate, one can write the function $U$ such that $\det \lambda=-U^2$ and this would give $\sqrt{-g}=1/\zeta$. 
 One then applies the equations of motions (EOMs) and the Hamiltonian constraint and then by integrating the EOM one can derive the \textit{``super angular momentum"} vector.
 
 First, one parameterizes the matrix $\lambda$ as
 \begin{gather}
 \lambda_{ab}= \left(
\begin{array}{cc}
X^0+X^1 & X^2\\
X^2 & X^0-X^1 
\end{array} \right ),
 \end{gather}
where  $\mathbf{X}=(X^0,X^1,X^2) $ is the $SO(1,2)$ vector.
 
 Then, one applies the reduced equation of motion and the Hamiltonian constraint as in \cite{Clement:2009gq}
 \begin{align}\label{eq:EOM}
&\mathbf{X}  \wedge ( \mathbf{X} \wedge  \mathbf{X}'''')+\frac{5}{2} \mathbf{X} \wedge (\mathbf{X}' \wedge \mathbf{X}''')+\frac{3}{2} \mathbf{X}' \wedge (\mathbf{X} \wedge \mathbf{X}'''')+\frac{9}{4} \mathbf{X}' \wedge (\mathbf{X}' \wedge \mathbf{X}'')  \nonumber\\  &
-\frac{1}{2} \mathbf{X}'' \wedge ( \mathbf{X} \wedge \mathbf{X}'')- \left[\frac{1}{8} (\mathbf{X}'^2)+\frac{m^2}{\zeta^2} \right] \mathbf{X}'' =  0,&
 \end{align}

\begin{align}\label{eq:hamiltonian}
H \ \ \ &  \equiv \ \ \ (\mathbf{X} \wedge \mathbf{X'} )\ . \ ( \mathbf{X} \wedge \mathbf{X}'''')-\frac{1}{2} (\mathbf{X} \wedge \mathbf{X}'')^2 +\frac{3}{2} (\mathbf{X} \wedge \mathbf{X}') \ . \ (\mathbf{X}' \wedge \mathbf{X}'')  \nonumber\\ & +\frac{1}{32} (\mathbf{X}'^2 )^2 +\frac{m^2}{2\zeta^2} (\mathbf{X'}^2)+\frac{2m^2 \Lambda} {\zeta^4} =0.
\end{align}  
 
 From these two equations one can find $\zeta^2$ and $\Lambda$. Then one can define the vector
 \begin{gather}
 \mathbf{L} \equiv \mathbf{X} \wedge \mathbf{X}',
 \end{gather}
 where prime denotes derivative with respect to $\rho$: $' \equiv \frac{d}{d\rho}$.
 
 Finally, the \textit{super angular momentum} of NMG $\mathbf{J}=(J^0, J^1, J^2)$ is
 \begin{gather}
 \mathbf{J}= \mathbf{L}+\frac{\zeta^2}{m^2} \left[ 2\mathbf{L} \wedge \mathbf{L'}+\mathbf{X}^2 \mathbf{L''}+\frac{1}{8} \big( \mathbf{X'}^2-4 \mathbf{X} \ .\  \mathbf{X''} \big) \mathbf{L} \right],
 \end{gather}
 where the products are defined as
 \begin{gather}
 \mathbf{A} . \mathbf{B}= \eta_{ij} A^i B^j, \ \ \ \ \ \ \ \ \ \ (\mathbf{A} \wedge {B})^i=\eta^{im} \epsilon_{mjk} A^j B^k, \ \  \ \ \ \ (\epsilon_{012}=1).
 \end{gather}

For the case of NMG one has \cite{Nam:2010ub}
\begin{align}\label{charge1}
 \eta  \left[\sigma Q_R^{\rho t}+\frac{1}{m^2} Q_K^{\rho t} \right]_{\zeta_T} \ &= \ \frac{1}{L} \left[ -\frac{\zeta^2}{2} \delta J^2 + \Delta_{\textit{Cor}} \right], \nonumber\\
  \eta  \left[ \sigma Q_R^{\rho t}+\frac{1}{m^2} Q_K^{\rho t}\right]_{\zeta_R} \ &= \ \frac{\zeta^2}{2} \delta (J^0-J^1),
  \end{align}

where $\eta$ and $\sigma$ are $\pm 1$, depending on the sign in the action. Based on Eq. (\ref{eq:action}), both of $\eta$ and $\sigma$ are positive in our case.
Also, for the case of NMG, $\Delta _{\textit{Cor}}$ which is the correction term to the mass is
\begin{gather}
\Delta _{\textit{Cor}}=\Delta_R+\Delta_K,
\end{gather}
 where
 
 \begin{align}
 &\Delta_R =  \frac{\zeta^2}{2} \left[ -(\mathbf{X} \ .\  \delta \mathbf{X'}) \right], \nonumber\\ &
 \Delta_K = \frac{\zeta^4}{m^2}\bigg [ -U^2 ( \mathbf{X''} \ . \ \delta \mathbf{X'} )+\frac{U^2}{2} \Big [ (U\delta U)'''-(\mathbf{X} \ . \  \delta \mathbf{X'})''-\frac{1}{2} (\mathbf{X'}\ . \ \delta \mathbf{X'} )' \Big]\nonumber\\ & -\frac{ U U'}{4}\Big [ (U \delta U)''-\frac{5}{2} (\mathbf{X}' \ . \ \delta \mathbf{X'} ) \Big]+ \Big[\mathbf{X'}^2-(U U')' \Big] ( U \delta U)'+ U U'(\mathbf{X''} \ . \ \delta \mathbf{X} )\nonumber\\ &
 + \Big[\frac{5}{4}(UU')'-\frac{21}{16} \mathbf{X'}^2 \Big] (\mathbf{X} \ . \ \delta \mathbf{X'} )+ \Big[-\frac{1}{2} (U U')''+\frac{9}{4} (\mathbf{X'} \ . \ \mathbf{X''} ) \Big ] U \delta U \bigg ].
 \end{align}
 Then the mass and angular momentum in NMG are
 \begin{gather} 
 \mathbf{M} \ = \ \frac{1}{4G} \sqrt{ - \text{det} \   g} \  \left [  Q_R^{rt} +\frac{1}{m^2} Q_K^{rt} \right]_{\zeta_T, r\to \infty}, \nonumber\\
 \mathbf{J} \ = \ \frac{1}{4G} \sqrt{ -\text{det} \ g} \  \left[  Q_R^{rt} +\frac{1}{m^2} Q_K^{rt} \right] _{\zeta_R , r\to \infty}.\label{charge2}
 \end{gather}

Finally, for calculating the entropy for any solution in NMG, one can use the following relation from \cite{Clement:2009gq}
\begin{gather}\label{entropy}
\mathbf{S}=\frac{A_h}{4 G} \left( 1+\frac{\zeta^2}{2m^2} \left[ (\mathbf{X\ .\ X''})-\frac{1}{4} (\mathbf{X'}^2) \right] \right).
\end{gather}
 
Note that there exists other methods of calculating the conserved charges. Specifically the methods developed in \cite{Hajian:2015xlp, Hajian:2016kxx, Ghodrati:2016vvf}, the so-called ``solution phase space method" is of our interest which is discussed in Appendix (\ref{sec-Higher d}).
 
 Now using these relations one can derive the charges, Gibbs free energies and the phase diagrams of several solutions of NMG.

 \subsection{ Examples of conserved charges of BHT solutions}\label{sec:exensemble}

First for the warped AdS black hole in the \textit{``grand canonical ensemble"} \cite{Detournay:2015ysa} as in the following form,

 \begin{gather}\label{metricensemble}
g_{\mu \nu}=\left(
\begin{array}{ccc}
 -\frac{r^2}{l^2}-\frac{H^2 (-r^2-4 l J+8 l^2 M)^2}{4l^3(l M-J)}+8M & 0 & 4J-\frac{H^2(4 l J-r^2)(-r^2-4 l J+8 l^2 M) }{4l^2(l M-J)}  \\  
0 & \frac{1}{\frac{16 J^2}{r^2}+\frac{r^2}{l^2}-8M } & 0 \\ 
4J-\frac{H^2(4 l J-r^2)(-r^2-4 l J+8 l^2 M) }{4l^2(l M-J)} & 0 & r^2-\frac{H^2 (4 J l -r^2)^2 }{4 l (l M-J)} 
\end{array} \right ),
\end{gather}

by reparametrizing the radial coordinate as $r^2 \to \rho$, and then by applying the equation of motion and Hamiltonian constraints in Eqs (\ref{eq:EOM}) and (\ref{eq:hamiltonian}) one can find
\begin{gather}
\zeta^2= \frac{8 l^2 m^2}{(1-2H^2)(17-42H^2)} , \ \ \ \ \ \  \ \ \ \ \Lambda=\frac{ m^2 (84H^4+60H^2-35) }{(17-42H^2)^2}.
\end{gather} 
 From the above relation one can see that the acceptable region for $\Lambda$ is
 \begin{gather}
 \frac{-35m^2}{289} < \Lambda < \frac{m^2}{21}.
 \end{gather}
Note that the special case of $\Lambda=\frac{m^2}{21}$ corresponds to the $\text{AdS}_2 \times S^1$.

Now for the metric \ref{metricensemble}, the components of the super angular momentum are

\begin{gather}
J^0= \frac{H^2 \left(1+l^2\right)}{4 l^3 (J-l M)},\ \ \ \ \ \ \ \ \  
J^1=  \frac{H^2 \left(1-l^2\right)}{4 l^3 (J-l M)}, \ \ \ \ \ \ \ \  
J^2=  \frac{H^2}{2 l^2 (J-l M)}.
\end{gather}

Then, using Eqs (\ref{charge1}) and (\ref{charge2}) one can find the charges 
\begin{gather}\label{charge1}
\mathbf{M}=\frac{16 \left(1-2 H^2\right)^{3/2} M}{G L\left(17-42 H^2\right) }, \ \ \ \ \ \ \ \ \ \ \ \  \ \ \  \mathbf{J}=\frac{16 \left(1-2 H^2\right)^{3/2} J}{G \left(17-42 H^2\right)}.
\end{gather}

One should note that $\Delta_{Cor}$ is zero here.

Now, for the above metric using Eq. (\ref{entropy}), the entropy is
 \begin{gather}\label{entropy1}
 \mathbf{S}=\frac{16 \pi \left(1-2 H^2\right)^{3/2} \text{  }}{G \left(17-42 H^2\right)} \sqrt{l^2 M+\sqrt{l^4 M^2-J^2 l^2}}.
 \end{gather}

We can then study this black hole solution in another ensemble. The asymptotically warped $\mathrm{AdS}_3$ black hole in NMG in the ADM form and thereby in the \textit{``quadratic/non-local ensemble"} takes the following form
\begin{equation}
\begin{split} \label{eq:WBTZ}
\frac{ds^2}{l^2}= dt^2+\frac{dr^2 }{ (\nu^2+3)(r-r_+)(r-r_-) }+(2\nu r-\sqrt {r_+ r_- (\nu^2+3)} ) dt d\varphi \\  
+\frac{r}{4} \Big[ 3(\nu^2-1)r +(\nu^2+3)(r_+ +r_-)-4\nu \sqrt {r_+ r_- (\nu^2+3)} \Big] d\varphi^2.
\end{split}
\end{equation}
So using Eqs (\ref{eq:EOM}) and (\ref{eq:hamiltonian}), one has
 \begin{gather}
 \zeta ^2=\frac{8 m^2}{l^4 \left(20 \nu ^2-3\right)}, \ \ \ \ \ \ \ \ \  \Lambda =\frac{m^2 \left(9-48 \nu ^2+4 \nu ^4\right)}{\left(3-20 \nu ^2\right)^2}.
 \end{gather}
and the components of the super angular momentum are
\begin{equation}
\begin{split}
\mathnormal{J^0}&= -\frac{l^4 \nu  (\nu ^2+3)\left (4-2 r_- \nu  \sqrt{r_+ r_- (\nu ^2+3)}-2 r_+\nu  \sqrt{r_+ r_- (\nu ^2+3)}+r_+ r_-\text{  }(5 \nu ^2+3)\right)}{2(20 \nu ^2-3)} ,  \nonumber\\ \mathnormal{J^1}&=  \frac{l^4 \nu  (\nu ^2+3) \left (-4-2 {r_-} \nu  \sqrt{r_+ r_-  (\nu ^2+3)}-2 r_+ \nu  \sqrt{r_+ r_-  (\nu ^2+3)}+r_+ r_-\text{  }(5 \nu ^2+3)\right)}{2(20 \nu ^2-3)}, \nonumber\\   \mathnormal{J^2}&=   -\frac{2 l^4 \nu  (\nu ^2+3) \left(({r_+}+{r_-}) \nu -\sqrt{r_+ r_-  (\nu ^2+3)}\right)}{20 \nu ^2-3}. 
\end{split}
\end{equation}
Next, by using Eq. (\ref{charge1}) and (\ref{charge2}) one can find the conserved charges \cite{Clement:2009gq, Donnay:2015iia}
\begin{align}
&\mathbf{M}=\frac{\nu  \left(\nu ^2+3\right) }{2 G \left(20 \nu ^2-3\right)}\left(\left(r_++r_-\right) \nu -\sqrt{r_+ r_- \left(\nu ^2+3\right)}\right), \nonumber\\
&\mathbf{J}= \frac{ \nu  \left(\nu ^2+3\right) }{4 G l\left(20 \nu ^2-3\right)}\left( \left(5 \nu ^2+3\right)r_+ r_- -2 \nu  \sqrt{ r_+ r_-\left(3+\nu ^2\right)}\left(r_++r_-\right)\right), \nonumber\\
& \mathbf{S}=\frac{4\pi  l \nu ^2 }{G \left(20 \nu ^2-3\right)}\sqrt{r_+ r_- \left(\nu ^2+3\right)+4 r_+ \nu  \left(r_+ \nu -\sqrt{r_+ r_- \left(\nu ^2+3\right)}\right)}.
\end{align}

As another example of solution of NMG with potentials for practical applications in condensed matter physics, one could also study the conserved charges of the Lifshitz geometry
\begin{gather}
ds^2=-\frac{r^{2z}}{l^{2z}} dt^2+\frac{l^2}{r^2} dr^2+\frac{r^2}{l^2} d {\vec{x}}^2.
\end{gather}
Here $\zeta ^2=-\frac{2 m^2l^{2+2 z} }{1+z(z-3)}$ and the vector of super angular momentum vainshes. The case of $z=3$ and $z=\frac{1}{3}$ could be a solution of the simple NMG with no matter content. For the case of $z=3$, one has $\zeta^2 = -2 l^8 m^2$.

Now considering the solution of the Lifshitz black hole \cite{Correa:2014ika},
\begin{gather}
ds^2=-\frac{r^{2z}}{l^{2z}} \left[ 1-M \left(\frac{l}{r}\right)^{\frac{z+1}{2}} \right] dt^2+\frac{l^2}{r^2} \left[1-M \left( \frac{l}{r}\right)^{\frac{z+1}{2}} \right]^{-1} dr^2 +\frac{r^2}{l^2} d\varphi^2,
\end{gather}
by taking $r \to \Big(  \rho (z+1)  \Big)^{\frac{1}{1+z}}$, one gets $\sqrt{-g}=1/\zeta=l^{-z}$ which results in  
\begin{gather}
\mathbf{M}=-\frac{\pi M^2 (z+1)^2 (3z-5)}{16 \kappa (z-1) (z^2-3z+1)}, \ \ \ \ \ \ \ \ \ \ \ \ \ \mathbf{J}=0, 
\end{gather}
in agreement with \cite{Correa:2014ika}. This leads us to the following Gibbs free energy
\begin{gather}
G_{\text{Lifshitz BH}}=\frac{M^2 \pi  z (z+1)^2 (3 z-5)}{16 k (z-1) (z(z-3)+1)}.
\end{gather}
Comparing this result with the free energy of the Lifshitz metric, one can see that in NMG the Lifshitz black hole is always the dominant phase.

 \section{Phase transitions of $\text{AdS}_3$ solution} \label{sec:AdSPhase}

The vacuum $\text{AdS}_3$ solution is
\begin{gather}
ds_{\text{AdS}_3}^2=l^2 (d\rho^2-\cosh^2 \rho \ dt^2+\sinh^2 \rho \ d\phi^2 ),
\end{gather}
where \cite{Grumiller:2009sn}
\begin{gather}
1/l^2=2m^2(1 \pm \sqrt{1+\frac{\Lambda}{m^2}}),
\end{gather}
and the boundary where the dual CFT is defined is located at $\rho \to \infty$.

For this case, to find the Gibbs free energy, we use the relation $G(T,\Omega) =TS[g_c]$, where $g_c$ is the Euclidean saddle and $\tau=\frac{1}{2\pi} (-\beta \Omega_E+i \frac{\beta}{l})$ is the modular parameter. We work in the regimes where the saddle-point approximation is valid.

First, we need to find the free energy of the vacuum solution. In \cite{Kraus:2005vz, Kraus:2006wn}, the authors derived a general result for the action of the thermal $\mathrm{AdS}_3$ in any theory as, 
\begin{gather}\label{Kraus}
S_E \big(AdS(\tau ,\tilde{\tau} )\big)=\frac{i \pi}{12 l} (c \tau-\tilde{c} \tilde{\tau}).
\end{gather}
Also, the modular transformed version of this equation gives the thermal action of the BTZ black hole. By changing the boundary torus as $\tau \to -\frac{1}{\tau}$, and then by using the modular invariance, one has 
\begin{gather}
ds^2_{\text{BTZ}}\left[-\frac{1}{\tau}\right]=ds^2_{\text{AdS}} [\tau],
\end{gather}
leading to the following result
\begin{gather}\label{KrausBH}
S_E \big(BTZ(\tau ,\tilde{\tau} )\big)=\frac{i \pi}{12 l} (\frac{c}{ \tau}- \frac{\tilde{c}}{ \tilde{\tau}}).
\end{gather}
In this equation the contributions of the quantum fluctuations of the massless field are neglected as they are suppressed for large $\beta$. 

One should notice that this equation and its modular transformed version are only true for the $\text{AdS}_3$ and not particularly for the ``warped $\text{AdS}_3$" or ``asymptotically warped AdS black holes". This equation is correct as in the Lorentzian signature, the thermal $\text{AdS}_3$ has the same form as in the global coordinates and also the global $\text{AdS}_3$ corresponds to NS-NS vacuum with zero Virasoro modes \cite{Kraus:2005vz}. These statements are not applicable for geometries with asymptotics other than AdS, including geometries with warped $\text{AdS}_3$ asymptotic backgrounds.

 In the next section, by redefining of the modular parameter $\tau$, and deriving the free energy by three different methods, we find a new expression for the thermal action of the warped $\text{AdS}_3$ in NMG case as well.

For now, inserting the central charges of the NMG \cite{Bergshoeff:2009aq, Liu:2009bk},
\begin{gather}
c_L=c_R=\frac{3l}{2G_N} \left( 1-\frac{1}{2m^2 l^2 }\right),
\end{gather}
and the modular parameter $\tau=\frac{1}{2\pi} (-\beta \Omega_E+i \frac{\beta}{l})$ in Eq. (\ref{Kraus}) results in
\begin{gather}
S_E=-\frac{1}{8 l T G_N} \Big( 1-\frac{1}{2m^2 l^2} \Big).
\end{gather}
This relation, unlike the corresponding equation in the TMG case, does not depend on the angular velocity $\Omega_E$. This is because the NMG has chiral symmetry and so the central charges are equal which causes that the terms containing $\Omega$ to cancel out.

Therefore, the Gibbs free energy is
\begin{gather}\label{eq:GADS}
G_{AdS}(T, \Omega)=-\frac{1}{8 l G_N} \Big( 1-\frac{1}{2m^2 l^2} \Big).
\end{gather}
Just by considering this equation, one can see that the stability condition of the vacuum $\text{AdS}_3$ in NMG is $m^2 l^2 >\frac{1}{2}$ which is different from the condition in the Einstein theory. \\

Additionally, the NMG theory also admits a general BTZ solution. The rotating BTZ black hole metric solution in this theory is of the following form
\begin{gather}
ds^2=(-\frac{2 \rho}{\tilde{l} ^2} +\frac{M }{2}) dt^2- j dt d\phi+(2\rho +\frac{M \tilde{l}^2}{2}) d\phi^2+\frac{ d\rho^2} {(\frac{4\rho^2}{\tilde{l}^{2}}- \frac{M ^2 \tilde{l}^2-j^2)}{4}) } ,
\end{gather}
where the AdS curvature comes from $l^{-2} =2m^2 \big[ 1 \pm \sqrt{1+\frac{\Lambda}{m^2} } \big]$ \cite{Clement:2009gq} and $M$ is the ADM mass of the black hole. 

If we aim to write the metric in the ADM form,
\begin{gather}
ds^2=-N(r)^2 dt^2+\frac{dr^2}{f(r)^2}+R(r)^2(N^\phi (r)dt+d\phi)^2,
\end{gather}
we need to go from the coordinate system $(t, \rho, \phi)$ to $(t, r, \phi)$, so we should change the radial coordinate as $ \rho=r^2/2 - M \tilde{l}^2 /4$, and then re-scale the three coordinates as $r\to\tilde{l} r$, $t\to -l t$ and  $\phi \to L \phi / \tilde{l}$. 

Then the metric becomes \cite{Nam:2010dd}
\begin{gather} \label{eq:BTZ}
ds^2=l^2 \Big[ -\frac{(r^2-r_+^2)(r^2-r_-^2)}{r^2}dt^2+\frac{r^2}{(r^2-r_+^2)(r^2-r_-^2)} dr^2+r^2(d\phi+\frac{r_+ r_-}{r^2} dt)^2 \Big].
\end{gather}
The Hawking temperature of this black hole is \cite{Nam:2010dd}
\begin{gather}
T_H=\frac{\kappa}{2\pi} =\frac{1}{2\pi l} \frac{\partial_r N}{\sqrt{g_{rr}} } \Big |_{r=r_+}=\frac{r_+}{2\pi l}\Big(1-\frac{r_-^2}{r_+^2} \Big),
\end{gather}
the entropy is
\begin{gather}
S_{BH}=\frac{\pi^2 l}{3} c( T_L+T_R),
\end{gather}
and the angular velocity at the horizon is defined as \cite{Nam:2010dd}
\begin{gather}
\Omega_H=\frac{1}{l} N^\phi (r_+)=\frac{1}{l} \frac{r_-}{r_+} .
\end{gather}
Also, the left and right temperatures are given by \cite{Maldacena:1998bw}
\begin{gather}
T_L=\frac{r_+ +r_-}{2\pi l}=\frac{T}{1- l \Omega}, \ \ \ \ \ \ \ \ \ \ \ \ \  T_R=\frac{r_+-r_-}{2\pi l}=\frac{T}{1+l \Omega},
\end{gather}
and the left and right energies can be defined as follows
\begin{gather}
E_L \equiv \frac{\pi^2 l}{6} c_L T_L^2,  \ \ \ \ \ \ \ \ \ \  \ \  \ \ E_R\equiv \frac{\pi^2 l}{6} c_R T_R^2.
\end{gather}
These parameters are related to the mass and angular momentum as \cite{Nam:2010dd}
\begin{gather}
M=E_L+E_R, \ \ \ \ \ \ \ \ \ J=l(E_L-E_R).
\end{gather}
The horizons of the BTZ black hole are located at
\begin{gather}
r_+=\sqrt{2\Big( \frac{M \tilde{l}^2 }{4} +\frac{\tilde{l}}{4} \sqrt{M^2 \tilde{l}^2 -j^2}  \Big) }=\frac{2\pi l T}{1-\Omega^2 l^2}, \ \ \ r_-=\sqrt{2 \Big( \frac{M \tilde{l}^2 }{4} -\frac{\tilde{l}}{4} \sqrt{M^2 \tilde{l}^2 -j^2} \Big ) }=\frac{2\pi \Omega l^2 T}{1-\Omega^2 l^2}.
\end{gather}
For the BTZ black hole in NMG which has an asymptotic AdS geometry, again the central charges are
\begin{gather}
c_L=c_R=\frac{3 l}{2G_N} \Big( 1-\frac{1}{ 2 m^2 l^2 } \Big).
\end{gather}
For having a physical theory, the central charge and the mass of the BTZ black hole should be positive which again leads to the condition of $m^2 l^2 >\frac{1}{2}$.

These parameters satisfy the first law of thermodynamics,
\begin{gather}
dM=T_H dS_{BH}+\Omega_H dJ,
\end{gather}
and the integral of it satisfies the Smarr relation \cite{Nam:2010dd},
\begin{gather}
M=\frac{1}{2}T_H S_{BH}+\Omega_H J.
\end{gather} 
Now one can read the Gibbs free energy from the following relation,
\begin{gather}
G=M-T_H S_{BH} -\Omega_H  J.
\end{gather}
Using all the above equations, the Gibbs free energy of the BTZ in NMG is
\begin{gather}\label{eq:GBTZ}
G_{BTZ} (T,\Omega)=-\frac{\pi ^2 T ^2\left(2 m^2 l^2-1 \right)}{4 G_N m^2l\left( 1 - l^2 \Omega ^2\right)}.
\end{gather}
This result can also be rederived by considering modular invariance. Hence, using the relation (\ref{KrausBH}) and $G(T, \Omega)=T S[g_c]$ again denotes the applicability of (\ref{Kraus}) for the $\text{AdS}_3$ case in NMG. 
From this relations one can see that for small rotations $\Omega$ as also explained in \cite{Myung:2015pua}, the thermal stability condition for BTZ black hole in NMG is $m^2 l^2 > \frac{1}{2}$, regardless of the size of the event-horizon. For this case, the Hawking-Page phase transition can occur between the BTZ black hole and the thermal solution, while for the case of $m^2 l^2 < \frac{1}{2}$  the fourth-order curvature terms is dominant  and in this case an inverse Hawking-Page phase transition between the BTZ black hole and the ground state massless BTZ black hole can occur  \cite{Myung:2015pua}. 

One can also discuss the interpolations and the continuous phase transitions between these phases such as the scenario in \cite{Zhang:2015wna}. 

We now extend these results to the higher angular momenta.

\subsection{The stability conditions}

For checking the local stability we find the Hessian, $H$, of the free energy $G(T,\Omega)$  of the BTZ metric as
\begin{gather}
H=\left(
\begin{array}{ll}
 \frac{\partial ^2G}{\partial T^2} & \frac{\partial ^2G}{\partial T \partial \Omega } \\ \\
 \frac{\partial ^2G}{\partial \Omega  \partial T} & \frac{\partial ^2G}{\partial \Omega ^2} \\
\end{array}
\right)= \left(
\begin{array}{ll}
 \frac{ \pi ^2 \left(2 m^2 l^2-1\right)}{2 G_N m^2 \left(\Omega ^2 l^2-1\right)} & \frac{\pi ^2 l^2 \left(1-2 l^2 m^2\right)  T \Omega }{G_N m^2 \left(\Omega ^2 l^2-1\right)^2} \\ \\
 \frac{\pi ^2 l^2 \left(1-2 m^2 l^2 \right)  T \Omega }{G_N m^2 \left(\Omega ^2 l^2-1\right)^2} & \frac{\pi ^2  T^2 \left(2 m^2 l^2-1 \right)  \left(l^2 +3 l^4 \Omega ^2\right)}{2 G_N m^2 \left(\Omega^2 l^2-1\right)^3} \\
\end{array}
\right).
\end{gather}
In the region where both of its eigenvalues are negative, the system is stable. By finding the eigenvalues of the above matrix and then by assuming $G_N=l=1$, the stable region is found to be $m^2 >1$ and $\Omega^2<1$ for any $T$, similar to the stability region of TMG discussed in \cite{Detournay:2015ysa}. 

Now for calculating the global stability, we calculate the difference of the free energies of AdS and BTZ backgrounds which gives
\begin{gather}
\Delta G=G_{AdS}-G_{BTZ}=\frac{2m^2 l^2-1}{4 G_N m^2 l} \left( \frac{\pi^2 T^2}{1- l^2 \Omega^2} -\frac{1}{4l^2} \right).
\end{gather}

\subsection{Phase diagrams}
When $\Delta G>0$, the BTZ black hole is the dominant phase and when $\Delta G<0$, the thermal $\text{AdS}_3$ is dominant. Assuming $G_N=l=1$, we show the phase diagrams in Figures (\ref{fig:p1}) and (\ref{fig:p2}). One can notice that, since in NMG unlike the TMG case, parity is conserved, the phase diagrams are symmetric.

\begin{figure}[ht!]
\centering 
\begin{minipage}{.5\textwidth}
  \centering  
\includegraphics[width=.7\linewidth]{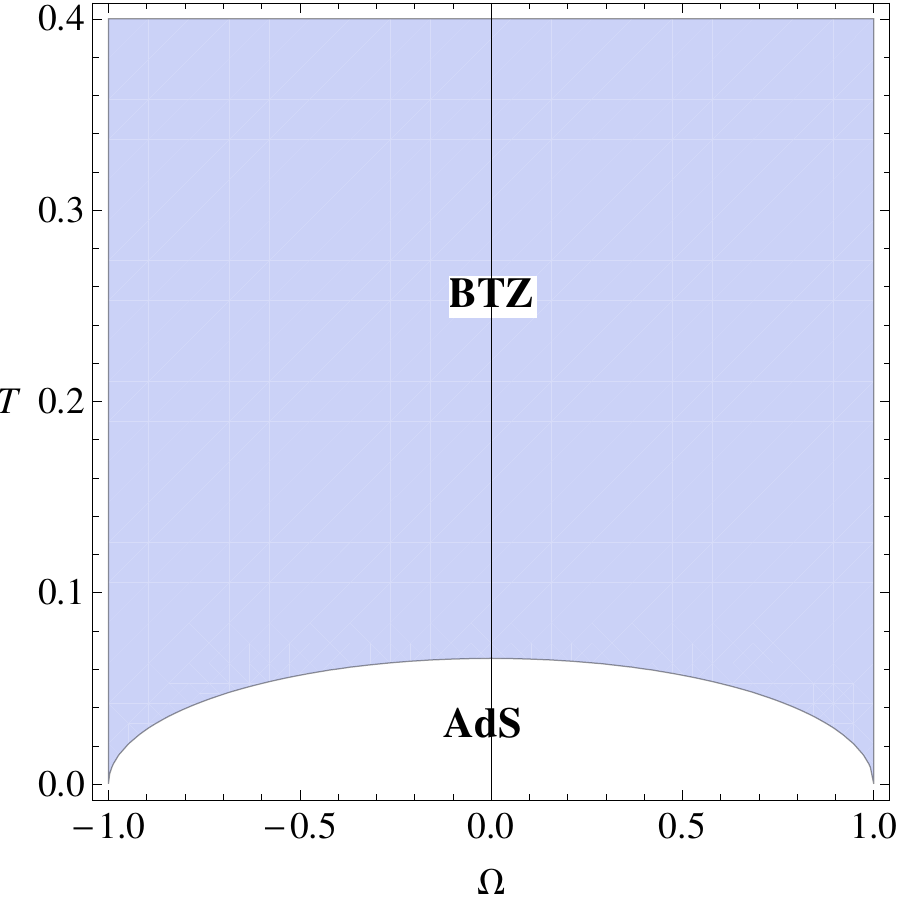} 
\caption{\label{fig:p1}  $m=1.05$.} 
\end{minipage}%
\begin{minipage}{.5\textwidth}
\centering  
\includegraphics[width=.7\linewidth]{p2}
\caption{\label{fig:p2}  $m=10$.}
\end{minipage}
\end{figure}

From the above diagrams, one can also notice that by decreasing $m$ the effect of the higher derivative correction terms to the Einstein-Hilbert action increases. This effect would make the BTZ black hole to form in a lower temperature. So, due to the fact that in NMG the modes are massive, forming a black hole in NMG is easier relative to pure Einstein gravity. On the other hand, increasing $m$ with a specific angular velocity causes the phase transition from $\text{AdS}_3$ to BTZ to occur at a higher temperature.

\section{Phase transitions of warped $\text{AdS}_3$ solution in quadratic ensemble} \label{sec:sec2}
In this section we first introduce the thermal $\text{WAdS}_3$ and Warped BTZ black hole solutions and then we present the phase diagrams. 

Both of these solutions have an asymptotic warped AdS geometry which are the squashed or stretched deformation of $\text{AdS}_3$ with a different symmetry algebra than the AdS case. This dissimilarity of the algebra makes the thermal properties different from the asymptotic AdS solution as well. We derive some relations for the thermal action of the warped solutions in NMG.

\subsection{G\"{o}del space-time}

The time-like $\text{WAdS}_3$ or the three-dimensional G\"{o}del space-time is the true vacuum of the $\text{WAdS}_3$ black hole \cite{Donnay:2015iia, Banados:2005da}.
The original G\"{o}del solution of the Einstein equations was four-dimensional. The non-trivial three-dimensional factor of G\"{o}del space-time which is within the family of deformed $\text{AdS}_3$ was first studied in \cite{Rooman:1998xf}. This metric is a constant curvature Lorentzian manifold with isometry group $U(1) \times SL(2, \mathbb{R})$ where the $U(1)$ factor is generated by a time-like Killing vector. As this metric can be embedded in the seven-dimensions flat space, it would possess time-like closed curves. However, it is still a solution of string theory which corresponds to the time-like generator of $SL(2,\mathbb{R})$ or a magnetic background \cite{Israel:2004vv}.

The metric is given by
\begin{gather}\label{eq:Godel}
ds^2=-dt^2-4 \omega r dt d\phi+\frac{\ell^2 dr^2}{(2r^2(\omega^2 \ell^2+1) +2\ell^2 r)}-\Big( \frac{2r^2}{ \ell^2} (\omega^2 \ell^2-1)-2r\Big) d\phi^2.
\end{gather}

In the special case of $\omega^2 \ell^2=1$, this metric corresponds to $\text{AdS}_3$. For this timelike solution we have
\begin{gather}\label{eq:LMG}
m^2=-\frac{(19\omega^2 \ell^2-2)}{2\ell^2},  \ \ \ \ \ \ \ \Lambda=-\frac{(11\omega^4 \ell^4+28 \omega^2 \ell^2-4)}{2\ell^2 (19 \omega^2 \ell^2-2)}.
\end{gather}

This metric were extensively studied in cosmological models although it contains closed timelike curves (CTCs) and is unstable with respect to the quantum fluctuations. As these causal pathologies are large scale deficiencies, some rotating objects with physical applications in cosmology or perhaps in condensed matter could be modeled by this metric surrounded by a more standard space-time \cite{Rooman:1998xf}, as sometimes by considering the  covering space, one avoids CTCs. Therefore, constructing the phase diagrams of this manifold could have interesting applications.

\subsection{Space-like warped BTZ black hole}

The warped $\mathrm{AdS}_3$ or warped BTZ black hole in NMG in the quadratic non-local ensemble in its ADM form can be written as
\begin{equation}
\begin{split} \label{eq:WBTZ}
\frac{ds^2}{l^2}= dt^2+\frac{dr^2 }{ (\nu^2+3)(r-r_+)(r-r_-) }+(2\nu r-\sqrt {r_+ r_- (\nu^2+3)} ) dt d\varphi \\  
+\frac{r}{4} \Big[ 3(\nu^2-1)r +(\nu^2+3)(r_+ +r_-)-4\nu \sqrt {r_+ r_- (\nu^2+3)} \Big] d\varphi^2.
\end{split}
\end{equation}
If $\nu^2=1$ the space is locally $\text{AdS}_3$, if $\nu^2 >1$ it is stretched and if $\nu^2<1$ it is a squashed deformation of $\text{AdS}_3$. 

For the space-like solution, the parameters are,
\begin{gather}\label{eq:mlBTZ}
m^2=-\frac{(20\nu^2-3)}{2 l^2}, \ \ \ \ \ \ \ \ \ \ \ \Lambda=-\frac{m^2 (4\nu^4-48\nu^2+9 ) }{(20\nu^2-3)^2}.
\end{gather}
If one employs the following relations
\begin{gather}
\omega=\frac{\nu}{l}, \ \ \ \ \ \ \ \ \   \  \omega^2 \ell^2+2=3 \ell^2/ l^2,
 \end{gather}
again one reaches to equation (\ref{eq:LMG}).

Notice that ``$l$" is the radius of space-like $\text{AdS}_3$ and ``$\ell$" is the radius of the warped time-like $\text{AdS}_3$. 
Similar to the way that by the global identifications one can derive BTZ black hole, one can also derive Eq. (\ref{eq:WBTZ}) from Eq. (\ref{eq:Godel}).

In order to have a real $m$ and a negative $\Lambda$ and therefore a physical solution, from Eq. (\ref{eq:LMG}) and Eq. (\ref{eq:mlBTZ}) the allowed ranges of $\nu$ and $\omega$ are
\begin{gather}
-\sqrt{\frac{2}{19}}<\omega \ell<\sqrt{\frac{2}{19}}, \ \ \ \ \ \ \ \ \ \ \ \  -\sqrt{\frac{3}{20} } <\nu < \sqrt{\frac{3}{20} }. 
\end{gather}

\subsection{The free energies and phase diagrams}\label{sub:free}
Now by using the thermodynamic quantities and conserved charges, we calculate the free energies of both of these space-times and then we proceed by making the phase diagrams.

Notice that the isometry group of the time-like $\text{WAdS}_3$ is $SL(2,\mathbb{R}) \times U(1)$ which is generated by four Killing vectors \cite{Donnay:2015iia}. By assuming a specific boundary condition, the authors in \cite{Donnay:2015iia} derived the asymptotic algebra of WAdS in NMG and then the central charge $c$ and the $\hat{u}(1)_k$ Ka$\check{\text{c}}$-Moody level $k$ as \cite{Donnay:2015joa}
\begin{gather}\label{eq:param}
c=\frac{48 \ell^4 \omega^3 }{G (19 \ell^4 \omega^4+17\ell^2 \omega^2-2)}=-\frac{96 l \nu^3}{G (20 \nu^4+57 \nu^2-9) }, \\ \nonumber\\   k=\frac{8 \omega (1+\ell^2 \omega^2)}{G(19 \ell^2 \omega^2-2)}=\frac{4\nu(\nu^2 +3)}{G l(20\nu^2-3)} .
\end{gather}

If we just simply assume that the relation (\ref{Kraus}) can be used here and the modular parameter is $\tau=\frac{1}{2\pi} (-\beta \Omega_E+i \frac{\beta}{l})$, then by using the above central charge one can find the free energy as
\begin{gather}\label{eq:G2}
G_{\text{timelike WAdS}}=-\frac{4 \ell^2 \omega^3 (\omega^2 \ell^2+2)}{3G(19\ell^4 \omega^4+ 17 \ell^2 \omega^2-2)}=-\frac{4\nu^2}{G(20\nu^2-3)}\times   \frac{\nu}{(\nu^2+3) l}.
\end{gather} 

We can also recalculate the free energy by using the conserved charges. These conserved charges of the timelike $\text{WAdS}_3$ in NMG have been calculated in \cite{Donnay:2015joa} for a \textit{``spinning defect"}. Using these relations one can take the limit of $\mu \to 0$ to find the mass and angular momentum as
\begin{gather}
\mathcal{M}=-\frac{4 \ell^2 \omega^2}{G(19 \ell^2 \omega^2-2)}, \ \ \ \ \ \ \ \ \ \ 
\mathcal{J}=-\frac{4 j \ell^4 \omega^3}{G(19\ell^2 \omega^2-2)}.
\end{gather}
For the time-like warped $\text{AdS}_3$, again the entropy and the temperature are zero. So the Gibbs free energy, $G=\mathcal{M}-\Omega \mathcal{J}$, is
\begin{gather}
G_\text{spinning defect }=\frac{4 \ell^2 \omega^2 \big((\mu-1)+\Omega j \ell^2 \omega \big)}{G(19\ell^2 \omega^2-2)}.
\end{gather}
Taking the limit of zero defect, the result is as follows 
\begin{gather}\label{eq:G1}
G_{\text{timelike WAdS}}=-\frac{4 \ell^2 \omega^2}{G (19 \ell^2 \omega^2 -2)}=-\frac{4\nu^2}{G (20\nu^2-3)}.
\end{gather}

Comparing (\ref{eq:G1}) with (\ref{eq:G2}), one can see that there is a factor of $N_1= \frac{\nu}{(\nu^2+3) l}$ difference. This factor can be introduced in the modular parameter to compensate for this discrepancy.

For re-deriving this factor we can also calculate the free energy in a third way. As the authors in \cite{Donnay:2015joa} found, the warped CFT version of the Cardy's formula of entropy 
\begin{gather}
S_{\text{WCFT}}=\frac{4\pi i}{k} \tilde{P}_0^{(\text{vac})} \tilde{P}_0+4\pi\sqrt{ -\tilde{L}_0^{+(\text{vac}) } \tilde{L}_0^+ }, 
\end{gather}
matches with the black hole entropy 
\begin{gather}
S_{BH}=\frac{8 \pi \nu^3}{ (20\nu^2-3) G_N } \left(r_+-\frac{1}{2\nu} \sqrt{ ( \nu^2+3) r_- r_+}\right).
\end{gather}
Now using the above equation and the relation (\ref{eq:radius}) of the next section, one can find the warped BTZ black hole entropy as 
\begin{gather}
S_{\text{WBTZ}}=\frac{8\pi \nu^2}{ G_N \Omega l (20\nu^2-3)}.
\end{gather}
If one uses the modular transformed equation for the BTZ black hole as
\begin{gather}
S=\frac{-i \pi c}{12 l} (\frac{1}{\tau}-\frac{1}{\tilde{\tau}}),
\end{gather}
then by using the central charge in Eq. (\ref{eq:centralcharge}), one can see that for matching the two relations, the modular parameter of the warped CFT should be defined as
\begin{gather}
\tau=\frac{2i \Omega \nu}{ (\nu^2 +3)}.
\end{gather}
One can see that again a similar factor is appearing here. The imaginary factor can point to the appearance of closed time-like curves (CTCs) in the bulk.  

 The factor of $\frac{\nu}{(\nu^2+3)}$ can actually be explained by studying the Killing vectors of the space-time. The orbiflold construction of warped $\text{AdS}_3$ preserves a $U(1) \times U(1)$ subgroup of $SL(2,\mathbb{R})  \times U(1)$ isometries and is generated by two Killing vectors
\begin{gather}
\xi^{(1)} =\partial_t,  \ \ \ \  \ \ \ \ \ \ \ \ \xi^{(2)}=\frac{2 l \nu }{(\nu^2+3) }\partial_t +\partial_{\varphi}. 
\end{gather}

The same factor of $\frac{2 l \nu }{(\nu^2+3) }$ is also in the construction of the manifold which changes the partition function and therefore the free energy. This factor is the normalization factor $N_1=\frac{2l\nu}{(\nu^2+3)}$ in \cite{Donnay:2015joa} which is being fixed by matching the asymptotic Killing vector $\ell_0$ with the vector $\xi^{(2)}$. In addition, in the WCFT context as in \cite{Detournay:2012pc}, this factor relates to the anomaly in the transformation operators $T$ and $P$ which generates the infinitesimal coordinate transformation in $x$ and the gauge transformation in the gauge bundle, respectively.

Due to the current anomaly $k$, the operators $T$ and $P$ mix with each other.  This can be seen like a ``tilt" $(\alpha)$ in the mapping from $x^-$ to $\phi$ coordinates as in \cite{Detournay:2012pc},
\begin{gather}
x_-=e^{i \phi}, \ \ \ \ \ \ \ \ \ \ \ \ \ \ \ x^+=t+2\alpha \phi.
\end{gather}

This spectral flow parameter, $\alpha$, which is a property of the specific theory on the cylinder can be related to the factor $N_1$ for any theory.
In general, for the warped $\text{AdS}_3$ solutions one cannot simply use the relation (\ref{Kraus}). However, for calculating the thermal action for space-times with warped $\text{AdS}_3$ asymptotes, one can redefine the modular parameter using the Killing vectors of the manifold or the normalization constant in the symmetry algebra of the asymptotic geometry.   

In fact, the redefinition of the modular parameters have been also seen in Kerr/CFT contexts \cite{Guica:2008mu}. Specifically the NHEK geometry has an enhanced $SL(2, \mathbb{R}) \times U(1)$ isometry group where a different normalization factor appears in the algebra and in the normalization factors between the Killing vectors, and therefore in the redefinition of the modular parameter. 

Now using the methods introduced in previous chapters for calculating the conserved charges, we calculate the thermodynamic properties and the Gibbs free energies of the black holes with asymptotic warped $\text{AdS}_3$ geometry which could be called ``warped BTZ black holes".  The thermodynamical quantities are \cite{Nam:2010dd, Donnay:2015iia}, 
\begin{gather}\label{Tomega}
T_H=\frac{\nu^2+3}{8\pi \nu l} \bigg( \frac{r_+-r_-}{r_+- \sqrt{ \frac{(\nu^2+3)r_+ r_-}{4 \nu^2} } } \bigg), \ \  \ \ \ \Omega_H=\frac{1}{\nu l} \bigg( \frac{1}{r_+- \sqrt{ \frac{( \nu^2+3)r_+ r_-}{4\nu^2} } } \bigg),  
\end{gather}

\begin{gather}
T_L=\frac{(\nu^2+3)}{8 \pi l^2} (r_+ +r_- -\frac{1}{\nu} \sqrt{(\nu^2+3) r_- r_+} )  , \ \ \ \ \ T_R=\frac{(\nu^2+3)}{8 \pi l^2} (r_+ -r_-),
\end{gather}
and the conserved charges, mass and angular momentum are
\begin{gather}\label{Mass}
\mathcal{M}=Q_{\partial _t}= \frac{\nu (\nu^2+3) }{G l (20 \nu^2-3) } \Big( (r_-+r_+)\nu -\sqrt{r_+ r_-(\nu^2+3) } \Big) , 
\end{gather}
\begin{gather}\label{angular}
\mathcal{J} = Q_{\partial_{\varphi} }= \frac{\nu (\nu^2+3) }{4 G l ( 20\nu^2-3)} \Big( (5\nu^2+3) r_+ r_- -2\nu \sqrt{r_+ r_-(\nu^2+3) }  (r_+ +r_-) \Big) .
\end{gather}
Also, the conditions for the existence of black hole are 
\begin{gather}
\mathcal{J}  \le \frac{G l (20\nu^2-3)}{4\nu (\nu^2+3) } \mathcal{M}^2,  \ \ \ \ \ \ \ \ \ \mathcal{M} \ge 0, 
\end{gather}
which specifically do not put any new constraint on $\nu$.

The entropy of warped BTZ black hole in NMG is again
\begin{gather}\label{eq:entropy}
S_{BH}=\frac{8 \pi \nu^3}{(20\nu^2-3) G } (r_+-\frac{1}{2\nu} \sqrt{(\nu^2+3) r_+ r_- } )  .
\end{gather}

These thermodynamical quantities satisfy the first law of thermodynamics and their integrals follow the Smarr-like relation
\begin{gather}
M=T_H S_{BH}+2 \Omega_H J.
\end{gather}
The central charge is \cite{Donnay:2015iia} 
\begin{gather}\label{eq:centralcharge}
c=-\frac{96 l \nu^3 }{G (20\nu^4+57\nu^2-9) }.
\end{gather}

One can study the behavior of cosmological constant $\Lambda$ and the central charge versus the warping parameter $\nu$ shown in Figures (\ref{fig:c1}) and (\ref{fig:c22}). In the region where there is no CTSs, it is a monotonically increasing function of $\nu$. Also at $\nu=0$ or $\nu\to \pm \infty$, one can see that the central charge is zero which indicates that for infinitely squashed or stretched space time the Casimir energy vanishes. Note that the central charge diverges at  $\nu= \pm  \sqrt{\frac{3}{20}}\sim \pm 0.387$. 

For a physical theory the central charge should be positive. So if one assumes that $G=l =1$, then the constraint on $\nu$ is $0<\nu <\sqrt{\frac{3}{20}}$.

\begin{figure}[ht!]
\centering 
\begin{minipage}{.5\textwidth}
  \centering
\includegraphics[width=.7\linewidth]{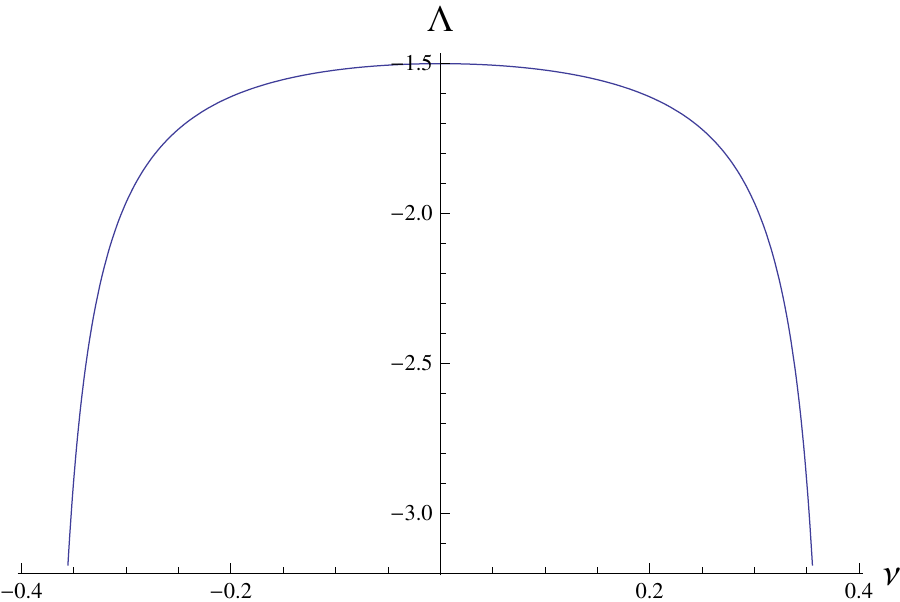} 
\caption{\label{fig:c1} The plot of cosmological \\ constant,  $\Lambda$ vs. $ -\sqrt{\frac{3}{20}}<\nu< \sqrt{\frac{3}{20}} $.} 
\end{minipage}%
\begin{minipage}{.5\textwidth}
\centering
\includegraphics[width=.7\linewidth]{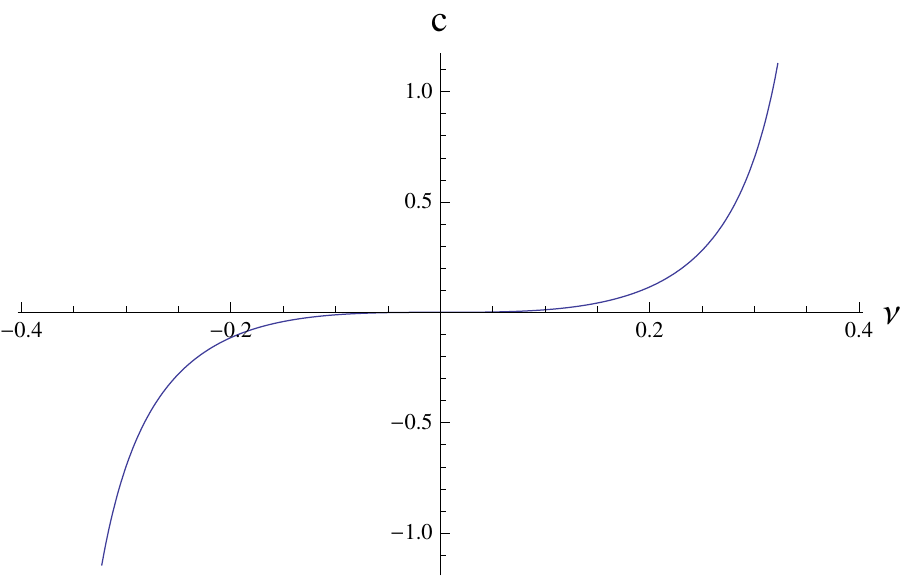} 
\caption{\label{fig:c22}  The central charge of NMG vs. $\nu$.}
\end{minipage}
\end{figure}

Defining
\begin{gather}
K=1+\frac{8 l \pi  T (l \pi  T-\nu ) \pm 4 \pi  l T \sqrt{4 \pi ^2 l^2 T^2-8 l \pi  T \nu +\nu ^2+3}}{\nu ^2+3}\text{  },
\end{gather}
then
\begin{gather}\label{eq:radius}
r_+=\frac{1}{\Omega \nu l \Big(1-\frac{1}{2 \nu} \sqrt{K(\nu^2+3)}\Big)  }, \ \ \ \ \ \ \ \ \ \  r_-=K r_+.
\end{gather}
The minus sign in $K$ above is acceptable which indeed could make $r_-$ smaller than $r_+$. Then the Gibbs free energy in terms of $\Omega ,T$, $K$ and $\nu$ is
\begin{equation}
\begin{split}
G_{WBTZ}&=\frac{1 }{G  \Omega l (20 \nu ^2-3)}\Bigg[- 8 \pi  T \nu^2 +\frac{ (\nu ^2+3) }{l \Big(1-\frac{1}{2\nu}\sqrt{K(\nu^2+3) }  \Big) }\Bigg ( (K+1)\nu -\sqrt{K(\nu^2+3) }  \nonumber\\&
 -\frac{(5\nu^2+3) K-2\nu(K+1)\sqrt{K(\nu^2+3) }    }{4\nu l \big(1-\frac{1}{2\nu}\sqrt{K(\nu^2+3)} \big )  } \Bigg)\Bigg].
\end{split}
\end{equation}

Notice that this Gibbs free energy only depends on $\Omega$, $T$ and $\nu$. One should also notice that the limit of the un-warped black hole is $\nu=\beta^2=1$ which corresponds to $m=1$ for $l=1$. In this limit the $G_{WBTZ}$ does not reach the limit of $G_{BTZ}$ in equation (\ref{eq:GBTZ}). This is specially obvious from the factors of $\Omega$ and $T$. Actually, this is not a real  problem as we should not expect to reach to the results of the previous section for the case of $\nu \to 1$ since these metrics have been written in two different coordinate systems. 

Now that we have found the free energies of the warped BTZ black hole and its vacuum, we can find the phase diagrams of temperature versus angular velocity as before and then we can compare them for different warping factors. Thus, we can study the effect of $\nu$ on the phase transitions in warped geometries.

We saw that the acceptable interval for $\nu$ is $0<\nu < \sqrt{\frac{3}{20} }\sim 0.3872 $. The phase diagram for $\nu=0.387$ is shown in Figure (\ref{fig:nu1}). The blue regions are where the warped BTZ black hole is the dominant phase and in the white regions the vacuum WAdS is dominant. If one increases $\nu$ until $\nu \ge \sqrt{\frac{3}{20}}$, the places of these two phases change with each other as it is also evident from the functionality of the central charge in terms of the warping factor $\nu$. 
 
 \begin{figure}[ht!]
\centering 
\begin{minipage}{.5\textwidth}
  \centering
\includegraphics[width=.7\linewidth]{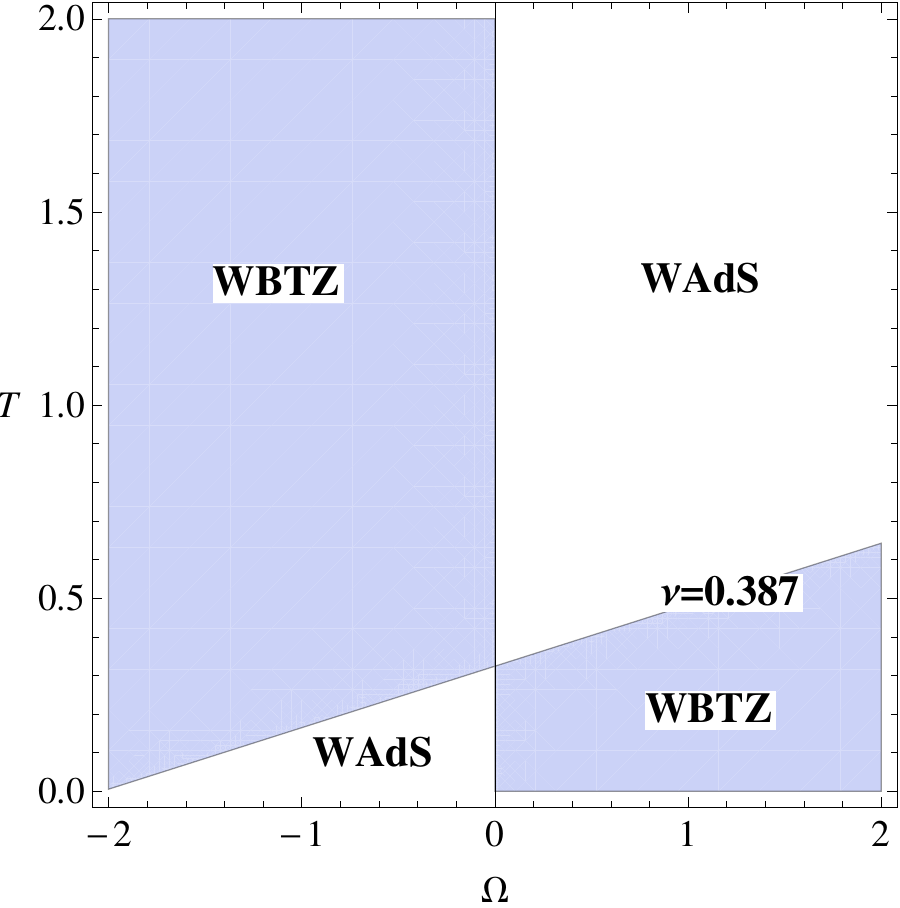}
 \caption{\label{fig:nu1}The phase diagram \\for $\nu=0.387$.} 
\end{minipage}%
\begin{minipage}{.5\textwidth}
\centering
\includegraphics[width=.7\linewidth]{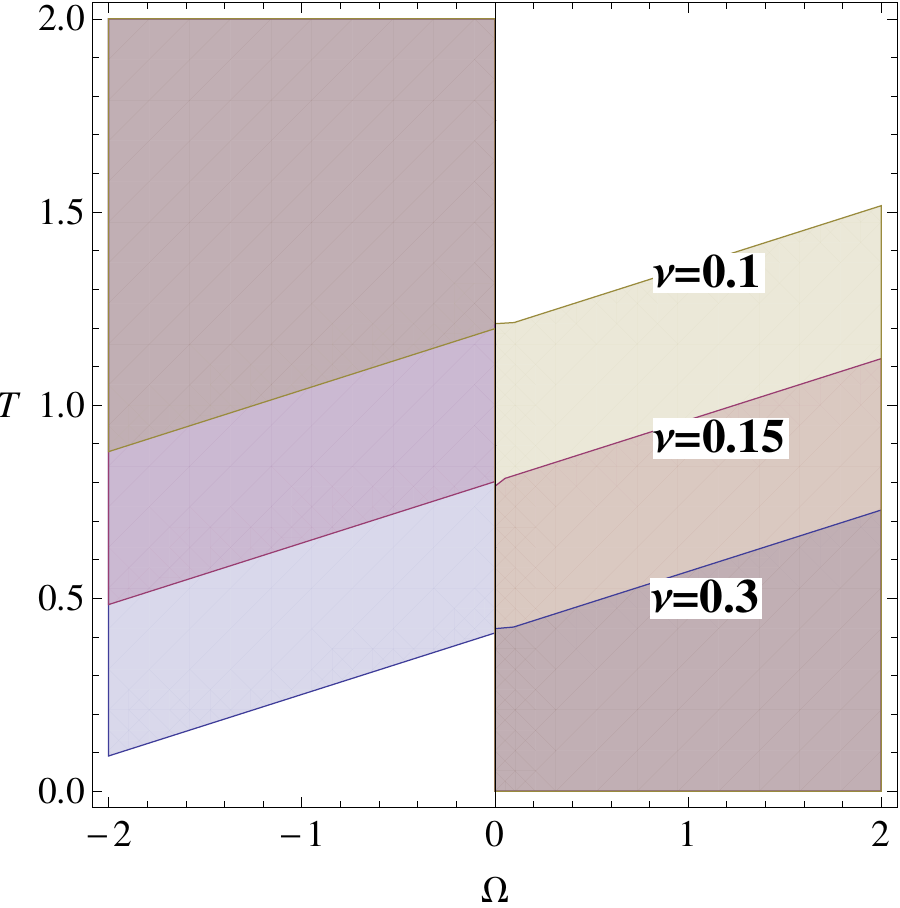}
\caption{\label{fig:warp} The phase diagram\\ for different $\nu$.}
\end{minipage}
\end{figure}

In these diagrams one can notice some other points. As one can see in the figures and also from the equations of the free energy of warped BTZ black hole, in the quadratic/non-local ensemble, unlike the grand canonical ensemble, the diagrams of warped AdS solutions are not symmetric although the NMG is parity preserving. One can notice that the behaviors for the positive and negative angular velocities are very different as the free energy is an odd function of $\Omega$ for the warped geometry unlike the previous case.

Also, if $\Omega>0$, the phase of warped black hole is dominant at lower temperatures and the thermal warped $\text{AdS}_3$ is dominant at higher temperatures.  However, in this case, a higher temperature would trigger the reverse Hawking-Page transition and the black hole phase transforms to the thermal warped $\text{AdS}_3$. So bigger $\Omega$ makes the black hole the dominant phase while bigger $T$ makes the vacuum phase dominant.

Furthermore, the effect of warping factor $\nu$ is shown in Figure (\ref{fig:warp}). If we define a critical temperature $T_c$ where the tilted line crosses the $\Omega=0$ axis, then one can see that increasing $\nu$ would decrease this critical temperature. So for $\Omega <0$, increasing the warping factor $\nu$ makes the black hole phase more dominant and for $\Omega>0$, it makes the vacuum $\text{AdS}_3$ the dominant phase.

\section{Phase diagram of warped $\text{AdS}_3$ solution in grand canonical ensemble}\label{sec:grand}

The WAdS black hole in the grand canonical solution is of the following form\\
\begin{gather}\label{metricg}
g_{\mu \nu}=\left(
\begin{array}{ccc}
 -\frac{r^2}{l^2}-\frac{H^2 (-r^2-4 l J+8 l^2 M)^2}{4l^3(l M-J)}+8M & 0 & 4J-\frac{H^2(4 l J-r^2)(-r^2-4 l J+8 l^2 M) }{4l^2(l M-J)}  \\  
0 & \frac{1}{\frac{16 J^2}{r^2}+\frac{r^2}{l^2}-8M } & 0 \\ 
4J-\frac{H^2(4 l J-r^2)(-r^2-4 l J+8 l^2 M) }{4l^2(l M-J)} & 0 & r^2-\frac{H^2 (4 J l -r^2)^2 }{4 l (l M-J)} 
\end{array} \right).
\end{gather}
The change of coordinates to go from this form to the form of Eq. (\ref{eq:WBTZ}) were derived in \cite{Detournay:2015ysa}. Also the phase diagram of this specific ensemble was just recently presented  in \cite{Detournay:2016gao}. Here we brought the phase diagram of this ensemble in Fig \ref{fig:grand} for the sake of comparison to the previous case.

Now using the charges and entropy derived in (\ref{charge1}) and (\ref{entropy1}), one can derive the Gibbs free energy as
\begin{gather}
G_{WBTZ} (T, \Omega)=\frac{-8 l^2 \pi^2 T^2 (1-2H^2)^{\frac{3}{2}} }{(17-42H^2)(1-l^2 \Omega^2) }, 
\end{gather}
and the vacuum corresponds to $M=-\frac{1}{8}$ and $J=0$. 

\subsection{local stability}
The Hessian matrix of the metric \ref{metricg} takes the following form \\
\begin{gather}
H=\left(
\begin{array}{cc}
 \frac{16 l^2 \pi ^2\left(1-2 H^2\right)^{3/2} }{\left(17-42 H^2\right) \left(l^2 \Omega ^2-1\right)} & \frac{-32 l^4 \pi ^2 T \Omega \left(1-2 H^2\right)^{3/2} }{\left(17-42 H^2\right) \left(l^2 \Omega ^2-1\right)^2} \\ \\
 \frac{-32l^4 \pi ^2 T \Omega  \left(1-2 H^2\right)^{3/2} }{\left(17-42 H^2\right) \left(l^2 \Omega ^2-1\right)^2} & \frac{16\pi ^2 T^2 \left(1-2 H^2\right)^{3/2}\left(l^4+3 l^6 \Omega ^2\right)}{\left(17-42 H^2\right) \left(l^2 \Omega ^2-1\right)^3} \\
\end{array}
\right).
\end{gather}

In order to have a locally stable solution, both of the eigenvalues of the Hessian should be negative. For the case of $\Omega=0$ the condition of making both eigenvalues negative is $H^2 <\frac{17}{42}$.

One can notice that unlike the previous ensemble, in the grand canonical ensemble the diagrams of warped BTZ black hole solution are symmetric. This could be just the result of the symmetry and parity preserving nature of this kind of solutions in BHT gravity. 

Also this could show us that the thermodynamical properties of these black holes and therefore the Hawking-Page phase diagrams could only really be meaningful in the grand canonical ensemble and not in other ensembles. The meaning and the dual interpretations of the phase diagrams in other thermodynamical ensembles is not particularly clear and deserves further study.

 \begin{figure}[htb!]
\centering 
  \centering
\includegraphics[width=.35\linewidth]{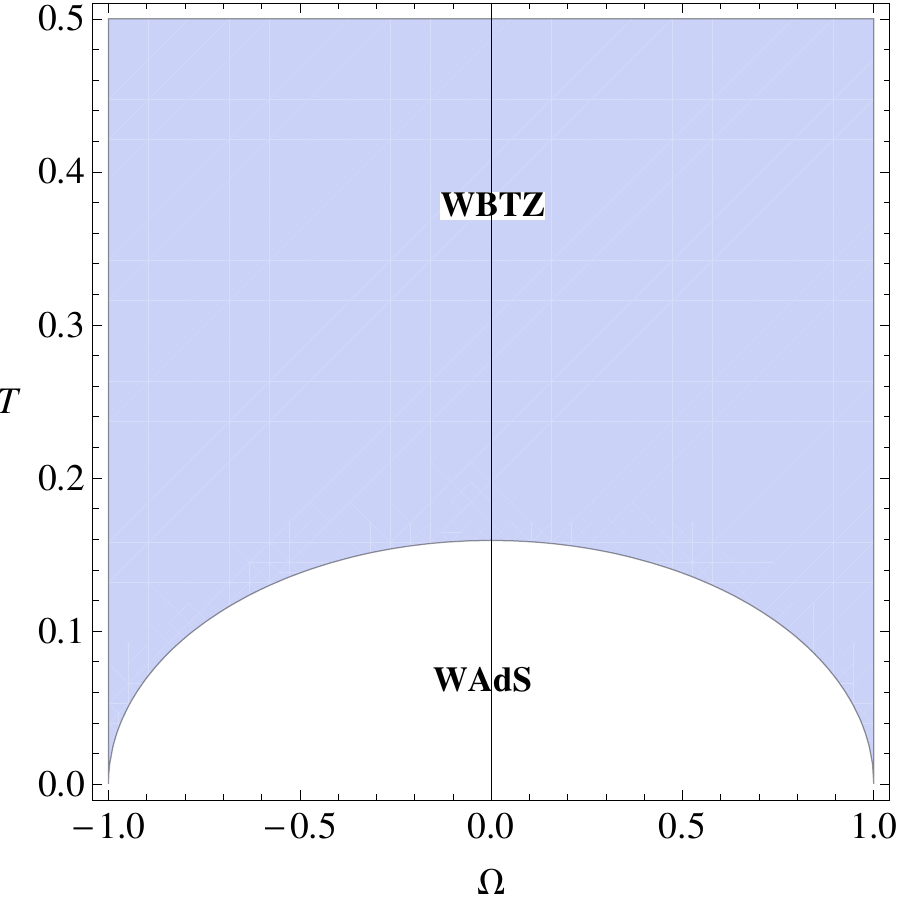}
\caption{\label{fig:grand}   Phase diagram for WAdS solution in grand canonical ensemble, $C=l=1$. } 
\end{figure}

\section{Phase diagram of the hairy black hole }\label{sec:hairy}

There exist yet another interesting black hole solution in the new massive gravity, the ``New Hairy black hole''. In this section, we are interested in studying the Hawking-Page phase transitions of this solution which first introduced in \cite{Oliva:2009ip} and later was studied more in \cite{Nam:2010dd}, \cite{Giribet:2009qz} and \cite{Mirza:2014xxa}. Its geodesics and entanglement entropy for the specific case of non-rotating solution were discussed recently in \cite{Hosseini:2015vya}. 

This hairy black hole solution exists for $m^2=\Lambda=-\frac{1}{2 l^2}$ as the parameters of the action (\ref{eq:action}). The form of its metric is as follows
\begin{gather}
ds^2=-N F dt^2+\frac{dr^2}{F}+r^2 (d\phi+N^\phi dt)^2,
\end{gather}
where
\begin{equation}
\begin{split}
N=&\Big( 1+\frac{b l^2}{4H} \big( 1-\Xi^{\frac{1}{2}} \big) \Big)^2,\ \ \ \ \ \ \ \ \ \  
N^\phi=-\frac{a}{2r^2} (4G_N M-bH),\nonumber\\
F=&\frac{H^2}{r^2} \Big( \frac{H^2}{l^2} +\frac{b}{2} \Big(1+ \Xi^{\frac{1}{2}} \Big)H+\frac{b^2 l^2}{16} \Big( 1-\Xi^ {\frac{1}{2}} \Big)^2- 4G_N M\ \Xi^{\frac{1}{2}} \Big),\nonumber\\
H=& \Big( r^2-2G_N M l^2 \big (1- \Xi^{\frac{1}{2}} \big) -\frac{b^2 l^4}{16} \big(1-\Xi^{\frac{1}{2}} \big)^2 \Big)^{\frac{1}{2}},
\end{split}
\end{equation}
and the definition of the parameter $\Xi$ is $\Xi :=1-\frac{a^2}{l^2}$ where $-l \le a \le l $. Now there are two conserved charges for this black hole which are $M$ and $J=M a$ and also a gravitational hair parameter which is $b$.
 
The thermodynamic parameters of this black hole are
\begin{equation}
\begin{split}
\Omega_+= & \frac{1}{a} \Big( \Xi^{\frac{1}{2}}-1 \Big),\ \ \ \ \ \ \ \ \ \ \ \
T= \frac{1}{\pi l} \Xi^{\frac{1}{2}} \sqrt{2 G_N \Delta M \Big (1+\Xi^{\frac{1}{2}} \Big)^{-1}},\nonumber\\
S= & \pi l \sqrt{\frac{2}{G_N} \Delta M \Big( 1+\Xi^{\frac{1}{2}}  \Big) },\ \ \ \ \ \ \ \ \ \ \ \ \
\Delta M = M+\frac{b^2 l^2}{16 G_N}.
\end{split}
\end{equation}

Then using all of these thermodynamic quantities one can read the Gibbs free energy.
We will see that the region where the black hole can be locally stable for any $b$ is $\Omega^2 l^2 <1$. So with this assumption we can simplify the relation as
\begin{gather}\label{GNBH}
G_{NBH} =\frac{l^2}{16G}\text{  }\left(\frac{16 \pi ^2 T^2 \left(5 l^2 \Omega ^2-1\right)}{\left(l^2 \Omega ^2-1\right)^2}-\frac{b^2 \left(3 l^2 \Omega ^2+1\right)}{l^2 \Omega ^2+1}\right). 
\end{gather}
We see that the Gibbs free energy, in addition to $\Omega$ and $T$,  depends also on the hair parameter $b$. One can also notice that there is no real $b$ which makes this free energy to vanish.

Now for studying the local stability we calculate the Hessian matrix which becomes
\begin{gather}
H=\left(
\begin{array}{ll}
 \frac{2 l^4 \pi ^2 \Omega ^2 \left(l^2 \Omega ^2+3\right)}{G \left(l^2 \Omega ^2-1\right)^2} & \ \ \ \  -\frac{4 l^4 \pi ^2 T \Omega  \left(5 l^2 \Omega ^2+3\right)}{G \left(l^2 \Omega ^2-1\right)^3} \\ \\
 -\frac{4 l^4 \pi ^2 T \Omega  \left(5 l^2 \Omega ^2+3\right)}{G \left(l^2 \Omega ^2-1\right)^3} & \ \ \ \ \frac{l^4 \left(b^2 \left(l^2 \Omega ^2-1\right)^4 \left(3 l^2 \Omega ^2-1\right)+24 \pi ^2 T^2 \left(l^2 \Omega ^2+1\right)^3 \left(1+5 l^2 \Omega ^2\left(2+l^2 \Omega ^2\right) \right)\right)}{4 G \left(l^2 \Omega ^2-1\right)^4 \left(l^2 \Omega ^2+1\right)^3} \\
\end{array}
\right).
\end{gather}

The region where both of the eigenvalues of the above matrix is negative would depend on the hair parameter $b$. The phase diagram for a specific value of $b=20$ is shown in Fig. (\ref{fig:localb1}).

For $G_N=l=1$ one can check that for any $b$ the angular velocity should be in the range of $-0.5< \Omega <0.5$, so that the black hole solution can be locally stable.  Increasing $\Omega$ can make the black hole locally unstable. Also the condition for $T$ depends on $b$. 

Increasing the hair parameter $b$ makes the locally stable region bigger. So basically the hair parameter makes the system more stable and $\Omega $ makes it more unstable. In condensed matter systems, one can also investigate the dual interpretation of this hair parameter and show how it makes the system stable.

 \begin{figure}[ht!]
\centering 
\begin{minipage}{.5\textwidth}
  \centering
\includegraphics[width=.7\linewidth]{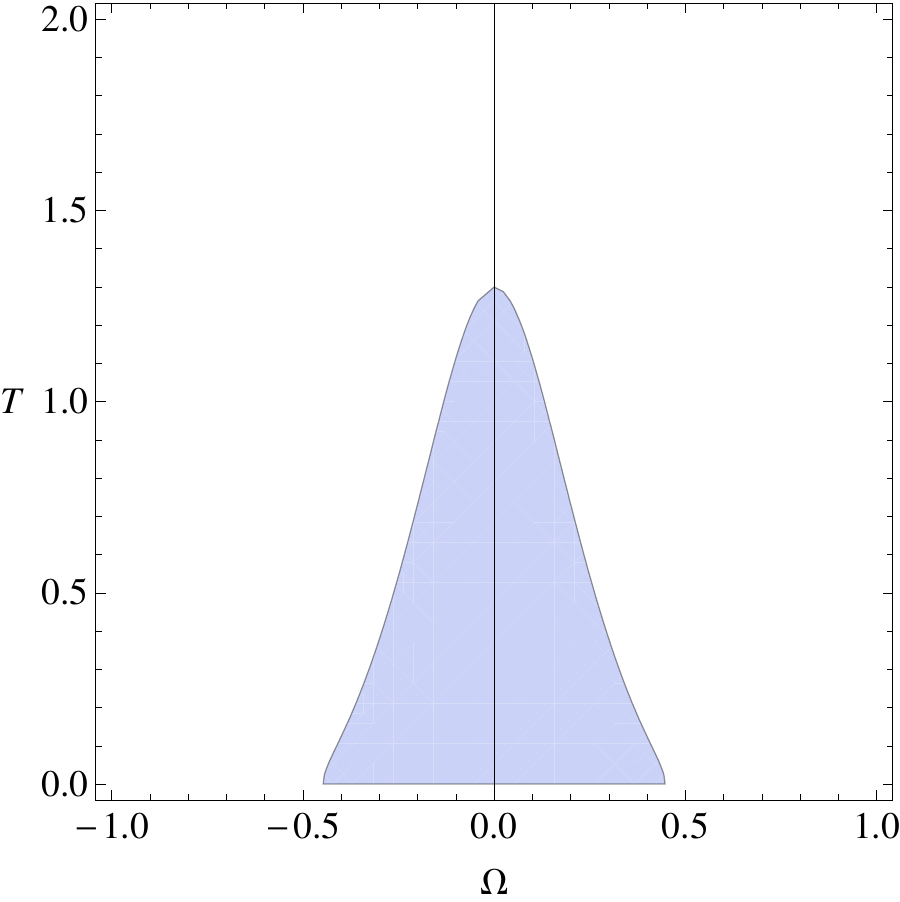}
 \caption{\label{fig:localb1}  The local stable region \\for $b=20$. } 
\end{minipage}%
\begin{minipage}{.5\textwidth}
\centering
\includegraphics[width=.7\linewidth]{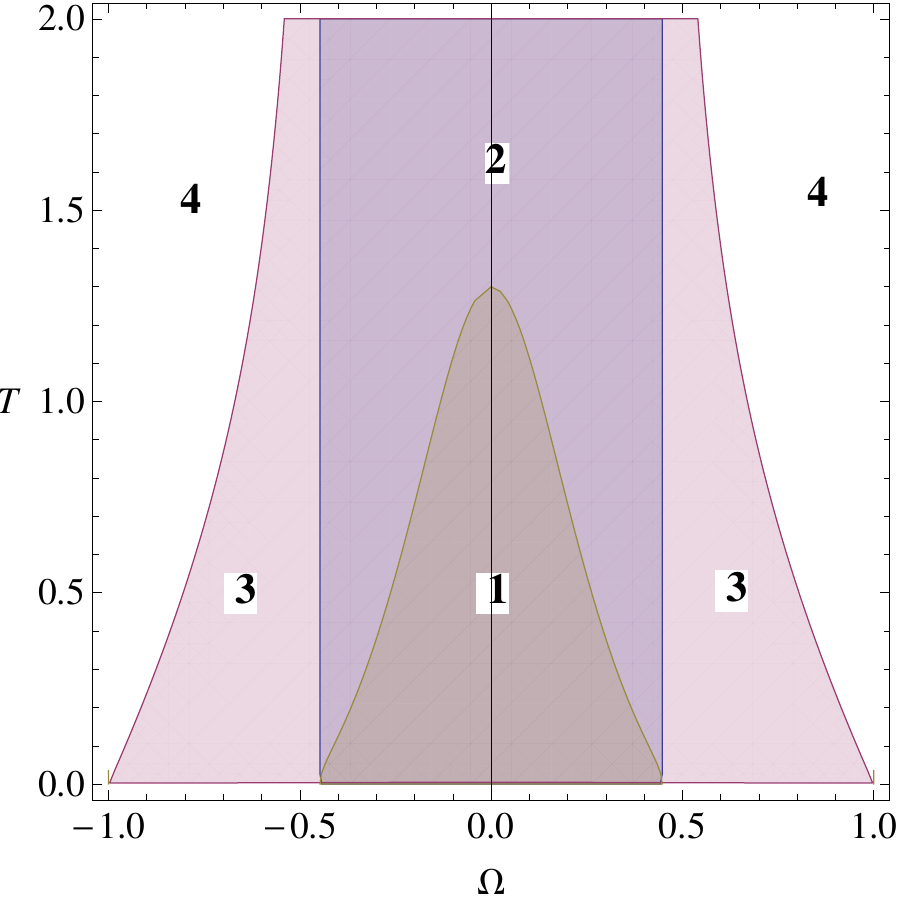}
\caption{ \label{fig:localb2}  The phase diagram\\ for $b=20$.} 
\end{minipage}
\end{figure}

To study the phase diagram of the system, we now can compare the Gibbs free energies of this black hole with the free energies of other solutions. The phase diagram for the region of local stability, the vacuum $\text{AdS}_3$ solution and the ground state of new hairy black hole is shown in figure (\ref{fig:localb2}). The region where $\Delta G_1=G_{\text{AdS}}-G_{\text{NBH}} >0$ is the union of regions 1, 2 and 3. By comparing the Gibbs free energy of this black hole with the the free energy of the vacuum AdS, one can see that for the region of locally stable, for any $b$, only the black hole phase is present. So the only phase that is both locally and globally stable is the black hole case. Outside of region 1, the phase would not be even locally stable. 

Also the case of $M=M_0=-\frac{b^2 l^2}{16 G}$ in this solution is the ground state which corresponds to an extremal case where both the left, right and the Hawking temperature and also the entropy vanish \cite{Giribet:2009qz}. The free energy is
\begin{gather}
{G_0}_{\text{NBH}}=-\frac{b^2 l^2}{16 G_N} \left(3-\frac{2}{1+l^2 \Omega ^2}\right),
\end{gather}
and the region where $\Delta G_2={G_0}_{\text{NBH}}-G_{\text{NBH}}>0$ is the union of 1 and 2. Again one can see that $-0.5 <\Omega <0.5$ is the region where the black hole can be stable.

One can also plot the diagram of $M$ versus $J$ and study the effect of other physical parameters or conserved charges on the phase transitions which could shed light on many other physical characteristics.

\section{The inner horizon thermodynamics}\label{sec:inner}
 It would also be useful to study the thermodynamics of black holes \textit{inside} the horizon as the quantities from \textit{outer} of the horizon $(r_+)$ combined with the \textit{inner} ones $(r_-)$ can provide additional information about the central charges or scattering data around the black hole. Also the relations between the thermodynamics of inner and outer of horizon would be of practical use in holographical applications, such as the examples in \cite{Detournay:2015ysa, Giribet:2015lfa}.
 
 One can first integrate the Wald's formula on $r=r_-$ to find the inner horizon entropy. Alternatively, one can use the following relations,
\begin{gather}
 T_{R,L}=\frac{T_- \pm T_+}{\Omega_- -\Omega_+}, \ \ \ \ \ \ S_{\pm}=S_R\pm S_L, \ \ \ \ \ S_{R,L}=\frac{\pi^2 l}{3} c \ T_{R,L}.
 \end{gather}
 So one gets,
 \begin{gather}
S_{-}=\frac{8 \pi \nu^3}{(20\nu^2-3) G } (r_- -\frac{1}{2\nu} \sqrt{(\nu^2+3) r_+ r_- } )  .
\end{gather}
The temperature at the inner horizon is
\begin{gather}
T_-=\frac{1}{2\pi l} \frac{\partial_r N} {\sqrt{g_{rr}} } \Big | _{r=r_-}=\frac{\nu^2+3}{8\pi \nu} \bigg (\frac{r_+-r_-}{r_- -\sqrt{\frac{(\nu^2+3)r_+ r_- }{4\nu^2} } } \bigg ),
\end{gather}
and the angular velocity at the inner horizon is
\begin{gather}
\Omega_-=\frac{1}{l} N^{\phi} (r_-)= \frac{1}{\nu  l} \bigg ( \frac{1}{ r_--\sqrt{\frac{ \left(\nu ^2+3\right)r_+ r_-}{4 \nu ^2}}}\bigg).
\end{gather}

As explained in \cite{Giribet:2015lfa}, the statement of inner horizons mechanics is that the product of all horizons' entropies should be proportional to the conserved charges that at the quantum level are quantized. In this case this charge is only $J$. The product of the inner and outer horizon entropies is
\begin{gather}
S_+ S_-=\frac{16 \pi^2 \nu^4}{ G^2 (20\nu^2-3)^2}\Big( (5\nu^2+3) r_+ r_--2\nu \sqrt{(\nu^2+3) r_+ r_-}(r_++r_-) \Big),
\end{gather}
which as expected can be written as a factor of $J$, i.e, $S_+ S_- \propto J $, and it is independent of $M$. This is because $S_+ S_-$ is holographically dual to the level matching condition of the warped $\text{CFT}_2$ and this is also consistent with the WAdS/WCFT picture. 

As explained in \cite{Chen:2012mh}, the mass-independence of the product of entropies are satisfied here as $T_+ S_+=T_- S_-$ which is also a result of the parity-preserving nature of this theory, i.e., the fact that the left and right central charges are equal. Also based on \cite{Castro:2013pqa}, as the Smarr-relation holds for the Einstein gravity with  the higher curvature corrections in the form of NMG, we expect such result. Moreover, the first law of black hole thermodynamics would be satisfied as
\begin{gather}
dM= \pm T_{\pm} dS_{\pm}+\Omega_{\pm} dJ.
\end{gather}
This was expected since, as explained in \cite{Chen:2012mh}, if the first law is satisfied in outer horizon, it should also be satisfied in the inner-horizon as well. Then, using both of these relations, one can derive the same results for the $M$ and $J$ in the form of equations (\ref{Mass}) and (\ref{angular}) which is consistent with WAdS/WCFT picture \cite{Donnay:2015vrb}.

\section{Entanglement entropy of WCFT in BHT}\label{entanglement}

Recently, using the Rindler method, the authors of \cite{Castro:2015csg} found a general expression for the entanglement and the Renyi entropy of a single interval in a $(1+1)$-dimensional WCFT. These authors also provided the geodesic and massive point particle description in a lower spin gravity as the dual of a WCFT. This way one could holographically study the results from the bulk geometry.

Their general result for the entanglement entropy of an interval in the warped CFT theories is as follows
\begin{gather}\label{eq:SEE}
S_{EE}= i P_0^{\text{vac}} \ell^* \left( \frac{\bar{L} }{L}-\frac{\bar{\ell^*} }{\ell^*} \right) -4 L_0^{\text{vac}} \log \left( \frac{L}{\pi \epsilon} \sin \frac{ \pi \ell^* }{L} \right ).
\end{gather}

In the above formula $\ell^*$ and $\bar{\ell^*}$ are the separation in space and time respectively and $L$ and $\bar{L}$ are related to the identification pattern of the circle that defines the vacuum of the theory \cite{Castro:2015csg}. 

The authors of \cite{Castro:2015csg} have explained that the second term is well-known and expected, but the first term seems exotic. With the motivation of finding the nature of the first term, we study this result for our specific case of NMG theory.

Since the theory is parity even, for the vacuum of $\text{WAdS}_3$ which is holographically dual to WCFT one can write the Virasoro and Kac-Moody operators as
\begin{gather}\label{eq:operatores}
\tilde{P}_0^{(\text{vac})}=\mathcal{M}^{\text{(vac)} }, \ \ \ \ \ \ \ \ \ \ \ \ \ \tilde{L}_0^{\text{(vac)}}=\frac{1}{k} \left (  \mathcal{M}^{\text{vac}}  \right)^2,
\end{gather}
where, as found in \cite{Donnay:2015iia}, the mass parameter is 
\begin{gather}\label{eq:M}
\mathcal{M}^{(\text{vac})}= i {\mathcal{M}}_{\text{God} }=-i \frac{4 \ell^2 \omega^2 }{G(19 \ell^2 \omega^2-2) }.
\end{gather}
Here, as expected, $P_0^{\text{(vac)}}$ has an overall factor of $i$ which makes the first term of (\ref{eq:SEE}) real.

It would also be interesting to write the Virasoro and $\text{U}(1)$ Kac-Moody operators of NMG in terms of its central charge as well. So using Eqs. (\ref{eq:operatores}) and (\ref{eq:M}), the expressions for the central charge and the level $k$ in Eq. (\ref{eq:param}), one can write
\begin{gather}
\tilde{P}_0^{\text{(vac)}}=-\frac{i  c }{12 l^2 \omega} (1+\ell^2 \omega^2 ), \ \ \ \ \ \ \ \ \ \  \tilde{L}_0^{\text{(vac)}}=-\frac{c}{24}.
\end{gather}
We can also compare it with the operators of TMG,
\begin{gather}
\tilde{P}_0^{\text{(vac)}}=-\frac{c_L}{24}, \ \ \ \ \ \ \ \ \ \ \ \ \ \ \tilde{L}_0^{\text{(vac)}}=-\frac{c_R}{24}.
\end{gather}

One can see that similar to the TMG case, the Virasoro part is only proportional to $c_R$. However, the Kac-Moody part does not only depend on $c$ but also on $\omega$, and therefore it cannot define a central charge. Still $c_L$ is called a central charge only by convention. These relations could be useful for studying the dual WCFT of time-like G$\ddot{\text{o}}$del or warped BTZ in these massive gravitational theories.

Now by using (\ref{eq:operatores}), the entanglement entropy of a strip in a WCFT theory dual to a NMG bulk could be found as
\begin{gather}\label{eq:SEE2}
S_{EE}=\frac{4\ell^2 \omega^2 }{G (19 \ell^2 \omega^2 -2) } \left(  \ell^* \Big( \frac{\bar{L} }{L}-\frac{\bar{\ell^*} }{\ell^*} \Big) + \frac{2\ell^2 \omega}{1+\ell^2 \omega^2}  \log \Big( \frac{L}{\pi \epsilon} \sin \frac{ \pi \ell^* }{L} \Big ) \right).
\end{gather}
Notice that $\ell$ here is just the length of AdS radius in the G$\ddot{\text{o}}$del space-time and is independent of the length of intervals $\ell^*$ and $\bar{\ell^*}$.

The plot of $S_{\text{EE}}$ versus $\omega$ is shown in Figure (\ref{fig:S1}). The lengths are considered to be constant and equal to one, so one can examine the effect of the warping factor of G$\ddot{\text{o}}$del space-time on the entanglement entropy.  As it is shown in Fig. (\ref{fig:S1}), the entanglement entropy is a monotonic function of $\omega$. One can also see that at $\omega= \pm \sqrt{\frac{2}{19}}$ the entanglement entropy diverges. 

 \begin{figure}[]
\centering 
\begin{minipage}{.5\textwidth}
  \centering 
\includegraphics[width=.7\linewidth]{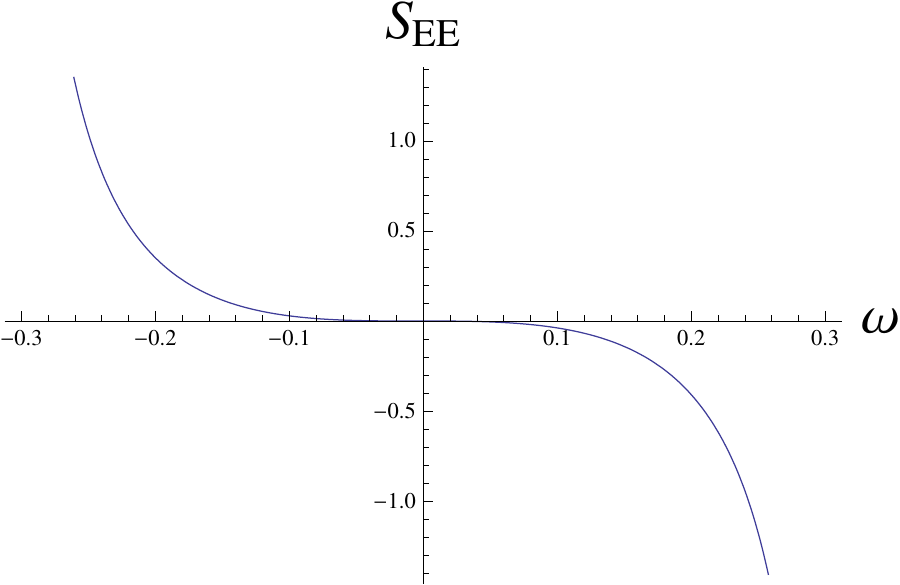}
 \caption{\label{fig:S1}  The plot of $S_{\text{EE}}$ versus $\omega$. } 
\end{minipage}%
\begin{minipage}{.5\textwidth}
\centering
\includegraphics[width=.7\linewidth]{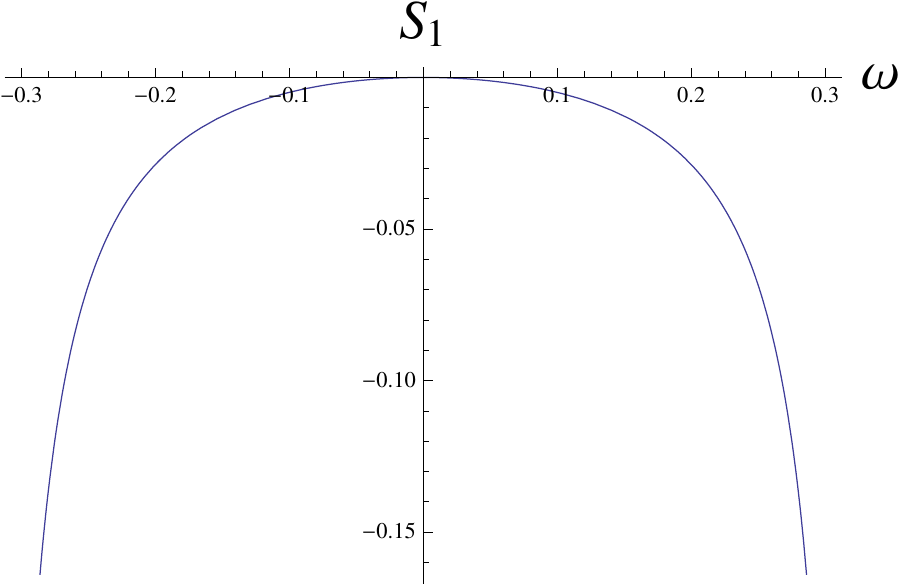} 
\caption{\label{fig:S2}  The plot of $S_1$ versus $\omega$.} 
\end{minipage}
\end{figure}

Also the plot of the peculiar first term contributing to the entanglement entropy for NMG versus the parameter $\omega$ is shown in Fig. (\ref{fig:S2}) where its  specific relation is
\begin{gather}
S_1=\frac{\omega k }{2(1+\ell^2 \omega^2) }  \Big( \frac{\bar{L} }{L}-\frac{\bar{\ell^*} }{\ell^*} \Big) \ell^2 \ell^*.
\end{gather}
  One can notice that, based on the sign of $\Big( \frac{\bar{L} }{L}-\frac{\bar{\ell^*} }{\ell^*} \Big)$, it can be a positive or negative term, has an extremum at $\omega=0$ and also it is a symmetric function. One can also write $\omega$ and $\ell$ in terms of the physical quantities $m$ and $\Lambda$ as
\begin{gather}\label{eq:relations}
\omega=\pm \sqrt{\frac{-2m^2}{7} \pm \frac{\sqrt{ 5m^4-7m^2 \Lambda} }{7 \sqrt{3}} }, \ \ \ \ \ \ \ \ \ \ \ 
\ell=\pm \sqrt{\frac{144m^2 \pm 38\sqrt {3m^2(5m^2-7 \Lambda) } } {11m^4-361m^2 \Lambda}  }.
\end{gather}

Using the above relations one can study the interplay between the entanglement entropy and physical parameters $m$ and $\Lambda$ as well.

As explained in \cite{Castro:2015csg}, in the holographic picture, the first term contributes to the black hole entropy. This term is a $\text{U}(1)$ contribution to the entanglement entropy and is not UV divergent like the second term. As a new observation in warped AdS geometries, one can see that $S_1$ is proportional to the volume of the interval although it corresponds to the vacuum states and not particularly to the mixed states. Also as noted in \cite{Castro:2015csg}, this term in WCFT, unlike CFT, is independent of the Renyi replica index and it is also a rather complicated function of physical parameter of the theory such as $m$ and $\Lambda$ as in relations (\ref{eq:relations}). These features could be compared with other solutions and other massive gravity theories in order to shed light on the nature of each term.

\chapter{Near horizon region behavior of $U(1)^4$ gauged supergravity black holes}\label{chap5}

As mentioned before, the Gauge/Gravity correspondence is a paradigm that relates a strongly coupled field theory in $d$ dimensions to a weakly coupled gravity theory in one dimension higher \cite{Maldacena:1997re}. In using this duality, one approach is, first, by using the symmetries of the desired field theory, one constructs a Lagrangian in the gravity theory and then solves the equations of motion to find the characteristics of the field theory side. This is the so called ``bottom-up" approach which has been used extensively to build some desired models. However, by choosing the parameters of the theory arbitrarily, this approach could give some unphysical or maybe unstable solutions; see our work in \cite{Ghodrati:2014spa} for an example. One way to resolve this issue is to use the ``top-down" approach. In this approach, by choosing a solution of supergravity for the bulk geometry from the beginning there would be a direct connection with string theory and supergravity, and therefore the CFT dual is completely known. This approach has been used among the others in \cite{DeWolfe:2012uv, DeWolfe:2013uba, DeWolfe:2011aa, DeWolfe:2015kma,Cosnier-Horeau:2014qya}.

 In this chapter we choose the extremal four-dimension $U(1)^4$ charged black brane solution of $\mathcal{N}=2$ gauged supergravity to holographically study a zero-temperature and non-zero density system of $3d$ $\mathcal{N}=2$ super-conformal field theory which has four distinct chemical potentials. The relevant field theories of these solutions were constructed in ABJM model \cite{Aharony:2008ug} and constitute a Chern-Simons theory with gauge group $U(N) \times U(N)$ and levels $k$ and $-k$, respectively. There is also a matter sector that leads to SUSY.  The main characteristic of these black holes/branes is that they have four charges and for a general extremal form of the theory, the near-horizon geometry is non-singular, the entropy is non-zero at zero temperature and also the near-horizon geometry has an $\mathrm{AdS}_2$ sector. 

For simplicity one can imagine that three of these four charges are equal, and therefore one calls the system  the regular extremal ``3+1-charge". One can choose one of these charges to be zero and ends-up with a ``3-charge" black brane, which has a singular near-horizon and zero entropy at zero temperature which is physically desirable since it is complying with thermodynamics expectations. 

As has been mentioned in \cite{DeWolfe:2013uba}, this singularity is a good type of singularity as it gets resolved in the one dimension higher uplifted theory. It can be shown that for $4d$ $U(1)^4$ gauged supergravity which is being uplifted to $5d$, and also $3d$ $U(1)^3$ gauged supergravity which is being uplifted to $4d$, the uplifted near-horizon geometries have an $\mathrm{AdS}_3$ or BTZ sector and the singularity is also being resolved in the uplifted backgrounds. 

We find that there are differences in the physical nature of the three equal charges and the other non-equal charge in the system of 3+1-charge black brane. We show that an oscillatory phase with the oscillatory momentum $k_{\text{osc}}$ can exist in the near horizon of extremal 3-charge black branes. So the near horizon can be unstable due to the effects of these charges. But for the 1-charge black brane, $k_{\text{osc}}$ cannot exist and even with a big 1-charge, the near horizon is always stable. The other difference between these two charges is that the gap can only exist when the 1-charge $q^\prime$ is turned off. By turning on even a small value of $q^\prime$ the gap would disappear \cite{DeWolfe:2013uba, DeWolfe:2015kma}, which is the result of non-uniformity of states and the discontinuity in the chemical potential $\mu_1$. Also in the uplifted geometry, only $q^\prime$ would associate to the Kaluza-Klein charge of reduction and the compact momentum. The source of the gauge field of 1-charge in the one dimension higher is the graviphoton, but the source of 3-charge gauge field $A_\mu$ can be understood by uplifting to $10d$ or $11d$ which is a two-form. 

By solving the Dirac equation for each specific mode, one can also study the linear response of probe spin-1/2 fermions in the background of $4d$ supergravity charged black brane. For the $5d$ case, this has been done in \cite{DeWolfe:2013uba, DeWolfe:2011aa, DeWolfe:2014ifa}. The authors mentioned that the main feature of the fermionic response in this top-down approach is that the mass term couples to the running dilaton and diverges at the singularity ($r \to 0$). This divergence of the mass term of fermions at the horizon is just a feature of the top-down solutions coming from supergravity which we aim to study here for the four dimensional case as well.

 In this chapter, we study the behavior of the Dirac equation and its solutions in the two mentioned near horizon limits and therefore different sectors of the theory.  The point is that as it has been mentioned in \cite{DeWolfe:2013uba}, for the extremal 2-charge case, there exists a gap of the order of the chemical potential where the fluctuations of the modes with the fluctuation energy of $\omega \equiv \omega_* -i \Gamma$ are stable and the ratio of the real part of the fluctuation energy to the imaginary part is constant, i.e. $\Gamma / \omega_{\ast} \to constant$, corresponding to a non-Fermi liquid behavior and the main result of our investigation is to establish that, in the four dimensional $3+1$ charged supergravity black hole case, such a gap and therefore such a behavior exists as well.

\section{ Charged black brane solution }
The Lagrangian of the $4d$ $U(1)^4$ gauged supergravity solution without axions which we are interested in is \cite{deBoer:2011zt}
\begin{eqnarray} \label{eq:lag}
e^{-1} \mathcal{L}_4&=R-\frac{1}{2}(\partial \vec{\varphi})^2+8g^2 (\text{cosh}\varphi_1+\text{cosh}\varphi_2+\text{cosh}\varphi_3)-\frac{1}{4}\sum\limits_{i=1}^4 e^{\vec{a}_i.\vec{\varphi}} \Big(F^i_{(2)}\Big)^2,
\end{eqnarray}
where
\begin{flalign} \label{eq:rel}
\vec{\varphi}&=(\varphi_1,\varphi_2,\varphi_3),\ \ \ \ 
\vec{a}_1=(1,1,1), \ \ \ \vec{a}_2=(1,-1,-1), \ \ \ \vec{a}_3=(-1,1,-1),  \ \ \ \vec{a}_4=(-1,-1,1). 
\end{flalign}

The special sector of this theory that we are going to study, is the one where three of the four gauge fields are equal to each other. Checking the equations of motion for the consistency of the theory implies that in this case three of four scalar fields should also set equal to each other. Using the supergravity constraints between the scalar fields $X_i=e^{\varphi_i}$, $ X_1 X_2 X_3 X_4=1$ and  Eq. (\ref{eq:rel}), we find that $\varphi_1=\varphi_2=-\varphi_3=\varphi$ and then by redefining the parameter $\varphi$ to $\varphi=\frac{\phi}{\sqrt{3}}$, the Lagrangian of 3+1-charge sector is
\begin{equation} \label{eq:lagmain}
\mathcal{L}_{(3+1)}=R-\frac{1}{2}(\partial \vec{\phi})^2+24g^2 \text{cosh} \frac{\phi}{\sqrt{3}}-\frac{3}{4} e^{\frac{\phi}{\sqrt{3}}} F^2-\frac{1}{4} e^{-\sqrt{3} \phi}f^2.
\end{equation}
Note that the kinetic term is not affected under rescaling.  By letting $F=F^\prime=0$ and for a constant and vanishing dilaton field $\phi=0$, corresponding to the $\mathrm{AdS_4}$ theory, we can find the coupling of the Lagrangian, i.e., $g=\frac{1}{2L}$. Also, one can find,  $X_1=X_2=X_3=e^{-\frac{\phi}{2\sqrt{3}}}$, and $X_4=e^{\frac{\sqrt{3}\phi}{2}}$. The dual of this theory is a $\mathrm{CFT_3}$, with the symmetry group of $\mathrm{SO(8)}$ and for our special case with the subgroup of $U(1)^4$. 

Just for comparison, the Lagrangian of $5d$ supergravity theory is \cite{DeWolfe:2013uba, DeWolfe:2014ifa}:
\begin{equation}
e^{-1}  \mathcal{L}=R-\frac{1}{2}(\partial\varphi)^2+\frac{8}{L^2}e^{\frac{\varphi}{\sqrt{6}}}+\frac{4}{L^2} e^{\frac{-2\varphi}{\sqrt{6}}}-e^{\frac{-4\varphi}{\sqrt{6}}}f_{\mu\nu}f^{\mu\nu}-2e^\frac{2\varphi}{\sqrt{6}}F_{\mu\nu}F^{\mu\nu}-2\epsilon^{\mu\nu\rho\sigma\tau}f_{\mu\nu}F_{\rho\sigma}A_{\tau},
\end{equation}
whose dual theory is a $\mathrm{CFT_4}$, with the symmetry group of $\mathrm{SO(6)}$ and the subgroup of $U(1)^3$. Note that for the $4d$ case, the Chern-Simons term is eliminated. 

Now the general form of the static $U(1)^4$ gauged black hole solution in $d=4$ is 

\begin{equation} \label{metric}
ds^2=-H^{-\frac{1}{2}} f(r) dt^2+H^\frac{1}{2}(\frac{dr^2}{f(r)}+r^2d\Omega_{2,k}^2),
\end{equation}
where
\begin{eqnarray}
H=H_1H_2 H_3 H_4,\ \ \ \ \ \
H_I=1+\frac{\mu\ \text{sinh}^2\beta_I}{k\ r}=1+\frac{q_I}{r}, \ \ \ \ \ \
f(r)=k-\frac{\mu}{r}+ \frac{r^2}{L^2} H,
\end{eqnarray}
$q_I=\frac{\mu\ {\text{sinh}}^2 \beta_I}{k}$ is a parameter of the solution, and $k$ can be 1, 0, -1, for the spherical, flat or hyperbolic $(S^2, T^2, H^2)$ foliating transverse space \cite{Cvetic:1999xp, Fareghbal:2008eh, Duff:1999gh}. Here we consider the case where $k=0$ corresponding to a black brane solution, since we want to study a dual non-compact CFT on a flat space in the boundary. 

The black brane solution of both $4d$ and $5d$ theories are in the family of extremal vanishing horizon (EVH) black holes. In these black holes, for the extremal case where the temperature goes to zero, $T \to 0$, then this leads to $A_h \to 0$ so $S\to 0$ while the ratio of $\frac{A_h}{T}$ remains constant. This is a desirable feature as the physical origin for a non-zero entropy at zero temperature has not been understood well. Note also that in their near horizon limit, these black hole solutions contain an $\mathrm{AdS}_3$ throat \cite{Fareghbal:2008ar, SheikhJabbaria:2011gc, deBoer:2010ac, deBoer:2011zt}.

The gauge fields and scalar fields of the theory are
\begin{gather}
A_I=\frac{Q_I}{q_I}(\frac{1}{H_I(r)}-\frac{1}{H_I(r_H)})dt=\frac{Q_I}{q_I+r_H} \left(1-\frac{q_I+r_H}{q_I+r}\right)dt, \ \ \ \ \ \  X_I=\frac{H^\frac{1}{4}}{H_I}.
\end{gather}

The difference between these gauge fields and the one in \cite{Fareghbal:2008eh} is that we chose a gauge where at $r=r_H$, $A_t=0$ so the chemical potentials become finite at the boundary. Also, $X_I$s  which parameterize the three scalars satisfy the relation $X_1 X_2 X_3 X_4 =1$.\\
For the case of $k=0$, by rescaling ${\text{sinh}}^2 \ \beta_I \to k\ {\text{sinh}}^2  \beta_I$ and considering the limit of $k$ to zero, \cite{Cvetic:1999xp} one gets
\begin{equation}
A_{(1)}^I=\frac{H_I^{-1}(r)-{H_I}^{-1}(r_H)}{\text{sinh}\beta_I} dt.
\end{equation}
In the above equation, $\mathrm{sinh}^2 \beta=\frac{q_i k}{q_i +\mu}$ where $q_I$ and $\mu$ are the parameters of the solution. These five parameters are related to each other by \cite{Fareghbal:2008eh}
 \begin{equation}
Q_I=\sqrt{q_I (q_I+\mu)}.
\end{equation}
The physical observables of these $4d$ black hole solutions are the ADM mass and the electric charges\cite{Fareghbal:2008eh} 
\begin{gather}
M=\frac{1}{2G_N^{(4)}}(2\mu+q_1+q_2+q_3+q_4), \ \ \ \ \ \ \  J_I=\frac{L}{2G_N^{(4)}} Q_I .
\end{gather}
At the horizon, $f(r=r_H)=0$, so 
\begin{equation}\label{eq:exsol}
-\mu r +\frac{1}{L^2}(r+q_1)(r+q_2)(r+q_3)(r+q_4)\ \big |_{(r=r_H)}=0.
\end{equation}
We want to study the 3+1-charge solution in the case where
\begin{equation}
q_1=q_2=q_3=q,\\  \ \ \ \ \ \ \ \ \ \ \ \ q_4=q^{\prime},
\end{equation}
corresponding to
\begin{equation}
Q_1=Q_2=Q_3=Q,\\  \ \ \ \ \ \ \ \ \ \ \ \ Q_4=Q^{\prime}.
\end{equation}
The prime are used just to define a new parameter here. So from (\ref{eq:exsol}), one would have
\begin{equation} \label{eq:mu}
\mu =\frac{ \left(q+r_H\right){}^3 \left(q'+r_H\right)}{L^2 r_H}.
\end{equation}
Then for the general regular case, using (\ref{eq:mu}), the horizon temperature is
\begin{equation}
T_H=\frac{1}{4\pi} \frac{\partial g_{tt}}{\partial r}\Big|_{ r=r_H}=\frac{  | 3 r_H{}^2+2 q' r_H- q q' |}{4 L^2 \pi  r_H }\sqrt{ \frac{q+r_H}{ q'+r_H}}.
\end{equation}
For the regular extremal 3+1-QBH where $T_H {(r=r_H)}=0$, there are two solutions
\begin{equation} \label{eq:extg}
 \mathrm{I.} \ \ \  r_H=-q=Q, \ \mbox{with any $q^\prime$ \ \ \ and}\ \ \ \ \ \  
 \mathrm{II.}\ \ \   3 r_H{}^2+2 q' r_H-q q'=0.
\end{equation}
For the first solution, $\mu=0$ and the solution is BPS displaying the expected mass to charge relation. As these black holes are solutions of $11d$ supergravity, for a BPS solution with $n$ number of non-zero charges $q_I$, $32/2^n$ number of super-symmetries are preserved. For the second solution, one has $\mu=\frac{(q+r_H)^4}{L^2 (q-2r_H)}$ and for $q>2r_H$, the condition on $\mu$ is $\mu>0$ and also the solution is non-BPS with no supersymmetry preserved \cite{Fareghbal:2008eh}. The point is that in $4d$, only the non-supersymmetric black holes with regular horizon with $\mu \ne 0$ can exist and therefore only the solution II is acceptable. The supersymmetric one with $\mu=0$, as has been studied in \cite{Leblond:2001gn}, are named superstars and are just naked singularities. \cite{Fareghbal:2008eh, Leblond:2001gn}

The entropy density in the general regular case is
\begin{equation}
s=\frac{\sqrt{(q+r_H)^3(q^\prime+r_H)}}{4G}.
\end{equation}
For the first solution of the extremal condition (\ref{eq:extg}), $q=-r_H$ and $s$ is degenerate to zero. However, for the second solution of the extremal regular 3+1-charge case, the entropy density which is the entropy of the ground state is 
\begin{equation}
s_{ext}=\frac{1}{4G}(q+r_H)^2({\frac{r_H}{q-2r_H}})^\frac{1}{2}=\frac{1}{4G}(q^\prime+r_H)^2 (\frac{3r_H}{q^\prime})^{\frac{3}{2}}.
\end{equation}
As we will demonstrate below, by considering the order of taking the limits to go from a regular 3+1-charge black brane to the extremal 3-charged black brane, a discontinuity in different parameters, such as in the entropy density, will show up which can be interpreted, in the dual $\mathcal{N}=4$ SYM theory, as a gap in the states. Within this gap the system displays the behavior of a Fermi liquid with long-lived quasiparticles and stable fluctuation modes while similar to the $5d$ case of reference \cite{DeWolfe:2013uba}, outside the gap the quasiparticles have short lives and the states show the behavior of non-Fermi liquids.

\subsection{3-charge black brane}
We first consider the case where $q^\prime= 0$ and $q\ne0$. Then for the general three-charge black brane solution we have
\begin{gather}
f(r)=\frac{1}{r L^2 }\big((r+q)^3-(r_H+q)^3\big), \ \ \ \ \ \ \ \ \ 
\mu=\frac{(q+r_H)^3}{L^2},\nonumber \\
A_t=\Big(\frac{q^2}{(q+r_H)^2}+\frac{q(q+r_H)}{L^2}\Big)^\frac{1}{2}\big(1-\frac{q+r_H}{q+r}\big),\ \ \ \ \ \ \ 
A_t^\prime=0,\ \ \ \ \ 
X=(1+\frac{q}{r})^{-\frac{1}{4}}, \ \ \ \ \ \ \nonumber\\   X^\prime=(1+\frac{q}{r})^\frac{3}{4},
T_H=\frac{3}{L^2 \pi  }\sqrt{r_H\left(q+r_H\right)}, \ \ \ \ \ \ \ \ \
s=\frac{\text{  }\sqrt{r_H\left(q+r_H\right){}^3}}{4G}.
\end{gather} 
For the extremal case, $T_H=0$, $S=0$ and it corresponds to the two solutions of (\ref{eq:extg}),
\begin{gather}\label{eq:ext1}
r_H=0,\  \mbox{or} \  q=-r_H  \ \ \text{extremal (3 QBHs)} \nonumber\\
\mu_{ext}=\frac{4q^3}{L^2} \  \mbox{or} \ \ 0
\end{gather}
Now we consider two scenarios for reaching from the regular non-extremal 3+1-charge to the extremal 3-charge black brane. If we first let $q'\to 0$ and then consider the extremality condition (\ref{eq:ext1}), then 
\begin{equation}
\frac{qq'}{r_H^2}\to 0.
\end{equation}  
On the other hand, by first applying the extremality condition (\ref{eq:extg}) and then letting $q' \to 0$, one gets 
\begin{flalign}
q=\frac{3r_H^2}{q'}+2r_H,\ \ \  \ \  \text{or} \ \ \ \ \ \ 
\frac{qq'}{r_H^2}\to 3.
\end{flalign}  
The difference between these two limits indicates the existence of a discontinuity and therefore a gap in the states of the dual field theory.
\subsection{1-charge black brane}
 If we assume $q=0$ and $q^\prime\ne 0$ which is a 1-charge black brane case, then
 \begin{gather}
 f(r)=\frac{1}{L^2}\Big( r(q^\prime+r)-\frac{r_H^2}{r}(q^\prime+r_H)\Big),\ \ \ \ \ 
 \mu=\frac{r_H^2(q^\prime+r_H)}{L^2},\nonumber\\
 A_t=0 , \ \ \ \ \ A_t^\prime=\Big(\frac{{q^\prime}^2}{(q^\prime+r_H)^2}+\frac{q^\prime }{q^\prime+r_H} \frac{r_H^2}{L^2} \Big)^\frac{1}{2}\big(1-\frac{q^\prime+r_H}{q^\prime+r}\big),\ \ \ \
 X=(1+\frac{q^\prime}{r})^{\frac{1}{4}},\ \ \ \ \ \nonumber\\  X^\prime=(1+\frac{q^\prime}{r})^{-\frac{3}{4}}, 
T_H=\frac{2 q'+3 r_H}{L^2 \pi  }\sqrt{\frac{r_H}{q'+r_H}},\ \ \ \ 
s=\frac{\sqrt{r_H{}^3\left(q'+r_H\right)}}{4G}.
\end{gather}
Again, for the extremal case $T_H=0$, $S=0$ and
\begin{gather}\label{eq:ext2}
r_H=0,\  \mbox{or} \ \  q^\prime=-\frac{3 r_H}{2},   \ \ \ \ \text{extremal  (1QBHs),}\nonumber\\
 \mu_{ext}=0 ,\  \mbox{or} \ \  \frac{-r_H^3}{2L^2}. 
\end{gather}
If we first let $q\to0$ and then apply the extremality condition, setting $\frac{qq^\prime}{r_H^2}$ to zero, and if we first use the condition (\ref{eq:extg}), then we find
\begin{equation}
q^\prime=\frac{3 r_H^2}{q-2r_H}, \ \ \ \ \ \ 
\frac{qq^\prime}{r_H^2}=\frac{3q}{q-2r_H}  \xrightarrow{q\to 0} 0.
\end{equation}
We can also use other ratios to check the results of the order of taking the limits, showing the discontinuity in the states. For instance, if first we apply $q\to0$ and then $r_H\to0$ we would find
\begin{equation}
\frac{q}{q^\prime}\to0, \ \ \ \  \ \  \ \ \ \ \ \  \frac{r_H}{q^\prime}\to0, \ \ \ \ \ \ \ \ \ \ \ \  \frac{q}{r_H}\to0.
\end{equation}
On the other hand, if we first apply (\ref{eq:extg}) and then $q\to0$, we would have
\begin{equation}
\frac{q}{q^\prime}=\frac{q(q-2r_H)}{3r_H^2}\to0, \ \ \ \ \ \ \  \frac{r_H}{q^\prime}=\frac{r_H(q-2r_H)}{3r_H^2}\to-\frac{2}{3}, \ \ \ \\ \ \ \ \ \ \ \ \frac{q}{r_H}\to 0.
\end{equation}
Once more the difference between the limits of these parameters indicates the existence of the gap.

\begin{figure}[ht!]
\centering
\parbox{15cm}{
\centering
\includegraphics[scale=1]{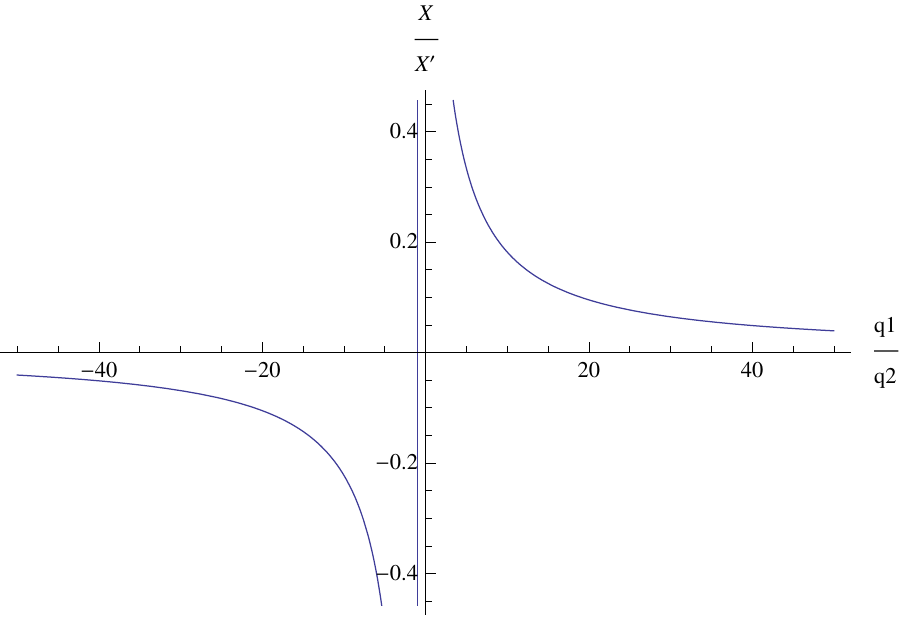}
\caption{The ratio of the dilaton fields $\frac{X}{X^\prime}=\frac{q^\prime+r}{q+r}$ for $r=1$ when $\frac{q}{q^\prime}\to\infty$, $\frac{X}{X^\prime}\to0$.\\Also for the specific $q$ and $q^\prime$, for $r\to\infty$ one gets $\frac{X}{X^\prime}\to1$. So the dialton fields are equal in the boundary.}
\label{fig:2figsA9}}
\end{figure} 

\begin{figure}[ht!]
\centering
\parbox{5cm}{
\includegraphics[scale=0.6]{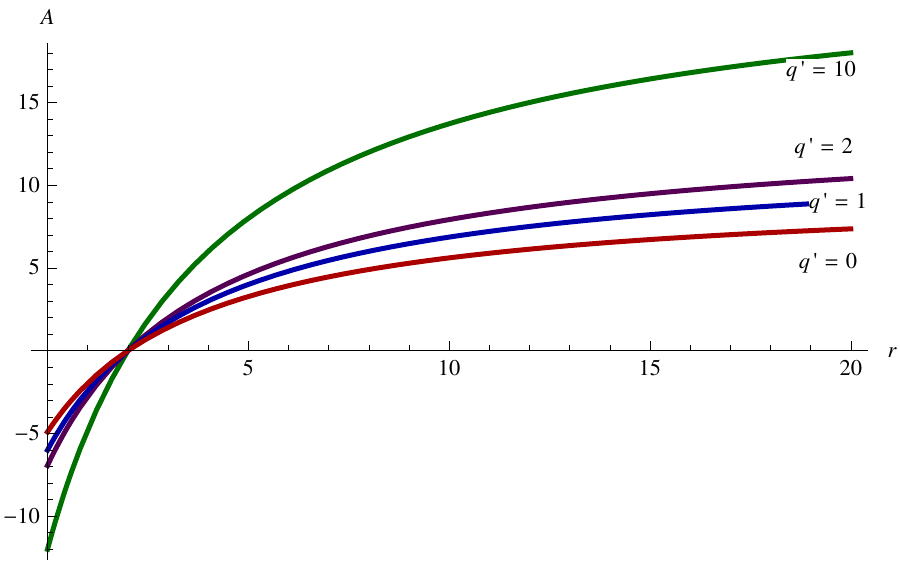}
\caption{Plot of $A$ v.s $r$ for different $q^\prime$. ($r_H=2,\ L=1 $). }
\label{fig:2figsA9}}
\qquad
~ ~ ~ ~ ~ ~ ~ ~ ~ ~ 
\begin{minipage}{7cm}
\includegraphics[scale=0.6]{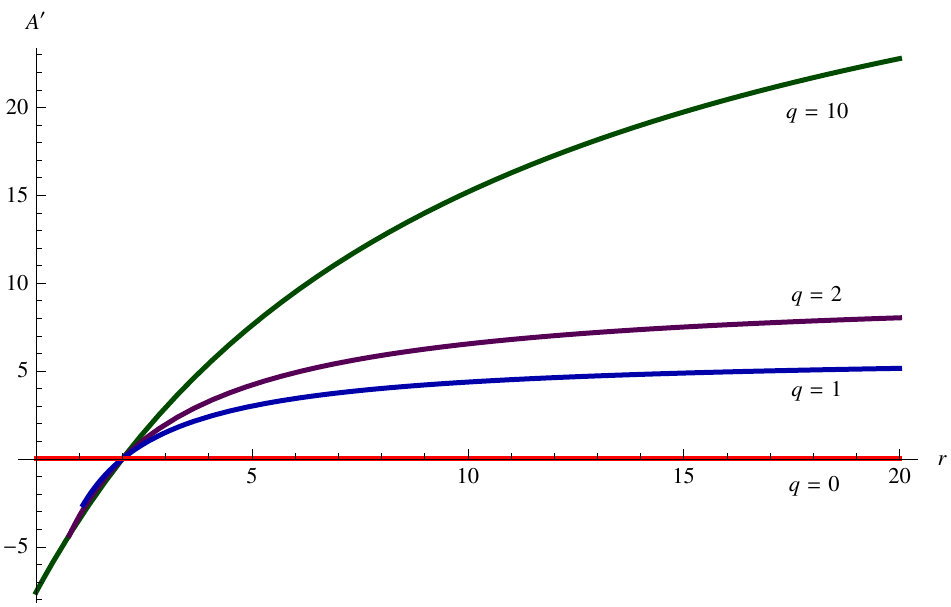}
\caption{Plot of $A^\prime$ v.s $r$ for different $q$. As $q\to0$, the curves become closer to the $x$ axis. At $q=0$, $A^\prime$ degenerates to a constant which is zero. ($r_H=2, \ L=1$) }
\label{fig:2figsB9}
\end{minipage}
\end{figure}

\subsection{Chemical Potentials}

The chemical potentials of these black hole solutions are

\begin{gather}
\mu_1=\frac{\sqrt{3}}{L} \left(\frac{q^2}{(q+r_H)^2}+\frac{q(q^\prime+r_H)(q+r_H)}{L^2 r_H}\right), \ \ \ \ \ 
\mu_2=\frac{\sqrt{3}}{L} \left(\frac{{q^\prime}^2}{(q^\prime+r_H)^2}+\frac{q^\prime(q+r_H)^3}{(q^\prime+r_H) L^2 r_H}\right).
\end{gather}

The factor of $\frac{\sqrt{3}}{L}$ is present to canonically normalize the chemical potentials. From these, it can be seen that, if we first let $q^\prime=0$ and then set $r_H\to 0$, then the limit of $\mu_1$ is $\mu_1 \to \frac{\sqrt{3} q^2}{L^3 r_H}$, but if first by letting $r_H \to 0$ and then $q^\prime \to 0$ we go to the extremal limit, then we see $\mu_1 \to 0$. This behavior of the chemical potential which relates to the boundary of the gauge field again indicates the gap in the CFT. For the $\mu_2$ case, one can see that by changing $q^\prime$ from a positive to a negative value, the chemical potential jumps from a positive finite value to a negative finite value and again this discontinuity indicates that some states are missing in between. 

As the chemical potential around $\rho_1=0$ does not change uniformly and there is a jump around the center, the concentration of states does not change continuously everywhere and a gap could appear. Within this gap, the chemical potential remains constant and is different from the values of the chemical potential around it. So this could be one hint that the degrees of freedom inside and outside of the gap should be decoupled.

From Fig. \ref{fig:2figsA9}, one can see that by increasing $\frac{q}{q^\prime}$ $\frac{\mu_c}{{\mu_c}^\prime}$ decreases, pointing to the fact that the susceptibility is negative and due to the perturbative instability in the spectrum of the bosonic fluctuations, the thermodynamic is unstable. 
\begin{figure}[ht!]
\centering
\parbox{15cm}{
\centering
\includegraphics[scale=1]{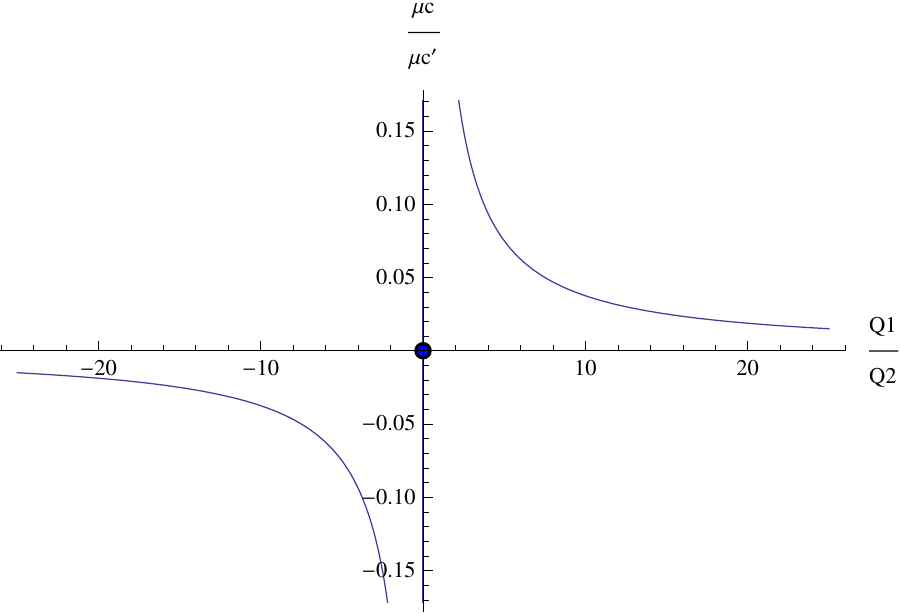}
\caption{The ratio of the chemical potentials $\frac{\mu_c}{{\mu_c}^\prime}$ v.s $\frac{Q}{Q^\prime}$, where $r_H=L=1$ and $Q=3$. }
\label{fig:2figsA9}}
\end{figure}

In the equilibrium, as the free energy is in its minimum, the sum of the chemical potentials is zero. During phase transitions or during any reaction, the chemical potential changes from higher values to the lower values and some free energies are being released. From this diagram, one can also see that when $\frac{\rho_1}{\rho_2}$ increases, the ratio of the chemical potentials, $\frac{\mu_1}{\mu_2}$, decreases which means that the system is going further away from equilibrium and so the thermodynamics is unstable. This is actually due to pair creation effects around this gap.
 
\subsection{The three regimes of energy}
 
 It has been shown that, as vortex lines play a crucial role in superconducting phase transitions, there is a type II regime (with no latent heat) where the superconducting phase transition is second order, a type I regime where the phase transition is first order (with latent heat) and these two regions are separated by a tricritical point. Consequently, in the phase diagram of superconducting materials, there exist three distinct regions corresponding to three different regimes of, oscillation, decaying to a finite gap and decaying to the zero gap. \\

\begin{figure}[ht!]
\centering
\parbox{10cm}{
\centering
\includegraphics[scale=0.3]{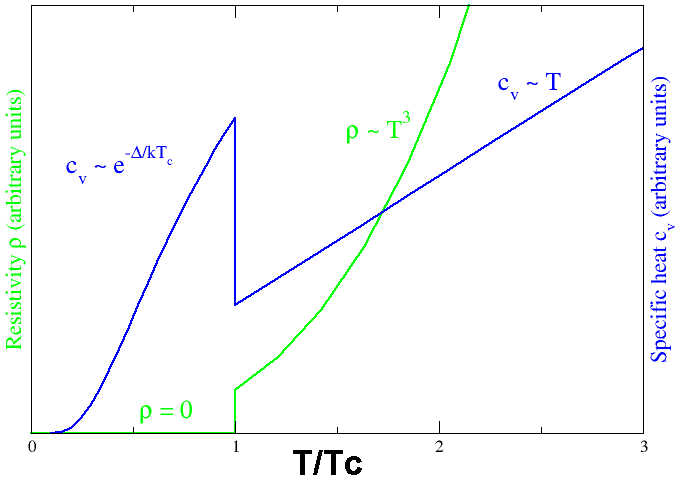}
\caption{Behavior of heat capacity ($c_v$, blue) and resistivity ($\rho$, green) at the superconducting phase transition (figure from wikipedia). }
\label{fig:2figsA9}}
\end{figure}

In the gravity side, by holography, we also expect to see three different regions. So we may consider three different limits to look at the black holes in the bulk which corresponds to three different regimes of energy in the dual CFT. Defining $\frac{r-r_H}{r_H}=\delta$ and $q=\tilde{q}\epsilon$, the three regimes are:
\begin{gather}
\frac{\delta}{\epsilon}>1\ \text{or} \ \frac{\tilde{q}}{r-r_H}>1 \to \text{system of 3 charged BH} \nonumber\\
\frac{\delta}{\epsilon}<1\ \text{or} \ \frac{\tilde{q}}{r-r_H}<1 \to \text{3+1 charged BH} \nonumber\\
\frac{\delta}{\epsilon}=1 \ \text{or} \ \frac{\tilde{q}}{r-r_H}=1 \to \text{excitations of $AdS_3$, BTZ}
\end{gather}

When in a system of fermions (or any charged particles), a strong electric field is turned on, one expects a form of symmetry breaking and then the appearance of different phases which may behave as a gap in the theory. Such behavior can be explored using a fermionic probe in the context of holography where the poles of Green's function can give us information about the behavior of the modes around this gap.

By uplifting the theory to $11d$ and then considering the black hole near horizon limit and also considering $q^\prime=0$, the near horizon geometry will decouple into $\text{AdS}_4 \times \text{M}^7$ where in the near horizon limit of 3-charge black hole, the AdS part is effectively an $\text{AdS}_3$. If $q^\prime$ is non-zero, but rather a small value, corresponding to a small momentum of the black hole, the near horizon geometry just close to the horizon decouples into another sector which is an excited $\text{AdS}_3$ or a rotating BTZ with energy $\epsilon_2$. In the $\text{UV}$ and far from the horizon, the geometry in $11d$ is $\text{AdS}_4 \times \text{S}^7$.

The decoupled regime in the bulk could be considered as a gap in the CFT with its own degrees of freedom not communicating with the degrees of freedom outside the gap. This $\text{CFT}_2$ could be considered as a fermionic $2d$ gas such as Luttinger systems. In the context of condensed matter physics there are many systems that can be approximately considered $1+1$. The flowing electrons in one spatial dimension (here due to the strong electric field), is a good approximation to describe quantum wires, carbon nanotubes, electrons along the edges of fractional quantum Hall effect systems, etc.

We can uplift the $4d$ theory to $5d$, and depending on how to fix the Killing spinors on the spherical part, the $5d$ spinors could be found. However, the interesting point is that the additional spatial dimension of the metric in not within these two geometries and is actually in the $S^7$ part. In the near horizon limit one of these dimensions from the $S^7$ part joins the $AdS_2$ sector and creates a warped $AdS_3$ geometry in the near horizon geometry of the black hole, which we will explain in more detail later.

\section{Dirac equation for $4d$ $U(1)^4$ gauged supergravity black brane}

Now for studying the dynamics of fermions in the background of this geometry, we consider the Dirac equation. 

The Dirac equation in this background is
\begin{equation}
\Big(i\gamma^\mu \nabla_\mu-m(\phi)+g q_1 \gamma^\mu A_\mu+g q_2 \gamma^\mu a_\mu+ip_1 e^{\frac{\phi}{2\sqrt{3}}} F_{\mu\nu} \gamma^{\mu\nu}+ip_2 e^{-\frac{\sqrt{3}}{2}\phi} f_{\mu\nu} \gamma^{\mu\nu}\Big) \chi=0.
\end{equation}

For the $\mathcal{N}=2$ truncation of gauged $\mathcal{N}=8$ supergravity without axions, the potential and super potential are 
\begin{gather}
V=-g^2 \sum\limits_{i<j} X_i X_j, \ \ \ \ \ \ \ W=\frac{1}{2} g \sum_i X_i,
\end{gather}


where $g=\frac{1}{2L}$. Using $X_1=X_2=X_3=e^{-\frac{\phi}{2\sqrt{3}} }$ and $X_4=e^{\frac{\sqrt{3}\phi}{2}}$, we get
\begin{gather}
V=\frac{-3}{4L^2}(e^{-\frac{\phi}{\sqrt{3}}}+e^{\frac{\phi}{\sqrt{3}}}), \ \ \ \ \ W=\frac{1}{4L}(3 e^{-\frac{\phi}{2\sqrt{3}}}+e^{\frac{\sqrt{3}\phi}{2} }).
\end{gather}

The fermion mass matrix can be found as
\begin{gather}
m_F^{ij}=\frac{\partial^2 W}{\partial \phi_i \partial \phi_j}.
\end{gather}

More explicitly one finds
\begin{gather}
m^{11}=m^{22}=m^{33}=\frac{1}{16L}\left(3e^{\frac{-\phi}{2\sqrt{3}} }+e^{ \frac{\sqrt{3}\phi}{2} } \right),\nonumber\\
m^{13}=m^{23}=-m^{12}=\frac{1}{16L}\left(e^{\frac{-\phi}{2\sqrt{3}} }-e^{\frac{\sqrt{3}\phi}{2}}\right), 
\end{gather}

with eigenvalues

\begin{gather}
\frac{1}{4 L}e^{-\frac{\phi }{2 \sqrt{3}}}, \ \ \  \frac{1}{4 L}e^{-\frac{\phi }{2 \sqrt{3}}},\ \ \ \frac{1}{16 L}\left(e^{-\frac{\phi }{2 \sqrt{3}}}+3 e^{\frac{\sqrt{3} \phi }{2}}\right).
\end{gather}

The scalar mass matrix also is

\begin{gather}
(m_s^2)^{ij}=\frac{\partial^2 V}{\partial \phi_i \partial \phi_j}=-\frac{1}{4L^2} (e^{-\frac{\phi}{\sqrt{3}} }+e^{\frac{\phi}{\sqrt{3}} }).
\end{gather}

The covariant derivative is   
 \begin{gather}
 \nabla_\mu=\partial_\mu-\frac{1}{4} \omega_{\hat{a}\hat{b}\mu} \Gamma^{\hat{a}\hat{b}}.
 \end{gather}

We choose the following Gamma matrices,
 
\[                                                          
\Gamma^{\hat{r}}=
  \begin{pmatrix}
    i\sigma_3 & 0 \\
0 & i\sigma_3 \\
  \end{pmatrix},
\]   \[
\Gamma^{\hat{t}}=
  \begin{pmatrix}
    \sigma_1 & 0 \\
0 & \sigma_1 \\
  \end{pmatrix},
\]
 \[
\Gamma^{\hat{i}}=
  \begin{pmatrix}
    i\sigma_2 & 0 \\
0 & -i\sigma_2 \\
  \end{pmatrix},
\]
which satisfy the standard relations:
\begin{gather}
 \{ \Gamma^\mu,\Gamma^\nu\}=2\eta^{\mu\nu}  \ \ \ \ \ \ \ \ \ \ \ \ 
 \Gamma^{\mu\nu}=\frac{1}{2}[\Gamma^\mu,\Gamma^{\nu}]. 
\end{gather}

The vierbeins $e_\mu^a$ are
\begin{gather}
e_t^a=\frac{1}{\sqrt{g_{tt}}} (\frac{\partial}{\partial t})^a=H(r)^{\frac{1}{4}} f(r)^{-\frac{1}{2}}(\frac{\partial}{\partial t})^a ,\nonumber\\
e_r^a=\frac{1}{\sqrt{g_{rr}}} (\frac{\partial}{\partial r})^a= H(r)^{-\frac{1}{4}} f(r)^{ \frac{1}{2} }  (\frac{\partial}{\partial r})^a , \nonumber\\
e_i^a=\frac{1}{\sqrt{g_{xx}}} (\frac{\partial}{\partial x^i})^a= H(r)^{-\frac{1}{4}}  r^{-1} (\frac{\partial}{\partial x^i})^a ,\nonumber\\
\end{gather}

and the spin connections, $(\omega_{\mu\nu})_a=(e_\mu)_b \Delta_a(e_\nu)^b$ are
\begin{gather}
(\omega_{tr})_t=-(\omega_{rt})_a=-\frac{  \partial_r \sqrt{g_{tt}}   }{\sqrt{g_{rr}}} (dt)_a=(\frac{1}{4} H^{-\frac{3}{2}}\frac{\partial H}{\partial r} f(r)-\frac{1}{2} H^{-\frac{1}{2}}\frac{\partial f}{\partial r})(dt)_a ,\nonumber\\
(\omega_{ir})_a=-(\omega_{ri})_a=\frac{\partial_r \sqrt{g_{xx}}}{\sqrt{g_{rr}}}(dx_i)_a=f^{\frac{1}{2}}(1+\frac{r}{4H})(dx^i)_a .
\end{gather}

A technical modification that effectively cancels the contributions of the spin connection follows from the following redefinition 
\begin{gather}\label{eq:chi}
\chi=\Big(\mathrm{Det}(g) \big | _{r=\mathrm{cte}}\Big)^{-\frac{1}{4}} e^{-i \omega t+i k x} \Psi=H ^{-\frac{1}{8}} r^{-1} f^{-\frac{1}{4}} e^{-i\omega t+ik x}\Psi.
\end{gather}

We define the projectors similarly to \cite{DeWolfe:2012uv} as, 
\begin{equation}\label{eq:projectors}
\Pi_{\alpha}\equiv \frac{1}{2}\Big(1-(-1)^\alpha i \gamma^{\hat{r}} \gamma^{\hat{t}} \gamma^{\hat{i}}\Big), \ \ \ \ \ \ \ \ P_{\pm} \equiv \frac{1}{2}\Big(1\pm i\gamma^{\hat{r}}\Big).
\end{equation}

Then, using Eq. (\ref{eq:projectors}), the definition for gamma matrices, and also Eq. (\ref{eq:chi}), we can split the Dirac equation into two equations for $\Psi_{\alpha+}$, and $\Psi_{\alpha-}$ as 
\begin{eqnarray}
\Big(\partial_r-H^{\frac{1}{4}} f(r)^{-\frac{1}{2}} m(\phi)\Big) \Psi_{\alpha+}&=&[-u(r)+(-1)^\alpha k_i r^{-1}  f(r)^{-\frac{1}{2}} -v(r)] \Psi_{\alpha -}, \nonumber\\  \nonumber\\
\Big(\partial_r+H^{\frac{1}{4}} f(r)^{-\frac{1}{2}} m(\phi)\Big) \Psi_{\alpha-}&=&[u(r)+(-1)^\alpha k_i r^{-1} f(r)^{-\frac{1}{2}} -v(r)] \Psi_{\alpha +}, 
\end{eqnarray}

where 
\begin{equation}
u(r)\equiv H^{\frac{1}{2}} f(r) ^{-1} (\omega +g q_1 \Phi_1+g q_2 \Phi_2), \ \ \ \ \ \ \  v(r) \equiv 2 \ H^{\frac{1}{4}} f(r) ^{-\frac{1}{2}} ( p_1 e^{\frac{\phi}{2\sqrt{3}}} \partial_r \Phi_1+ p_2 e^{-\frac{\sqrt{3}}{2} \phi} \partial_r \Phi_2).\nonumber\\
\end{equation}

Then, the Dirac equation becomes two decoupled second order equations as below
\begin{equation}\label{eq:diracsplit}
\begin{aligned}
\Psi^{\prime\prime}_{\alpha\pm}-F_{\pm} \Psi^\prime_{\alpha\pm} + \Big[ \mp \partial_r \big(m \ H^{\frac{1}{4}} f(r)^{-\frac{1}{2}}\big) -m^2 H^{\frac{1}{2}}f(r)^{-1} + & \\  
 \big( u(r)^2-\big(v(r)-(-1)^\alpha k_i r^{-1}  f(r)^{-\frac{1}{2}}\big)^2\big) \pm m \ H^{\frac{1}{4}} f(r)^{-\frac{1}{2}} F_{\pm} \Big] \Psi_{\alpha \pm}=& 0,
\end{aligned}
\end{equation}

where
\begin{equation}
F_{\pm}=\partial_r \mathrm{log} \big[u(r) \mp (-1)^\alpha k_i r^{-1}  f(r)^{-\frac{1}{2}} \pm v(r)\big].
\end{equation}

Similar to \cite{DeWolfe:2012uv}, one can see that the solution of Eq. (\ref{eq:diracsplit}) is invariant under
\begin{equation}
p_i\to -p_i, \ \ \ \ \ \  q_i \to -q_i, \ \ \ \ \ \ \omega \to -\omega, \ \ \ \ \ \ \ \  k \to -k.
\end{equation}
So the solution of conjugate fermions does not change if we take $(k,\omega)\to (-k,-\omega)$.

Our geometry in the near boundary limit $r\to \infty$ is $\mathrm{AdS}_4$, where the mass term becomes dominant there, and if we consider $ \big | mL\big | \neq 1/2$, the solution similar to \cite{DeWolfe:2012uv} is of the following form,

\begin{equation}
\Psi_{\alpha +} \sim A_{\alpha}(k) r^{mL} + B_{\alpha} (k) r^{-mL-1}, \ \ \ \ \ \ \  \Psi_{\alpha-}\sim C_{\alpha}(k)r^{mL-1}+D_{\alpha}(k) r^{-mL}.
\end{equation}

Now, similarly to \cite{DeWolfe:2012uv}, the relations between the coefficients of the solutions are

\begin{equation}
 C_\alpha= \frac{L^2 (\omega +(-1)^\alpha k)}{2mL-1} A_\alpha, \ \ \ \ \ \ B_\alpha=\frac{L^2(\omega-(-1)^\alpha k)}{2mL+1} D_\alpha,
 \end{equation}
where $mL\equiv 2(m_1+m_2)$. If $m>0$, $A$ is the source term and $D$ is the response and vise versa for $m<0$ for a dual fermion of opposite chirality \cite{DeWolfe:2013uba}. 

For $mL=\frac{1}{2}$ we have
\begin{equation}
 C_\alpha= L^2 (\omega +(-1)^\alpha k) A_\alpha, \ \ \ \ \ \ \ \ \ \ \  B_\alpha=\frac{L^2(\omega-(-1)^\alpha k)}{2mL+1} D_\alpha,
 \end{equation}
and then at the horizon, the retarded Green's function for the dual fermionic operator for the fluctuations is the ratio of the response to the source, $(G_R)_{\alpha\beta}=\frac{D_\alpha}{A_{\beta}}$.

Since similar to \cite{DeWolfe:2012uv}, our Eq. (\ref{eq:diracsplit}), decouples for the $\Psi_{\alpha\pm}$, then our Green's function is also diagonal leading to the following relation, $G_{22}(\omega, k)=G_{11} (\omega, -k)$.

Also, by finding the poles of the Green's function at Fermi energy one can find the Fermi surfaces, (where $k=k_F$), which at zero frequency is 
\begin{gather}
A_\beta(\omega=0, k=k_F)\equiv 0.
\end{gather}

\subsection{ Near-horizon analysis: 3+1-charge case}
In this section we study the near horizon limit of the Dirac equation first for the general ``regular" 3+1-charge case and then we look at the near horizon solution of the extremal 3-charge and finally we consider the 1-charge black brane case.

By defining the parameters, 
\begin{gather}\label{eq:parameter4}
\beta _1=\frac{1}{L\left(q+r_H\right){}^2}\sqrt{\frac{q \big(\left(q+r_H\right){}^4+L^2 q \left(q-2 r_H\right)\big)}{q-2 r_H}}, \nonumber\\
\beta _2=\frac{\left(q-2r_H\right)\sqrt{3 \left(\left(q+r_H\right){}^4+3L^2r_H{}^2\right)}}{L r_H\left(q+r_H\right){}^2}, \nonumber\\
{k_0}^2=\frac{\left(q+r_H\right){}^2 \sqrt{r_H}}{\sqrt{q-2r_H}}, \ \ \ \ \ \ \ \ \  (L_2)^2=\frac{ L^2 \sqrt{r_H (q-2r_H)}}{3 (q-r_H)},
\end{gather}
the leading terms for each component of the metric and the gauge fields of the 3+1-charge ``extremal" supergravity black brane, where one can eliminate $q^\prime$ by the extremality condition (\ref{eq:extg}), would be as follows
\begin{gather}
g_{tt} =-\frac{1}{ (L_2)^2} (r-r_H)^2+\Theta \left(r-r_H\right){}^3,\nonumber   \\ \nonumber 
g_{rr} = (L_2)^2 (r-r_H)^{-2}+\Theta \left(r-r_H\right){}^{-1}, \ \ \ \ \ \ \  g_{ii} = {k_0}^2+\Theta \left(r-r_H\right){}^1.
\end{gather}
As one can see, in the general 3+1-charge case, an $\mathrm{AdS}_2$ factor is appeared in the near horizon geometry.

The leading terms of the gauge fields are
\begin{gather}
\Phi_1 =\beta _1 \left(r-r_H\right)+\Theta \left(r-r_H\right){}^2,\ \ \ \ \ \ \ \ \ \ 
\Phi_2= \beta _2 \left(r-r_H\right)+\Theta \left(r-r_H\right){}^2,
\end{gather}
which are linear in terms of $r$ and similar to the 2+1-charge case of $5d$ \cite{DeWolfe:2012uv, DeWolfe:2013uba} have a single zero. Notice that the parameters $\beta_1$ and $\beta_2$ and also $p_1$ and $p_2$ are dimensionless.
Also, near the horizon, the leading term of the scalar field $\phi=\sqrt{3}\  \mathrm{log} X$ which is a constant and independent of $r$ is in the following form
\begin{gather}
\phi = \phi_0=\frac{\sqrt{3}}{4} \log \left(\frac{r_H}{q-2 r_H}\right)+\Theta (r-r_H)^2.
\end{gather}

Now we study the near horizon limit of the Dirac equation. For doing so, first in the near horizon limit ($(r\to r_H)$), we assume $\omega=0$ and then we study small $\omega$ limit while $\frac{\omega}{r-r_H}$ is constant and then we comment on how these two limits do not commute.

First, for $ \omega=0$ in the near horizon limit one finds
\begin{gather}
u = \frac{\left(L_2\right){}^2}{2 L }\left(q \beta _1+\frac{3r_H{}^2}{q-2 r_H} \beta _2\right)\frac{1}{r-r_H}+\Theta \left(r-r_H\right){}^0 ,\nonumber\\  \nonumber\\ 
 v =2 L_2\left(\text{  }p_1 e^{\frac{ \phi }{2\sqrt{3}}} \beta _1+ p_2 e^{-\frac{\sqrt{3} \phi }{2}} \beta _2\right)\frac{1}{\text{   }r-r_H}+\Theta \left(r-r_H\right){}^0.
 \end{gather}
 
Then, one finds the near horizon limit of $F_{\pm}\to F_{NH}$ as
\begin{gather}
F_{ NH}=-\frac{1}{r-r_H}.
\end{gather}

Taking the near-horizon limit of the function $F$ makes the form of Dirac equation much simpler as similar to \cite{DeWolfe:2012uv} the Dirac equation simplifies to
\begin{gather}
\partial_r ^2 \Psi _{\alpha \pm} +\frac{1}{r-r_H}\Psi _{\alpha \pm}^\prime -\frac{\nu_k ^2}{(r-r_H)^2} \Psi_{\alpha\pm}=0,
\end{gather}
where $\nu_k$ is
\begin{equation} \label{eq:nu}
\begin{split}
\nu _k=\sqrt{\left( m^2(\phi_0 ) +\left(\frac{\tilde{k}}{k_0}\right){}^2\right)\left(L_2\right)^2-\frac{\left(L_2\right){}^4 }{4 L^2} \left( q_1 \beta _1+\frac{3 r_H{}^2 }{ q_1-2r_H} \beta _2\right)^2},
\end{split}
\end{equation}
with 
\begin{gather}
\tilde{k} \equiv k_i-(-1)^{\alpha }2 k_0\left( p_1 e^{\frac{ \phi_0 }{2\sqrt{3}}}\beta _1+\text{  }p_2 e^{-\frac{\sqrt{3}}{2} \phi_0 }\beta _2\right).
\end{gather}

Note that we tried to choose the most similar way of defining the parameters for the sake of comparison to the five dimensional case of \cite{DeWolfe:2012uv}. However, although the equations look similar, the definitions of the quantities (\ref{eq:parameter4}) are different from those in \cite{DeWolfe:2012uv}. There are other differences between the above equation and the corresponding one in \cite{DeWolfe:2012uv}; for example, $\frac{L_2} {\tau_0}$ is replaced with ${L_2}^4$ and also the definition of $L_2$ here is different from the one in \cite{DeWolfe:2012uv}. Although the equations show similar behaviors in four and five dimensions, the physics that we extract from these formulas have some similarities to \cite{DeWolfe:2012uv} which we mention a few of those results here.

One can check that the effect of Pauli terms is just to shift the origin of the 3-momentum $k_i$. However, for $\alpha_1$ and $\alpha_2$, due to the different signs, the shifts are in the opposite directions.  Similarly the solution to the Dirac equation is of the form
\begin{gather}
\Psi \sim (r-r_H)^{\pm \nu_k},
\end{gather}
where this solution has a hidden near horizon region approaching $\text{AdS}_2 \times \mathbb{R}^2$. We can see this by defining
\begin{gather}
r-r_H= \lambda (L_2)^2\frac{1}{\zeta}, \ \ \ \ \ t=\frac{1}{\lambda} \tau .
\end{gather} 
By taking the limit $\lambda \to 0$ and by keeping $\zeta$ and $\tau$ fixed, we find the near horizon metric as
\begin{gather}
ds^2=\frac{(L_2)^2 }{ \zeta^2} (-d\tau^2 +d\zeta^2)+ {k_0}^2 d\vec{x}^2.
 \end{gather} 
The radius of this $AdS_2 \times \mathbb{R}^2$ metric is 
\begin{gather}
L_{\mathrm{AdS}_2}=L_2=\frac{L \big(r_H (q-2r_H)\big)^{\frac{1}{4}}}{\sqrt{3(q-r_H)}},
\end{gather}
and the near-horizon gauge fields are
\begin{gather}
A_\mu dx^\mu=\frac{\beta_1 (L_2)^2}{\zeta} d\tau, \ \ \ \ \ \ \ \ \ \ \ {A_\mu}^\prime dx^\mu=\frac{\beta_2 (L_2)^2}{\zeta} d\tau.
\end{gather}

Matching the boundary conditions in and out of $\mathrm{AdS_2}$ geometry requires one to choose the positive sign of $\nu$, i.e., $\Psi \sim (r-r_H)^{ +\nu_k}$. 

The oscillatory region is where $\nu_{k_{osc}}$ is imaginary \cite{DeWolfe:2012uv}. This would happen when the effective electric coupling 
\begin{gather}
(qe)_{\text{eff}} \equiv \frac{\left(L_2\right){}^2 }{2 L} \left( q_1 \beta _1+\frac{3 r_H{}^2 }{ q_1-2r_H} \beta _2\right),
\end{gather}
is stronger than the effect of mass and the shifted momentum $\tilde{k}$. This is intuitively correct, since near the horizon, a higher mass of the scalar field is in accordance with bigger chemical potentials $\mu_1$ and $\mu_2$ which can make the near-horizon region more stable. Also a higher momentum $k_i$ of infalling waves is associated to the higher stability of states. Putting it in another way, the bigger electric couplings near the horizon makes this region more unstable which this leads to particle creation in the near horizon $\text{AdS}_2$ region. 

It is worth noticing that, as mentioned in \cite{DeWolfe:2012uv}, the fact that the mass term here is a function of the scalar field and as a result, a function of the chemical potential $\mu_i$, is specifically a feature of the solutions coming from the gauged supergravity theories and from a top-down approach. 

Now, by using Eq. (\ref{eq:nu}) and solving the following equation, one can find where the oscillatory region appears, 
\begin{gather}
\nu_{k_{\text{osc}}}=0.
\end{gather}

Similar to \cite{DeWolfe:2012uv}, for the four dimensional case this happens where
\begin{gather}
\tilde{k}^2_{\text{osc}}=\left(\frac{k_0}{L_2}\right)^2\left((qe)^2_{\text{eff}}-m^2(\phi) L^2_2\right).
 \end{gather}
Note that this oscillatory momentum can exist only when the effective near-horizon electric field coupling is bigger than the effective mass term.

We now investigate the existence of the oscillatory region in the near horizon of two cases of 3-charge $(q^\prime \to 0)$ and 1-charge $(q \to 0)$ black brane solution.
The square of the near horizon oscillatory momentum, ${k^2}_{\text{osc}}$, for the $3$-charge black brane (i.e. $q^\prime \to 0$) is
\begin{gather}
k^2{}_{\text{osc}}=\frac{\sqrt{3q'} \ q^{3/2} }{36 L^2}\left( L^2+ q^2 \right)- \nonumber\\ \frac{q^{5/4}}{2L^2}\Big( m_1 m_2\text{  }\left(\frac{q'}{3}\right)^{1/4}\big(\frac{5}{2}\sqrt{\frac{q'}{3}}+\sqrt{q}\big)+\frac{m_2{}^2 q^{3/4}}{2}+\frac{q^{1/4}\sqrt{3q'}}{6 }\left(m_1{}^2+2m_2{}^2\right)\Big) + 
\Theta({q^\prime})^1.
\end{gather}

One can see that for the existence of the oscillatory region, $m_1$ and particularly $m_2$ should be smaller than $q$. The effect of $m_2$ is more important than $m_1$ due to the third and forth terms. So in this case, for the positivity of the right hand side and therefore the existence of $k_{\text{osc}}$,  $q$ and $m_2$ are competing with each other.
 
The oscillatory region cannot exist for the 1-charge black hole as the leading term in the expansion of $k_{\text{osc}}$ for the case of $q\to 0$ is negative,
\begin{gather}
k^2{}_{\text{osc}}= -\frac{3 m_1{}^2q q'}{8 L^2}-\frac{3 \sqrt{3q'} q^{3/2} m_1 m_2}{4\sqrt{2} L^2}+\frac{q^2 }{432 L^2}\left(169 L^2+81 m_1{}^2-243 m_2{}^2\right)+\Theta (q)^{5/2}.   
\end{gather}
As one can see, the right hand side is always negative and so for the four dimensional 1-charge black brane, the near horizon is always stable. 

Now we examine the case where $\omega$ is non-zero to study its effects on the Dirac equation and the behavior of the dispersion relation. For doing so, we turn to the notation of \cite{DeWolfe:2013uba}. By defining $U_{\pm}$ one can write the four-dimension near horizon Dirac equation as 

\begin{gather}
U^{\prime \prime}+ \left(\frac{1}{r-r_H}+...\right)U^{\prime} + \left(\frac{L^4 \left(q-2 r_H\right) r_H \omega ^2}{9 \left(q-r_H\right){}^2 \left(r-r_H\right){}^4}+\frac{n\omega}{(r-r_H)^3}-\frac{\nu^2}{(r-r_H)^2}+...    \right)U=0,
\end{gather}
where 
\begin{gather}
n=\frac{i L^2\text{  }\sqrt{r_H\left(q-2 r_H\right) }}{3 \left(q-r_H\right) }+\frac{L^3r_H\left(q \beta _1 \left(q-2 r_H\right)+3 \beta _2 r_H{}^2\right)}{9 \left(q-r_H\right){}^2 }.
\end{gather}

One can see that $\nu^2$ is a complicated number which depends on $\omega^2$, $\omega$ and a term which depends only on $r_H$, $L$ and $p$. For zero $\omega$, only the $\nu^2$ term remains and the term containing $\frac{1}{r-r_H}$ is dominant.

\section{Uplift of the metric to five dimensions }
First by taking the near-horizon limit of the $4d$ extremal 3-charge black hole metric, one gets

\begin{gather}\label{eq:nearh}
ds^2= \frac{q^{\frac{5}{2}}}{ r^{\frac{1}{2}}L^2} dt^2-\frac{r^{\frac{1}{2}}L^2 }{q^{\frac{5}{2}}} dr^2+ q^{\frac{1}{2}} r^{\frac{3}{2}} d \vec{x}^2.
\end{gather}
The scalar field near the horizon behaves as 
\begin{gather}
e^{\frac{\phi}{2\sqrt{3}}}=\left(\frac{q}{r}\right)^{\frac{1}{4}}.
\end{gather} 

After reduction, the reduced Kaluza-Klein Ansatz of a $5d$ metric turns to be (\ref{eq:nearh})
\begin{gather}
d {\hat{s}}^2=e^{2\alpha \phi} ds^2+ e^{2\beta \phi} (dz+\mathcal{A})^2=e^{\frac{-5}{\sqrt{3}}\phi} ds^2+e^{\frac{-4}{\sqrt{3}}\phi} (L d \varphi_3+\mathcal{A})^2.
\end{gather}
Then the $5d$ metric is as follows,
\begin{gather}
d\hat{s}^2=\frac{r^2}{L^2}dt^2-\frac{L^2}{r^2} dr^2+\frac{r^2}{q^2}L^2d {\varphi_3}^2+\frac{r^4}{q^2} d {\vec{x}_{2,k}}^2.
\end{gather}
This has a null warped $\mathrm{AdS}_3$ sector which by itself is interesting.  Thus, it would be illuminating to find the central charges of the specific dual $\text{CFT}_2$ of this metric and study the corresponding Virasoro algebra in order to find the behavior of the gap and the near horizon modes.

\section{The $11d$ uplifted metric}

As has been studied in \cite{DeWolfe:2013uba}, the $S^7$ reduction of eleven dimensional supergravity gives rise to a $SO(8)$ gauged $\mathcal{N}=8$ supergravity in $4d$. After further truncation it leads to $\mathcal{N}=2$ gauged supergravity, which the bosonic sector of the theory comprises four commuting $U(1)$ gauge potential fields, three dilatons, three axions and our desired metric \ref{metric}. 

In $11d$, if one does not consider the axions, the full theory would not be consistent, as the $U(1)$ gauge fields source the term $\epsilon^{\mu\nu\rho\sigma} F_{\mu\nu} F_{\rho\sigma}$. However, if we wish to consider the $4d$ non-rotating electrically charged black holes, we can consider axions to be zero in $11d$.
So we can uplift the four dimensional theory consistently by considering the $11d$ metric Ansatz as in \cite{DeWolfe:2013uba},

\begin{equation}
ds_{11}^2=\widetilde{\Delta}^{2/3} ds_4^2+g^{-2} \widetilde{\Delta}^{-1/3} \sum\limits_{i} X_i^{-1}\bigg({d\mu_i}^2+{\mu_i}^2(d\phi_i+g {A^i}_{(1)})^2\bigg),
\end{equation}
where
\begin{gather*}\label{eq:11d}
\widetilde{\Delta}=\sum\limits_{i=1}^4 X_i \mu_i^2, \ \ \ \ \ \ \ \ \ \ \ 
X_i=e^{-\frac{1}{2} \vec{a_i}.\vec{\varphi}},\nonumber\\
\mu_1=\text{sin}\theta, \ \ \ \ \ \ \mu_2=\text{cos}\theta\ \text{sin}\varphi, \ \  \ \ \ \mu_3=\text{cos}\theta\ \text{cos}\varphi \ \text{sin}\psi, \ \ \ \ \ \mu_4=\text{cos}\theta\ \text{cos}\varphi \ \text{cos}\psi.
\end{gather*}

Note that these quantities satisfy the following relations,
\begin{equation}
\sum\limits_i \mu_i^2=1, \ \ \ \ \  \ \ \ \ X_1X_2 X_3 X_4=1, \  \ \ \ \ \ \ \ \ \vec{a}_i.\vec{a}_j=4\delta_{ij}-1.
\end{equation}
A consistent choice for the parameters is given in (\ref{eq:lag}).

As has been suggested in \cite{DeWolfe:2013uba}, the 4-form field strength Ansatz in $11d$ is
 \begin{gather}
 F_{(4)}= 2g \sum\limits_{i} \big(X_i^2 \mu_i^2-\tilde{\Delta} X_i\big) \epsilon_{(4)}+ \nonumber\\ \frac{1}{2g} \sum\limits_{i} X_i^{-1} \bar{\ast} dX_i \wedge d(\mu_i^2)-\frac{1}{2g^2}\sum\limits_{i} X_i^{-2} d(\mu_i^2)\wedge(d\phi_i+g A_{(1)}^i )\wedge \bar{\ast} F^i_{(2)} .
 \end{gather}
 It is worth mentioning that in the eleven dimension theory, the angular dependence coming from  $\mu_i$ will be cancelled out in the equation of motion due to the contribution of both the metric and the 4-field strength in the euqation. 
 
\subsection{Near horizon limit in $11d$}

The geometry of 3+1-charge black brane in $11d$ is
\begin{gather}
ds_{11}^2=\Delta^{\frac{2}{3}}\Big[-\frac{f}{H} dt^2+\frac{dr^2}{f}+r^2 (dx^2+dy^2)\Big]+{ds_7^2}\nonumber\\
ds_7^2=\Delta^{-\frac{1}{3}}\bigg[L^2 \mathcal{H} \Big(\sum\limits_{i=1}^3 d\mu_i^2+\mu_i^2\big(d\phi_i+A_i\frac{dt}{L}\big)^2\Big)+L^2 \mathcal{H}^\prime\Big(d\mu_4^2+\mu_4^2\big(d\phi_4+A^\prime\frac{dt}{L}\big)^2\Big)\bigg],
\end{gather}
where
\begin{gather}
H_1= H_2=H_3=\mathcal{H}=1+\frac{q_1}{r}, \ \ \ \ \ \ \ \ \
H_4=\mathcal{H^\prime}=1+\frac{q_2}{r},\ \ \ \ \ \ \ \ \ \ \
H=\mathcal{H}^3 \mathcal{H^\prime}.
\end{gather}

 If in Eq. (\ref{eq:11d}), one considers $\mu_4=\text{cos}\ \theta$, the general form of $\Delta$ could be written as
\begin{gather}
\Delta= \mathcal{H^\prime}\mathcal{H}^3\Big(\frac{1-\mu_4^2}{\mathcal{H}}+\frac{\mu_4^2}{\mathcal{H}^\prime}\Big)= \mathcal{H}^2\big[\mathcal{H}^\prime+\mu_4^2\big(\mathcal{H}-\mathcal{H^\prime}\big)\big]=
\mathcal{H}^2\big[\mathcal{H}^\prime+\mu_4^2\big(\frac{q_1-q_2}{r}\big)\big].
\end{gather}

The three form-field strength $C_3$ satisfying $F_4=dC_3$ is \cite{Fareghbal:2008ar}
\begin{gather}
C^{(3)}=-\frac{r^3}{2} \Delta dt \wedge d^2 \sigma_2 -\frac{L^2}{2} \sum\limits_{i=1}^3 Q_i \mu_i^2 \big(d\phi_i-\frac{q_i}{Q_i}\frac{dt}{L}\big)\wedge d \Omega_2-\frac{L^2}{2} \Big(Q_4\mu_4^2\big(d\phi_4-\frac{q_4}{Q_4}\frac{dt}{L}\big)\wedge d\Omega_2\Big),
\end{gather}
where $d\Omega_2$ is the volume form on a unit radius two sphere.

 Now we consider two scenarios of taking the near horizon limits. In both of these two cases $q_1$ is a finite non-zero value. 
In the first scenario, we first consider $q_2 \to 0$, and then we take the near horizon limit $r\to r_H$. As it has been shown in the equation (\ref{eq:ext1}), for the non-BPS black hole, the horizon is at $r_H=0$. So in this case, we will get the near horizon geometry of a 3-charge black hole in $11d$ at the end. We call this limit $L_1$.

Then for the other case, while keeping $q_2$ finite, which corresponds to a finite horizon $r_H$, we take the near horizon limit $r \to r_H$ which will lead to the near horizon geometry of the 3+1-charge black hole in $11d$. In the next step, we can consider $q_2 \to 0$ while we are in the near horizon geometry. We call this limit $L_2$.

 So in brief the order of the limits are as follows \\ 
$L_1$: \ \ $q_2 \to 0$, \ \ \ \ \ \ \           $r\to r_H$,  \ \ \ \ $r_H\to 0,$\\
$L_2$: \ \ $q_2=\text{finite}$, \ \ \ $r\to r_H$, \ \ \ \ $r_H=\text{finite}$, \ \ \ \ $q_2 \to 0.\\$ 

Now we study each scenario in more details.

\subsubsection{$L_1$ Limit}
For the $L_1$ case we would find
\begin{gather}
f=-\frac{\mu}{r}+\frac{r^2}{L^2}\big(1+\frac{q_1}{r}\big)^3 \big(1+\frac{q_2}{r}\big),\nonumber\\\simeq \frac{3{q_1}^2}{L^2}+\frac{1}{r}(-\mu+\mu_{cr}+\frac{3{q_1}^2 q_2}{L^2})+\frac{q^3 q_2}{r^2 L^2}\ :\text{Finite} \ \ \ \ \ \ \  \  \mu_{cr}=\frac{{q_1}^3}{L^2}.
\end{gather}
 In order to keep $f$ finite, we need to consider
 \begin{gather}
 r\sim r_H \sim \epsilon,\ \ \ \ \ \ \ \ \ \ \ \ \ \ \
 q_2 \sim \epsilon^2, \ \ \ \ \ \ \ \ \ \ \ \ \ \ \ 
 (\mu-\mu_{cr})\sim \epsilon.
 \end{gather}
 Therefore,
\begin{gather}
f\simeq \frac{3{q_1}^2}{L^2}-\frac{\mu-\mu_{cr}}{r}+\frac{{q_1}^3 q_2}{r^2 L^2},\nonumber\\ 
\mathcal{H}=1+\frac{{q_1}}{r} \simeq\frac{{q_1}}{r},\ \ \ \ \ \ \ \ \ 
\mathcal{H^\prime}=1+\frac{q_2}{r} \simeq 1+O(\epsilon), \ \ \ \ \ \ \ \  \Delta\simeq \mu_4^2\frac{{q_1}^3}{r^3},  \nonumber\\
A=\frac{Q_1}{{q_1}}\big(\frac{1}{\mathcal{H}}-1\big)\simeq -\frac{Q_1}{{q_1}}   \simeq -\sqrt{1+\frac{{q_1}^2}{L^2}},\ \ \ \ \ \ \ \ \ 
a=\frac{Q_2}{q_2}\big(\frac{1}{\mathcal{H}^\prime}-1\big)\simeq -\frac{Q_2}{r}   \simeq -\sqrt{\frac{{q_1}^3}{L^2}}.\nonumber\\
\end{gather}

The above limit in the parameter space should be accompanied by the following near-horizon limit for the coordinates,
\begin{gather}
(r-r_H)\sim \epsilon, \ \ \ \ \ \ \ \ \ \ \ \ \ \ 
 t\sim \epsilon ^{\frac{-1}{2}}, \ \ \ \ \ \ \ \ \ \ \ \ \  
 \phi_4\sim \epsilon^{\frac{-1}{2}}.
\end{gather}

Now using these, we can find the limits of the components of the metric,
\begin{gather}
r^2\ \Delta^{\frac{2}{3}}\ d\Omega_2^2={q_1}^2 \  \mu_4^{\frac{4}{3}},\nonumber\\
-\Delta^{\frac{2}{3}}\frac{f}{H} \ dt^2=-\mu_4^\frac{4}{3}\ \frac{r}{{q_1}}\ f \ dt^2,\nonumber\\
\frac{\Delta^{\frac{2}{3}}}{f}\ dr^2=\mu_4^{\frac{4}{3}}\  \frac{{q_1}^2}{f} \ \frac{dr^2}{r^2}\ : \text{Finite}.
\end{gather}
Note that $\frac{dr}{r}$ is finite, because if we consider $r=r_H+\rho$, then $\frac{d\rho}{(r_H+\rho)}$ and therefore $g_{rr}=\mu_4^{\frac{4}{3}} \frac{{q_1}^2}{f} \Big(\frac{d\rho}{\rho+r_H}\Big)^2$ are finite. 

Now, considering 

\begin{equation}
d\phi_i+A_i\frac{dt}{L}=d\phi_i-\sqrt{1+\frac{{q_1}^2}{L^2}} \frac{dt}{L}=d\varphi_i,
\end{equation}

the components for the $ds_7$ part are
\begin{gather}
\mu_i, \phi_i: \ \ \ \ \ \ \mu_4^{-\frac{2}{3}} L^2 \sum\limits_{i=1}^3 (d\mu_i^2+\mu_i^2 d\varphi_i^2),\nonumber\\ 
\mu_4, \phi_4: \ \ \ \ \ \ \mu_4^{-\frac{2}{3}}\ \frac{r}{{q_1}}\ L^2  \Big(d\mu_4^2+\mu_4^2 \big(d\phi_i^4-\sqrt{\frac{{q_1}^3}{L^2}}\frac{dt}{L}\big)^2\Big).
\end{gather}

So, finally the near horizon geometry in this case is
\begin{gather}
ds^2=\mu_4^{\frac{4}{3}} \Big[-\frac{r}{{q_1}}\ f \ dt^2 +\frac{{q_1}^2}{r^2} \ \frac{dr^2}{f}+\frac{r}{{q_1}} \ L^2 \big(d\phi_4-\sqrt{\frac{{q_1}^3}{L^2}} \ \frac{dt}{L}\big)^2 +{q_1}^2\ (dx^2+dy^2) \Big]\\ \nonumber
+\mu_4^{-\frac{2}{3}} L^2 \sum\limits_{i=1}^3 \big(d\mu_i^2+\mu_i^2\ d\varphi_i^2\big).
\end{gather}
As it can be seen, in this order of limits, the near horizon geometry is $\mathrm{AdS}_3 \times T^2\times T^6$ or $\mathrm{BTZ} \times T^2\times T^6$.  The BTZ sector is rotating with the angular momentum $\frac{\sqrt{{q_1}^3}}{L^2}$. For the case of ${q_1}=-r_H$ and therefore, $\mu=0$ (BPS point), the three dimensional part describes a global $\mathrm{AdS}_3$ space \cite{Fareghbal:2008ar}.

Also the scalar fields are
\begin{gather}
X\simeq \Big(\frac{r}{{q_1}}\Big)^{\frac{1}{4}}\simeq 0, \ \ \ \ \ \ \ \ \ \ \ \ \ \ \ \ \ \ 
X^ \prime \simeq \Big(\frac{{q_1}}{r}\Big)^{\frac{3}{4}} \simeq {q_1}^{\frac{3}{4}} \epsilon^{-\frac{3}{4}}.
\end{gather}
By considering the limit of $q_2 \to 0$ and then taking the near horizon limit, in the near horizon geometry only one scalar field would still be present which is $\mathrm{AdS}_3 \times T^2\times T^6$.

\subsubsection{$L_2$ Limit}

Now for the other case ($\text{L}_2$) both $q_2$ and $r_H$ are finite. Going to near horizon limit $r\to r_H$, the components of metric could be found as

\begin{gather}
g_{tt}=-\frac{3 \left({q_1}-r_H\right)\text{  }}{L^2 r_H }\left(\frac{r_H+\mu _4{}^2\left({q_1}-3 r_H \right)}{{q_1}-2 r_H}\right)^{\frac{2}{3}}\left(r-r_H\right){}^2+\Theta(r-r_H)^3,\nonumber\\ \nonumber\\
g_{rr}=\frac{L^2 \left({q_1}-2 r_H\right){}^{\frac{1}{3}} \left(r_H+\mu _4{}^2\left({q_1}-3 r_H \right)\right){}^{\frac{2}{3}}}{3 \left({q_1}-r_H\right) \left(r-r_H\right){}^2}+\frac{1}{\Theta\left (r-r_H\right)},\nonumber\\ \nonumber\\
g_{ii}=\left(\frac{r_H+\mu _4{}^2\left({q_1}-3 r_H \right)}{{q_1}-2 r_H }\right){}^{\frac{2}{3}}\left({q_1}+r_H \right){}^2+\Theta \left(r-r_H\right){}^1.
\end{gather}

For taking the limit of the $ds_7$ part, one first needs to know the limit of the gauge fields which are as below

\begin{gather}
\Phi_1=-\sqrt{\frac{{q_1} \left(L^2 {q_1} \left({q_1}-2 r_H\right)+\left({q_1}+r_H\right){}^4\right)}{L^2 \left({q_1}-2 r_H\right)\left({q_1}+r_H\right){}^2}}+\Theta \left(r-r_H\right){}^1,\nonumber\\ \nonumber\\
\Phi_2=-\frac{\sqrt{3\left(\left({q_1}+r_H\right){}^4+3L^2r_H{}^2\right)}}{ L({q_1}+r_H)}+\Theta \left(r-r_H\right){}^1.
\end{gather}

These are constant in the leading term and thus are independent of $r$. So one can simply rename the angular component as $d\varphi_i=d \phi_i+A_i \frac{dt}{L}$ and finds the limit of the $ds_7$ part as
\begin{gather}
\Delta^{-\frac{1}{3}} L^2  \mathcal{H} \sum\limits_{i=1}^3 d\mu_i^2=\frac{L^2 \left({q_1}-2 r_H\right){}^{1/3} }{ \left(r_H+\mu _4{}^2\left({q_1}-3 r_H \right)\right){}^{1/3}}+O\left[r-r_H\right]{}^1,\nonumber\\  
\Delta^{-\frac{1}{3}} L^2  \mathcal{H}^\prime \sum\limits_{i=1}^3 d\mu_i^2=\frac{L^2 r_H}{\left({q_1}-2 r_H\right){}^{2/3} \left(r_H+\mu _4{}^2\left({q_1}-3 r_H \right)\right){}^{1/3}}+O\left[r-r_H\right]{}^1.
\end{gather}
All the components of $ds_7$ and also $g_{ii}$ are independent of $r$ and so the near horizon geometry is $\mathrm{AdS}_2 \times T^2 \times T^7$. Now considering $q_2 \to 0$ imposes $r_H \to 0$.

One can see that, generally, for applications of AdS/CFT, for example for a $3d$ condensed matter system, instead of using the gravity theory in a 4d bulk, embedding the theory into the general $11d$ M-theory leads to being able to extract more information about the boundary CFT.

\begin{appendices}

\chapter{Solution phase space method of calculating conserved charges}\label{sec-Higher d}
The solution phase space method (SPSM) introduced in \cite{Hajian:2015eha,Hajian:2016kxx} has reproduced successfully the conserved charges and also the first law(s) for the standard (black hole) solutions of Einstein-Hilbert gravitational theories. Explicit examples can be found in Refs. \cite{Hajian:2015eha}. The goal of this appendix is utilizing the SPSM for the gravitational theories with higher curvature terms, specifically the ``New Massive Gravity'' introduced in (\ref{eq:action}).

 Explicitly, the Lagrangian which we will focus on, has the metric $g_{\alpha\beta}$, some gauge fields $A^a_\mu$, and some scalar fields $\phi^I$, in arbitrary dimension $d$,
\vspace*{-0.2cm}

{\small
\begin{flalign}\label{Lagrangian scalar}
\mathcal{L}=\frac{1}{16\pi G}\Big(&f(R,\phi)\!+\mathrm{a}(\phi) R_{\mu\nu}R^{\mu\nu}\!+\mathrm{b}(\phi) R_{\mu\nu\alpha\beta}R^{\mu\nu\alpha\beta}\!-\mathrm{c}_{ab}(\phi)F_{\mu\nu}^a F^{b\, \mu\nu}\!-\!2\mathrm{d}_{_{IJ}}(\phi)\nabla^\mu\phi^I\nabla_\mu\phi^J\Big).\!\!\!\!&&
\end{flalign}}

\vspace*{-0.2cm}
\noindent The $R^\mu_{\,\,\nu\alpha\beta}$, $R_{\mu\nu}$, and $R$ are Riemann tensor, Ricci tensor, and Ricci scalar respectively. The $F^a=\mrd A^a$ are the field strengths. The coefficients $\mathrm{a}(\phi)$, $\mathrm{b}(\phi)$, $\mathrm{c}_{ab}(\phi)$, and $\mathrm{d}_{_{IJ}}(\phi)$ are some functions of $\phi^I$. Notice that the $f(R,\phi)$ term covers the Einstein-Hilbert gravity with cosmological constant.  Lagrangian $d$-form is the Hodge dual of \eqref{Lagrangian scalar}, $\mathbf{L}=\star \mathcal{L}$,
\begin{equation}\label{Lagrangian top form}
\mathbf{L}=\frac{\sqrt{-g}}{d!}\,\,\epsilon_{\mu_1\mu_2\cdots \mu_d}  \,\mathcal{L}\,\,\mrd x^{\mu_1}\wedge \mrd x^{\mu_2}\wedge\cdots\wedge \mrd x^{\mu_d}\,.
\end{equation}
The $\epsilon_{\mu_1\mu_2\cdots \mu_d}$ is the Levi-Civita symbol, \emph{i.e.} $\epsilon_{_{012\cdots d-1}}=+1$ and changes sign by the odd permutations of indices. We will use the following conventions
\begin{equation}
h^{\mu\nu}\equiv \delta g^{\mu\nu}=g^{\mu\alpha}g^{\nu\beta}\delta g_{\alpha\beta},\qquad \delta F^{\mu\nu}\equiv g^{\mu\alpha}g^{\nu\beta}(\delta \mrd A)_{\alpha\beta}=g^{\mu\alpha}g^{\nu\beta}(\mrd \delta A)_{\alpha\beta}\,.
\end{equation}
Hence, the indices for the perturbed fields can be raised and lowered similar to other tensors. Let us label the terms in the Lagrangian \eqref{Lagrangian scalar} by $f$, a, b, c, and d respectively. The equations of motion (e.o.m) for the chosen Lagrangian, considering variations with respect to the metric, gauge fields, and scalar fields, are respectively \cite{Stelle:1977ry}
\vspace*{-0.2cm}

{\small \begin{flalign}
&E_{f\,\mu\nu}+E_{\text{a}\,\mu\nu}+E_{\text{b}\,\mu\nu}+E_{\text{c}\,\mu\nu}+E_{\text{d}\,\mu\nu}=0\,, \label{eom g}\\
&\hspace{2cm}E_{f\,\mu\nu}=\frac{1}{2}f g_{\mu\nu}-f' R_{\mu\nu}+\nabla_{\mu}\nabla_\nu f' -\Box f' g_{\mu\nu}\nonumber\\
&\hspace{2cm}E_{\text{a}\,\mu\nu}=\mathrm{a}\big(\frac{1}{2}R_{\alpha\beta}R^{\alpha\beta}g_{\mu\nu}+\nabla^\alpha(\nabla_\mu R_{\alpha\nu}+\nabla_\nu R_{\alpha\mu})-\nabla_\alpha\nabla_\beta R^{\alpha\beta} g_{\mu\nu}-\Box R_{\mu\nu}-2R_{\mu\alpha}R^{\alpha}_\nu\big) \nonumber\\
&\hspace{2cm}E_{\text{b}\,\mu\nu}=\mathrm{b}\big( \frac{1}{2}R_{\rho\sigma\alpha\beta}R^{\rho\sigma\alpha\beta}g_{\mu\nu}-2R_{\mu\gamma \alpha\beta}R_\nu^{\,\,\gamma \alpha\beta}-2\nabla^\alpha\nabla^\beta(R_{\mu\alpha\nu\beta}+R_{\nu\alpha\mu\beta})\big)\nonumber\\
&\hspace{2cm}E_{\text{c}\,\mu\nu}=2\text{c}_{ab}\big(F_{\mu\alpha}^a F_\nu^{b\,\alpha}-\frac{1}{4}F_{\alpha\beta}^a F^{b\, \alpha\beta}g_{\mu\nu} \big)\nonumber\\
&\hspace{2cm}E_{\text{d}\,\mu\nu}=2\text{d}_{_{IJ}}\big(\nabla_\mu \phi^I \nabla_\nu \phi^J -\frac{1}{2}\nabla^\alpha\phi^I\nabla_\alpha\phi^J g_{\mu\nu} \big)\,, \nonumber\\
&\nabla_\nu \big(\mathrm{c}_{ab} F^{b\, \mu\nu}\big)=0\,,\\
&4\nabla_\alpha \big(\mathrm{d}_{_{IJ}}\nabla^\alpha \phi^J\big)+\frac{\partial f}{\partial \phi} +\frac{\partial\mathrm{a}}{\partial \phi} R_{\mu\nu}R^{\mu\nu}\!+\frac{\partial \mathrm{b}}{\partial\phi} R_{\mu\nu\alpha\beta}R^{\mu\nu\alpha\beta} \!-\!\frac{\partial\mathrm{c}_{ab}}{\partial \phi}F_{\mu\nu}^a F^{b\, \mu\nu}\!-\!2\frac{\partial\mathrm{d}_{_{IJ}}}{\partial \phi}\nabla^\mu\phi^I\nabla_\mu\phi^J=0, \!\!\!\!&&
\end{flalign}}
 
\vspace*{-0.2cm}
\noindent where the notation $f'\equiv \frac{\partial f}{\partial R}$ is used. We need to find the $\mathbf{\Theta}_{_\text{LW}}$, $\mathbf{Q}_\epsilon$, and most importantly, the $\boldsymbol{k}_\epsilon$ for this theory. Their derivation and final results are standard practice in the literature. Hence we only report the final results here. 

By variation of Lagrangian $\delta \mathbf{L}$ and imposing e.o.m, the surface $d\!-\!1$-form $\mathbf{\Theta}_{_\text{LW}}$ can be read to be $\mathbf{\Theta}_{_\text{LW}}=\star \Theta$, \emph{i.e.}
\begin{equation}
\mathbf{\Theta}_{_\text{LW}}=\frac{\sqrt{-g}}{(d-1)!}\,\,\epsilon_{\mu\mu_1\cdots \mu_{d-1}} \,(\Theta_{f}^\mu+\Theta_{\text{a}}^\mu+\Theta_{\text{b}}^\mu+\Theta_{\text{c}}^\mu+\Theta_{\text{d}}^\mu)\,\,\mathrm{d}x^{\mu_1}\wedge \cdots\wedge \mathrm{d}x^{\mu_{d-1}}, 
\end{equation}
in which
\vspace*{-0.2cm}

{\small \begin{flalign}
&\Theta_{f}^{\mu}(\delta\Phi,\Phi)=\frac{1}{16\pi G}\big(f'(\nabla_\alpha h^{\mu\alpha}-\nabla^\mu h)-\nabla_\alpha f'h^{\mu\alpha}+\nabla^\mu f' h\big)\,, \nonumber\\
&\Theta_{\text{a}}^\mu(\delta\Phi,\Phi)=\frac{\text{a}}{16\pi G}\big(2R_{\alpha\beta}\nabla^\alpha h^{\beta \mu}-2\nabla_\alpha R^{\mu}_{\,\,\beta}h^{\alpha\beta}\!-\!R^\mu_{\,\,\alpha}\nabla^\alpha h+\nabla^\alpha R^\mu_{\,\,\alpha}h-R_{\alpha\beta}\nabla^\mu h^{\alpha\beta}\!+\!\nabla^\mu R_{\alpha\beta}h^{\alpha\beta}\big)\,,\nonumber\\
&\Theta_{\text{b}}^\mu(\delta\Phi,\Phi)=\frac{\text{b}}{4\pi G}( \nabla^\nu R^{\mu}_{\,\,\alpha\nu\beta}h^{\alpha\beta}-R^{\mu}_{\,\,\alpha\nu\beta}\nabla^\nu h^{\alpha\beta})\,,\nonumber\\
&\Theta_{\text{c}}^\mu(\delta\Phi,\Phi)=\frac{-1}{4\pi G}\mathrm{c}_{ab}\,F^{a\,\mu\nu}\,\delta A^b_{\nu}\,,\nonumber\\
&\Theta_{\text{d}}^\mu(\delta\Phi,\Phi)=\frac{-1}{4\pi G}\mathrm{d}_{_{IJ}}\,\nabla^\mu\phi^I\delta\phi^J\,,&&
\end{flalign}}

\vspace*{-0.2cm}
\noindent where $h\equiv h^\alpha _{\,\,\alpha}$.  Having the $\mathbf{\Theta}$ in our hand, for a generic $\epsilon=\{\xi,\lambda^a\}$, and imposing the e.o.m Eq.\eqref{eom g}, the Noether-Wald $d\!-\!2$-form $\mathbf{Q}_\epsilon$ can be read as 
\begin{equation}
\mathbf{Q}_\epsilon=\frac{\sqrt{-g}}{(d-2)!\,2!}\,\,\epsilon_{\mu\nu\mu_1\cdots \mu_{d-2}} \,(\mathrm{Q}_{f\,\epsilon}^{\mu\nu}+\mathrm{Q}_{\mathrm{a}\,\epsilon}^{\mu\nu}+\mathrm{Q}_{\mathrm{b}\,\epsilon}^{\mu\nu}+\mathrm{Q}_{\text{c}\,\epsilon}^{\mu\nu}+\mathrm{Q}_{\mathrm{d}\,\epsilon}^{\mu\nu})\,\,\mathrm{d}x^{\mu_1}\wedge \cdots\wedge \mathrm{d}x^{\mu_{d-2}}, 
\end{equation}
 in which 
\vspace*{-0.3cm}

{\small
\begin{flalign}
&\mathrm{Q}_{f\,\epsilon}^{\mu\nu}=\frac{1}{16\pi G}\big(2\nabla^\mu f' \xi^\nu-f' \nabla^\mu \xi^\nu \big)- [\mu\leftrightarrow\nu]\,,\nonumber\\
&\mathrm{Q}_{\text{a}\,\epsilon}^{\mu\nu}=\frac{\text{a}}{8\pi G}\big(\nabla^\mu R^\nu_{\,\,\alpha}\xi^\alpha+ R^{\nu}_{\,\,\alpha}\nabla^\alpha\xi^\mu - \nabla^\alpha R^\nu_{\,\,\alpha}\xi^\mu\big)-[\mu\leftrightarrow\nu]\,,\nonumber\\
&\mathrm{Q}_{\text{b}\,\epsilon}^{\mu\nu}=\frac{\text{b}}{4\pi G}\big(\nabla^\alpha R^{\mu\nu}_{\,\,\,\,\,\,\alpha\beta}\,\xi^\beta-R^{\mu\alpha\nu\beta}\nabla_\alpha\xi_\beta\big)-[\mu\leftrightarrow\nu]\,,\nonumber\\
& \mathrm{Q}_{\text{c}\,\epsilon}^{\mu\nu}=\frac{-1}{4\pi G}\mathrm{c}_{ab}F^{a\,\,\mu\nu}(A^b_\rho\xi^\rho+\lambda^b)\,,\nonumber\\
&\mathrm{Q}_{\text{d}\,\epsilon}^{\mu\nu}=0\,.&&
\end{flalign}}

\vspace*{-0.2cm}
\noindent By variation of the $\mathbf{Q}_\epsilon$ with respect to all dynamical fields, utilizing the textbook relations
\vspace*{-0.3cm}

{\small \begin{flalign}
&\delta \sqrt{-g}=\frac{\sqrt{-g}}{2}h^\alpha_{\,\,\alpha}\,, \quad \quad \qquad \delta \Gamma ^\lambda_{\mu\nu}= \frac{1}{2}g^{\lambda \sigma}(\nabla _\mu h_{\sigma \nu}+\nabla_\nu h_{\sigma \mu}-\nabla _\sigma h_{\mu\nu}),\quad \qquad \qquad \delta \epsilon_{\mu\nu\mu_1\cdots \mu_{d-2}}=0\,,\nonumber\\
&\delta R_{\mu\nu\alpha\beta}=\frac{1}{2}\big(2R_{\mu\nu\alpha\gamma}\,h^\gamma_{\,\,\beta}-\nabla_\mu\nabla_\alpha h_{\beta\nu}\!+\!\nabla_\mu\nabla_\beta h_{\alpha\nu}\!-\!\nabla_\mu\nabla_\nu h_{\alpha\beta}\!+\!\nabla_\nu\nabla_\alpha h_{\beta\mu}\!-\!\nabla_\nu\nabla_\beta h_{\alpha\mu}\!+\!\nabla_\nu\nabla_\mu h_{\alpha\beta}\big),\nonumber\\
&\delta R_{\mu\nu}=\frac{1}{2}\left( \nabla_\alpha \nabla_\mu h^\alpha_{\,\,\nu}+\nabla_\alpha \nabla_\nu h^\alpha_{\,\,\mu}-\Box h_{\mu\nu}-\nabla_\mu\nabla_\nu h \right),   \qquad \delta R=\nabla_\mu \nabla_\nu h^{\mu\nu}-\Box h-R_{\mu\nu}h^{\mu\nu}\,,&&
\end{flalign}}

\vspace*{-0.2cm}
\noindent and putting the results into the general formula for $\boldsymbol{k}_\epsilon$\cite{Hajian:2015xlp,Ghodrati:2016vvf}, the final applicable tensor for calculation of conserved charges can be derived as
\begin{equation}\label{k epsilon general}
\boldsymbol{k}_\epsilon(\delta\Phi,\Phi)=\frac{\sqrt{-g}}{(d-2)!\,2!}\,\,\epsilon_{\mu\nu\mu_1\cdots \mu_{d-2}} \,(k_{f\,\epsilon}^{\mu\nu}+k_{\text{a}\,\epsilon}^{\mu\nu}+k_{\text{b}\,\epsilon}^{\mu\nu}+k_{\text{c}\,\epsilon}^{\mu\nu}+k_{\text{d}\,\epsilon}^{\mu\nu})\,\,\mathrm{d}x^{\mu_1}\wedge \cdots\wedge \mathrm{d}x^{\mu_{d-2}}, 
\end{equation}
where
\vspace*{-0.2cm}

{\small
\begin{flalign}
&k_{f\,\epsilon}^{\mu\nu}(\delta\Phi,\Phi)=\dfrac{1}{16 \pi G}\Big[\Big( h^{\mu\alpha}\nabla_\alpha\xi^\nu-\nabla^\mu h^{\nu\alpha}\xi_\alpha-\frac{1}{2}h \nabla^\mu\xi^\nu\Big)f'+2\Big(R^{\mu\alpha}\nabla_\alpha h-\nabla_\alpha R h^{\mu\alpha}-R^\mu_{\,\,\alpha}\nabla_\beta h^{\alpha\beta}\nonumber\\
&\hspace*{2.9cm}-\Box \nabla^\mu h +\nabla_\alpha\nabla^\mu\nabla_\beta h^{\alpha\beta}-\nabla^\mu(R_{\alpha\beta} h^{\alpha\beta})+\frac{1}{2}\nabla^\mu R\, h \Big)\xi^\nu f''\!+\!\Big(R_{\alpha\beta}h^{\alpha\beta}-\nabla_\alpha\nabla_\beta h^{\alpha\beta}\nonumber\\
&\hspace*{2.9cm}+\Box h\Big)(\nabla^\mu\xi^\nu f''-2\nabla^\mu R \,\xi^\nu f''')\Big]+\frac{\delta \mathrm{Q}_{f\,\epsilon}^{\mu\nu}}{\delta\phi^I}\delta\phi^I-\Theta^\mu_f \xi^\nu-[\mu\leftrightarrow\nu], \\
&k_{\text{a}\,\epsilon}^{\mu\nu}(\delta\Phi,\Phi)=\dfrac{\text{a}}{16 \pi G}\Big[\Big(\nabla^\alpha R_\alpha^{\,\,\mu} h-\nabla_\alpha R h^{\mu\alpha}\!-\!\nabla^\mu(R_{\alpha\beta}h^{\alpha\beta})+\nabla^\mu\nabla_\alpha\nabla_\beta h^{\alpha\beta}-\nabla^\mu \Box h\Big)\xi^\nu \!+\! \Big( 2\nabla_\beta R^\mu_{\,\,\alpha}h^{\beta\nu}\nonumber\\
&\hspace*{2.9cm}-2 R^{\mu\beta}\nabla_\beta h^\nu_{\,\, \alpha}-2\nabla^\mu R_{\alpha\beta}h^{\nu\beta}- \nabla^\mu (\nabla_\alpha\nabla^\nu h -\nabla_\beta \nabla_\alpha h^{\nu\beta} +\Box h^\nu_{\,\,\alpha}-\nabla^\beta \nabla^\nu h_{\alpha\beta}) \nonumber\\
&\hspace*{2.9cm} +\nabla^\mu R^\nu_{\,\,\alpha} h +2 R^{\mu\beta} \nabla^\nu h_{\alpha\beta}\Big) \xi^\alpha +\Big(\nabla_\alpha \nabla^\mu h - \nabla_\beta \nabla_\alpha h^{\mu\beta}-\nabla^\beta \nabla^\mu h_{\alpha\beta}+\Box h^{\mu}_{\,\,\alpha}\nonumber\\
&\hspace*{2.9cm} +2(R_{\alpha\beta}h^{\mu\beta}+R^{\mu\beta}h_{\alpha\beta})-R^{\mu}_{\,\,\alpha} h\Big) \nabla^\alpha \xi^\nu \Big]\!+\!\frac{\mathrm{Q}_{\text{a}\,\epsilon}^{\mu\nu}}{\text{a}}\frac{\partial \text{a}}{\partial\phi^I}\delta \phi^I\!-\!\Theta^\mu_{\text{a}} \xi^\nu\!-\![\mu\leftrightarrow\nu],\!\!\!\!\! 
\end{flalign} 
\begin{flalign}
&k_{\text{b}\,\epsilon}^{\mu\nu}(\delta\Phi,\Phi)=\dfrac{\text{b}}{8 \pi G}\Big[\Big(2(R^\mu_{\,\,\alpha\beta\gamma}\!\!-\!R^\mu_{\,\,\beta\alpha\gamma})h^{\nu\gamma}\!\!+\! R^{\mu\,\,\,\nu}_{\,\, \alpha\,\, \beta}h \!-\! R^{\mu\,\,\,\nu}_{\,\,\alpha\,\,\gamma}h_\beta^{\,\,\gamma}\!-\! R^{\mu\,\,\,\nu}_{\,\,\beta\,\,\gamma}h_\alpha^{\,\,\gamma} \!-\! \nabla^\mu \nabla_\alpha h^\nu_{\,\,\beta}\!+\!\nabla^\mu \nabla_\beta h^\nu_{\,\,\alpha}\Big)\!\nabla^\beta\xi^\alpha\nonumber\\
&\hspace*{2.9cm} +\Big(R^{\mu\beta}(\nabla_\beta h^\nu_{\,\, \alpha}-\nabla_\alpha h^\nu_{\,\,\beta})+R^{\mu\,\,\,\nu}_{\,\,\beta\,\,\gamma} \nabla^\gamma h_\alpha^{\,\,\beta}+\frac{1}{2}R^{\mu\nu}_{\,\,\,\,\,\alpha\gamma}(\nabla_\beta h^{\beta\gamma}-\nabla^\gamma h)\nonumber\\
&\hspace*{2.9cm} +2(\nabla_\beta R^\mu_{\,\,\alpha}-\nabla^\mu R_{\alpha\beta})h^{\nu\beta}+\nabla^\mu \nabla_\beta\nabla_\alpha h^{\nu\beta}-\nabla^\mu\Box h^\nu_{\,\,\alpha}+\nabla^\mu R^\nu_{\,\,\alpha}h + \nabla^\mu R^\nu_{\,\,\beta} h_\alpha^{\,\,\beta}\nonumber\\
&\hspace*{2.9cm} -\nabla^\mu (R^\nu _{\,\,\beta\alpha\gamma}h^{\beta\gamma})\Big) 2\xi^\alpha\Big]+\frac{\mathrm{Q}_{\text{b}\,\epsilon}^{\mu\nu}}{\text{b}}\frac{\partial \text{b}}{\partial\phi^I}\delta \phi^I -\Theta^\mu_{\text{b}} \xi^\nu-[\mu\leftrightarrow\nu], \\
&k_{\text{c}\,\epsilon}^{\mu\nu}(\delta\Phi,\Phi)=\frac{1}{8 \pi G}\Big[\big(\frac{-h}{2} \,\mathrm{c}_{ab}\, F^{a\,\mu\nu}\!+\!2\,\mathrm{c}_{ab}\,F^{a\,\mu\alpha}h_\alpha^{\;\;\nu}-\mathrm{c}_{ab}\,\delta F^{a\,\mu\nu}\!-\!\frac{\partial\,\mathrm{c}_{ab}}{\partial \phi^I}\,F^{a\,\mu\nu}\delta\phi^I\big)({\xi}^\alpha A^b_\alpha+\lambda^b)-\nonumber\\
 & \hspace*{2.9cm}\,\mathrm{c}_{ab}\,F^{a\,\mu\nu}\xi^\alpha \delta A^b_\alpha-2\,\mathrm{c}_{ab}\,F^{a\,\alpha\mu}\xi^\nu \delta A^b_\alpha\Big]-[\mu\leftrightarrow\nu]\,, \\
 &k_{\text{d}\,\epsilon}^{\mu\nu}(\delta\Phi,\Phi)=\frac{1}{8\pi G}\Big[\xi^\nu\,\mathrm{d}_{_{IJ}}\,\nabla^\mu\phi^I\,\delta\phi^J\Big]-[\mu\leftrightarrow\nu]\,. &&
\end{flalign} 
}

\vspace*{-0.2cm}
\noindent Having the $\boldsymbol{k}_\epsilon$, and being equipped with the parametric variations $\hat \delta \Phi$, calculation of conserved charges associated with exact symmetries $\eta=\{\zeta,\lambda^a\}$ of the (black hole) solutions $\hat \Phi(x^\mu;p_j)$ to the Lagrangian \eqref{Lagrangian scalar} can be performed by using SPSM.

To exemplify, in this section we will work out conserved charges and first law(s) of thermodynamics for some black hole solutions to the Lagrangian \eqref{Lagrangian scalar}.

\section{Example 1: $\mathbf{z=3}$ Lifshitz black hole in $\boldsymbol{d}\mathbf{=3}$} \label{sec-example}
Consider
\begin{equation}\label{coefficients ex 1}
f=R+\frac{13}{l^2}-\frac{3l^2}{4}R^2, \qquad \text{a}=2l^2, \qquad \text{b}=\text{c}=\text{d}=0\,,
\end{equation}
\emph{i.e.} the new massive gravity (NMG) Lagrangian\cite{Bergshoeff:2009hq}
\begin{equation}\label{NMG}
\mathcal{L}=\frac{1}{16\pi G}\left(R-2\Lambda+\frac{1}{\mathfrak{m}^2}(R_{\mu\nu}R^{\mu\nu}-\frac{3}{8}R^2)\right),
\end{equation}
in which $\Lambda=-\frac{13}{2l^2}$ and $\mathfrak{m}^2=\frac{1}{2l^2}$. 

For $z=3$ and in 3-dim, we can have a family of black holes $\hat g_{\alpha\beta}(x^\mu;m)$ as solution to this theory  \cite{AyonBeato:2009nh,AyonBeato:2010tm} in the form
\begin{equation}
\mrd s^2=-(\frac{r}{l})^{2z}(1-\frac{ml^2}{r^2})\,\mrd t^2 +\frac{\mrd r^2}{\frac{r^2}{l^2}(1-\frac{ml^2}{r^2})}+r^2 \mrd \varphi^2.
 \end{equation} 
 
Let us analyze the thermodynamics of this family of black holes using SPSM. Putting Eq.  \eqref{coefficients ex 1} into the general result  Eq. \eqref{k epsilon general}, $k_\epsilon^{\mu\nu}$ can be read for our specific theory.  Then if for simplicity we choose $\partial \Sigma$ to be surfaces of constant $(t,r)$, the conserved charge variations for an exact symmetry $\eta$ can be simply read through
\begin{align}\label{k epsilon tr}
\hat\delta H_{\eta}= \oint_{\partial \Sigma}\boldsymbol{k}_\eta(\hat\delta g_{\alpha\beta},\hat g_{\alpha\beta}) =\int_0^{2\pi}\sqrt{-\hat g}\,k_\eta^{tr}(\hat\delta g_{\alpha\beta},\hat g_{\alpha\beta}) \,\mathrm{d}\varphi\,,
\end{align}
in which $k_\eta^{tr}$ is the $tr$ component of the $k_\eta^{\mu\nu}$. 

Inserting parametric variations $\hat\delta g_{\alpha\beta}=\frac{\partial \hat g_{\alpha\beta}}{\partial m}\delta m$ in it, conserved charges can be calculated, irrespective of the asymptotic Lifshitz behavior.\\
\underline{Mass}: We can choose the stationarity Killing $-\partial_t$ as the generator to which the mass is associated to. The minus sign has been adopted to make the mass and entropy positive. Hence, by $\eta_{_M}=\{-\partial_t,0\}$ the result of calculating Eq. \eqref{k epsilon tr} is
\begin{equation}
\hat \delta M\equiv \hat \delta H_{\eta_{_M}}=\frac{m}{2G}\delta m= \delta(\frac{m^2}{4G})  \qquad \Rightarrow \qquad M=\frac{m^2}{4G}.
\end{equation}
The reference point (constant of integration) was chosen $M\!=\!0$ for the geometry with $m=0$.\\
\noindent\underline{Angular momentum}: Choosing $\eta_{_J}=\{\partial_\varphi,0\}$, by similar analysis as the mass, angular momentum turns out to be
\begin{equation}
\hat \delta J\equiv -\hat \delta H_{\eta_{_J}}=0\times \delta m \qquad \Rightarrow \qquad J=0.
\end{equation}
\noindent\underline{Entropy}: The surface gravity for the horizon of this solution is $\kappa_{_\mathrm{H}}=\frac{r_{_\mathrm{H}}^3}{l^4}$ in which $r_{_\mathrm{H}}=\sqrt{ml^2}$. The entropy of the event horizon is defined to be the conserved charge associated with the horizon Killing vector $\zeta_{_\mathrm{H}}$ normalized by the Hawking temperature $T_{_\mathrm{H}}=\frac{\kappa_{_\mathrm{H}}}{2\pi}$. Therefore, by $\eta_{_\mathrm{H}}=\frac{2\pi}{\kappa_{_\mathrm{H}}}\{\zeta_{_\mathrm{H}},0\}$ and the identity $\zeta_{_\mathrm{H}}= -\partial_t$, entropy attributed to the horizon, via a similar integration as other conserved charges, is calculated to be
\begin{equation}
\hat \delta S_{_\mathrm{H}}\equiv \hat \delta H_{\eta_{_\mathrm{H}}}=\frac{\pi l}{G \sqrt{m}}\delta m=\delta (\frac{2\pi r_{_\mathrm{H}}}{G}) \qquad \Rightarrow\qquad     S_{_\mathrm{H}}=\frac{2\pi r_{_\mathrm{H}}}{G}\,.
\end{equation}
The reference point is chosen to be $S_{_\text{H}}\!=\!0$ for the geometry identified by $m=0$.
Notice that the entropy is proportional to the area (here the length) of the horizon, but without the usual factor of $\frac{1}{4}$. The results above are in agreement with the results of quasi-local method \cite{Gim:2014nba}.\\
\noindent\underline{First law:} Having made the entropy free of being calculated on the horizons, the first law of thermodynamics would be the simple identity $\eta_{_\mathrm{H}}=\frac{1}{T_{_\mathrm{H}}}\{-\partial_t,0\}$. The proof follows from the linearity of the generic charge variation $\delta H_\epsilon$ in terms of its generator $\epsilon$. Mathematically,
\begin{equation}
\begin{cases}
\delta M\equiv \delta H_{\eta_{_M}}\,, \,\delta S_{_\mathrm{H}}\equiv \delta H_{\eta_{_\mathrm{H}}}\\
\eta_{_\mathrm{H}}=\frac{1}{T_{_\mathrm{H}}}\eta_{_M}\\
\end{cases}
\quad \xrightarrow{\text{linearity of $\delta H_\epsilon$ in $\epsilon$\,\,\,}}
\qquad \delta S_{_\mathrm{H}}=\frac{1}{T_{_\mathrm{H}}}\delta M\,.
\end{equation}
Notice that the $\delta$ in the proof is a generic perturbation which satisfies linearized e.o.m. So, it is not restricted to the parametric variations. Moreover, integration over the horizon or asymptotics does not play any role in this proof.

\section{Example 2: Warped BTZ black hole}
As mentioned in \ref{sec:exensemble}, the warped BTZ solution is a $3$-dimensional black hole identified with two parameters $p_{_1}=m$ and $p_{_2}=j$, 

{\small
\begin{align}\label{WBTZ}
\mrd s^2= &\Big( \frac{-r^2}{l^2}-\frac{H^2 (-r^2+4lj+8l^2 m)^2}{4l^3(lm+j)}+8m\Big) \mrd t^2 -2\Big( 4j-\frac{H^2(4lj+r^2)(-r^2+4lj+8l^2m)}{4l^2(lm+j)}\Big)\mrd t\,\mrd \varphi\nonumber\\
&+\frac{1}{\frac{16 j^2}{r^2}+\frac{r^2}{l^2}-8m}\mrd r^2 +\Big(r^2-\frac{H^2(4lj+r^2)^2}{4l(lm+j)}\Big) \mrd \varphi^2\,.
\end{align}
}

After reading the $k^{\mu\nu}_\epsilon$ for this theory from the general result in Eq.\eqref{k epsilon general}, and equipped with the parametric variations
\begin{equation}
\hat\delta g_{\alpha\beta}=\frac{\partial \hat g_{\alpha\beta}}{\partial m}\delta m+\frac{\partial \hat g_{\alpha\beta}}{\partial j}\delta j\,,
\end{equation}
 one can find the conserved charges by an integration similar to Eq.\eqref{k epsilon tr}. Notice that because of the linearity of $\delta H_\eta (\delta \Phi,\Phi)$ in $\delta \Phi$, the parametric variations can be inserted term by term into the calculations. This makes the calculations simpler. \\ 
\underline{Mass}: By $\eta_{_M}=\{\partial_t,0\}$, it turns out that
\begin{equation}
\hat \delta M\equiv \hat \delta H_{\eta_{_M}}=\frac{16 (1-2H^2)^{\frac{3}{2}}}{G(17-42\,H^2)}\delta m + 0\times \delta j  \qquad \Rightarrow \qquad M=\frac{16 (1-2H^2)^{\frac{3}{2}}m}{G(17-42\,H^2)}.
\end{equation}
\noindent\underline{Angular momentum}: Choosing $\eta_{_J}=\{\partial_\varphi,0\}$, one finds
\begin{equation}
\hat \delta J\equiv -\hat \delta H_{\eta_{_J}}=0\times \delta m+ \frac{16 (1-2H^2)^{\frac{3}{2}}}{G(17-42\,H^2)}\delta j\qquad \Rightarrow \qquad J=\frac{16 (1-2H^2)^{\frac{3}{2}}j}{G(17-42\,H^2)}.
\end{equation}
\noindent\underline{Entropies}: There are two horizons in the Warped BTZ geometry \eqref{WBTZ}. So, we would find two entropies attributed to them. The horizons are situated at $r_\pm^2=4(l^2m\pm \sqrt{l^4m^2-l^2j^2})$, collectively denoted by $r_{_\mathrm{H}}$. The surface gravities, angular velocities, and the Killing vectors of the horizons are
\begin{equation}
\kappa_{_\mathrm{H}}=\frac{r_{_\mathrm{H}}^4-16\, l^2j^2}{l^2 r_{_\mathrm{H}}^3}\,, \qquad \Omega_{_\mathrm{H}}=\frac{4j}{r_{_\mathrm{H}}^2}\,, \qquad \zeta_{_\mathrm{H}}=\partial_t+\Omega_{_\mathrm{H}}\partial_\varphi\,,
\end{equation}
respectively. 

By integrations over arbitrary surfaces of constant time and radius, the entropies as conserved charges associated with the exact symmetries $\eta_{_\mathrm{H}}=\frac{2\pi}{\kappa_{_\mathrm{H}}}\{\zeta_{_\mathrm{H}},0\}$, are calculated to be
\begin{equation}
\hat \delta S_{_\mathrm{H}}= \frac{\partial (\frac{8\pi (1-2H^2)^{\frac{3}{2}}r_{_\mathrm{H}}}{G(17-42\,H^2)})}{\partial m}\delta m+\frac{\partial (\frac{8\pi (1-2H^2)^{\frac{3}{2}}r_{_\mathrm{H}}}{G(17-42\,H^2)})}{\partial j}\delta j \qquad \Rightarrow\qquad     S_{_\mathrm{H}}=\frac{8\pi (1-2H^2)^{\frac{3}{2}}r_{_\mathrm{H}}}{G(17-42\,H^2)}\,.
\end{equation}
The reference poinst for all of the charges above have been chosen to vanish on the geometry identified by $m=j=0$. Our results match exactly with the results of other methods \cite{Ghodrati:2016ggy,Detournay:2016gao}.\\
\noindent\underline{First law:} For any generic perturbation which satisfies linearized e.o.m, the first laws follow:
\begin{equation}
\hspace*{-0.1cm}\begin{cases}
\delta M\equiv \delta H_{\eta_{_M}}\,, \,\delta J\equiv -\delta H_{\eta_{_J}}\,, \,\delta S_{_\mathrm{H}}\equiv \delta H_{\eta_{_\mathrm{H}}}\\
\eta_{_\mathrm{H}}=\frac{1}{T_{_\mathrm{H}}}(\eta_{_M}+\Omega_{_\mathrm{H}}\eta_{_J})\\
\end{cases}
\quad \xrightarrow{\text{linearity of $\delta H_\epsilon$ in $\epsilon$\,\,\,}}
\quad \delta S_{_\mathrm{H}}=\frac{1}{T_{_\mathrm{H}}}(\delta M-\Omega_{_\mathrm{H}}\delta J)\,.
\end{equation}

So for both of these examples the first law of thermodynamics is satisfied.

\end{appendices}
\pagebreak

\medskip

\addcontentsline{toc}{chapter}{Bibliography}
\bibliographystyle{JHEP}
\bibliography{Wref}

\end{document}